\documentclass{aa}
\usepackage[multidot]{grffile}
\usepackage{graphicx}
\usepackage{txfonts}
\usepackage{natbib}
\usepackage[colorlinks,allcolors=blue,bookmarks=false]{hyperref} 
\pdfoutput=1
\bibpunct{(}{)}{;}{a}{}{,}
\begin{document}

\title
{
Multiscale, multiwavelength extraction of sources and filaments\\ 
using separation of the structural components: \textsl{getsf}
}


\author
{
A.~Men'shchikov 
}


\institute
{
AIM, IRFU, CEA, CNRS, Universit{\'e} Paris-Saclay, Universit{\'e} Paris Diderot, Sorbonne Paris Cit{\'e}, F-91191 Gif-sur-Yvette, 
France\\\email{alexander.menshchikov@cea.fr}
}

\date{Received 16 November 2020 / Accepted 22 February 2021 }

\offprints{Alexander Men'shchikov}
\titlerunning{Multiscale, multiwavelength extraction of sources and filaments: \textsl{getsf}}
\authorrunning{A.~Men'shchikov}


\abstract
{ 
High-quality astronomical images delivered by modern ground-based and space observatories demand adequate, reliable software for
their analysis and accurate extraction of sources, filaments, and other structures, containing massive amounts of detailed
information about the complex physical processes in space. The multiwavelength
observations with highly variable angular resolutions across wavebands require extraction tools that preserve and use the
invaluable high-resolution information. Complex fluctuating backgrounds and filamentary structures appear differently on various
scales, calling for multiscale approaches for complete and reliable extraction of sources and filaments. The availability of many
extraction tools with varying qualities highlights the need to use standard model benchmarks for choosing the most reliable and
accurate method for astrophysical research.

This paper presents \textsl{getsf}, a new method for extracting sources and filaments in astronomical images using separation of
their structural components, designed to handle multiwavelength sets of images and very complex filamentary backgrounds. The method
spatially decomposes the original images and separates the structural components of sources and filaments from each other and from
their backgrounds, flattening their resulting images. It spatially decomposes the flattened components, combines them over
wavelengths, detects the positions of sources and skeletons of filaments, and measures the detected sources and filaments, creating
the output catalogs and images. The fully automated method has a single user-defined parameter (per image), the maximum size of the
structures of interest to be extracted, that must be specified by users. This paper presents a realistic multiwavelength set of
simulated benchmark images that can serve as the standard benchmark problem to evaluate qualities of source- and
filament-extraction methods.

This paper describes \textsl{hires}, an improved algorithm for the derivation of high-resolution surface densities from
multiwavelength far-infrared \emph{Herschel} images. The algorithm allows creating the surface densities with angular resolutions
that reach $5.6{\arcsec}$ when the $70$\,$\mu$m image is used. If the shortest-wavelength image is too noisy or cannot be used for
other reasons, slightly lower resolutions of $6.8{-}11.3{\arcsec}$ are available from the $100$ or $160$\,$\mu$m images. These high
resolutions are useful for detailed studies of the structural diversity in molecular clouds.

The codes \textsl{getsf} and \textsl{hires} are illustrated by their applications to a variety of images obtained with ground-based
and space telescopes from the X-ray domain to the millimeter wavelengths.
} 
\maketitle


\section{Introduction}
\label{introduction}

Multiwavelength far-infrared and submillimeter dust continuum observations with the large space telescopes \emph{Spitzer},
\emph{Herschel}, and \emph{Planck} in the past decades greatly increased the amount and improved the quality of the available data
in various areas of astrophysical research. Observed images with diffraction-limited angular resolutions and high sensitivity
reveal an impressive diversity of the enormously complex structures, covering orders of magnitude in intensities and spatial
scales. The images feature foremost the bright fluctuating backgrounds, omnipresent filaments, and huge numbers of sources of
different physical nature, all blended with each other, whose appearance and resolution are often markedly different at short and
long wavelengths. The massive amount of information that is coded in the fine structure of the observed images must contain clues
to the complex physical processes taking place in space, but these clues are extremely difficult to decipher. It is quite clear
that the era of these high-quality data from space telescopes and large ground-based interferometers, such as the Atacama Large
Millimeter/submillimeter Array (\emph{ALMA}), requires more sophisticated tools for their accurate analysis and correct
interpretation than those developed for the lower-quality images of the past. Adequate extraction methods must be explicitly
designed for the multiwavelength imaging observations with highly dissimilar angular resolutions across wavebands. They must also
be able to handle the bright filamentary backgrounds that vary on all spatial scales, whose fluctuation levels differ by several
orders of magnitude across the observed images.

The source- and filament-extraction methods are growing in numbers. In the area of star formation, a new method was published every
year or two within the seven-year period after the launch of \emph{Herschel}. \cite{Rosolowsky_etal2008} devised
\textsl{dendrograms} to describe the hierarchical structure of clumps observed in the data cubes from molecular line observations.
The method carries out topological analysis of image structures by isophotal contours at varying intensity levels and represents
them graphically as a tree. \cite{Molinari_etal2011} created \textsl{cutex} to extract sources in star-forming regions observed
with \emph{Herschel}. The method analyzes multidirectional second derivatives of the observed image to detect sources, and it
measures them by fitting elliptical Gaussians on a planar background to their peaks. \cite{Men'shchikov_etal2012} developed
\textsl{getsources}, the multiwavelength source extraction method for the \emph{Herschel} observations of star-forming regions. The
method spatially decomposes images, combines them into wavelength-independent detection images, subtracts the backgrounds of
detected sources, and measures the sources, deblending them when they overlap. \cite{Kirk_etal2013} presented \textsl{csar} for the
\emph{Herschel} images. The method analyzes areas of connected pixels that are bound by closed isophotal contours, descending to a
predefined background level and partitioning peaks at their lowest isolated contours into sources. \cite{Berry2015} created
\textsl{fellwalker} to identify clumps in submillimeter data cubes. The method finds image peaks by tracing the line of the
steepest ascent and identifies sources as the hill with the highest value found for all pixels in its neighborhood.
\cite{Sousbie2011} produced \textsl{disperse} to identify structures in the large-scale distribution of matter in the Universe. The
method applies the computational topology to trace filaments and other structures. \cite{Men'shchikov2013} developed
\textsl{getfilaments} to improve the source extraction with \textsl{getsources} on the filamentary backgrounds observed with
\emph{Herschel}. The method separates filaments from sources in spatially decomposed images and subtracts them from the detection
images, thereby reducing the rate of spurious sources. \cite{Schisano_etal2014} devised a Hessian matrix-based approach to extract
filaments in \emph{Herschel} observations of the Galactic plane. The method analyzes multidimensional second derivatives to
identify filaments and determine their properties. \cite{Clark_etal2014} presented \textsl{rht} to characterize fibers in the
interstellar \ion{H}{i} medium. The method has been applied to various observations of diffuse \ion{H}{i}, revealing alignment of
the fibers along magnetic fields. \cite{KochRosolowsky2015} published \textsl{filfinder} to identify filaments in the
\emph{Herschel} images of star-forming regions. the method applies a mathematical morphology approach to isolating filaments in
observed images. \cite{Juvela2016} presented \textsl{tm} to trace filaments in observed images. The method matches a predefined
template (stencil) of an elongated structure at each pixel of an image by shifting and rotating the template and analyzing the
parameters of the matches.

These extraction methods all have several important issues. Sources and filaments are handled completely independently by these
methods, although numerous \emph{Herschel} observations have demonstrated that there is a tight physical relation between them.
Most sources are found in filamentary structures, and the corresponding starless, prestellar, and protostellar cores are thought to
form inside the structures that are created by dynamical processes, magnetic fields, and gravity within a molecular cloud. All
major structural components of the observed images, that is, the background cloud, filamentary structures, and sources, are heavily
blended with each other; curved filaments are even blended with themselves. The degree of their blending increases at longer
wavelengths with lower angular resolutions, which increases the inaccuracies in their detections, measurements, and interpretations.

Most of the extraction methods focus on detecting structures, whereas the most important and difficult problem is measuring them
accurately. Numerous algorithms only partition the image between sources and do not allow them to overlap, although deblending of
the mixed emission of the structural components is an indispensable property of an accurate extraction method. For best detection
and measurement results, source-extraction methods must be able to separate underlying filamentary structures and
filament-extraction methods must be able to separate sources. The existing source- and filament-extraction methods use completely
different approaches, and the quality of their results is expected to be very dissimilar.

\begin{figure*}
\centering
\centerline{
  \resizebox{0.77\hsize}{!}{\includegraphics{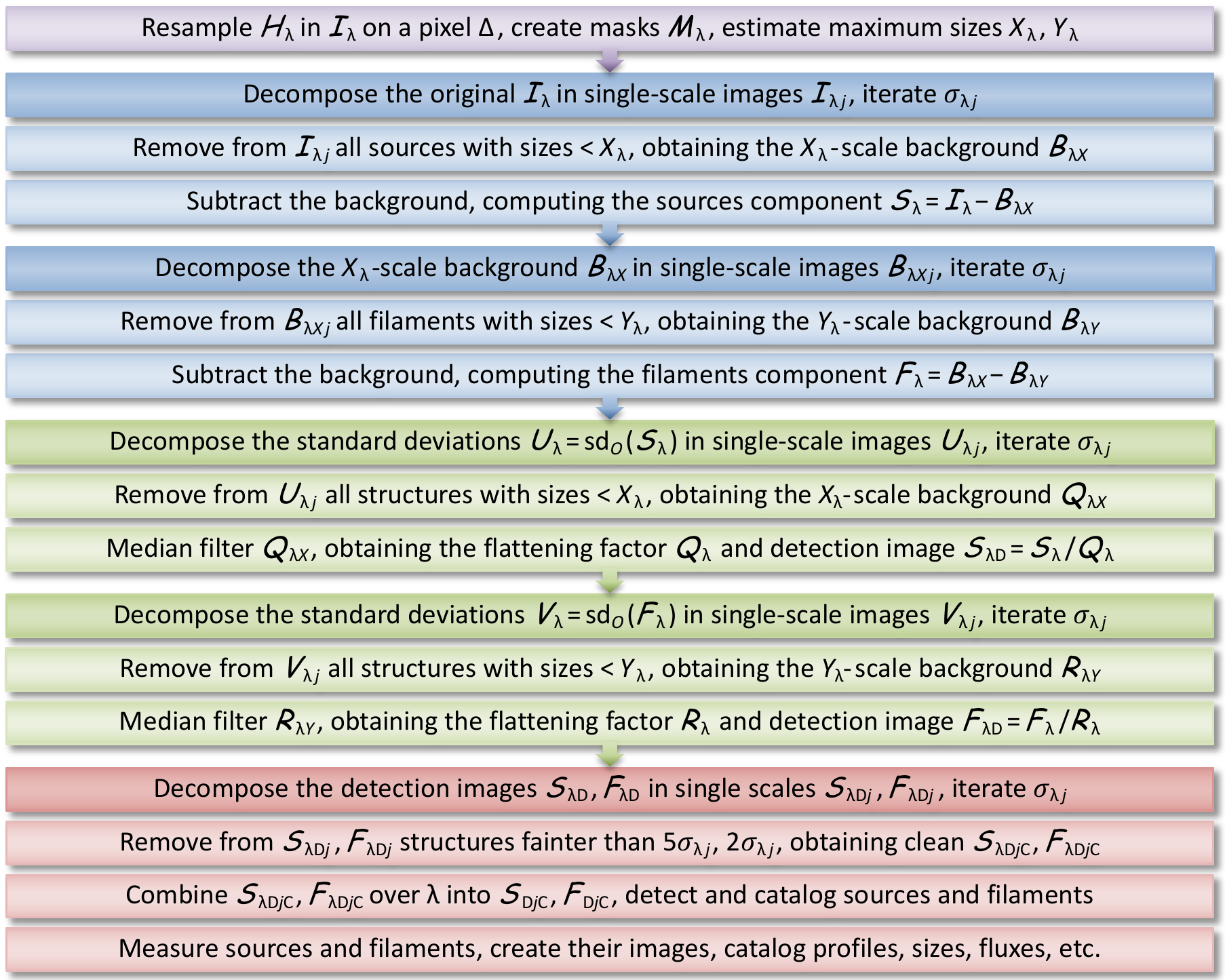}}}
\caption
{ 
Flowchart of the image processing steps in \textsl{getsf}. The colored blocks represent preparation (\emph{purple}), background
subtraction (\emph{blue}), image flattening (\emph{green}), and extraction of sources and filaments (\emph{red}). 
} 
\label{flowchart}
\end{figure*}

It seems unlikely that methods that are based on very different approaches would give consistent results in terms of detection
completeness, number of false-positive detections, and measurement accuracy. In practice, various methods do perform very
differently, as can be shown quantitatively on simulated benchmarks for which the properties of all components are known. This
highlights the need of systematic comparisons of different methods in order to understand their qualities, inaccuracies, and
biases. The danger is real that numerous uncalibrated methods are applied for the same type of star formation studies, which would
give inconsistent, contradictory results and incorrect conclusions. This would create serious, lasting problems for the science.

Source- and filament-extraction methods are the critically important tools that must be calibrated and validated using a standard
set of benchmark images with fully known properties of all components before they are applied in astrophysical contexts. It would be
desirable to use the same extraction tool to exclude any biases or dissimilarities that are caused by different methods. If a new
extraction method is to be used, it must be tested on standard benchmarks to ensure that its detection and measurement qualities
are consistent or better. This approach is usually practiced within research consortia, but this does not solve the global problem
that the results obtained from the same data by different consortia or research groups using different tools may still be affected
by the uncalibrated (or suboptimal) tools that were used.

This paper presents \textsl{getsf}, a new multiwavelength method for extracting sources and filaments. It also describes a
realistic simulated benchmark, resembling the \emph{Herschel} images of star-forming regions, which is used below to illustrate the
method and in a separate paper (Men'shchikov 2021, submitted) to quantitatively evaluate its performance. The multiwavelength
benchmark simulates the images of a dense cloud with strong nonuniform fluctuations, a wide dense filament with a power-law
intensity profile, and hundreds of radiative transfer models of starless and protostellar cores with wide ranges of sizes, masses,
and profiles. The simulated benchmark with fully known parameters allows quantitative analyses of extraction results and conclusive
comparisons of different methods by evaluating their extraction completeness, reliability, and goodness, along with the detection
and measurement accuracies. The multiwavelength images can serve as the standard benchmark problem for other source- and
filament-extraction methods, allowing researchers to perform their own tests and choose the most reliable and accurate extraction
method for their studies. Instead of publishing benchmarking results for some of the existing methods, it seems a better idea to
provide researchers with the benchmark\footnote{\url{http://irfu.cea.fr/Pisp/alexander.menshchikov/\#bench}} and a quality
evaluation system (Men'shchikov 2021, submitted) to enable comparisons of the methods of their choice. In practice, this approach
of having own experience is much more convincing and it allows a consistent evaluation of newly developed methods.

The new source and filament extraction method \textsl{getsf} represents a major improvement over the previous algorithms
\textsl{getsources}, \textsl{getfilaments}, and \textsl{getimages} \citep[][hereafter referred to as Papers I, II, and
III]{Men'shchikov_etal2012,Men'shchikov2013,Men'shchikov2017}; throughout this paper, the three predecessors are collectively
referred to as \textsl{getold}. The new method (Fig.~\ref{flowchart}) consistently handles two types of structures, sources and
filaments, that are important for studies of star formation, separating the structural components from each other and from their
backgrounds. All major processing steps of \textsl{getsf} employ spatial decomposition of images into a number of finely spaced
single-scale images to better isolate the contributions of structures with various widths. The method produces accurately
flattened detection images with uniform levels of the residual background and noise fluctuations. To detect sources and filaments,
\textsl{getsf} combines independent information contained in the multiwaveband single-scale images of the structural components,
preserving the higher angular resolutions. Then \textsl{getsf} measures and catalogs the detected sources and filaments in their
background-subtracted images. The fully automated method needs only one user-defined parameter, the maximum size of the structures
of interest to extract, constrained by users from the input images on the basis of their research interests.

This work follows Papers I--III in advocating a clear distinction between the words \emph{source} and \emph{object}, unlike many
publications in which it is implicitly assumed that the two are completely equivalent. "Source" used in the context of the source
extractions and statistical analysis of their results, and "object" is only used in the context of the physical interpretation of
the extracted sources. In this paper, the sources are defined as the emission peaks (mostly unresolved) that are significantly
stronger than the local surrounding fluctuations, indicating the presence of the physical objects in space that produced the
observed emission. The implicit assumption that an unresolved far-infrared source on a complex fluctuating background contains
emission of just one single object is invalid in general. Too often, an emission peak is actually a blend of many components,
produced by different physical entities. This is illustrated by the recent images of the massive star-forming cloud
\object{W43-MM1} \citep{Motte_etal2018}, obtained with the \emph{ALMA} interferometer. This object appears as a single source in
the \emph{Herschel} images, even with the $5.6$ and $11.3${\arcsec} resolutions at $70$ and $160$\,$\mu$m. However, the \emph{ALMA}
image (Sect.~\ref{alma}) displays a rich cluster of much smaller sources that are unresolved or just slightly resolved even at the
$0.44${\arcsec} resolution.

\begin{figure*}                                                               
\centering
\centerline{
  \resizebox{0.328\hsize}{!}{\includegraphics{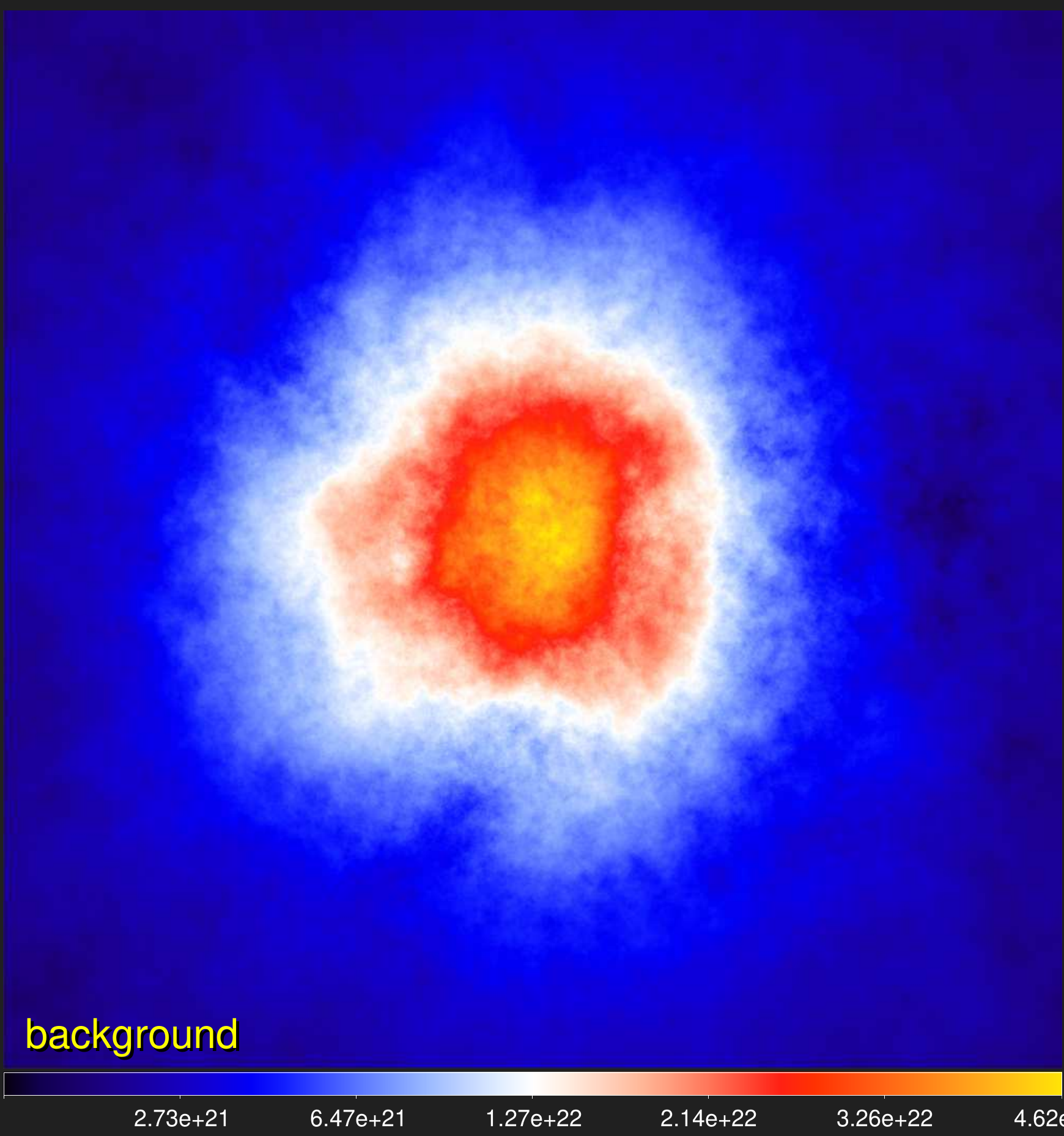}}
  \resizebox{0.328\hsize}{!}{\includegraphics{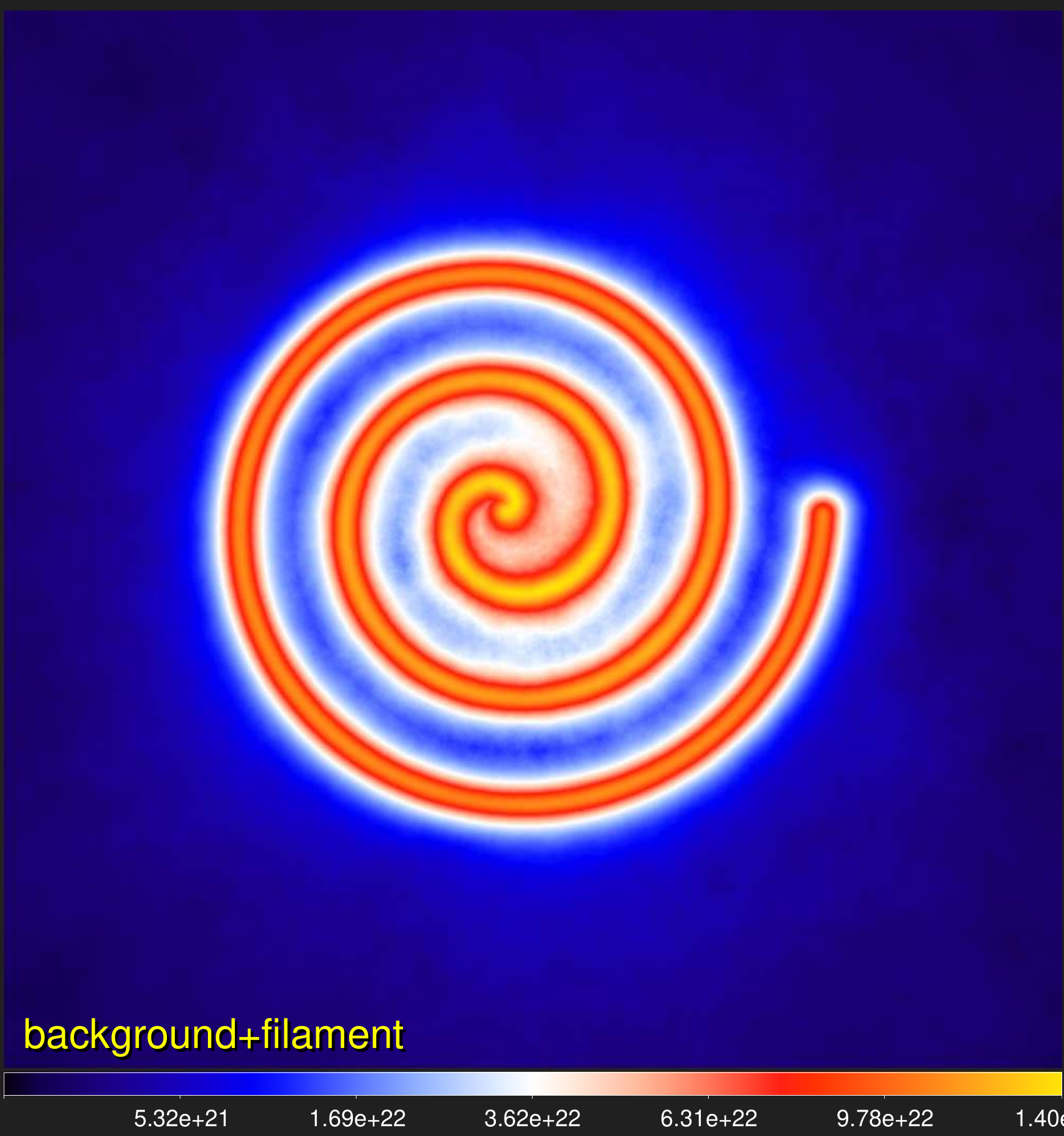}}
  \resizebox{0.328\hsize}{!}{\includegraphics{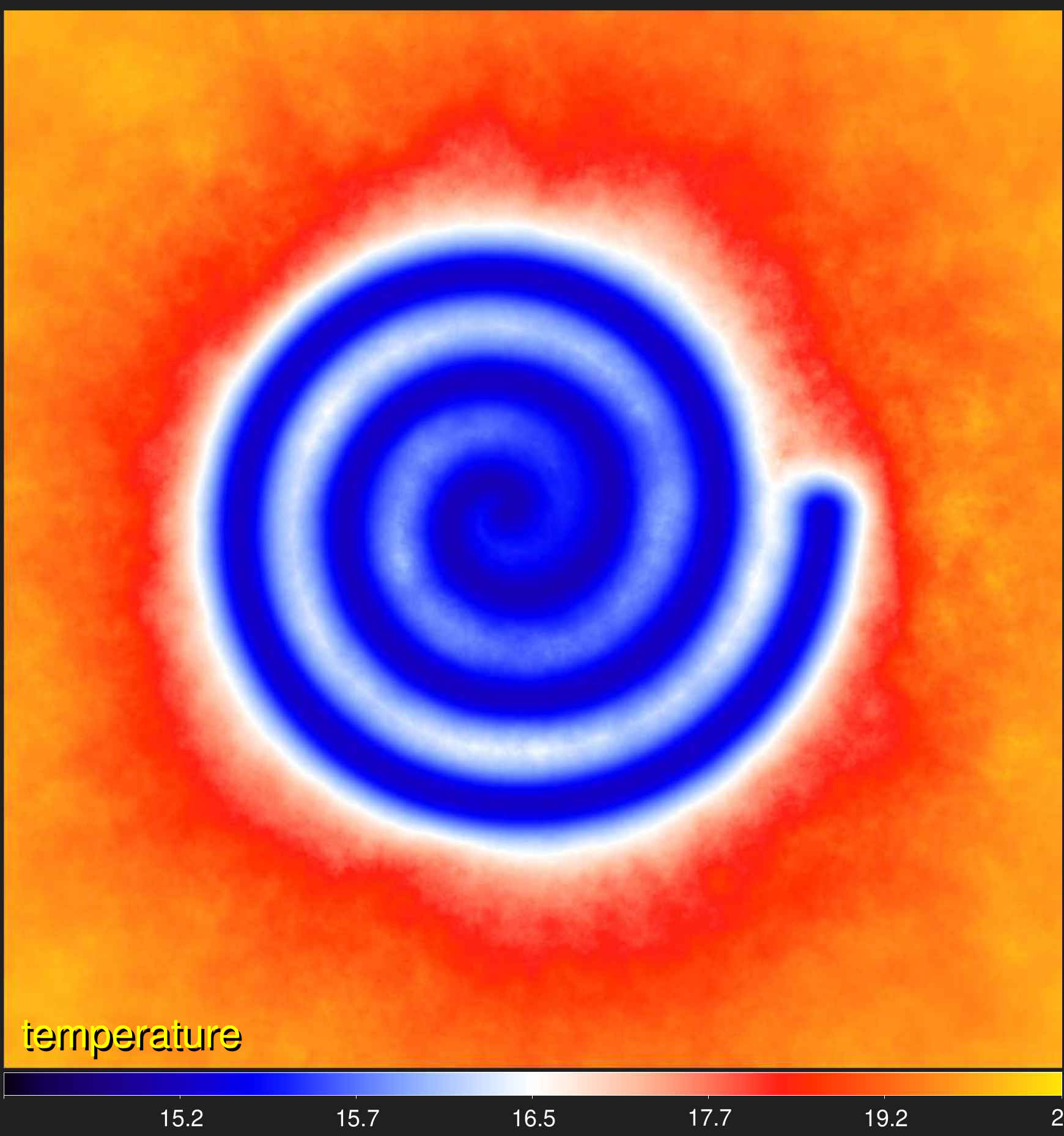}}}
\caption
{ 
Background surface densities ($\mathcal{D}_{\rm B}, \mathcal{D}_{\rm C}$) and average line-of-sight dust temperatures
($\mathcal{T}_{\!\rm C}$) used to compute the simulated \emph{Herschel} images $\mathcal{C}_{\lambda}$ of the filamentary cloud
from Eq.~(\ref{simformula}). Square-root color mapping.
} 
\label{bgimage}
\end{figure*}

Section~\ref{skybench} describes the new multiwavelength benchmark for source- and filament-extraction methods, resembling the
\emph{Herschel} observations of star-forming regions. Section~\ref{getsf} presents \textsl{getsf}, the new source- and
filament-extraction method, employing separation of the structural components. Section~\ref{applications} illustrates the
performance of \textsl{getsf} on a large variety of images that were obtained with different telescopes in a wide spectral range,
from X-rays to millimeter wavelengths. Section~\ref{strenlimits} describes all strengths and limitations of \textsl{getsf}.
Section~\ref{conclusions} presents a summary of this work. Appendix~\ref{hiresinacc} discusses inaccuracies of the surface
densities and temperatures, derived by spectral fitting of the images. Appendix~\ref{decomposition} describes the single-scale
spatial decomposition that is used by \textsl{getsf} in its processing steps. Appendix~\ref{getsfdetails} gives details on the
software.

In this paper, images are represented by capital calligraphic characters (e.g., $\mathcal{A}, \mathcal{B}, \mathcal{C}$) and
software names and numerical methods are typeset slanted (e.g., \textsl{getsf}) to distinguish them from other emphasized words.
The curly brackets $\{\}$ are used to collectively refer to either of the characters, separated by vertical lines. For example,
$\{a|b\}$ refers to $a$ or $b$ and $\{A|B\}_{\rm \{a|b\}c}$ expands to $A_{\rm \{a|b\}c}$ or $B_{\rm \{a|b\}c}$, as well as to
$A_{\rm ac}$, $A_{\rm bc}$, $B_{\rm ac}$, or $B_{\rm bc}$. 


\section{Benchmark for source and filament extractions}
\label{skybench}

Realistic multiwavelength, multicomponent images of a simulated star-forming region were computed to present \textsl{getsf} in
this paper and to compare its performance with the previous benchmark that was used in Papers I and III.
The benchmark images were created for all \emph{Herschel} wavebands (at $\lambda$ of $70$, $100$, $160$, $250$, $350$, and
$500$\,${\mu}$m). They consist of independent structural components: a background cloud $\mathcal{B}_{\lambda}$, a long filament
$\mathcal{F}_{\lambda}$, round sources $\mathcal{S}_{\lambda}$, and small-scale instrumental noise $\mathcal{N}_{\lambda}$:
\begin{equation} 
\mathcal{H}_{\lambda} = \mathcal{B}_{\lambda} + \mathcal{F}_{\lambda} + \mathcal{S}_{\lambda} + \mathcal{N}_{\lambda} =
\mathcal{C}_{\lambda} + \mathcal{S}_{\lambda} + \mathcal{N}_{\lambda},
\label{skycomponents}
\end{equation} 
where $\mathcal{C}_{\lambda}{\,=\,}\mathcal{B}_{\lambda}{\,+\,}\mathcal{F}_{\lambda}$ is the emission intensity of the filamentary
background. All simulated images were computed on a $2${\arcsec} pixel grid with $2690{\,\times\,}2690$ pixels, covering
$1.5{\degr}{\times\,}1.5${\degr} or $3.7$\,pc at a distance $D{\,=\,}140$\,pc of the nearest star-forming regions (e.g., those in
Taurus or Ophiuchus).


\subsection{Simulated filamentary background}
\label{simfilback}

An image of the background surface density was computed from a purely synthetic scale-free background $\mathcal{D}_{\rm A}$ (cf.
Paper I), with $N_{{\rm H}_2}{\,\sim\,}2.7{\,\times\,}10^{20}$ to $5{\,\times\,}10^{22}$\,cm$^{-2}$ that had uniform fluctuations
across the entire image. To simulate complex astrophysical backgrounds with strongly nonuniform fluctuations
\citep[e.g.,][]{Ko"nyves_etal2015}, $\mathcal{D}_{\rm A}$ was multiplied by a circular shape $\mathcal{P}$ with a radial profile
defined by Eq.~(\ref{moffatfun}) below (with $\Theta{\,=\,}1500${\arcsec} and $\zeta{\,=\,}2$), normalized to unity and centered on
the image; finally, a constant value of $1.5{\,\times\,}10^{21}$\,cm$^{-2}$ was added to increase the minimum value. The surface densities of the resulting
background cloud image $\mathcal{D}_{\rm B}$ (Fig.~\ref{bgimage}) are $1.5{\,\times\,}10^{21}$ to
$4.8{\,\times\,}10^{22}$\,cm$^{-2}$ and the fluctuations differ by approximately two orders of magnitude. The total mass of the
cloud is $M_{\rm B}{\,=\,}1.78{\,\times\,}10^{3}$\,$M_{\sun}$.

\begin{figure*}                                                               
\centering
\centerline{
  \resizebox{0.328\hsize}{!}{\includegraphics{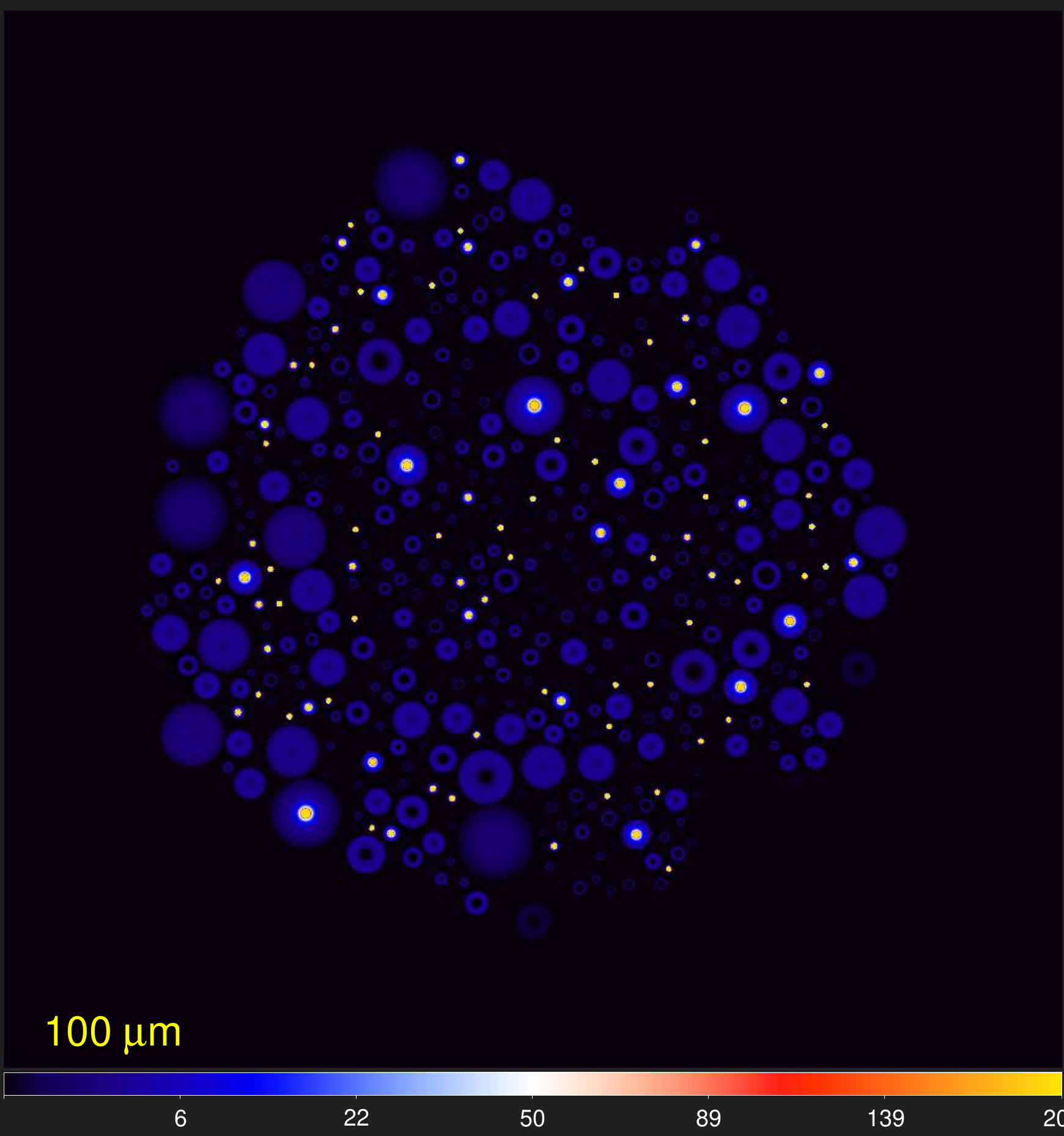}}
  \resizebox{0.328\hsize}{!}{\includegraphics{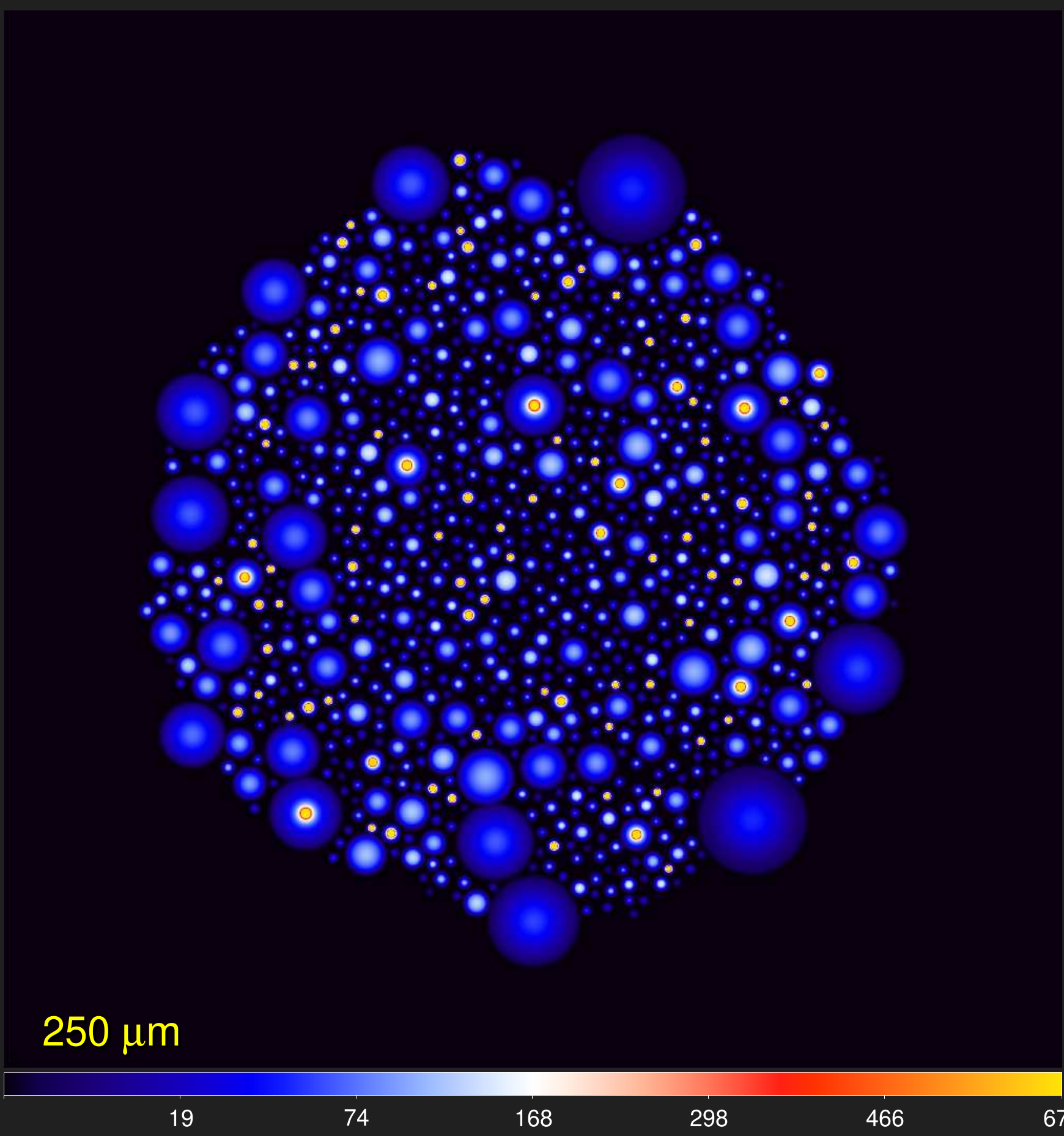}}
  \resizebox{0.328\hsize}{!}{\includegraphics{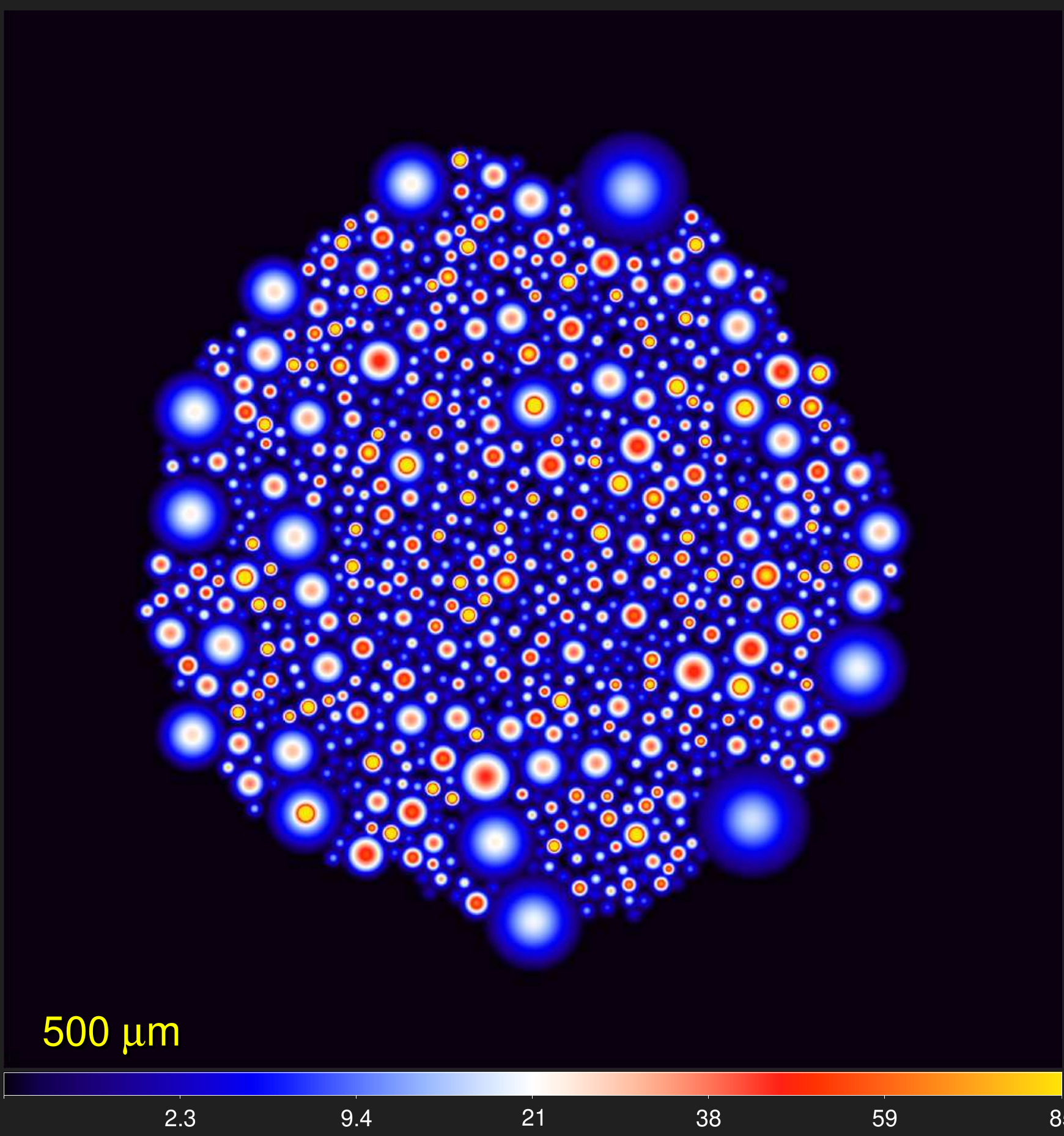}}}
\caption
{ 
Component of sources $\mathcal{S}_{\lambda}$ that is composed of the images of radiative transfer models of 828 starless and 91
protostellar cores and convolved to the \emph{Herschel} resolutions $O_{\lambda}$ (cf. Sect.~\ref{skycomponents}), shown at three
selected wavelengths. Only the bright unresolved emission peaks of the protostellar cores, clearly visible at $100$\,{${\mu}$m},
appear in the $70$\,{${\mu}$m} image (not shown). Square-root color mapping.
} 
\label{simcores}
\end{figure*}

To simulate filamentary backgrounds, a long spiral filament was added to the background cloud $\mathcal{D}_{\rm B}$. The spiral
shape was chosen so that the filament occupied various areas of the cloud with very different surface densities and to cause the
filament to be blended (with itself) to some extent. The spiral filament image $\mathcal{D}_{\rm F}$ has a crest value of
$N_{0}{\,=\,}10^{23}$\,cm$^{-2}$, a full width at half-maximum (FWHM)
$W{\,=\,}0.1$\,pc, and a radial profile similar to those observed with
\emph{Herschel} in star-forming regions \citep[e.g.,][]{Arzoumanian_etal2011,Arzoumanian_etal2019},
\begin{equation} 
N_{{\rm H}_2}(\theta) = N_{0} \left(1 + (2^{1/\zeta}\!- 1) \,(\theta / \Theta)^2\right)^{-\zeta},
\label{moffatfun}
\end{equation} 
where $\theta$ is the angular distance, $\Theta$ is the structure half-width at half-maximum, and $\zeta$ is a power-law exponent. With
$\Theta{\,=\,}75${\arcsec} (or $0.05$\,pc at $D{\,=\,}140$\,pc) and $\zeta{\,=\,}1.5$, this Moffat (Plummer) function approximates
a Gaussian of $0.1$\,pc (FWHM) in its core and it transforms into a power-law profile ${N_{{\rm
H}_2}(\theta){\,\propto\,}\theta^{\;\!-3}}$ for ${\theta{\,\gg\,}\Theta}$. The filament mass $M_{\rm
F}{\,=\,}3.04{\,\times\,}10^{3}$\,$M_{\sun}$ and length $L_{\rm F}{\,=\,}10.5$\,pc correspond to the linear density $\Lambda_{\rm
F}{\,=\,}290$\,$M_{\sun}$\,pc$^{-1}$. The resulting surface densities $\mathcal{D}_{\rm C}{\,=\,}\mathcal{D}_{\rm
B}{\,+\,}\mathcal{D}_{\rm F}$ of the filamentary cloud are in the range of $1.7{\,\times\,}10^{21}$ to
$1.4{\,\times\,}10^{23}$\,cm$^{-2}$ (Fig.~\ref{bgimage}), and its total mass is $M_{\rm C}{\,=\,}4.82{\,\times\,}10^{3}$\,$M_{\sun}$.

To approximate the nonuniform line-of-sight dust temperatures of the star-forming clouds observed with \emph{Herschel}
\citep[e.g.,][]{Men'shchikov_etal2010,Arzoumanian_etal2019}, an image of average line-of-sight temperatures was improvised as
\begin{equation} 
\mathcal{T}_{\!\rm C} = 200 \left(10^{-20} \mathcal{D}_{\rm C} + 20\right)^{-1}{\!+\,}15\,{\rm K}.
\label{skytdust}
\end{equation} 
The pixel values of the resulting temperature image $\mathcal{T}_{\!\rm C}$  range between $15$\,K in the innermost areas of the
filamentary cloud and $20$\,K in its outermost parts (Fig.~\ref{bgimage}). The temperatures from Eq.~(\ref{skytdust}) were used to
simulate the cloud images $\mathcal{C}_{\lambda}$ in all \emph{Herschel} wavebands, assuming optically thin dust emission:
\begin{equation} 
\mathcal{C}_{\nu} = B_{\nu}(\mathcal{T}_{\rm C})\,\mathcal{D}_{\rm C}\,\kappa_{\nu} \eta \mu m_{\rm H},
\label{simformula}
\end{equation} 
where $B_{\nu}$ is the blackbody intensity, $\kappa_{\nu}$ is the dust opacity, $\eta{\,=\,}0.01$ is the dust-to-gas mass ratio,
$\mu{\,=\,}2.8$ is the mean molecular weight per H$_2$ molecule, and $m_{\rm H}$ is the hydrogen mass. The dust opacity was parameterized as
a power law $\kappa_{\nu}{\,=\,}\kappa_{0}\left(\nu/\nu_{0}\right)^{\,\beta}$ with $\kappa_0{\,=\,}9.31$\,cm$^{2}$g$^{-1}$ (per
gram of dust), $\lambda_0{\,=\,}300$\,${\mu}$m, and $\beta{\,=\,}2$.


\subsection{Simulated starless and protostellar cores}
\label{simulcores}

To populate the filamentary cloud with realistic sources, 156 radiative transfer models were computed by a numerical solution of
the dust continuum radiative transfer problem in spherical geometry \citep[using \textsl{modust},][]{Bouwman2001}. The models
adopted tabulated absorption opacities $\kappa_{\rm abs}$ for dust grains with thin ice mantles \citep{OssenkopfHenning1994},
corresponding to a density $n_{\rm H}{\,=\,}10^6$\,cm$^{-3}$ and coagulation time $t{\,=\,}10^5$\,yr. The opacity values at
$\lambda{\,>\,}160$\,${\mu}$m were replaced with a power law $\kappa_{\lambda}{\,\propto\,}\lambda^{-2}$, consistent with the
parameterization used in Eq.~(\ref{simformula}).

The models of three populations of starless cores and one population of protostellar cores cover wide ranges of masses (from $0.05$
to $2$\,$M_{\sun}$) and half-maximum sizes (from ${\sim\,}0.001$ to $0.1$\,pc). Density profiles of the critical Bonnor-Ebert
spheres were adopted for starless cores, whereas the protostellar cores have power-law densities $\rho(r){\,\propto\,}r^{-2}$.
Starless cores consist of low-, medium-, and high-density subpopulations, following the $M{\,\propto\,}R$ relation for the
isothermal Bonnor-Ebert spheres (with $T_{\rm BE}{\,=\,}7, 14, 28$\,K) in the area of the mass-radius diagram occupied by
prestellar cores observed in the Ophiuchus and Orion star-forming regions \citep{Motte_etal1998,Motte_etal2001}.

Both types of cores were embedded in background spherical clouds with a uniform surface density of
$3{\,\times\,}10^{21}$\,cm$^{-2}$ and outer radius of $1.4{\,\times\,}10^{5}$\,AU ($1000${\arcsec} or $0.68$\,pc). In an isotropic
interstellar radiation field \citep{Black1994} with the strength parameter $G_{0}{\,=\,}10$ \citep[e.g.,][]{Parravano_etal2003},
the embedding clouds acquired temperatures of $T{\,\approx\,}22$\,K at their edges, consistent with the highest values of
$\mathcal{T}_{\rm C}$ from Eq.~(\ref{skytdust}). The embedding clouds lowered $T(r)$ toward the interiors of both starless and
protostellar cores. Accreting protostars in the centers of the protostellar cores, however, produced luminosity $L_{\rm
A}{\,\propto\,}M$ and thus sharply peaked temperature distributions deeper in their central parts.


\subsection{Complete simulated images}
\label{simcomplete}

Individual surface density images of the models of 828 starless and 91 protostellar cores were distributed in the dense areas
($N_{{\rm H}_2}{\,\ge\,}5{\,\times\,}10^{21}$\,cm$^{-2}$) of the filamentary cloud $\mathcal{D}_{\rm C}$. They were added
quasi-randomly, without overlapping, to the $\mathcal{D}_{\rm C}$ image at positions, where their peak surface density exceeded
that of the cloud $N_{{\rm H}_2}$ value. An initial mass function (IMF)-like broken power-law mass function with a slope ${\rm d}N/{\rm d}M$ of $-1.3$ for
$M{\,\le\,}0.5$\,{$M_{\sun}$} and $-2.3$ for $M{\,>\,}0.5$\,{$M_{\sun}$} was used to determine the numbers of models per mass bin
$\delta{\log_{10}\!M}{\,\approx\,}0.1$ in each of the four populations. This resulted in the surface densities $\mathcal{D}_{\rm S}$,
the intensities $\mathcal{S}_{\lambda}$ of sources (Fig.~\ref{simcores}), and in the complete simulated images
$\mathcal{C}_{\lambda}{\,+\,}\mathcal{S}_{\lambda}$.

\begin{figure*}                                                               
\centering
\centerline{
  \resizebox{0.328\hsize}{!}{\includegraphics{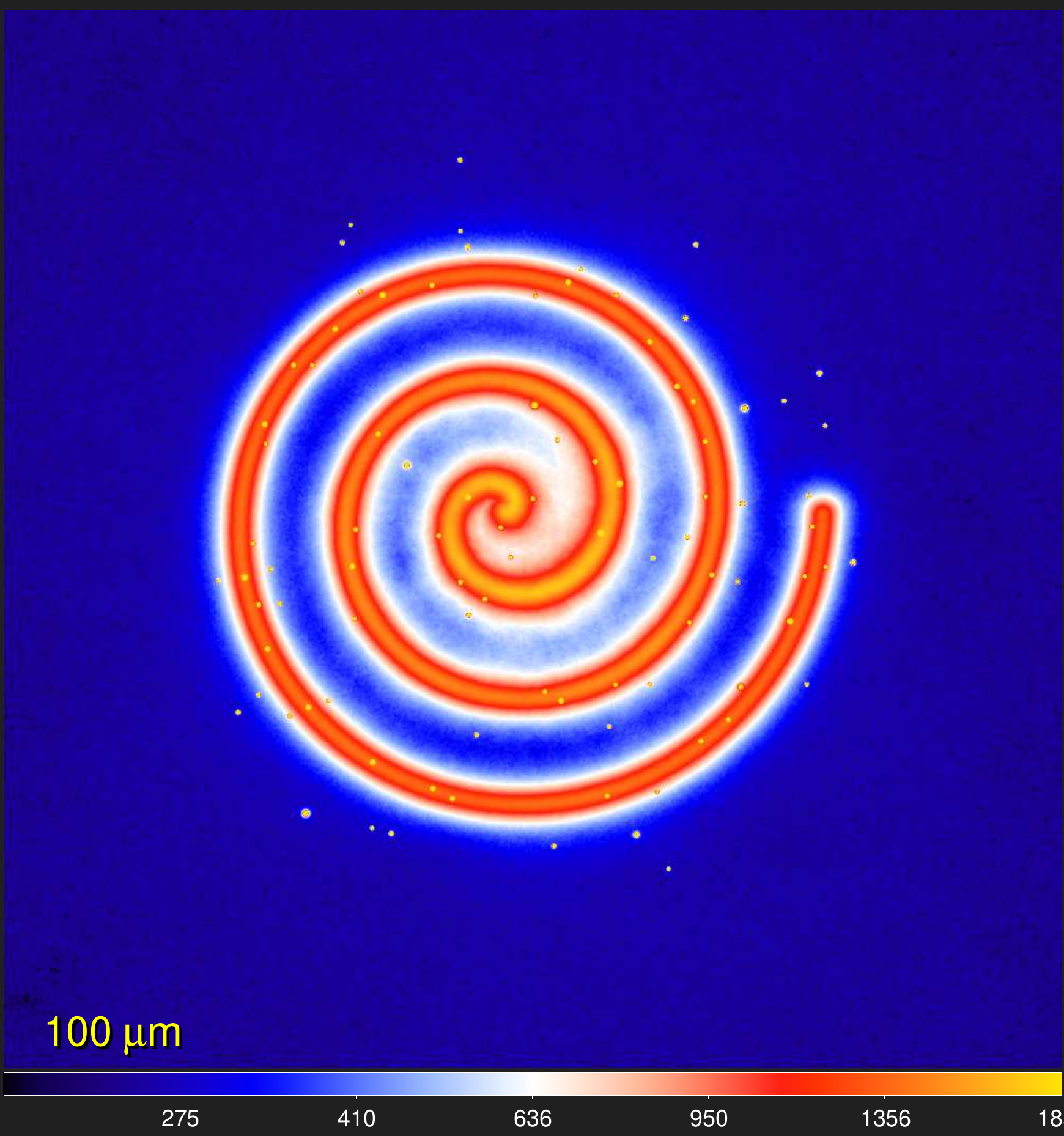}}
  \resizebox{0.328\hsize}{!}{\includegraphics{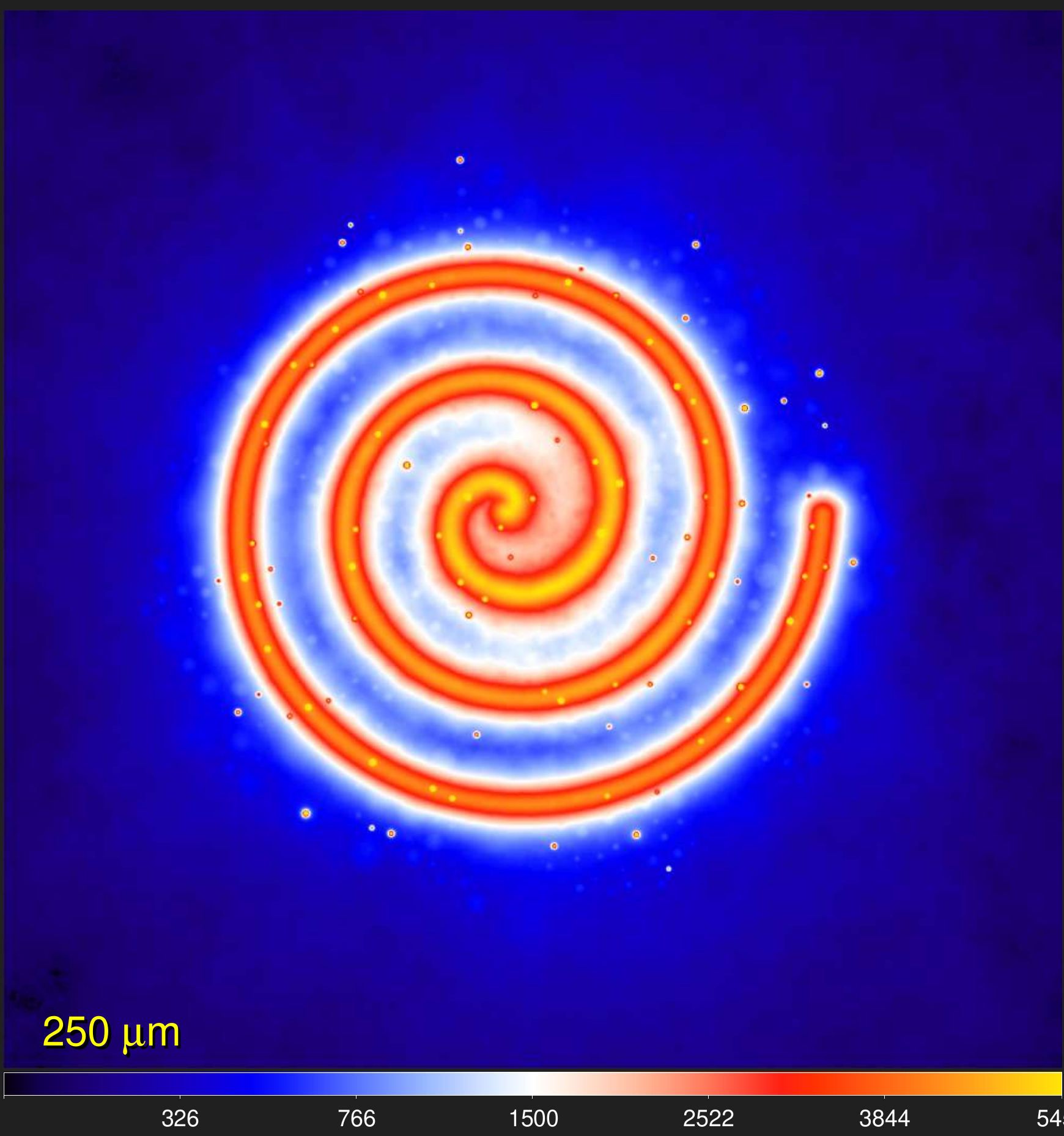}}
  \resizebox{0.328\hsize}{!}{\includegraphics{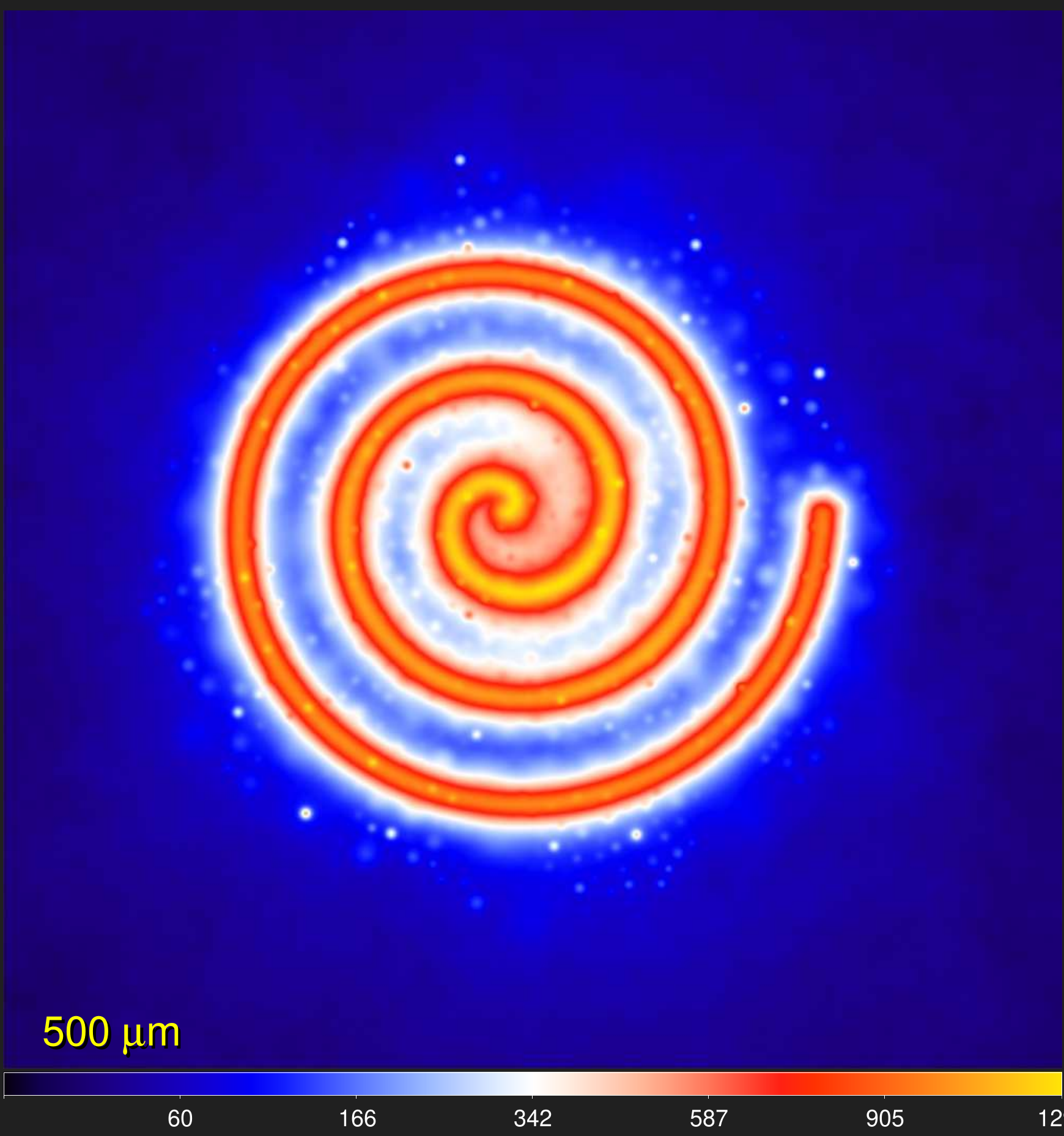}}}
\caption
{ 
Images $\mathcal{H}_{\lambda}$ of the simulated star-forming region, defined by Eq.~(\ref{skycomponents}), shown at three selected
wavelengths. The benchmark images are a superposition of four structural components: the background $\mathcal{B}_{\lambda}$, the
filament $\mathcal{F}_{\lambda}$, the sources $\mathcal{S}_{\lambda}$, and the noise $\mathcal{N}_{\lambda}$. Two simpler variants
of this benchmark are also available: without the filament and without the background. Square-root color mapping.
} 
\label{simimages}
\end{figure*}

The final simulated \emph{Herschel} images $\mathcal{H}_{\lambda}$ from Eq.~(\ref{skycomponents}) of the modeled star-forming
region were obtained by adding different realizations of the random Gaussian noise $\mathcal{N}_{\lambda}$ at $70$, $100$, $160$,
$250$, $350$, and $500$\,{${\mu}$m} and convolving the resulting images to the angular resolutions $O_{\lambda}$ of $8.4$, $9.4$,
$13.5$, $18.2$, $24.9$, and $36.3${\arcsec}, respectively (Fig.~\ref{simimages}). The resulting images $\mathcal{H}_{\lambda}$ have
$\sigma$ noise levels of $6$, $6$, $5.5$, $2.5$, $1.2$, and $0.5$\,MJy\,sr$^{-1}$, resembling the actual noise measured in the
\emph{Herschel} images of the Rosette molecular complex \citep[][]{Motte_etal2010}.

\section{Source- and filament-extraction method}
\label{getsf}

The main processing steps of \textsl{getsf} are outlined in Fig.~\ref{flowchart}, where several major blocks of the algorithm are
highlighted. The method may be summarized as follows: (1) preparation of a complete set of images for an extraction, (2) separation
of the structural components of sources and filaments from their backgrounds, (3) flattening of the residual noise and background
fluctuations in the images of sources and filaments, (4) combination of the flattened components of sources and filaments over
selected wavebands, (5) detection of sources and filaments in the combined images of the components, and (6) measurements of the
properties of the detected sources and filaments.

Like its predecessors, \textsl{getsf} has just a single, user-definable parameter: the maximum size (width) of the structures of
interest to extract. Internal parameters of \textsl{getsf} have been carefully calibrated and verified in numerous tests using
large numbers of diverse images (both simulated and real-life observed images) to ensure that \textsl{getsf} works in all cases.
This approach rests on the conviction that high-quality extraction methods for scientific applications must not depend on the human
factor. It is the responsibility of the creator of a numerical method to make it as general as possible and to minimize the number
of free parameters as much as possible. An internal multidimensional parameter space of complex numerical tools must never be
delegated to the end user to explore if the aim is to obtain consistent and reliable scientific results.


\subsection{Preparation of images for extraction}
\label{preparation}

The multiwavelength extraction methods must be able to use all available information contained in the observed images across
various wavebands with different angular resolutions. It is usually beneficial to collect all available images for a
specific region of the sky under study.


\subsubsection{Original observed set of images}
\label{obsimages}

To prepare multiwavelength $\mathcal{H}_{\lambda}$ for processing with \textsl{getsf}, it is necessary to convert them into the
images $\mathcal{I}_{{\!\lambda}}$, all on the same grid of pixels. To this end, \textsl{getsf} resamples all images \citep[using
\textsl{swarp},][]{Bertin_etal2002} on a pixel size, chosen to be optimal for the highest-resolution images available. It is very
important to carefully verify alignment of the resampled $\mathcal{I}_{{\!\lambda}}$ and correct it (if necessary) to ensure that
all unresolved intensity peaks remain on the same pixel across all wavebands. To reveal possible misalignments, it is sufficient to
open each pair of prepared images in \textsl{ds9} \citep{JoyeMandel2003} and blink the two frames, going from the
highest-resolution to the lowest-resolution images.

Most astronomical images have irregularly shaped coverage and limited usable areas that differ between wavebands. To include only
the ``good'' parts of the $\mathcal{I}_{{\!\lambda}}$ coverage in the image processing, it is necessary to create masks
$\mathcal{M}_{\lambda}$ (with pixel values 1 or 0). With these masks, \textsl{getsf} can process only the good areas of
$\mathcal{I}_{{\!\lambda}}$ that have a mask value of 1. To facilitate the image preparation, \textsl{getsf} always creates default
masks ${\mathcal{M}_{\lambda}{\,=\,}1}$. However, for most real observations, the masks must be prepared very carefully and
independently for each image. To manually create the masks, one can use \textsl{imagej} \citep{Abramoff_etal2004} or
\textsl{gimp}\footnote{\url{http://www.gimp.org/}} that allows users to create a polygon over an image, convert the polygon into a
mask, and save it in the FITS format.

\begin{figure*}                                                               
\centering
\centerline{
  \resizebox{0.328\hsize}{!}{\includegraphics{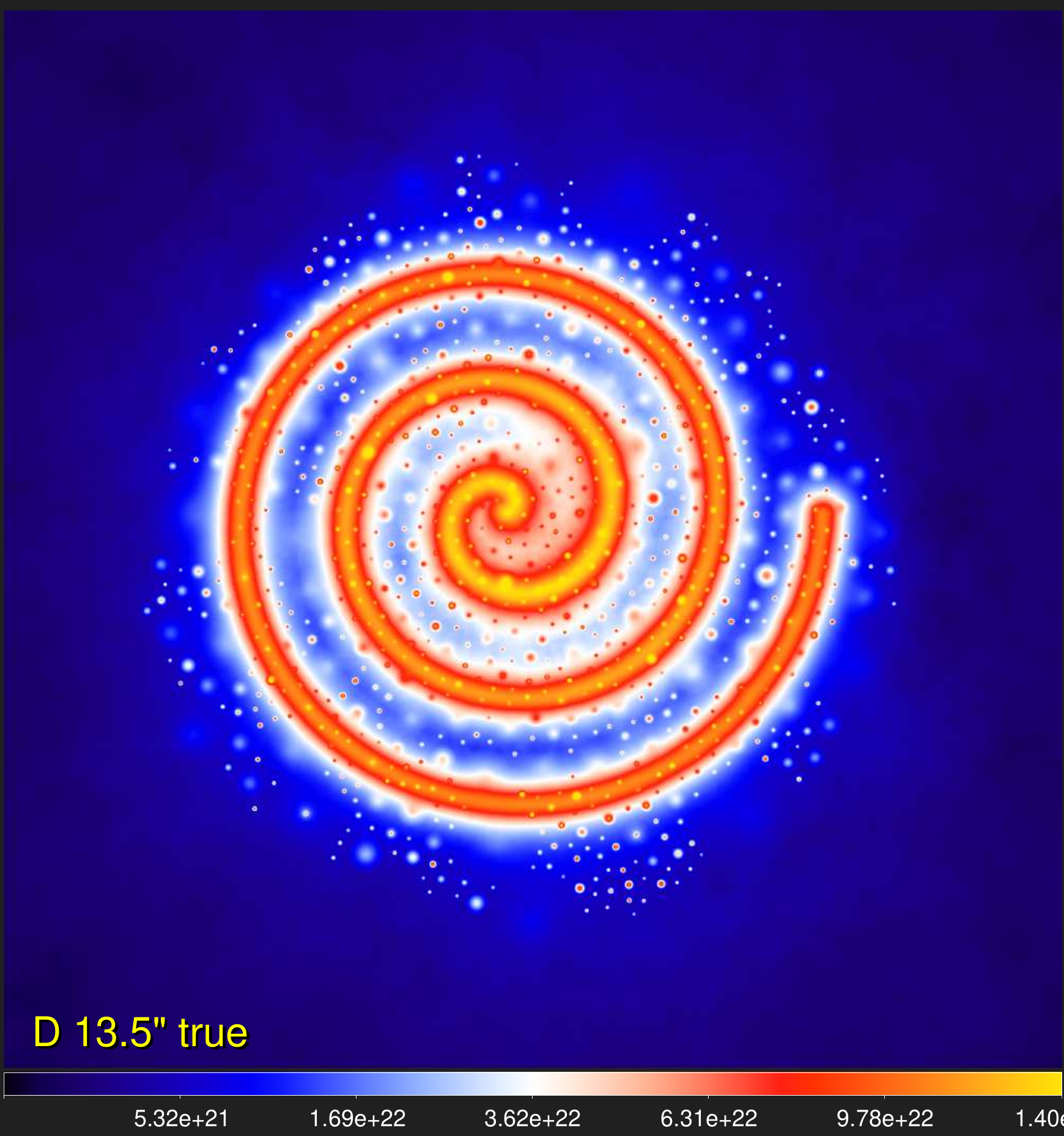}}
  \resizebox{0.328\hsize}{!}{\includegraphics{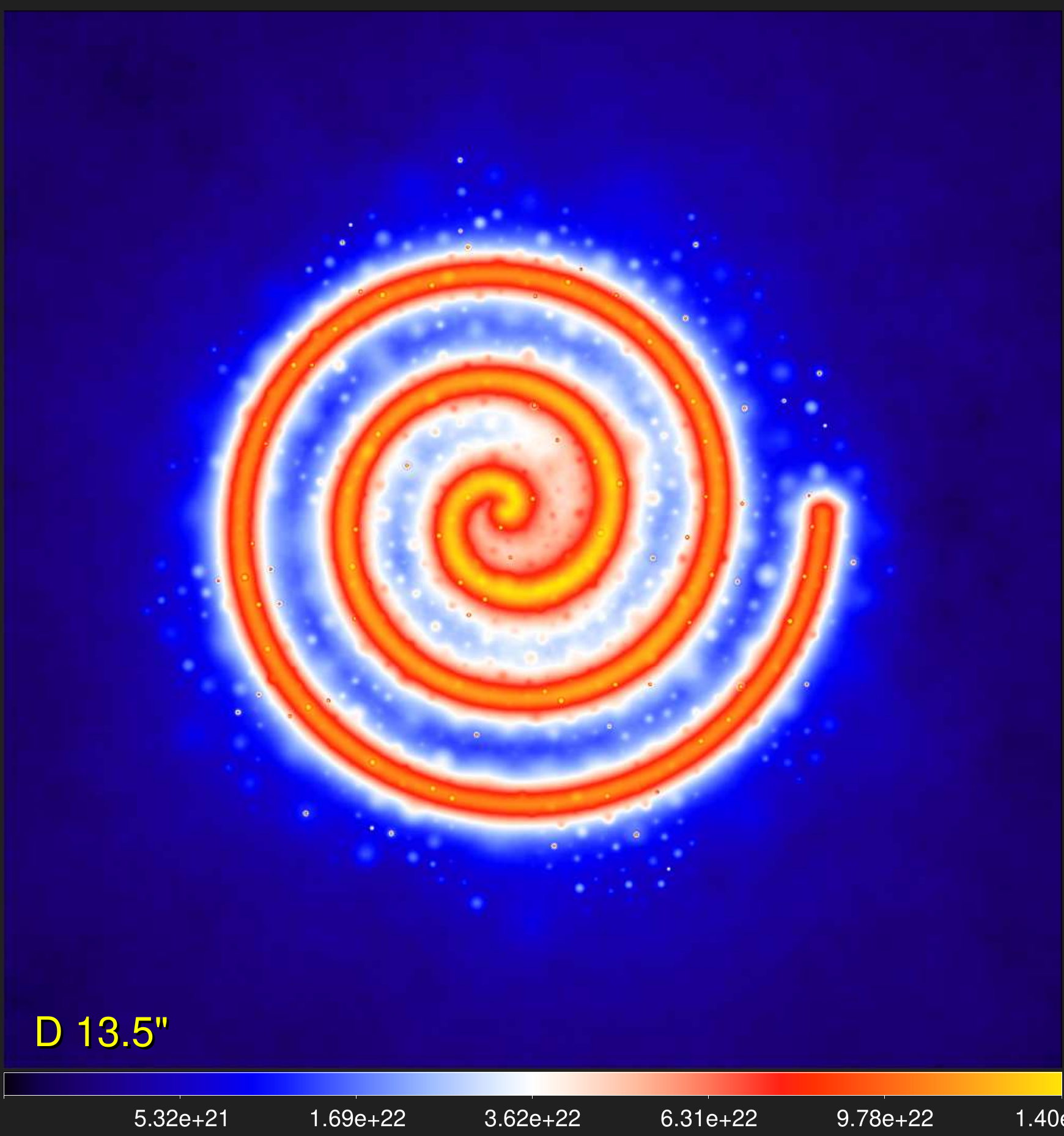}}
  \resizebox{0.328\hsize}{!}{\includegraphics{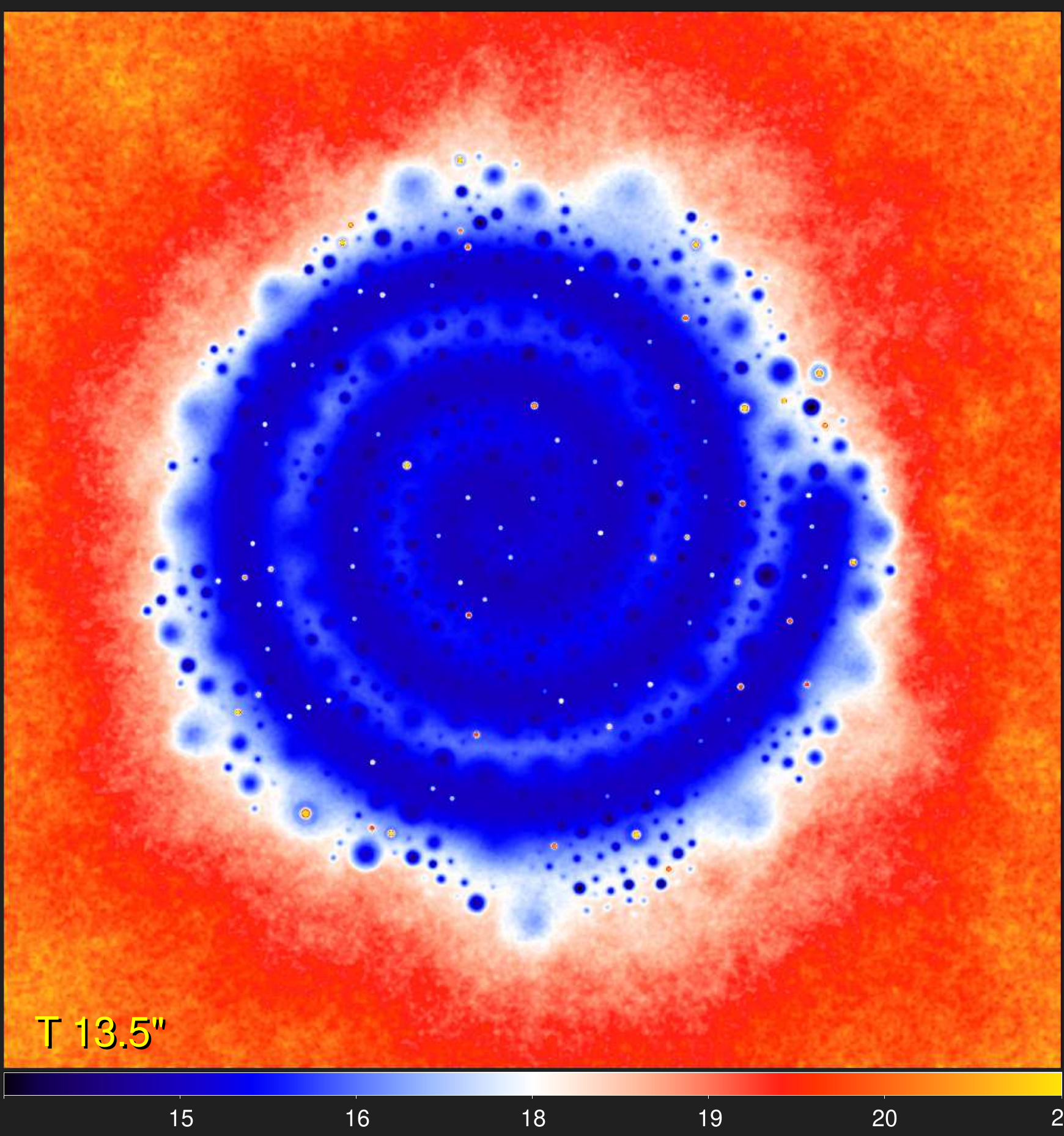}}}
\caption
{ 
Derived surface densities and temperatures (Sect.~\ref{hiresimages}). The true model image $\mathcal{D}_{\rm
C}{\,+\,}\mathcal{D}_{\rm S}$ and the \textsl{hires} surface density $\mathcal{D}_{13{\arcsec}}$ and temperature
$\mathcal{T}_{\!13{\arcsec}}$ derived from Eq.~(\ref{superdens}) with $\lambda_{\rm H}{\,=\,}160$\,${\mu}$m ($O_{{\rm
H}}{\,=\,}13.5${\arcsec}) are shown. Many of the sources, clearly visible in the true image (\emph{left}), are not discernible in
the derived surface density (\emph{middle}) because of the inaccuracies in the temperatures from fitting spectral shapes
$\Pi_{\lambda}$. Square-root color mapping, except the \emph{right} panel with linear mapping.
} 
\label{addhires}
\end{figure*}


\subsubsection{Derived high-resolution images}
\label{hiresimages}

The multiwavelength far-infrared \emph{Herschel} images open the possibility of computing maps of surface density and dust
temperature by fitting the spectral shapes $\Pi_{\lambda}$ of the image pixels. The standard procedure assumes that (1) the
original images represent optically thin thermal emission of dust grains with a power-law opacity
$\kappa_{\lambda}{\,\propto\,}\lambda^{-\beta}$ and a constant $\beta$ value, (2) the dust temperature is constant along the lines
of sight passing through each pixel of the images, and (3) the lines of sight are not contaminated by unrelated radiation at either
end, in front of the observed structures and behind them. Unfortunately, one or more of the assumptions are likely to be invalid,
especially the stipulation of the opacity law and the constant line-of-sight temperatures \citep[e.g.,][]{Men'shchikov2016}. The
values of the derived surface densities and temperatures therefore must be considered as fairly unreliable and implying large error
bars.

When we assume that the observations include the \emph{Herschel} images, the spectral shapes $\Pi_{\lambda}$ of each pixel can be fit
at several wavelengths ($160{-}500$\,{${\mu}$m}) and resolutions ($18.2{-}36.3${\arcsec}), which results in three sets of surface
densities and dust temperatures. The highest-resolution derived images are the least reliable because they are obtained from
fitting only two images (at $160$ and $250$\,${\mu}$m), whereas the lowest-resolution maps are the most accurate because
they come from fitting four independent images (at $160$, $250$, $350$, and $500$\,${\mu}$m).

In an attempt to combine the higher accuracy of the lower-resolution images with the higher angular resolutions of the less
accurate images, \cite{Palmeirim_etal2013} published a simple algorithm that uses complementary spatial information contained in
the observed images to create a surface density image with the resolution $O_{\rm P}{\,=\,}18.2{\arcsec}$ of the $250$\,${\mu}$m
image. When this approach is extended to temperatures, the sharper images can be computed by adding the higher-resolution information to
the low-resolution images as differential terms,
\begin{equation} 
\mathcal{\{D|T\}}_{\rm P}{\,=\,}\mathcal{\{D|T\}}_{4}{\,+\,}\delta\mathcal{\{D|T\}}_{3}{\,+\,}\delta\mathcal{\{D|T\}}_{2},
\label{hiresdentem}
\end{equation} 
where the base surface density and temperature $\mathcal{\{D|T\}}_{4}$ are derived by fitting the $160$, $250$, $350$, and
$500$\,${\mu}$m images at the lowest resolution $O_{500}{\,=\,}36.3{\arcsec}$. The additional terms, containing the
higher-resolution contributions, are produced by unsharp masking,
\begin{equation} 
\delta\mathcal{\{D|T\}}_{\{2|3\}}{\,=\,}{\mathcal{\{D|T\}}_{\{2|3\}}{\,-\,}
{\mathcal{G}_{\{3|4\}}{\,*\,}\mathcal{\{D|T\}}_{\{2|3\}}}},
\label{diffterms}
\end{equation} 
where $\mathcal{\{D|T\}}_{3}$ are computed by fitting the three images at $160$, $250$, and $350$\,${\mu}$m at the resolution
$O_{350}{\,=\,}24.9{\arcsec}$ , and $\mathcal{\{D|T\}}_{2}$ are obtained by fitting the two images at $160$ and $250$\,${\mu}$m at
the resolution $O_{250}{\,=\,}18.2{\arcsec}$; the Gaussian kernels ${\mathcal{G}_{\{3|4\}}}$ convolve the images to the next lower
resolutions $O_{\{350|500\}}$.

The following generalization of the above algorithm allows deriving surface densities and temperatures with any (arbitrarily
high) angular resolution existing among the observed $\mathcal{I}_{\!{\lambda}}$. The three independently derived maps of
temperatures $\mathcal{T}_{\{2|3|4\}}$ with the resolutions of $18.2{-}36.3{\arcsec}$ and six observed \emph{Herschel} images
with their native resolutions $O_{{\lambda}}$ of $8.4{-}36.3{\arcsec}$ define 18 surface densities,
\begin{equation} 
\mathcal{D}_{O_{\lambda}{\{2|3|4\}}} = \frac{\mathcal{I}_{\nu}}{B_{\nu}(\mathcal{T}_{\{2|3|4\}})\,\kappa_{\nu} 
\eta \mu m_{\rm H}},
\label{mosurfden}
\end{equation} 
with the assumptions and parameterizations of Eq.~(\ref{simformula}). It is required that the resolution of temperatures must not
be higher than $O_{\lambda}$, which excludes $\mathcal{D}_{O_{350}{2}}$ and $\mathcal{D}_{O_{500}{\{2|3\}}}$ from the algorithm and
provides 15 independently computed variants of the surface densities of the observed region, with different resolutions. The
high-resolution surface density image is computed as
\begin{equation} 
\mathcal{D}_{{O}_{{\rm H}}}\!= \mathcal{D}_{O_{500}}{\,+}\sum^{500}_{\lambda=\lambda_{\rm H}}
\max\left(\delta\mathcal{D}_{O_{\lambda}{2}},\delta\mathcal{D}_{O_{\lambda}{3}},\delta\mathcal{D}_{O_{\lambda}{4}}\right),
\label{superdens}
\end{equation} 
where $\lambda_{\rm H}$ denotes the wavelength of the image $\mathcal{I}_{\!{\lambda_{\rm H}}}$ with the desired angular resolution 
$O_{{\rm H}}{\,\equiv\,}O_{{\lambda}_{\rm H}}$ and the differential terms with higher-resolution information are obtained by the 
same unsharp masking,
\begin{equation} 
\delta\mathcal{D}_{O_{\lambda}{\{2|3|4\}}} = {\mathcal{D}_{O_{\lambda}{\{2|3|4\}}} -
{\mathcal{G}_{O_{\lambda+}}{*\,}\mathcal{D}_{O_{\lambda}{\{2|3|4\}}}}},
\label{superterms}
\end{equation} 
where $\mathcal{G}_{O_{\lambda+}}$\! is the Gaussian kernel (regarded as the delta function at $500$\,${\mu}$m), convolving
$\mathcal{D}_{O_{\lambda}{\{2|3|4\}}}$ to a lower resolution of the next longer wavelength. For images at
$\lambda{\,<\,}250$\,${\mu}$m, only the positive values of $\delta\mathcal{D}_{O_{\lambda}{\{2|3|4\}}}$ are used in
Eq.~(\ref{superdens}) to circumvent the problem of creating artificial depressions and negative pixels around strong peaks due to
the resolution mismatch ($O_{\lambda}{\,<\,}O_{250}$) between $\mathcal{I}_{\!\lambda}$ and the lower-resolution
$\mathcal{T}_{\{2|3|4\}}$ in Eq.~(\ref{mosurfden}).

The problem is caused by the sharp radial temperature gradients toward the unresolved centers of protostellar cores
\citep[cf.][]{Men'shchikov2016}. They are smeared out by the low resolutions of $18.2{-}36.3{\arcsec}$, hence the fitting of
$\Pi_{\lambda}$ leads to underestimated temperatures $\mathcal{T}_{\{2|3|4\}}$ and overestimated values (within an order of
magnitude) of peak surface densities at higher resolutions $O_{\lambda}{\,<\,}O_{250}$. This means that unsharp masking of the
overestimated peaks could create negative annuli in $\delta\mathcal{D}_{O_{\lambda}{\{2|3|4\}}}$ and negative pixels in
$\mathcal{D}_{{O}_{{\rm H}}}$. Fortunately, the surface densities are quite accurate outside the unresolved peaks
(Appendix~\ref{hiresinacc}).

A slight modification of Eq.~(\ref{superdens}) allows deriving the high-resolution surface densities with an enhanced contrast
of all unresolved or slightly resolved structures,
\begin{equation} 
\mathcal{D}^{+}_{{O}_{{\rm H}}}\!= \mathcal{D}_{{O}_{{\rm H}}}{\,+}\sum^{500}_{\lambda=\lambda_{\rm H}}\sum^{4}_{n=2}
\max\left(\delta\mathcal{D}_{O_{\lambda}{n}},0\right),
\label{superdenshc}
\end{equation} 
where the positive parts of the differential high-resolution terms from Eq.~(\ref{superterms}) are added to $\mathcal{D}_{{O}_{{\rm
H}}}$. These high-contrast images may be useful for detection of unresolved structures because the latter are usually diluted by
the observations with insufficient angular resolution; a higher contrast improves their visibility.

\begin{figure*}                                                               
\centering
\centerline{
  \resizebox{0.328\hsize}{!}{\includegraphics{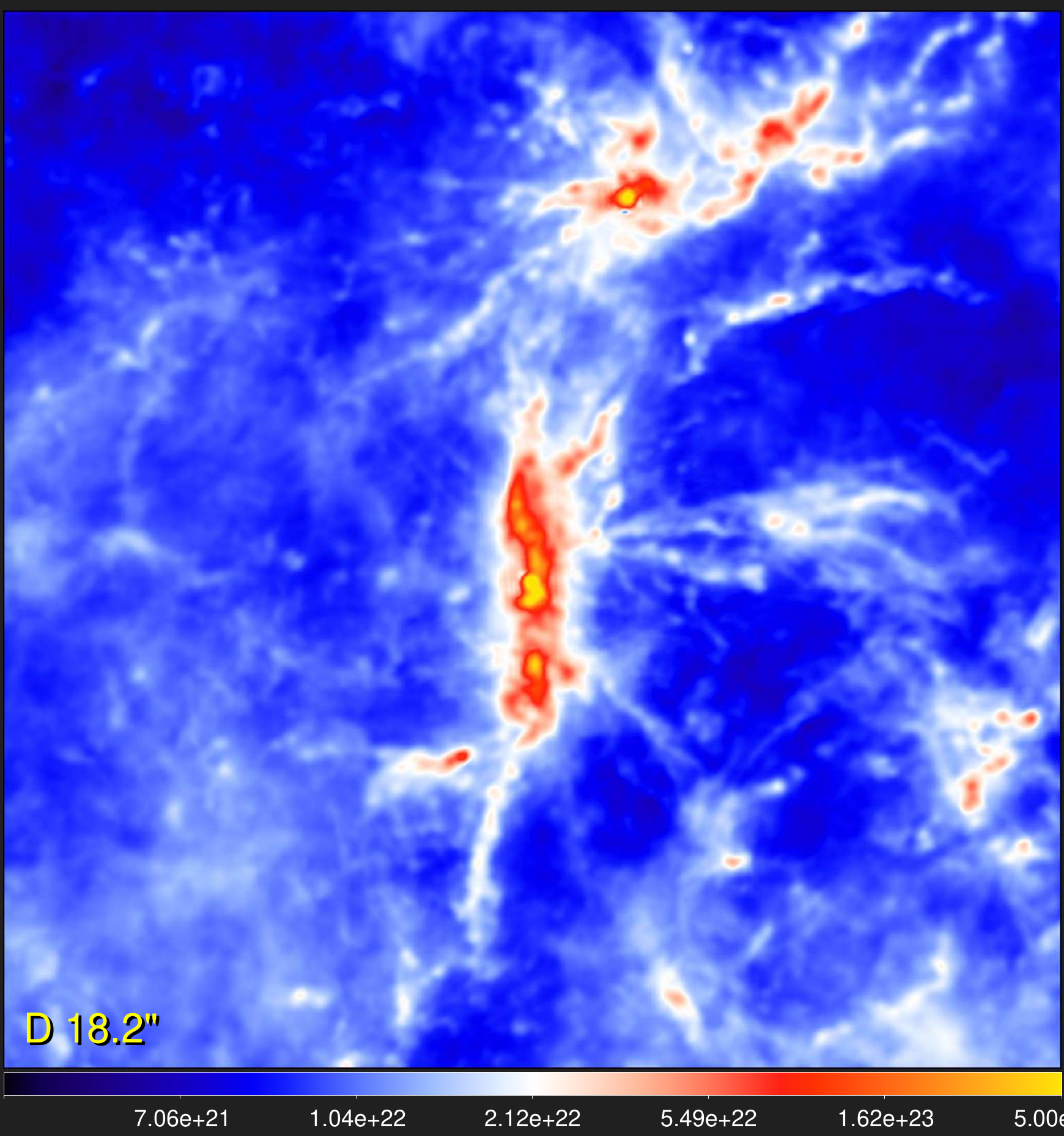}}
  \resizebox{0.328\hsize}{!}{\includegraphics{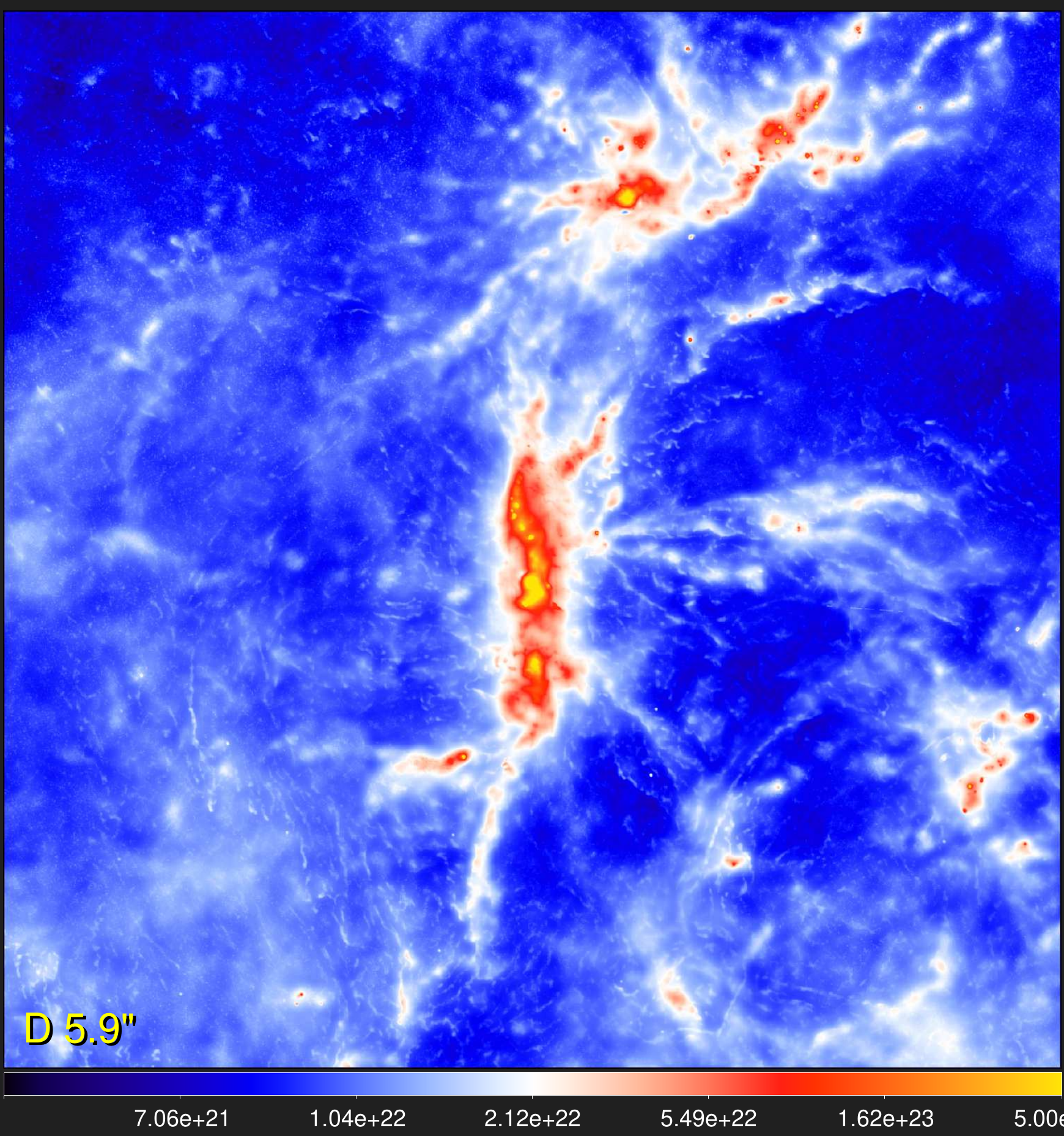}}
  \resizebox{0.328\hsize}{!}{\includegraphics{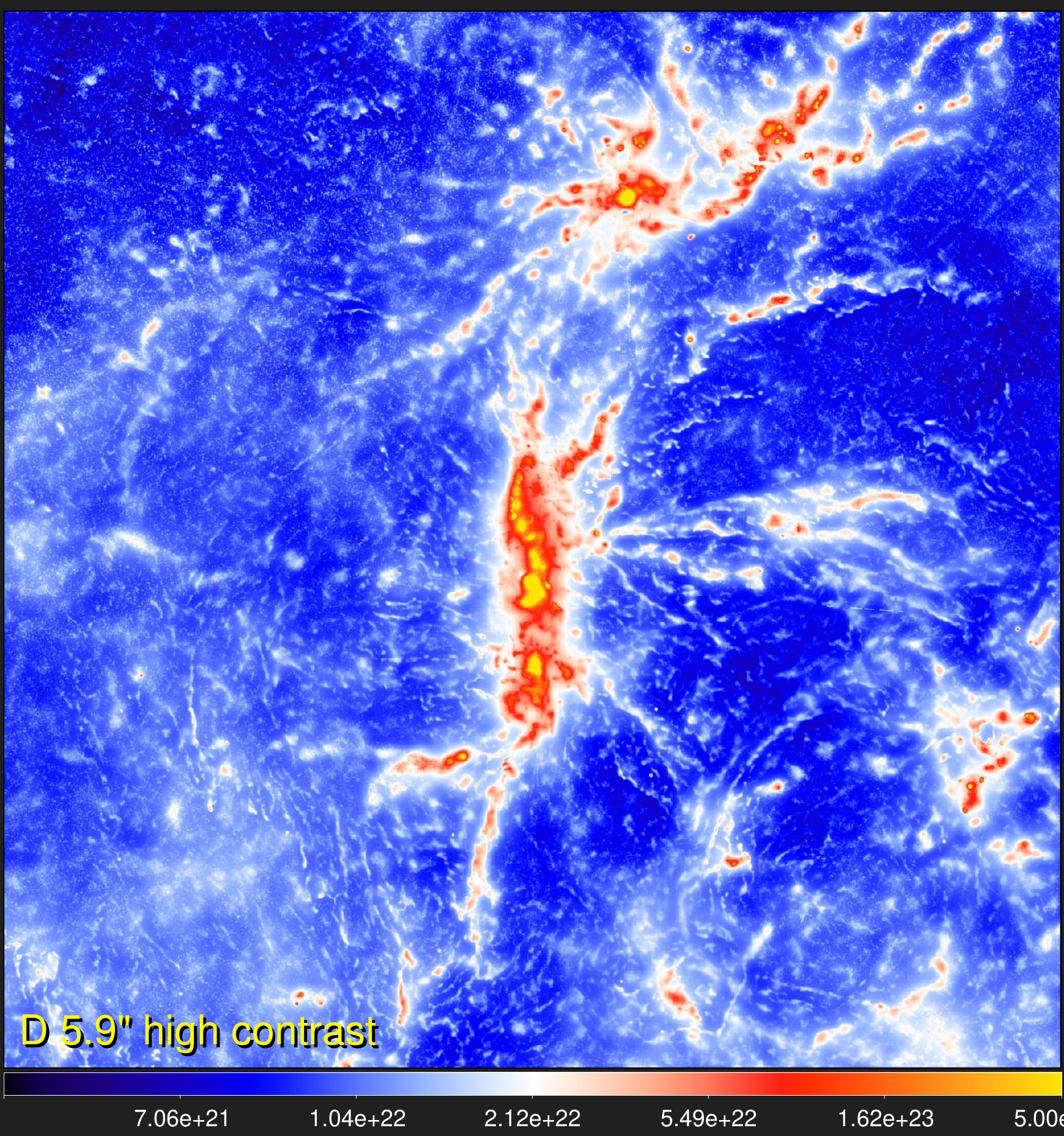}}}
\vspace{0.5mm}
\centerline{
  \resizebox{0.328\hsize}{!}{\includegraphics{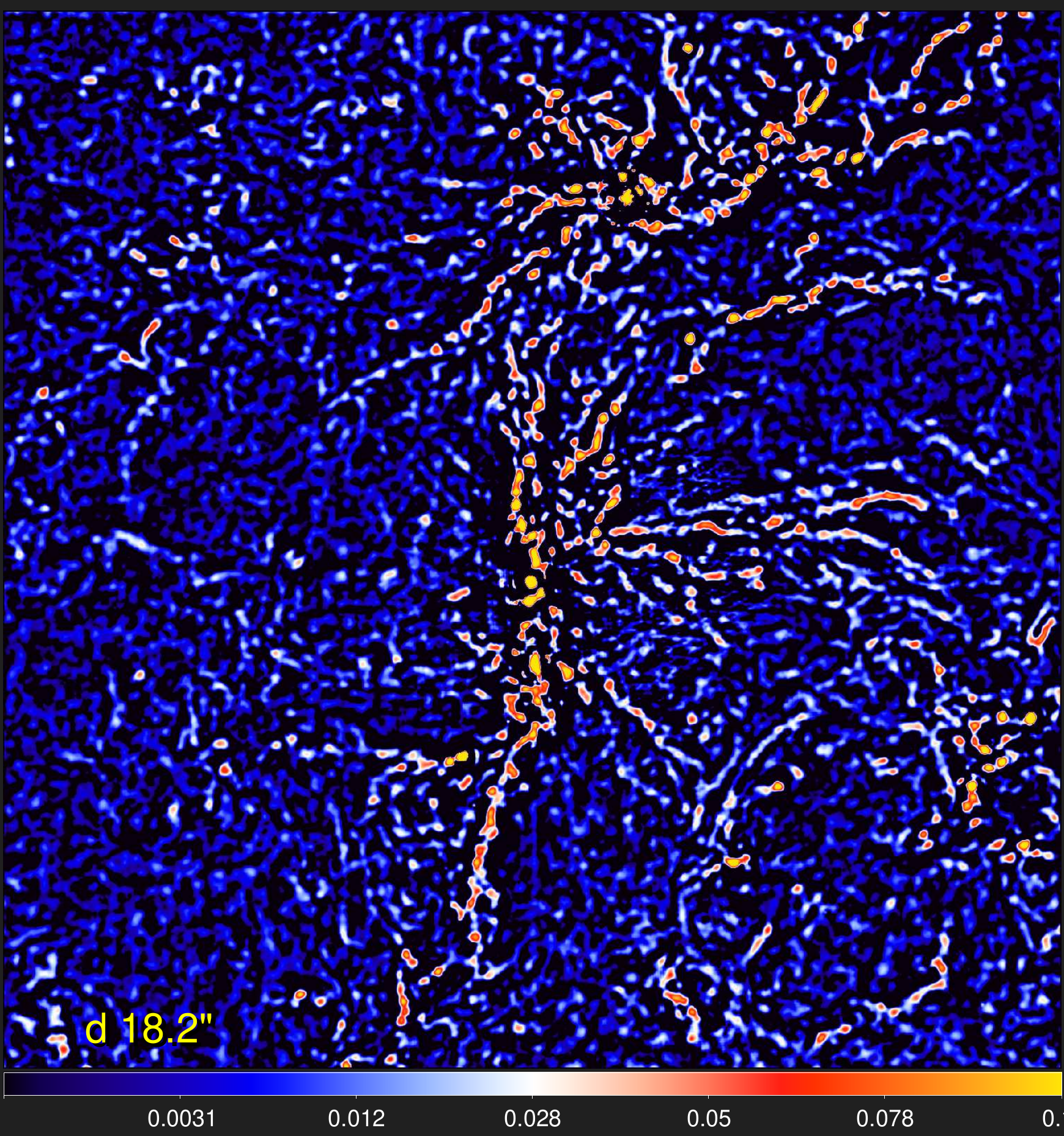}}
  \resizebox{0.328\hsize}{!}{\includegraphics{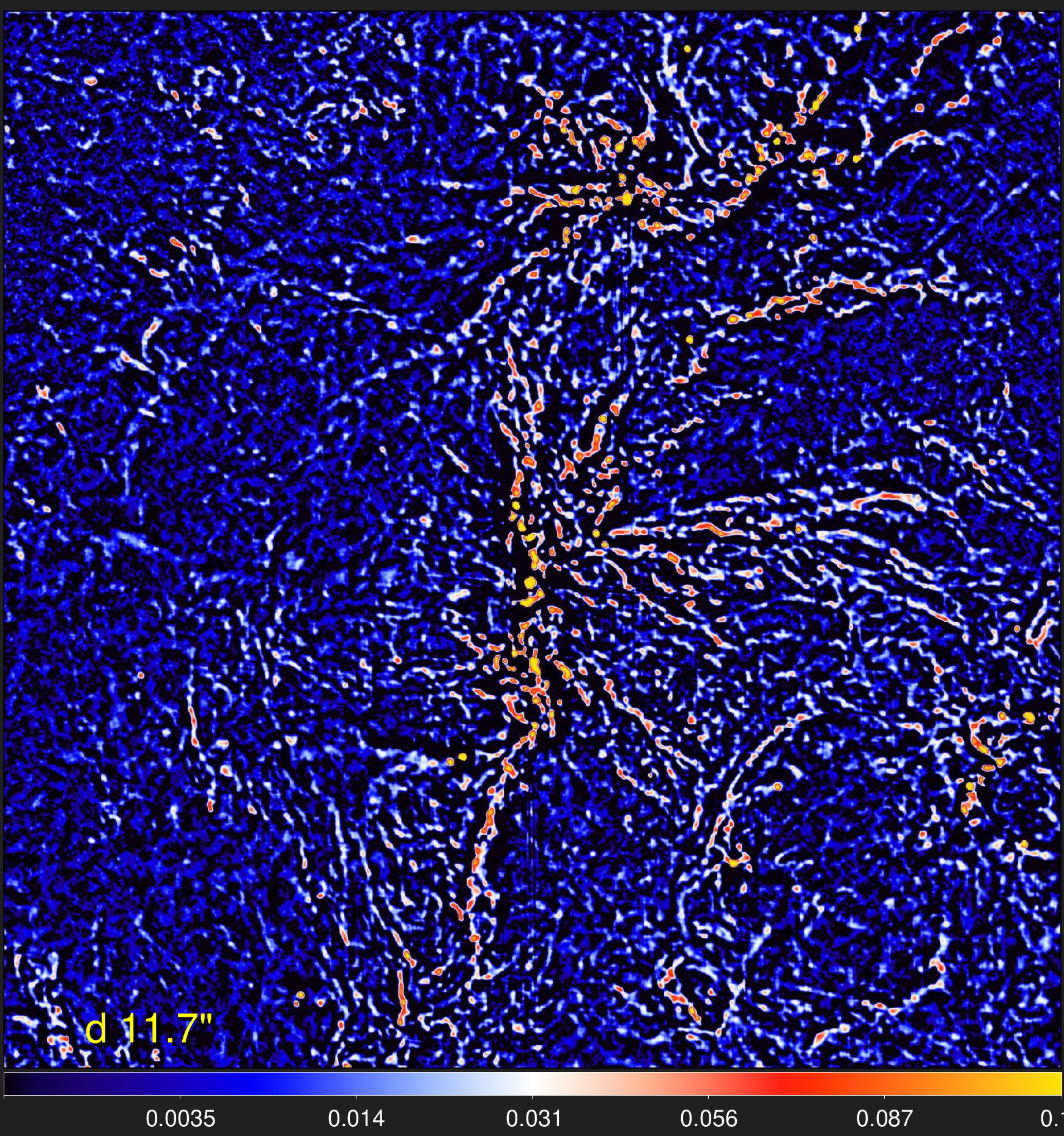}}
  \resizebox{0.328\hsize}{!}{\includegraphics{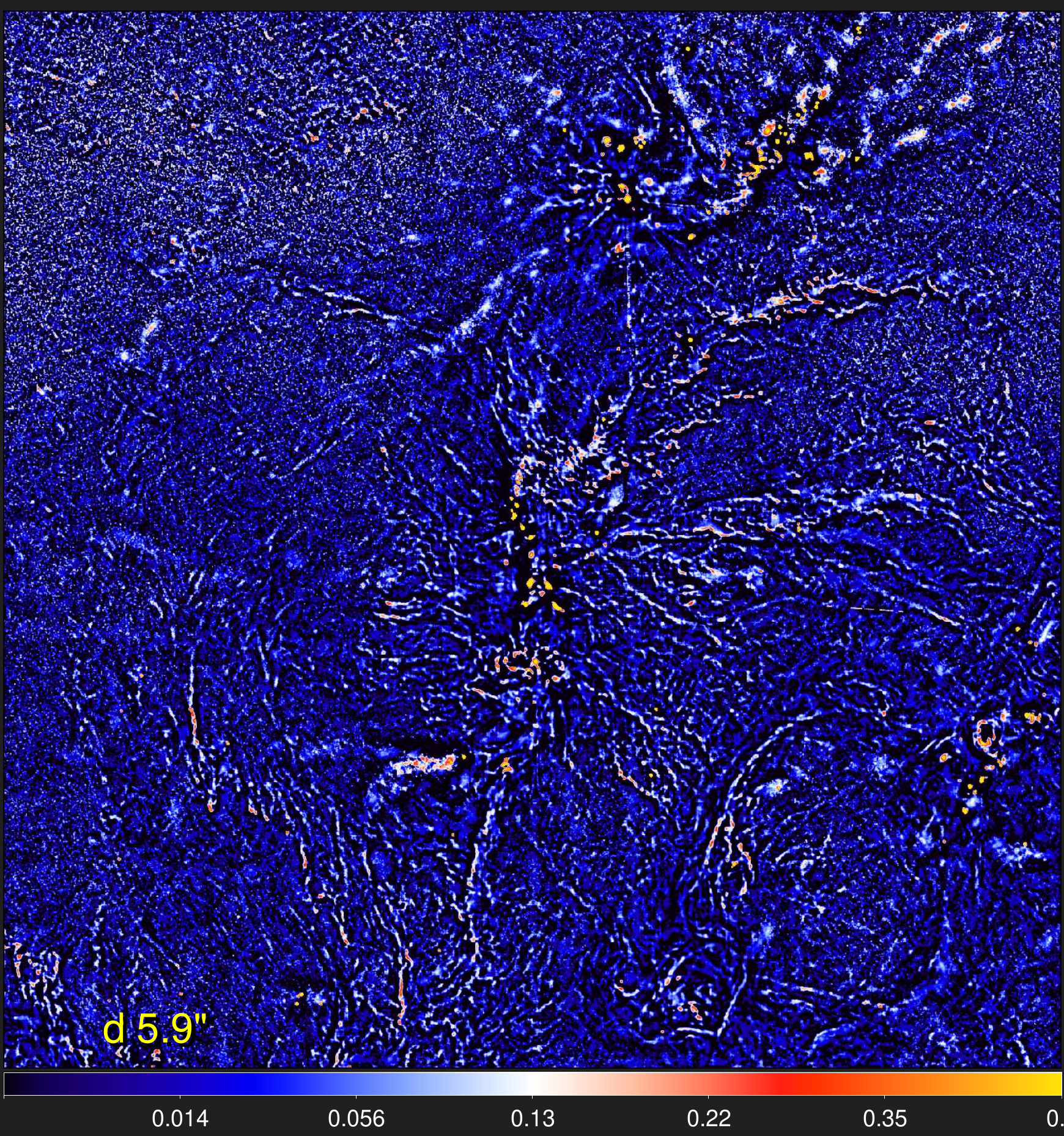}}}
\caption
{ 
High-resolution surface densities obtained for the \emph{Herschel} images of \object{Cygnus X} \citep[HOBYS project,][Bontemps et
al., in prep.]{Motte_etal2010,Hennemann_etal2012}. The \emph{top} row shows the \textsl{hires} surface densities
$\mathcal{D}_{{O}_{{\rm H}}}$ from Eq.~(\ref{superdens}) with $O_{\rm H}{\,=\,}18.2{\arcsec}$ and $5.9{\arcsec}$ resolutions, and
the high-contrast ${{\mathcal{D}_{{O}_{\rm H}}}^{\!\!\!\!\!\!\!\!\!+}}$\,\,\, from Eq.~(\ref{superdenshc}) with $O_{\rm
H}{\,=\,}5.9{\arcsec}$. The \emph{bottom} row displays the relative differences of $\mathcal{D}_{18{\arcsec}}$,
$\mathcal{D}_{12{\arcsec}}$, and $\mathcal{D}_{6{\arcsec}}$ with respect to the next lower-resolution surface densities
$\mathcal{D}_{25{\arcsec}}$, $\mathcal{D}_{18{\arcsec}}$, and $\mathcal{D}_{12{\arcsec}}$, respectively. Logarithmic and
square-root color mapping in the \emph{top} and \emph{bottom} rows, correspondingly.
} 
\label{cygxdr21}
\end{figure*}

A high-resolution temperature $\mathcal{T}_{\!{O}_{{\rm H}}}$, consistent with the high-resolution surface density
$\mathcal{D}_{{O}_{{\rm H}}}$, is computed by numerically inverting the Planck function,
\begin{equation} 
\mathcal{T}_{\!{O}_{{\rm H}}}\!= B^{-1}_{{\nu}_{\rm H}}
\left(\frac{\mathcal{I}_{{\nu}_{\rm H}}}{\mathcal{D}_{{O}_{{\rm H}}} \kappa_{{\nu}_{\rm H}} \eta \mu m_{\rm H}}\right),
\label{supertemp}
\end{equation} 
with $\nu_{\rm H}{\,=\,}c \lambda^{-1}_{\rm H}$, where $c$ is the speed of light. The high-resolution images
$\mathcal{\{D|T\}}_{13{\arcsec}}$ are shown in Fig.~\ref{addhires} along with the true simulated $\mathcal{D}_{\rm
C}{\,+\,}\mathcal{D}_{\rm S}$ (Sect.~\ref{simcomplete}). A comparison demonstrates that the pixel-fitting procedure reduces
visibility of many unresolved or slightly resolved starless cores, which is the manifestation of the invalid assumption of the
uniform line-of-sight temperatures. Starless (prestellar) cores have lower temperatures in their centers, and their smearing by an
insufficient resolution leads to overestimated temperatures that suppress the surface density peaks (cf. Fig.~\ref{hiresacc}).

The new \textsl{hires} algorithm, outlined by Eqs.~(\ref{mosurfden})--(\ref{supertemp}), brings the benefits of a resolution
${O}_{{\rm H}}{\,\approx\,}8{\arcsec}$, twice better than $O_{\rm P}$ and four times better than $O_{500}$, if the image quality at
the shortest wavelengths permits this. Moreover, the angular resolutions of the \emph{Herschel} images at $70$, $100$, and
$160$\,${\mu}$m, obtained with a slow scanning speed of $20{\arcsec}$s$^{-1}$, are even higher: $6$, $7$, and $11{\arcsec}$,
respectively. These observations, illustrated in Fig.~\ref{cygxdr21}, allow deriving the surface densities and temperatures
with ${O}_{{\rm H}}{\,\approx\,}6{\arcsec}$, a three times better resolution than when using Eq.~(\ref{hiresdentem}). If the
$70$\,${\mu}$m image is too noisy or there is evidence of its strong contamination by emission unrelated to that of the adopted
dust grains (e.g., polycyclic aromatic hydrocarbons or transiently heated very small dust grains), then the derived images may
still have the $7{-}13{\arcsec}$ resolution of the $100$ or $160$\,${\mu}$m wavebands. In addition to the high resolution, the images from
Eq.~(\ref{superdens}) also have a better quality than those from Eq.~(\ref{hiresdentem}) because they accumulate all
available high-resolution information from the (up to 15) independently computed images $\mathcal{D}_{O_{\lambda}{\{2|3|4\}}}$ that
use all three temperatures $\mathcal{T}_{\{2|3|4\}}$ with each original $\mathcal{I}_{\!{\lambda}}$.

The \textsl{hires} algorithm works with any number $2{\,\le\,}N{\,\le\,}6$ of \emph{Herschel} wavebands. If the $160\,{\mu}$m image
is unavailable or disabled, then the temperature $\mathcal{T}_{2}$ at the resolution $O_{250}$ is removed from
Eq.~(\ref{mosurfden}) and $\mathcal{T}_{\{3|4\}}$ at the resolutions $O_{\{350|500\}}$ are obtained from fitting of only the $250$,
$350$, and $500\,{\mu}$m images. If the $250\,{\mu}$m image is also unavailable or disabled, then only the single temperature
$\mathcal{T}_{4}$ at the lowest resolution $O_{500}$ remains in Eq.~(\ref{mosurfden}), obtained from the $350$ and $500\,{\mu}$m
images. Although the algorithm is unaffected by the changes, the reduction in the number of independently derived temperatures
would lower the angular resolution and accuracy of the resulting surface density image.
The improved algorithm can use the realistic, non-Gaussian point-spread functions (PSF) published by \cite{Aniano_etal2011}.
However, the surface densities are largely determined by the SPIRE bands with nearly Gaussian PSFs, whereas only the PACS
160\,$\mu$m band is used in the pixel fitting. The benchmark tests have shown that effects of the realistic PSFs on surface
densities are very small, at percent levels, much smaller than the general uncertainties of the pixel-fitting methods
(Appendix~\ref{hiresinacc}). It is therefore sufficient to use the Gaussian PSFs when surface densities are derived.

For some studies, it might be useful to have all images at the same wavelength-independent angular resolution. With the
high-resolution surface densities and temperatures from Eqs.~(\ref{superdens}) and (\ref{supertemp}), it is straightforward to
obtain such images:
\begin{equation} 
\mathcal{J}_{{\nu}{O}_{{\rm H}}} = 
B_{\nu}(\mathcal{T}_{\!{O}_{{\rm H}}})\,\mathcal{D}_{{O}_{{\rm H}}} \kappa_{\nu} \eta \mu m_{\rm H}, 
\label{hiresinten}
\end{equation} 
with the assumptions and parameterizations of Eq.~(\ref{simformula}). For example, the intensities
$\mathcal{J}_{\!{\lambda}{13\arcsec}}$ at $250$, $350$, and $500$\,${\mu}$m would be sharper than $\mathcal{I}_{\!\lambda}$ by the
factors $1.3$, $1.8$, and $2.7$, respectively.

When the available original set of images $\mathcal{I}_{\!\lambda}$ allows creation of $\mathcal{D}_{{O}_{{\rm H}}}$, it is
advantageous to have it complement the original data set, handling it as an image $\mathcal{I}_{{\!\lambdabar}}$ ``observed'' in a
fictitious waveband ${\lambdabar}$. In the multiwavelength extractions with \textsl{getsf}, it may be recommended to use
$\mathcal{D}_{{O}_{{\rm H}}}$ for better detections and deblending of dense structures. The surface densities are not accurate
enough for source measurements, as demonstrated in Appendix~\ref{hiresinacc} and \cite{Men'shchikov2016}.

The following presentation and discussion of \textsl{getsf} implicitly assumes that the additional detection images are contained
in the set of images $\mathcal{I}_{{\!\lambda}}$. In other words, all supplementary wavebands are included in the set of $\lambda$
prepared for extraction. The latter was done for a multiwavelength data set that included all images in the \emph{Herschel}
wavebands (Fig.~\ref{simimages}) and the high-resolution surface density $\mathcal{I}_{{\!\lambdabar}} \equiv
\mathcal{D}_{13{\arcsec}}$ (Fig.~\ref{addhires}), a total of $N_{\rm W}{\,=\,}7$ wavelengths.

\begin{figure*}                                                               
\centering
\centerline{
  \resizebox{0.328\hsize}{!}{\includegraphics{hi.surface.density.r13p5.pdf}}
  \resizebox{0.328\hsize}{!}{\includegraphics{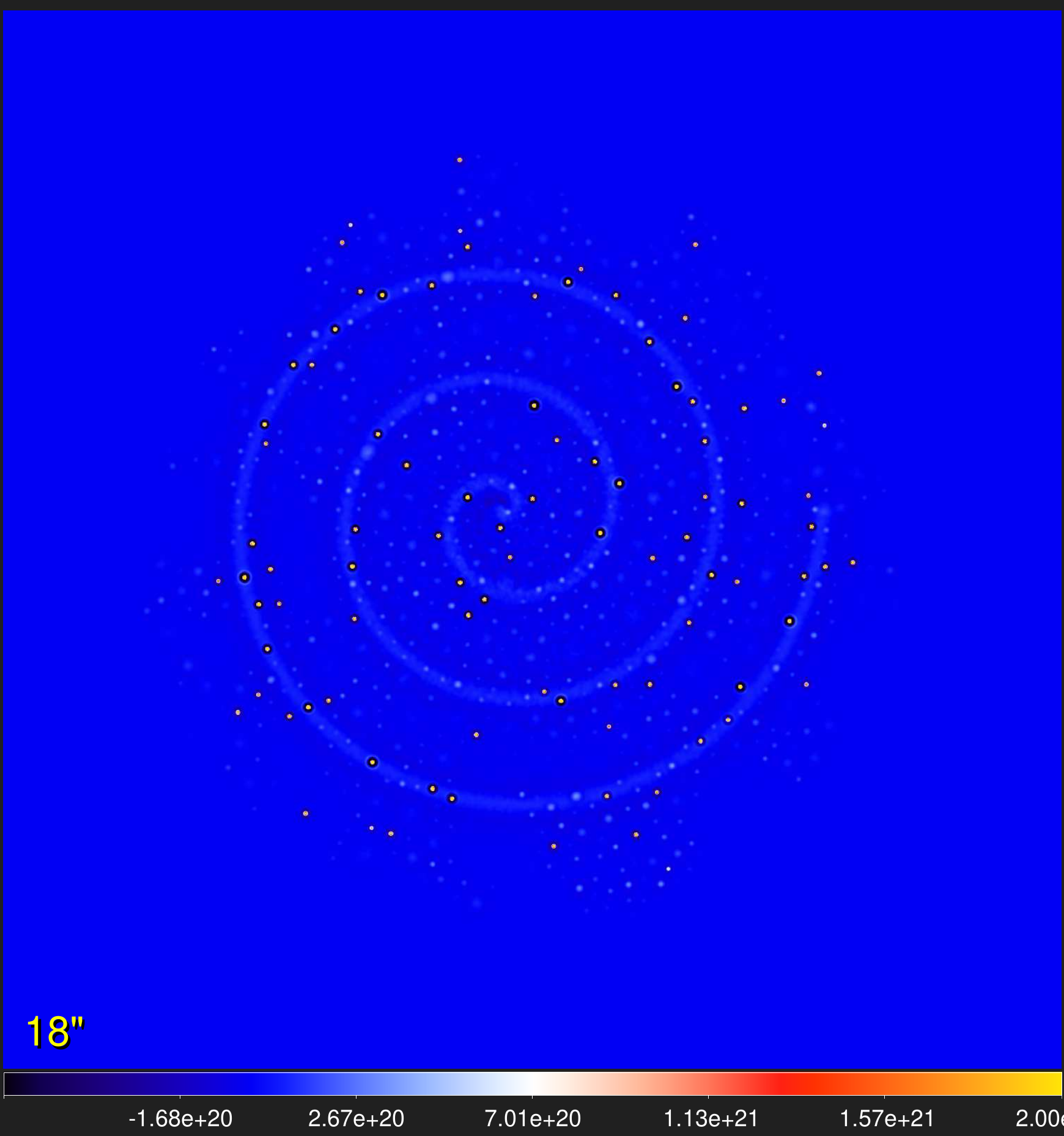}}
  \resizebox{0.328\hsize}{!}{\includegraphics{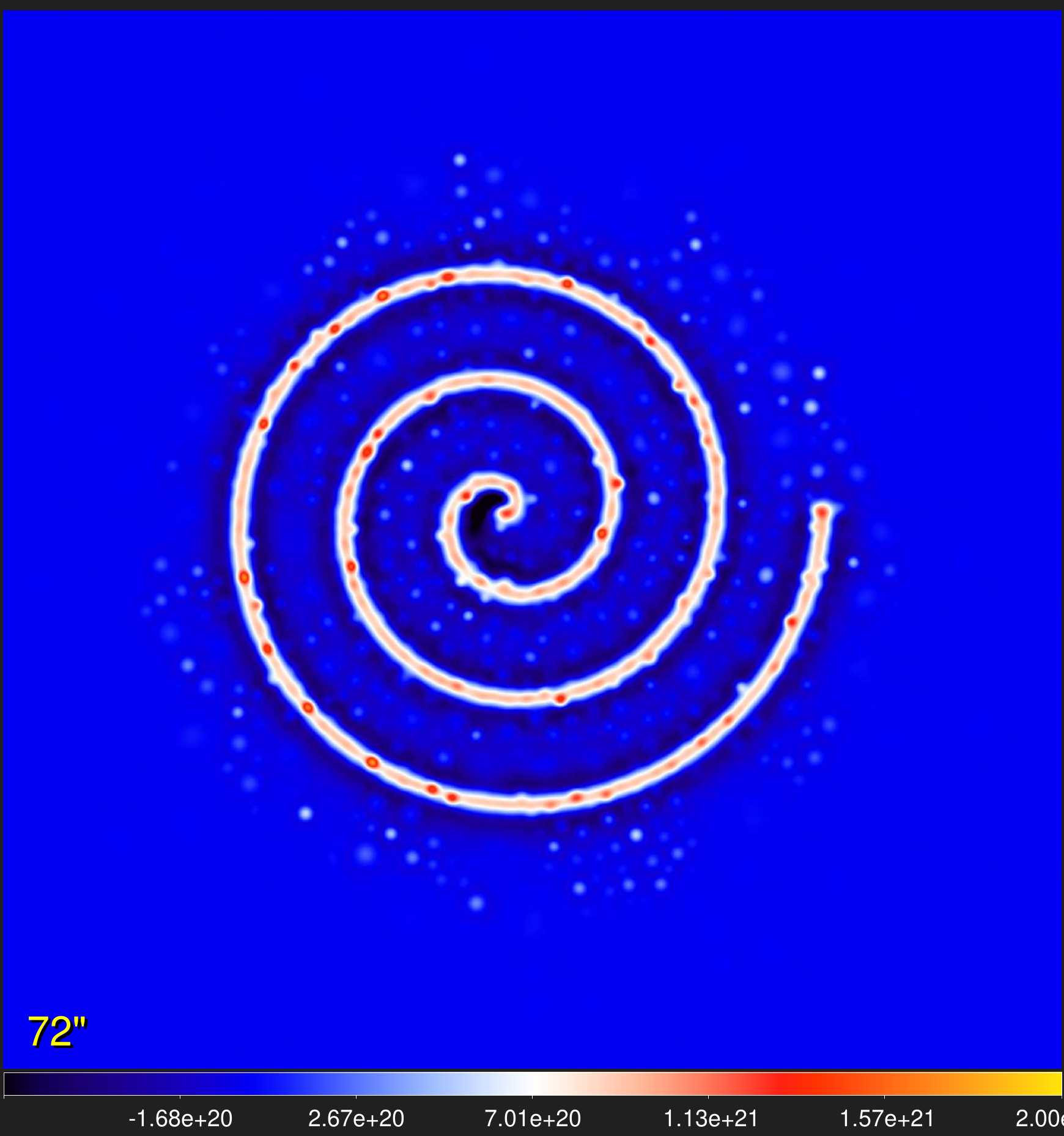}}}
\vspace{0.5mm}
\centerline{
  \resizebox{0.328\hsize}{!}{\includegraphics{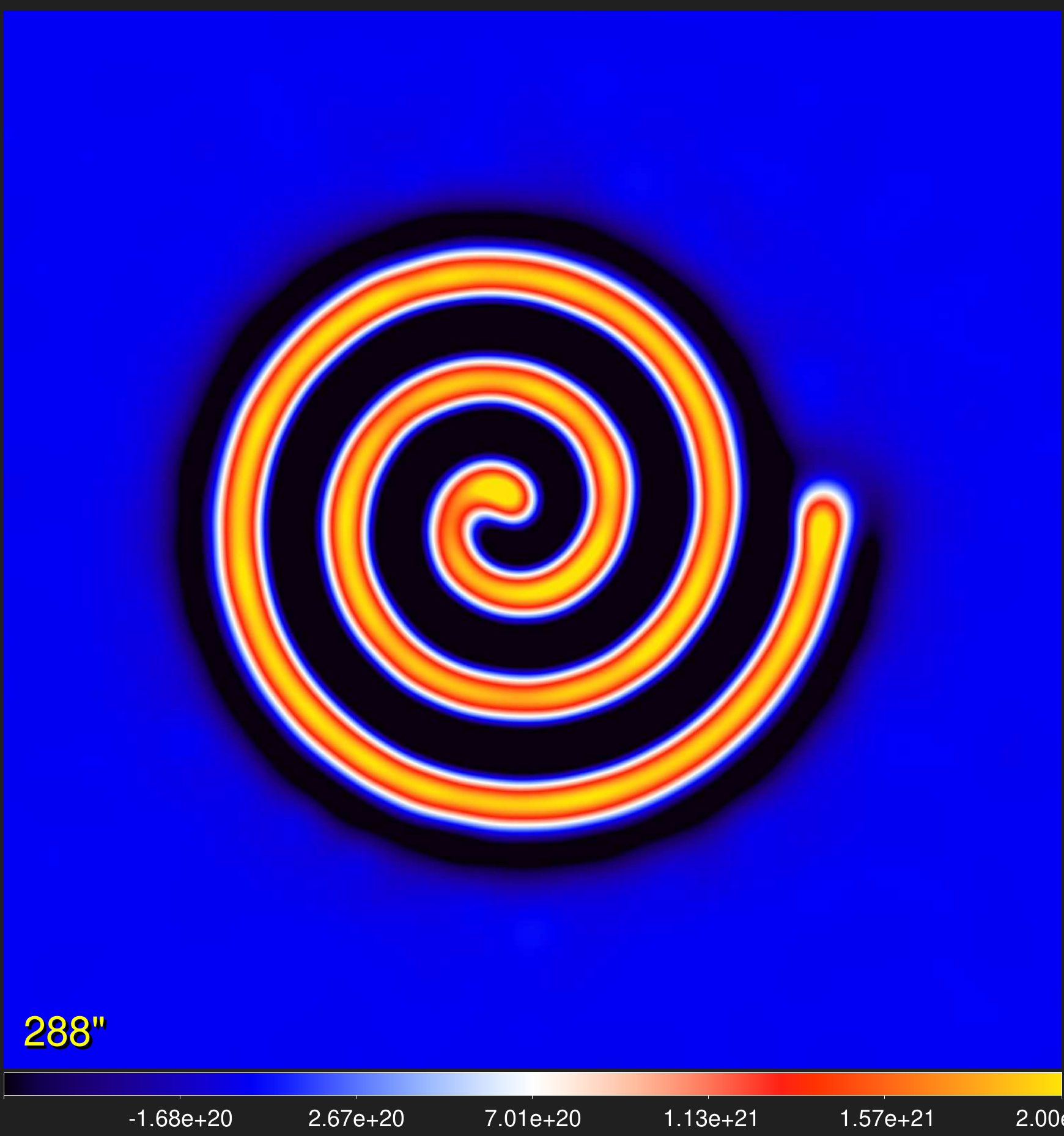}}
  \resizebox{0.328\hsize}{!}{\includegraphics{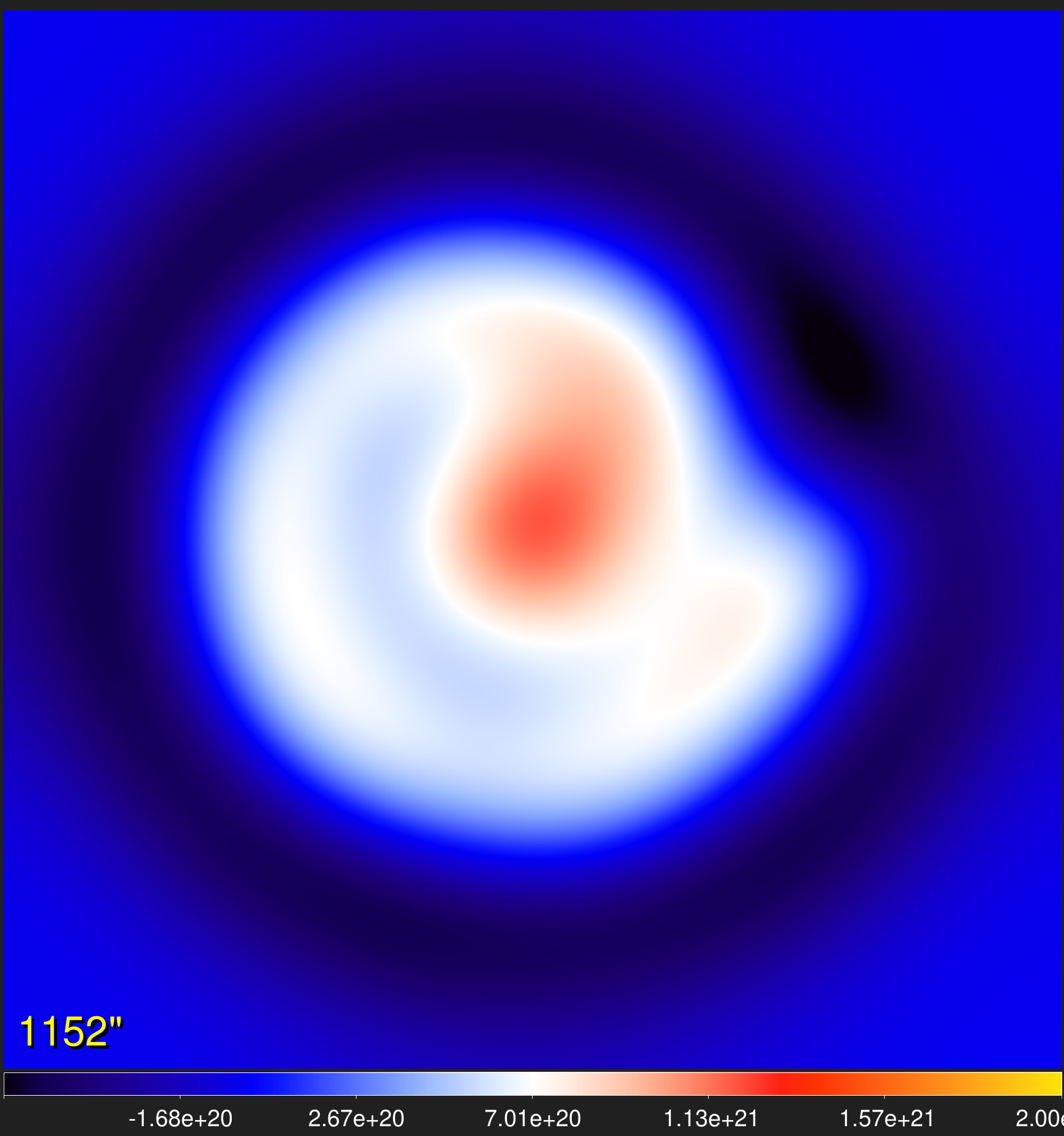}}
  \resizebox{0.328\hsize}{!}{\includegraphics{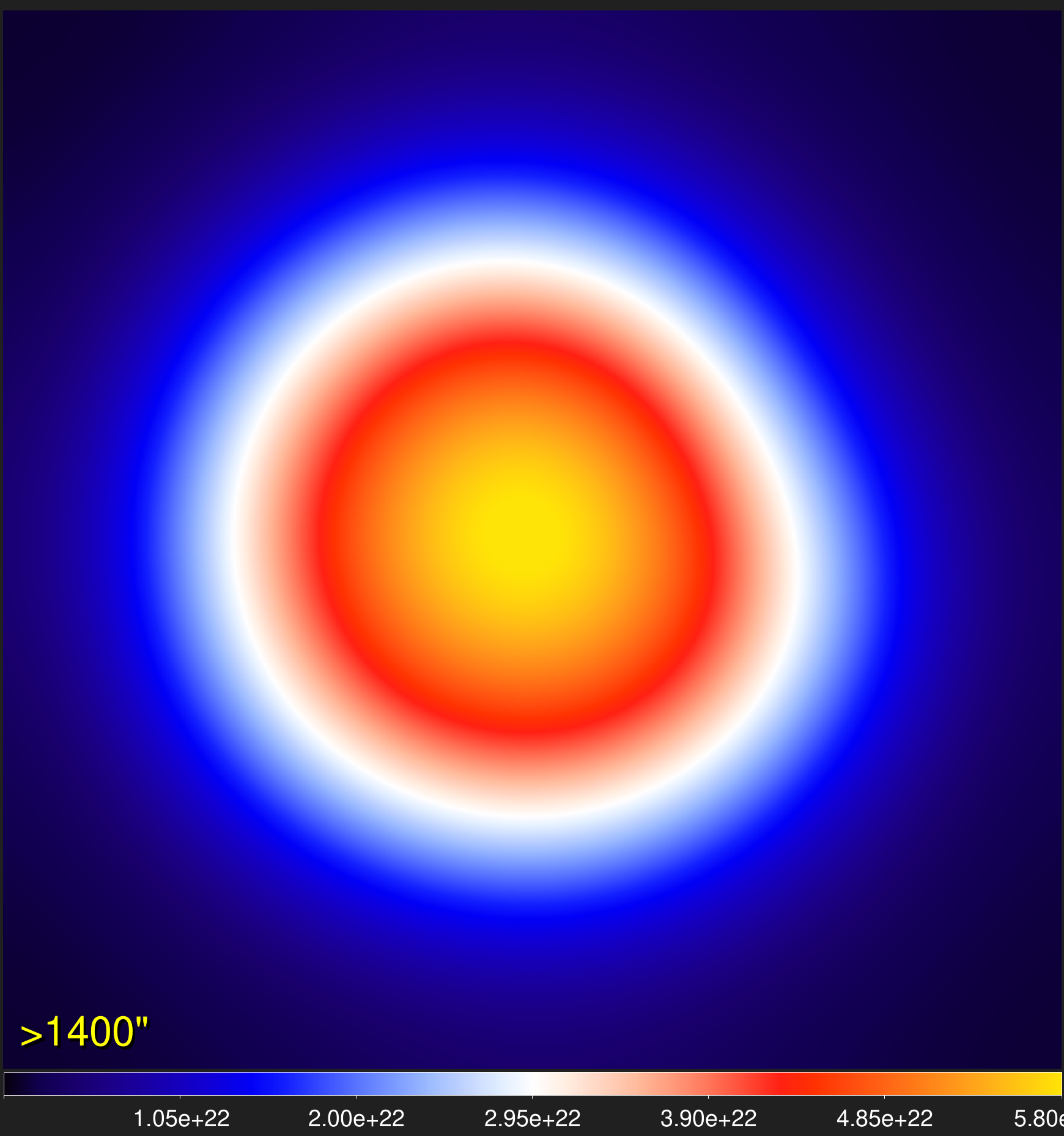}}}
\caption
{ 
Spatial decomposition (Sect.~\ref{decompos}, Appendix~\ref{decomposition}) for $\mathcal{I}_{{\!\lambdabar}} \equiv
\mathcal{D}_{13{\arcsec}}$ from Eq.~(\ref{superdens}) in single scales between $4$ and $1400${\arcsec}. The original
\textsl{hires} surface density (\emph{top left}) and decomposed $\mathcal{I}_{{\!\lambdabar}{j}}$ on selected scales $S_{\!j}$  that
differ by a factor of 4 are plotted. The remaining largest scales $\mathcal{G}_{\!N_{\rm S}}{*\,}\mathcal{I}_{{\!\lambdabar}}$
(\emph{bottom right}) are outside the decomposition range. Linear color mapping.
} 
\label{singlescales}
\end{figure*}


\subsubsection{Practical definition of maximum size}
\label{maxsizes}

Before starting any extraction with \textsl{getsf}, it is necessary to formulate the aim of the study and determine what structures
of interest are to be extracted. The method knows and is able to separate three types of structures: sources, filaments, and
backgrounds. To separate the structural components with \textsl{getsf}, the maximum size $\{X|Y\}_{\lambda}$ of the sources
($X_{\lambda}$) and filaments ($Y_{\lambda}$) of interest needs to be manually (visually) estimated from the prepared
$\mathcal{I}_{{\!\lambda}}$ independently for each waveband, which can be accomplished by opening an image in \textsl{ds9} and
placing a circular region fully covering the width of the largest structure. The maximum size of structures is the single physical
parameter that the method needs to know for each observed image. Being a function of the type of structures (sources, filaments)
and the waveband $\lambda$, it is split into $X_{\lambda}$ and $Y_{\lambda}$ in this paper for convenience.

The maximum size $\{X|Y\}_{\lambda}$ is defined as the footprint radius (in arcsec) of the largest source and the widest filament
to be extracted. A footprint size has the meaning of a full width at zero (background) level: the largest two-sided extent from a
source peak or filament crest at which this structure is still visible in $\mathcal{I}_{{\!\lambda}}$ against its background. For a
Gaussian intensity distribution, the footprint radius is slightly larger than the half-maximum width $H_{\lambda}$ of a structure.
For a power-law intensity profile, the footprint radius may become much larger than $H_{\lambda}$. If the widest filaments of
interest are blended (overlapping each other with their footprints), $Y_{\lambda}$ must be increased accordingly to approximate the
full extent of the blend. In contrast, it is not necessary to adjust $X_{\lambda}$ for blended sources because their final
background will be determined from their footprints at the measurement step (Sect.~\ref{srcmeasurement}).

It is not necessary (also not possible) to evaluate the maximum size parameter $\{X|Y\}_{\lambda}$ very precisely, a $50${\%}
accuracy is quite sufficient. Its purpose is to set a reasonable limit to the spatial scales when separating the structural
components, and to the size of the structures to be measured and cataloged. The method works with spatially decomposed images, and
it needs to know the maximum scale. It makes no sense to perform the decomposition up to a very large scale if the extraction is
aimed at much smaller sources or narrower filaments. The method has no limitations with respect to the sizes (widths) of the
structures to extract. However, it is important to avoid detecting, measuring, and cataloging the peaks that are unnecessarily too
wide because they would likely overlap with other sources of interest, which potentially would make their measurements less
accurate.

To extract all structures in the benchmark images presented in Sect.~\ref{skybench}, the estimated $X_{\lambda}$ values for sources
are $16$, $25$, $30$, $150$, $150$, and $150${\arcsec}, whereas the estimated $Y_{\lambda}$ values for the filament are
$350${\arcsec} in all six \emph{Herschel} wavebands (Fig.~\ref{simimages}); in the additional surface density image
$\mathcal{I}_{{\!\lambdabar}} \equiv \mathcal{D}_{13{\arcsec}}$ (Fig.~\ref{addhires}), the $\{X|Y\}_{\lambdabar}$ values are the
same as those for the $250{-}500$\,{${\mu}$m} images.


\subsection{Backgrounds of the structural components}
\label{deriveback}

Complex observed images may be radically simplified by subtracting backgrounds on spatial scales much larger than the maximum size
$\{X|Y\}_{\lambda}$. The independent largest sizes for sources and filaments effectively define two different backgrounds for the
two scales. The $X_{\lambda}$-scale background $\mathcal{B}_{{\lambda}X}$ is derived to separate the component of sources
$\mathcal{S}_{{\lambda}}$, whereas the $Y_{\lambda}$-scale background $\mathcal{B}_{{\lambda}{Y}}$ is obtained to separate the
component of filaments $\mathcal{F}_{{\lambda}}$. The backgrounds are collectively referred to as
$\mathcal{B}_{{\lambda}\{{X}|{Y}\}}$.

The true background under the observed structures is fundamentally unknown, and it is a major source of large uncertainties and
measurement inaccuracies, especially for the faintest structures. In practice, the backgrounds $\mathcal{B}_{{\lambda}\{{X}|{Y}\}}$
are defined in \textsl{getsf} as the smooth intensity distributions on spatial scales $S_{\!j}$ larger than $4\{X|Y\}_{\lambda}$
that remain in $\mathcal{I}_{{\!\lambda}}$ after a complete removal of all sources or filaments with the maximum size of
$\{X|Y\}_{\lambda}$. In contrast to the background derivation by median filtering (\textsl{getimages}, Paper III), which may become
extremely slow for very large images and wide structures, \textsl{getsf} employs a more direct, precise, and effective clipping
algorithm to separate the structures.


\subsubsection{Decomposition of the original images}
\label{decompos}

In general, observed images are very complex blends of various structural components on different spatial scales, and
great advantages are obtained when a spatial decomposition is used to simplify the images (cf. Papers I and II). Following the \textsl{getold}
approach, \textsl{getsf} employs successive unsharp masking (Appendix~\ref{decomposition}) to decompose the original images
$\mathcal{I}_{{\!\lambda}}$ into a set of single-scale images $\mathcal{I}_{{\!\lambda}{j}}$ (Fig.~\ref{singlescales}). It also
uses an iterative algorithm (Appendix~\ref{decomposition}) to determine a single-scale standard deviation
$\sigma_{{\!\lambda}{j}}$, as well as its total value $\sigma_{{\!\lambda}}$, which are used to separate the structural components
present in $\mathcal{I}_{{\!\lambda}}$.


\subsubsection{Separation of the structural components}
\label{clipping}

The backgrounds $\mathcal{B}_{{\lambda}\{{X}|{Y}\}}$ are computed by cutting small round peaks and elongated structures off the
decomposed images $\mathcal{I}_{{\!\lambda}{j}}$ and recovering the full images using Eq.~(\ref{recovered}). It is important to
note that the appearance of the structures in the decomposed images depends on both the spatial scale and intensity level.

To remove the structural components, \textsl{getsf} slices $\mathcal{I}_{{\!\lambda}{j}}$ by a number $N_{\rm L}$ of intensity
levels $I_{{\lambda}{j}{l}}$, spaced by ${\delta\ln{I_{{\lambda}{j}}}{\,=\,}0.05}$ from the image maximum down to
$\sigma_{{\!\lambda}{j}}$ for sources and to $0.3\sigma_{{\!\lambda}{j}}$ for filaments. Each slice $l$ cuts through all the
structures present in $\mathcal{I}_{{\!\lambda}{j}}$ on that intensity level, producing various shapes of connected pixels,
\begin{equation} 
\mathcal{I}_{{\!\lambda}{j}{l}} = \min\left(\max\left(\mathcal{I}_{{\!\lambda}{j}}, I_{{\lambda}{j}{l}}\right), 
I_{{\lambda}{j}{l}}\right), \,\,\, {l{\,=\,}1,2,\dots, N_{\rm L}}.
\label{levels} 
\end{equation} 

Relatively round source-like peaks in $\mathcal{I}_{{\!\lambda}{j}}$ may be effectively distinguished from elongated structures by
the number of connected pixels $N_{{\lambda}{j}{l}}$ that their shapes occupy in the slice $\mathcal{I}_{{\!\lambda}{j}{l}}$ (cf. Papers
I and II). The single-scale images indeed most clearly show the structures with matching sizes
(${H_{\lambda}{\,\approx\,}S_{\!j}}$), whereas the signals from much narrower and much wider structures are suppressed. As a
consequence, the source-like shapes occupy relatively small areas of connected pixels in ${\mathcal{I}}_{{\!\lambda}{j}{l}}$ that
are comparable to the area ${{\pi}{S_{\!j}}^{\!\!2}}$ of the convolution kernel $\mathcal{G}_{\!j}$. In contrast to the round
peaks, elongated shapes in $\mathcal{I}_{{\!\lambda}{j}{l}}$ have greater lengths $L_{\lambda}$  than widths
${W_{\lambda}{\,\approx\,}S_{\!j}}$, which means that the filamentary shapes in slices ${\mathcal{I}}_{{\!\lambda}{j}{l}}$ extend
over much larger areas than ${{\pi}{S_{\!j}}^{\!\!2}}$.

In addition to $N_{{\lambda}{j}{l}}$, \textsl{getsf} uses two more quantities to discriminate between sources and filaments: elongation
$E_{{\lambda}{j}{l}}$ and sparsity $S_{{\!\lambda}{j}{l}}$. They are defined by the major and minor sizes ($a_{{\lambda}{j}{l}}$
and $b_{{\lambda}{j}{l}}$) of each cluster of connected pixels, obtained from intensity moments (cf. Appendix F in Paper I),
\begin{equation} 
{E_{{\lambda}{j}{l}} \equiv \frac{a_{{\lambda}{j}{l}}}{b_{{\lambda}{j}{l}}}}, \,\,\,\, 
{S_{{\!\lambda}{j}{l}} \equiv \frac{\pi a_{{\lambda}{j}{l}} b_{{\lambda}{j}{l}}} {N_{{\lambda}{j}{l}}\,\Delta}},
\label{elongspars} 
\end{equation} 
where $\Delta$ is the pixel size. Only simple and relatively straight filamentary shapes can be identified in
$\mathcal{I}_{{\!\lambda}{j}{l}}$ by their elongation. Most of the actually observed filaments in space are shaped quite
irregularly on different scales and intensity levels. The elongation $E_{{\lambda}{j}{l}}$ alone cannot be used to quantify
strongly curved, not very dense clusters of connected pixels that meander around (e.g., a spiral structure). Although
$E_{{\lambda}{j}{l}}$ may well be close to unity for sparse shapes, high values of $S_{{\!\lambda}{j}{l}}$ for these structures
would indicate that they do not belong to sources.

\begin{figure*}                                                               
\centering
\centerline{
  \resizebox{0.328\hsize}{!}{\includegraphics{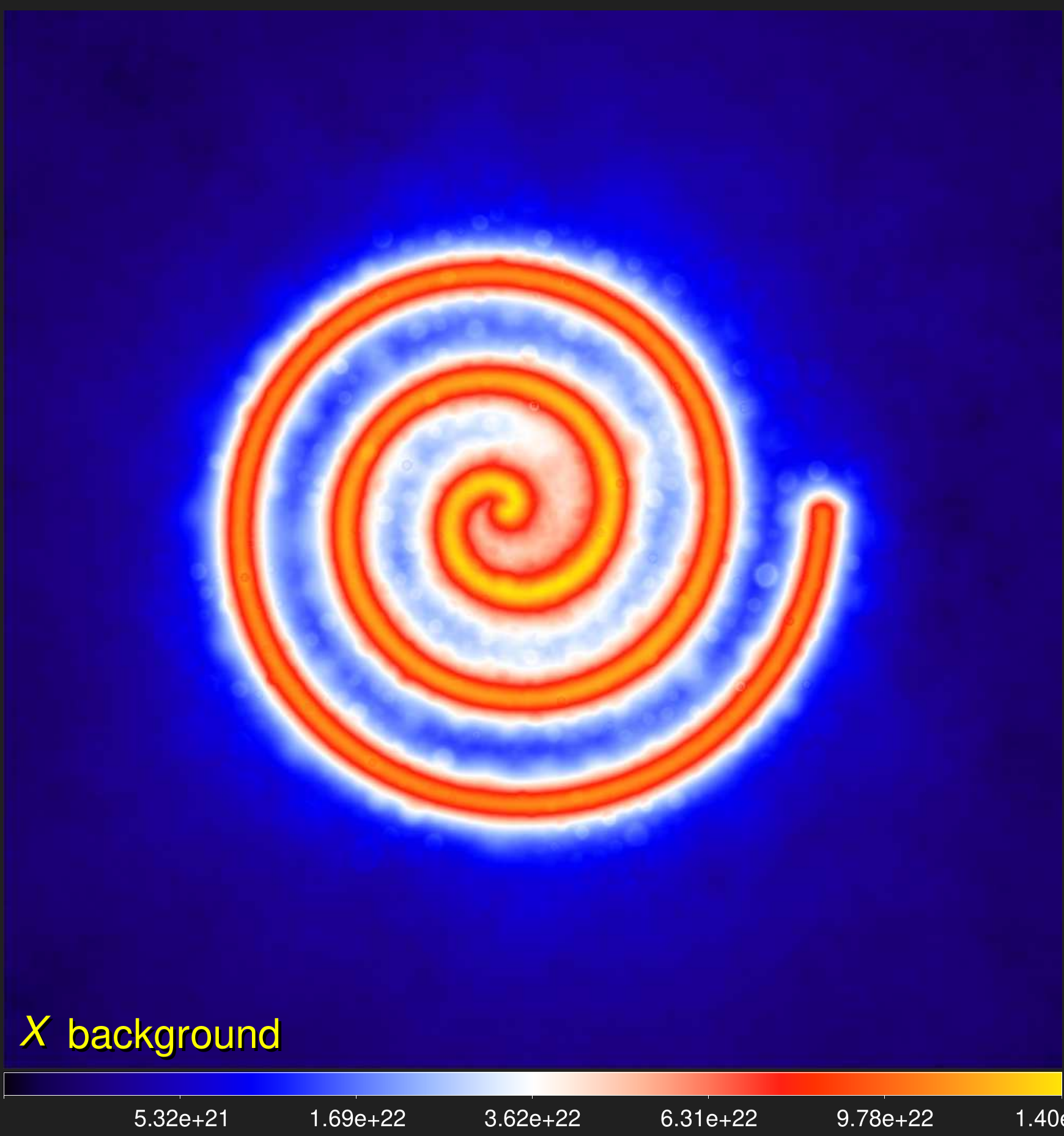}}
  \resizebox{0.328\hsize}{!}{\includegraphics{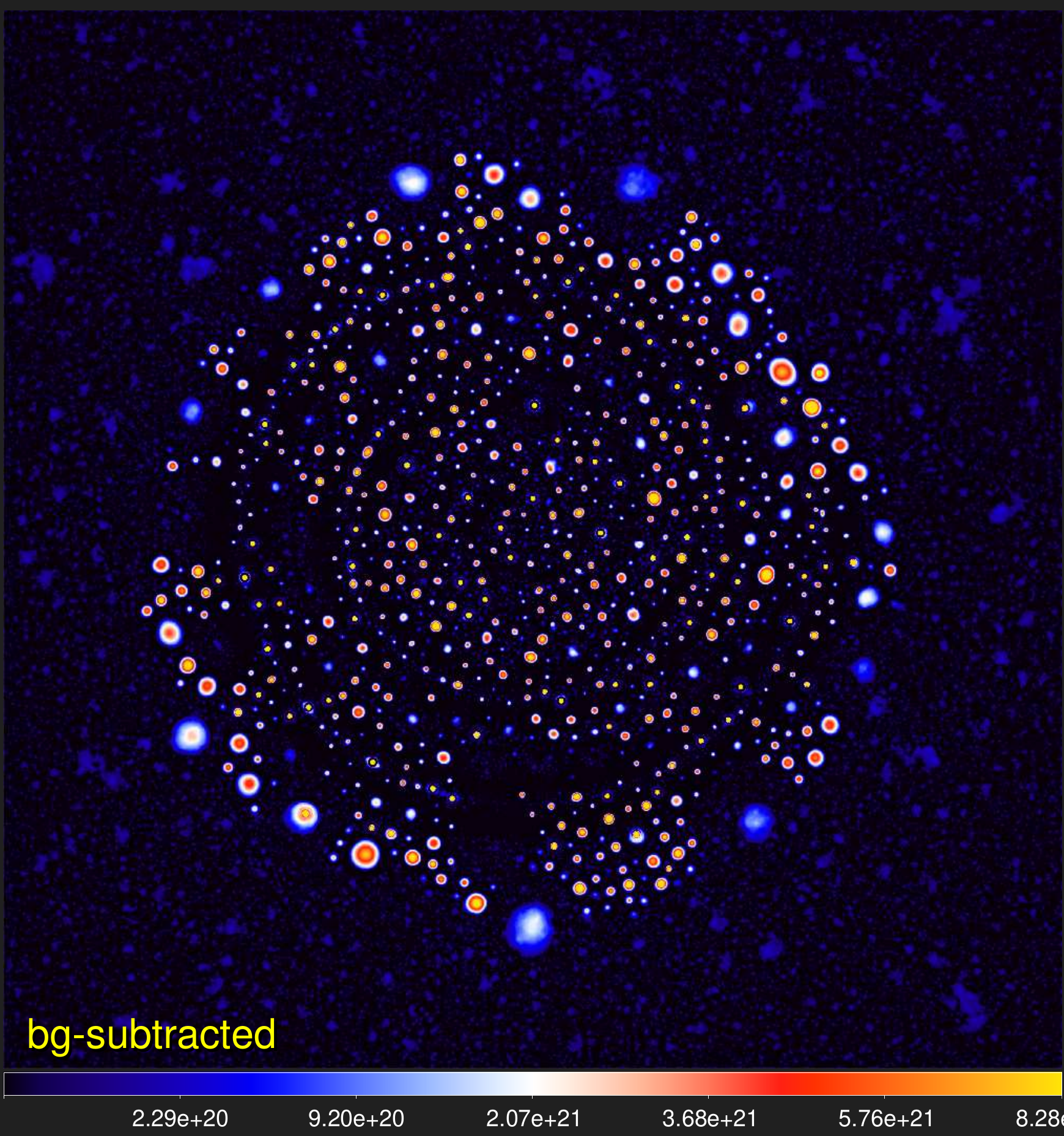}}
  \resizebox{0.328\hsize}{!}{\includegraphics{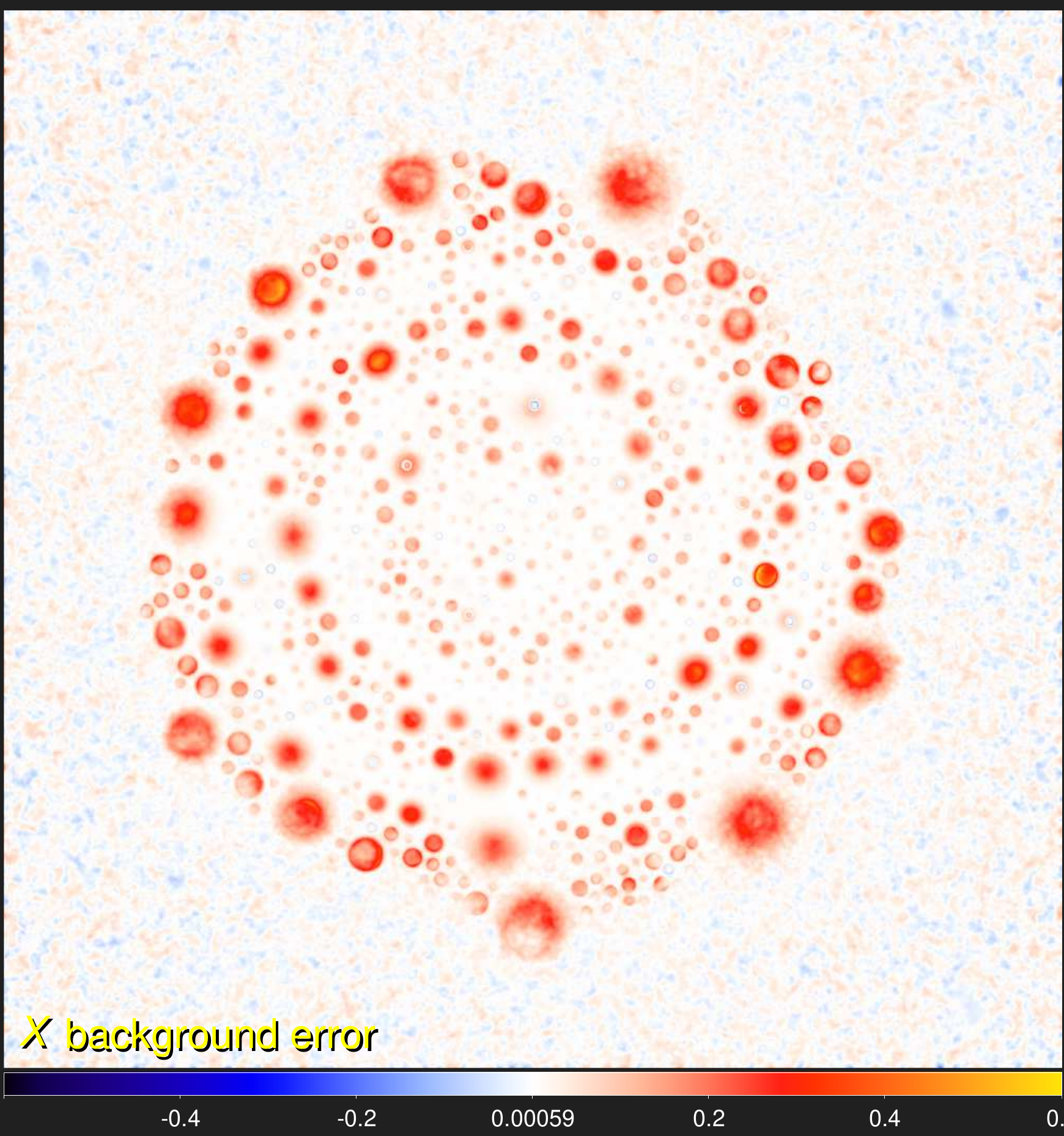}}}
\vspace{0.5mm}
\centerline{
  \resizebox{0.328\hsize}{!}{\includegraphics{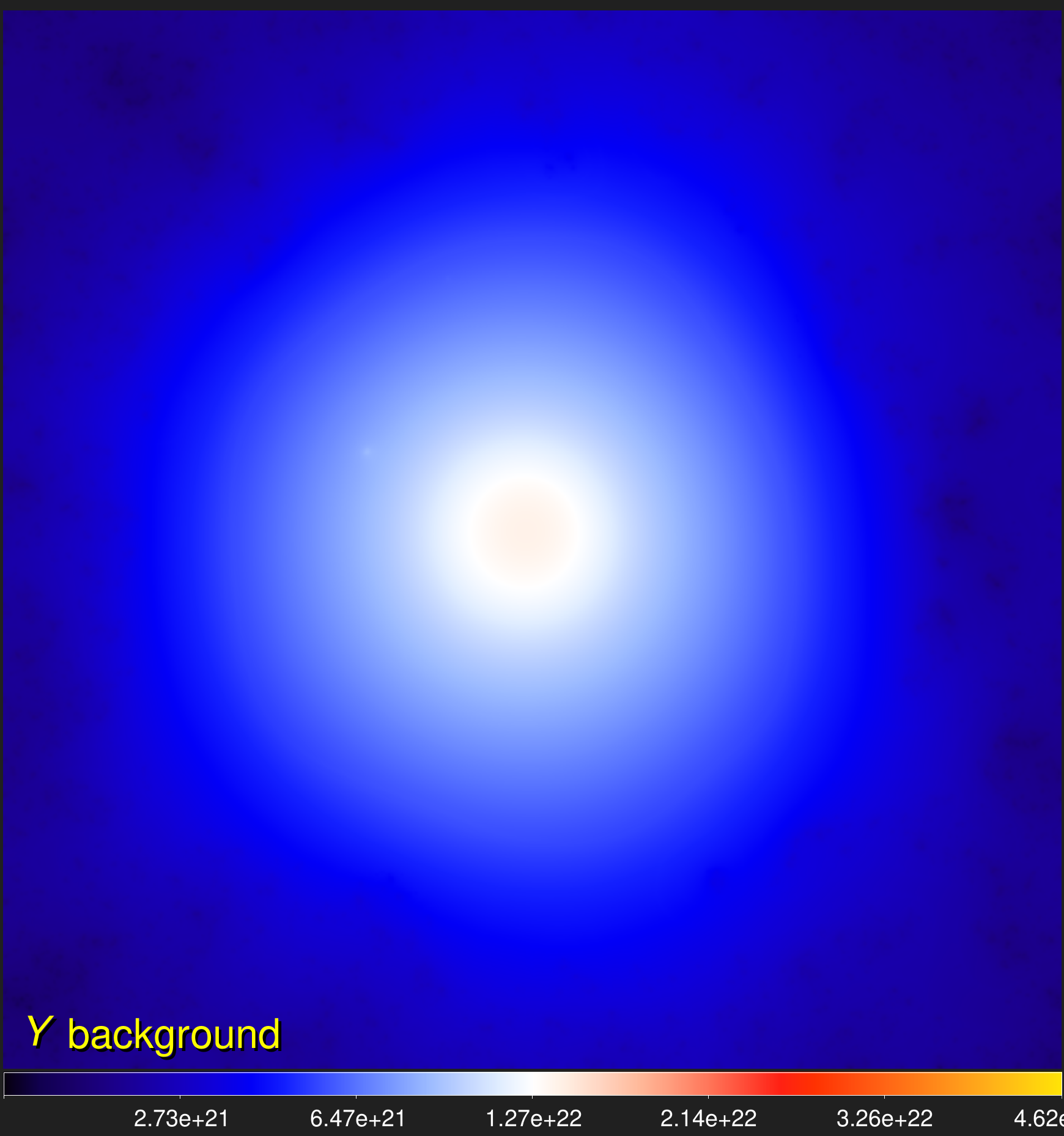}}
  \resizebox{0.328\hsize}{!}{\includegraphics{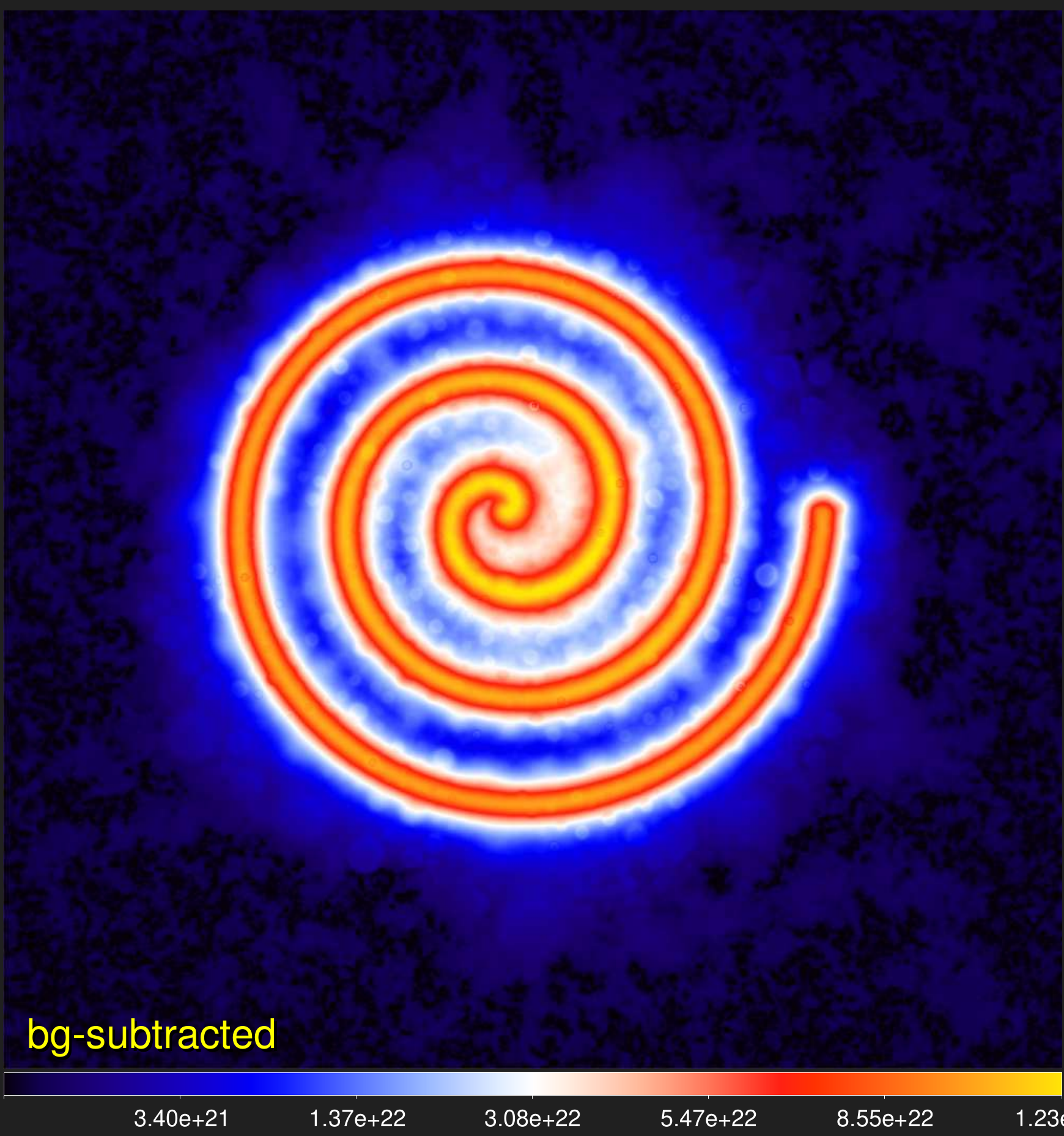}}
  \resizebox{0.328\hsize}{!}{\includegraphics{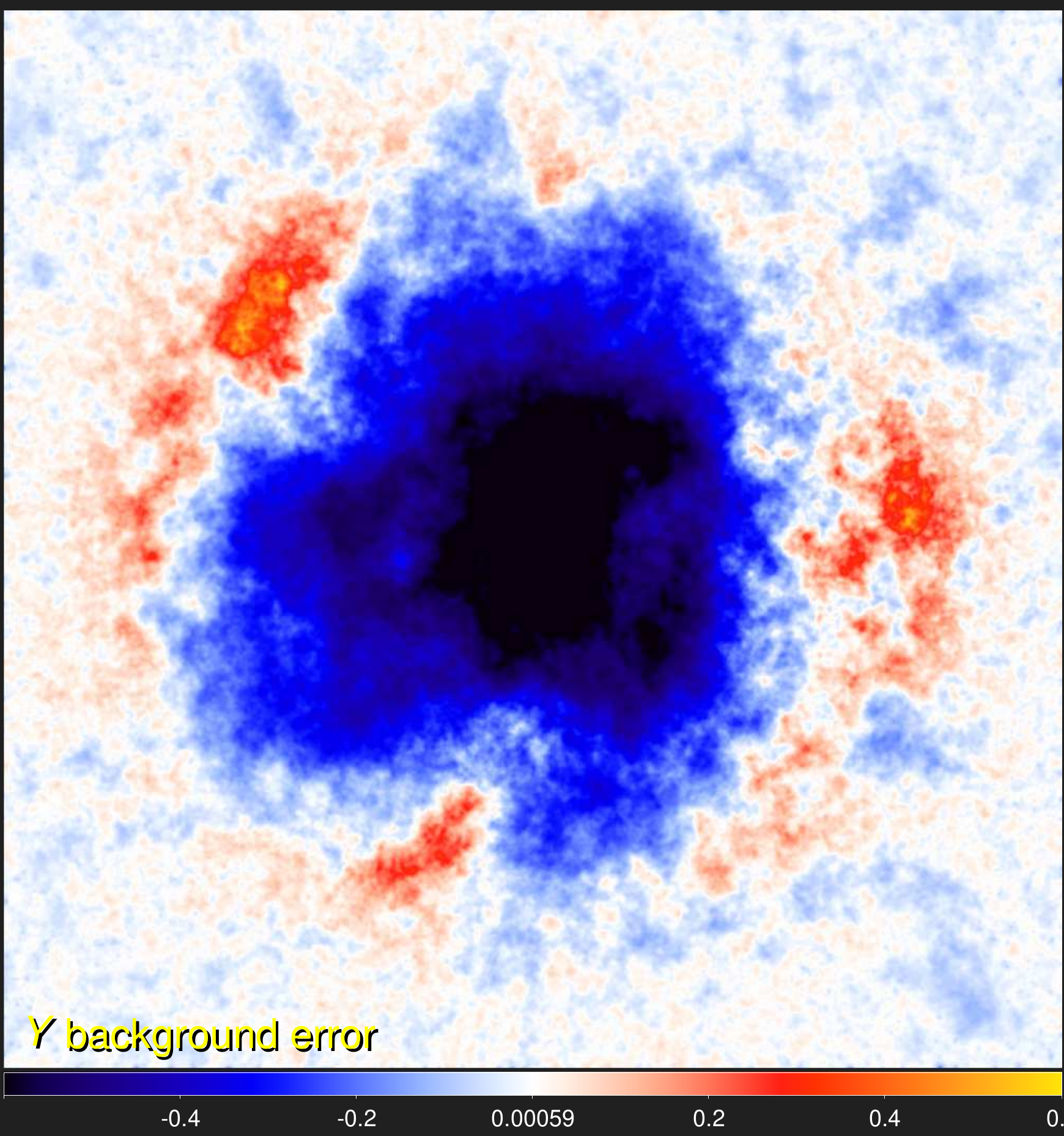}}}
\caption
{ 
Background derivation (Sect.~\ref{deriveback}) for $\mathcal{I}_{{\!\lambdabar}} \equiv \mathcal{D}_{13{\arcsec}}$ from
Eq.~(\ref{superdens}). The \emph{left} panels show the backgrounds $\mathcal{B}_{{\lambdabar}{X}}$ and
$\mathcal{B}_{{\lambdabar}{Y}}$, obtained using the procedure described by Eqs.~(\ref{background})--(\ref{convergence}). The
\emph{middle} panels show the corresponding background-subtracted $\mathcal{S}_{{\lambdabar}}$ and $\mathcal{F}_{{\lambdabar}}$
from Eq.~(\ref{bgsubtracted}). The \emph{right} panels show the relative errors of $\mathcal{B}_{{\lambdabar}{X}}$ and
$\mathcal{B}_{{\lambdabar}{Y}}$ with respect to the true model backgrounds $\mathcal{D}_{\rm C}$ and $\mathcal{D}_{\rm B}$
(Fig.~\ref{bgimage}), convolved to the same resolution. The filament is heavily blended with itself in the central area, therefore
its background is systematically underestimated there (\emph{lower right}). Square-root color mapping, except in the \emph{right}
panels, which show linear mapping.
} 
\label{bgderivation}
\end{figure*}

The structural components are separated in single scales $\mathcal{I}_{{\!\lambda}{j}}$ using the three quantities described above.
The shapes produced by sources in a slice $\mathcal{I}_{{\!\lambda}{j}{l}}$ are not very elongated, not very sparse, and not very
large. In contrast, the shapes produced by filaments in a slice $\mathcal{I}_{{\!\lambda} {j}{l}}$ are elongated or sparse. Hence,
these definitions for the source-like and filament-like shapes are written as
\begin{eqnarray} 
\left.\begin{aligned}
&{E_{{\lambda}{j}{l}} \le 1.47} \,\land\, {S_{{\!\lambda}{j}{l}} \le 1.39} \,\land\,
{N_{{\lambda}{j}{l}} \le \,\pi \left( \xi_{{\lambda}{j}}\,S_{\!j} \right)^{2}\!{\Delta^{-2}}}, \\
&{E_{{\lambda}{j}{l}} > 3.00} \,\lor\, {S_{{\!\lambda}{j}{l}} > 1.39},
\end{aligned}\right.
\label{defsrcfil} 
\end{eqnarray} 
where the limiting values of elongation and sparsity were determined empirically from numerous benchmark extractions. The
$\xi_{{\lambda}{j}}$ factor accounts for the fact that the area of a decomposed unresolved peak increases nonlinearly toward the
smallest spatial scales ${S_{\!j}{\,\la\,}O_{\lambda}}$. The factor may be determined empirically by decomposing an unresolved peak
$\mathcal{P}$ in single scales $\mathcal{P}_{\!j}$ (Fig.~\ref{gaussdecomposed}) and finding the distances $\theta$, where the
one-dimensional profile $P_{\!{j}}(\theta)$ through the peak has ${{\rm d}P_{\!j}/{\rm d}\theta{\,=\,}0}$ for $P_{\!j}{\,<\,}0$,
\begin{equation} 
{\xi_{{\lambda}{j}}} = 0.47 \left( {O_{\lambda}\,S^{-1}_{\!j}} \right)^{1.34}\! + 0.83.
\label{xifactor} 
\end{equation} 
The $\xi_{{\lambda}{j}}$ factor ensures that $N_{{\lambda}{j}{l}}$ has appropriate values and that single-scale peaks are clipped
cleanly on all spatial scales.

Various shapes formed by connected pixels are identified and analyzed in each single-scale slice using the \textsl{tintfill}
algorithm \citep{Smith_1979}\footnote{\url{http://portal.acm.org/citation.cfm?id=800249.807456}}, previously employed by
\textsl{getold} to detect sources and filaments (Papers I and II). Deriving the background $\mathcal{B}_{{\lambda}{X}}$ of sources,
\textsl{getsf} decomposes $\mathcal{I}_{{\!\lambda}}$ and removes all source-like shapes from $\mathcal{I}_{{\!\lambda}{j}{l}}$,
according to their definition in Eq.~(\ref{defsrcfil}), in an iterative procedure (Sect.~\ref{iterateback}). Deriving the
background $\mathcal{B}_{{\lambda}{Y}}$ of filaments, \textsl{getsf} decomposes $\mathcal{B}_{{\lambda}{X}}$ and removes all
filament-like shapes from $\mathcal{B}_{{\lambda}{X}{j}{l}}$, according to their definitions in Eq.~(\ref{defsrcfil}), in the same
iterative procedure. The shapes are erased from each slice $l$ by setting all their pixels to zero.


\subsubsection{Reconstruction of the backgrounds}
\label{iterateback}

When we denote with $\mathcal{B}_{{\lambda}\{{X}|{Y}\}{j}{l}{\rm C}}$ either of the single-scale background slices
${\mathcal{B}}_{{\lambda}{X}{j}{l}{\rm C}}$ or ${\mathcal{B}}_{{\lambda}{Y}{j}{l}{\rm C}}$ after the shape removal, the
backgrounds on scale $j$ are reassembled from the clipped slices as
\begin{equation} 
\mathcal{B}_{{\lambda}\{{X}|{Y}\}{j}{\rm C}}{\,=\,}\sum\limits_{l=1}^{N_{\rm L}} \mathcal{B}_{{\lambda}\{{X}|{Y}\}{j}{l}{\rm C}}.
\label{reassembly}
\end{equation} 
To properly reconstruct the complete backgrounds $\mathcal{B}_{{\lambda}\{{X}|{Y}\}}$ from
$\mathcal{B}_{{\lambda}\{{X}|{Y}\}{j}{\rm C}}$, it is not sufficient to just sum them over scales. The single-scale processing
scheme requires that it must be done indirectly in several steps by reconstructing the complete images of sources and
filaments.

In the first step, \textsl{getsf} recomputes the single-scale sources and filaments that have been clipped, removing all negative
values from the reassembled single-scale backgrounds,
\begin{equation} 
\{\mathcal{S}|\mathcal{F}\}_{{\lambda}{j}}{\,=\,}\{\mathcal{I}_{{\!\lambda}}|\mathcal{B}_{{\lambda}{X}}\} - 
\max\left(\mathcal{B}_{{\lambda}\{{X}|{Y}\}{j}{\rm C}},0\right).
\label{erasedstruc}
\end{equation} 
In the second step, \textsl{getsf} computes the full images of the sources and filaments over all scales, recursively summing
the clipped structures from the largest to the smallest scales and removing all negative values from each partial sum,\begin{equation} 
\{\mathcal{S}|\mathcal{F}\}_{{\lambda}}{\,=\,}\max\left(\{\mathcal{S}|\mathcal{F}\}_{{\lambda}} +
\{\mathcal{S}|\mathcal{F}\}_{{\lambda}{j}},0\right), \,\,\, {j{\,=\,}J_{{\lambda}\{{X}|{Y}\}},\dots,2,1},
\label{clippedstrucs}
\end{equation} 
where $J_{{\lambda}\{{X}|{Y}\}}$ is the number of the largest spatial scales ${4\{X|Y\}}_{\lambda}$ for the backgrounds
$\mathcal{B}_{{\lambda}{\{{X}|{Y}\}}}$ and the initial value of the recursive sum is set to zero. The complete backgrounds are
obtained by subtracting the structures from the original images,
\begin{equation} 
\mathcal{B}^{\,0}_{{\lambda}\{{X}|{Y}\}}{\,=\,}\{\mathcal{I}_{{\!\lambda}}|\mathcal{B}_{{\lambda}{X}}\} - 
\{\mathcal{S}|\mathcal{F}\}_{{\lambda}}.
\label{background}
\end{equation} 

\begin{figure*}                                                               
\centering
\centerline{
  \resizebox{0.328\hsize}{!}{\includegraphics{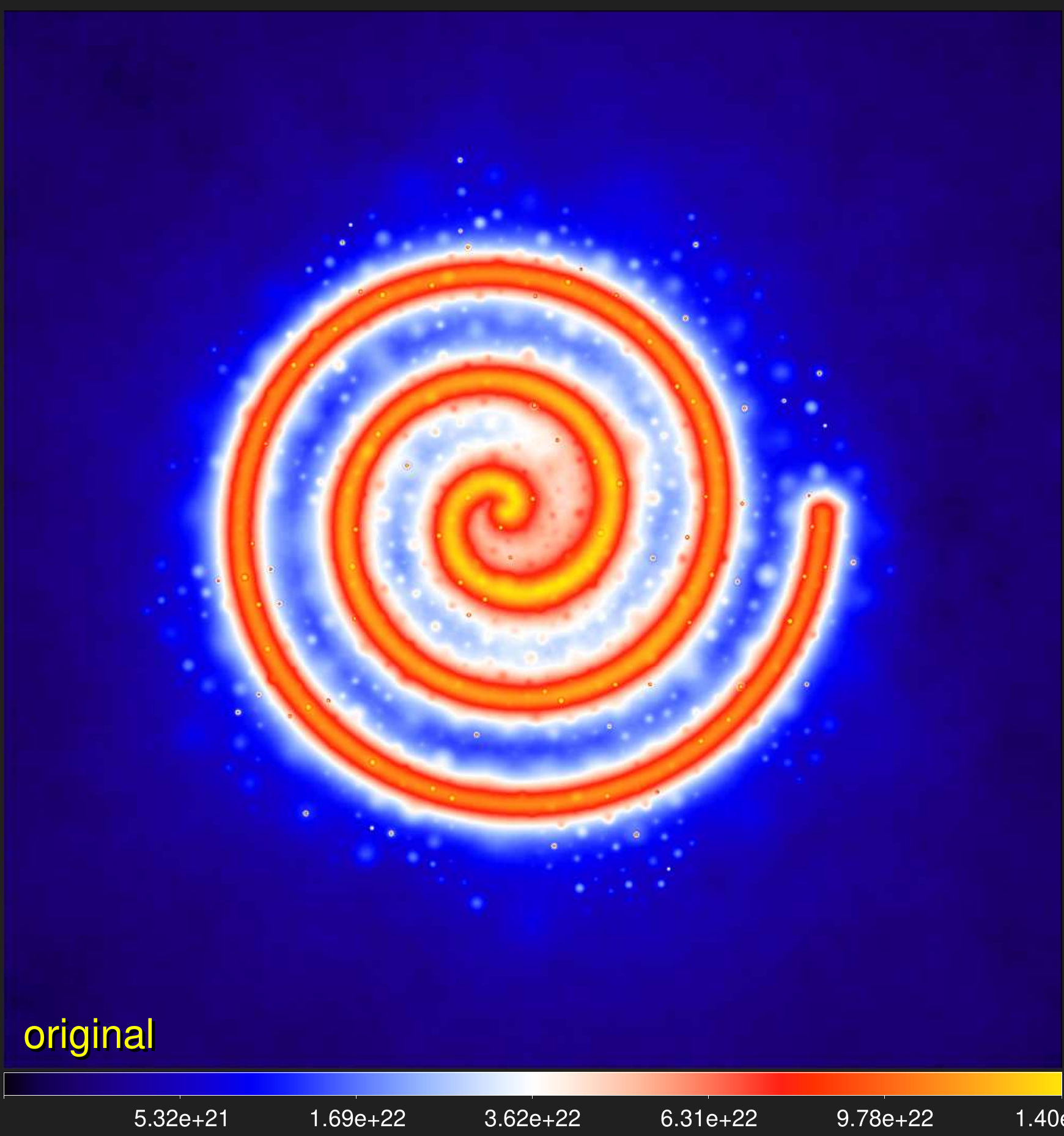}}
  \resizebox{0.328\hsize}{!}{\includegraphics{bench.165.obs.bgs.s150as.pdf}}
  \resizebox{0.328\hsize}{!}{\includegraphics{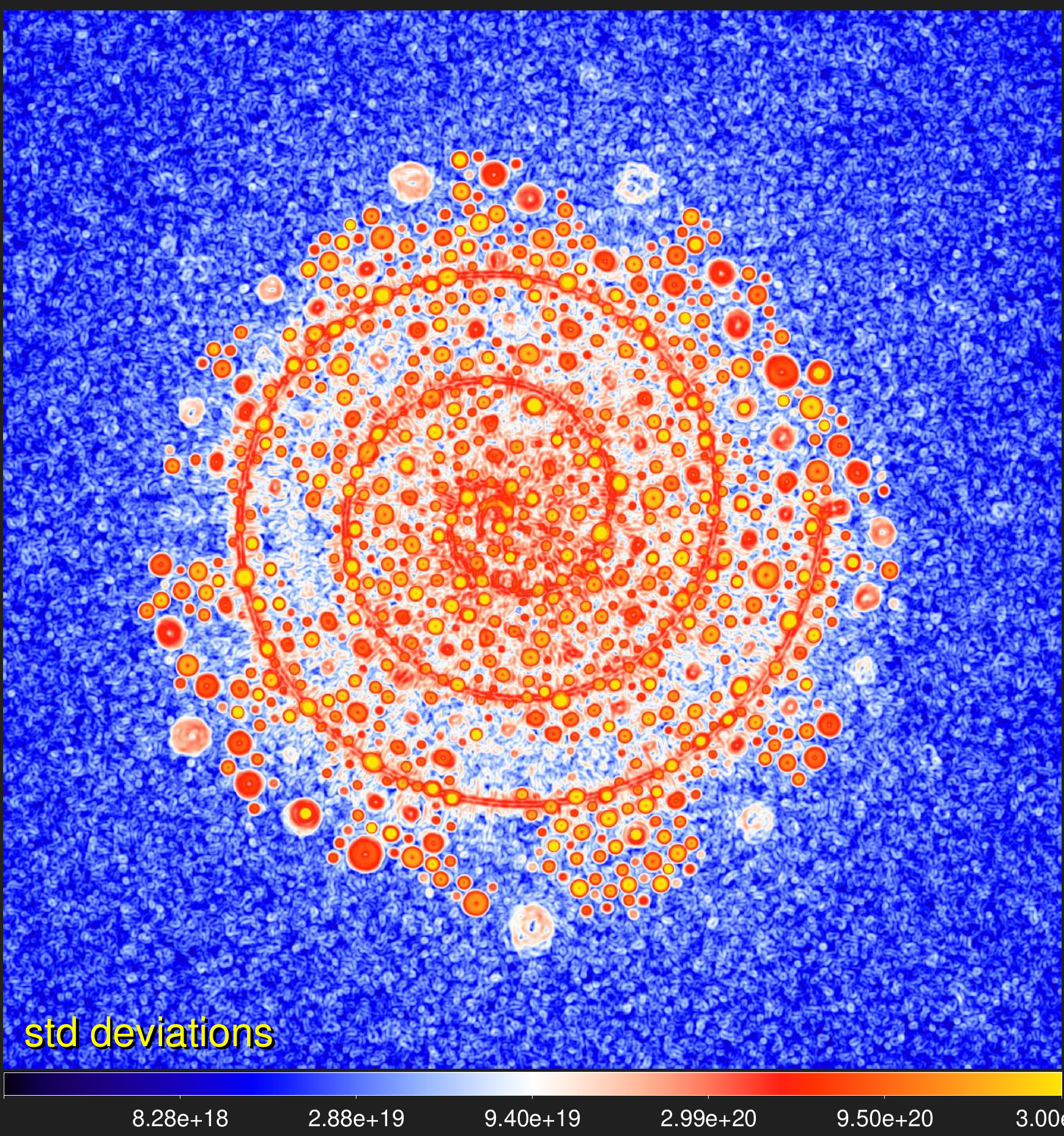}}}
\vspace{0.5mm}
\centerline{
  \resizebox{0.328\hsize}{!}{\includegraphics{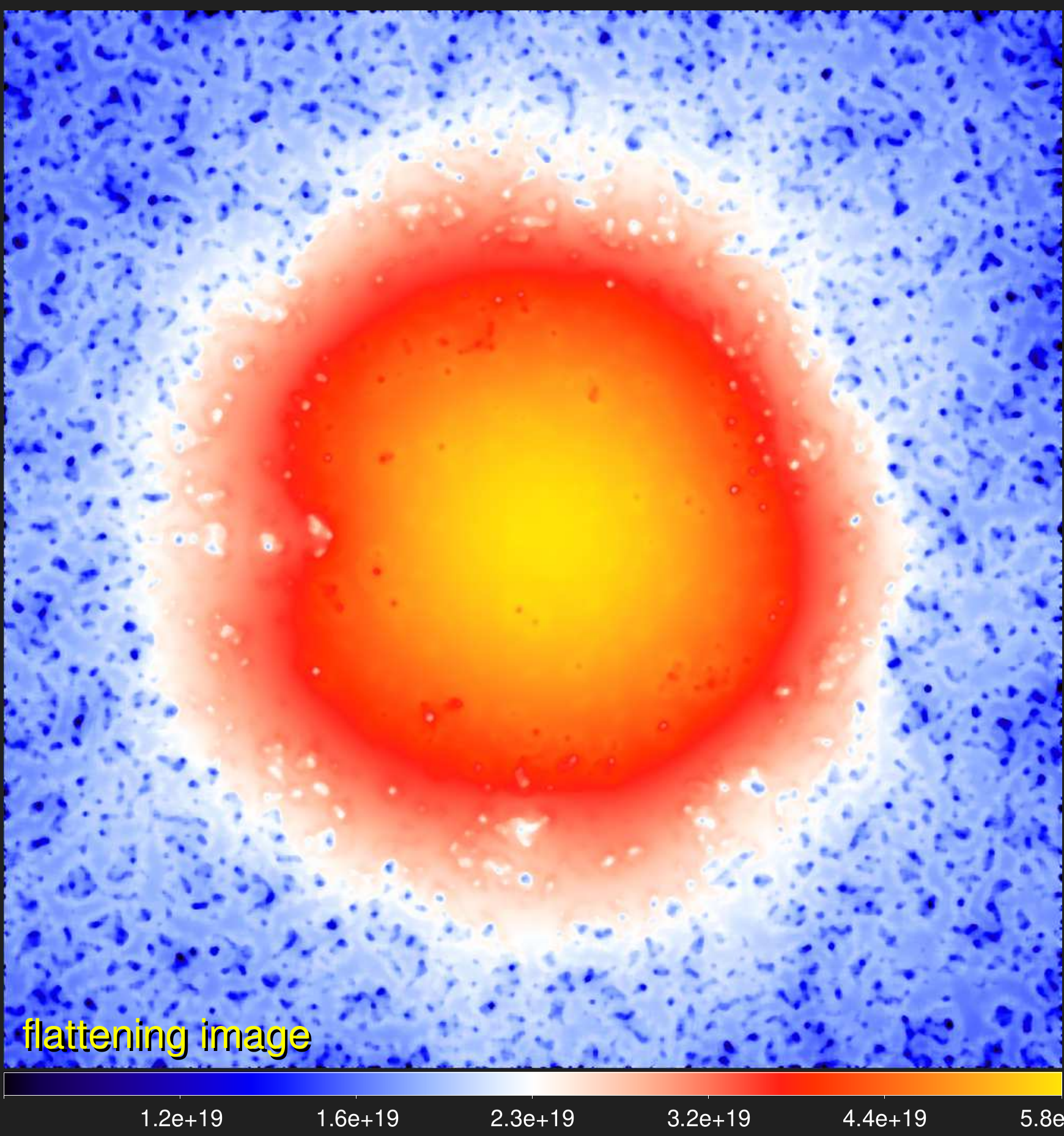}}
  \resizebox{0.328\hsize}{!}{\includegraphics{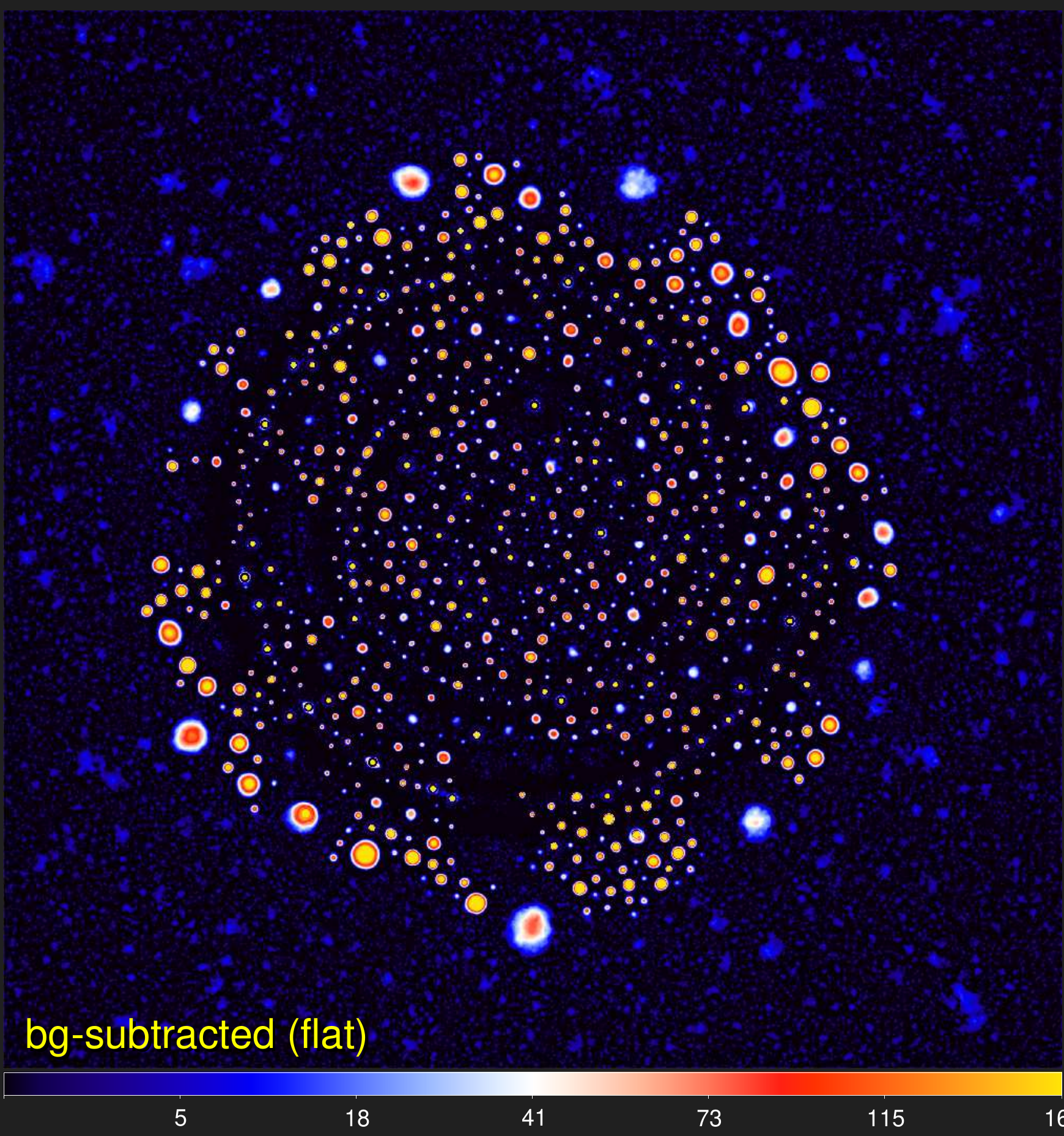}}
  \resizebox{0.328\hsize}{!}{\includegraphics{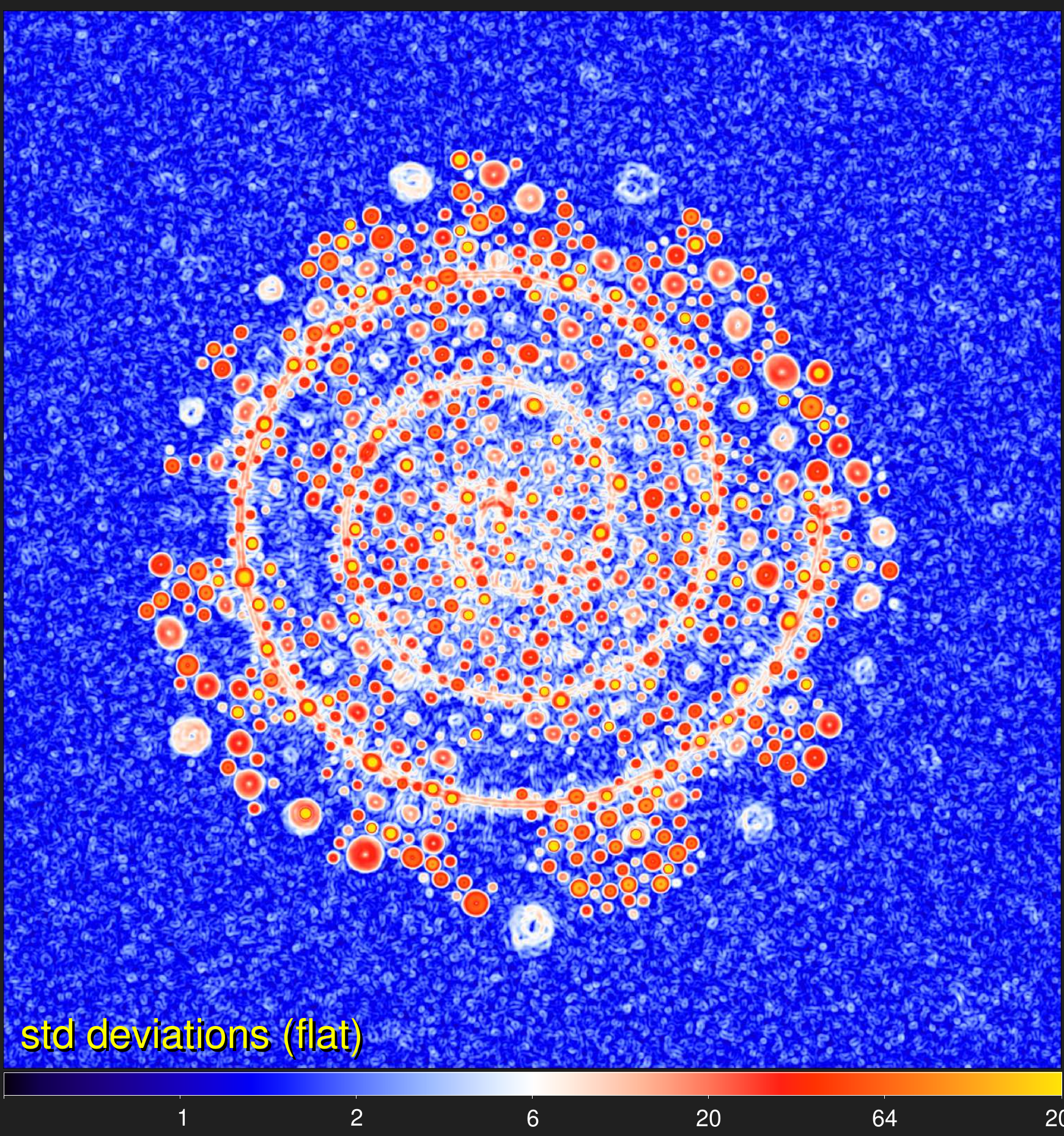}}}
\caption
{ 
Flattening for the component $\mathcal{S}_{{\lambdabar}}$ (Sect.~\ref{flattening}) for $\mathcal{I}_{{\!\lambdabar}} \equiv
\mathcal{D}_{13{\arcsec}}$ from Eq.~(\ref{superdens}). The \emph{top} row shows the original $\mathcal{I}_{\!\lambdabar}$, the
background-subtracted $\mathcal{S}_{{\lambdabar}}$ from Eq.~(\ref{bgsubtracted}), and the standard deviations
$\mathcal{U}_{{\lambdabar}}$ from Eq.~(\ref{stdflatten}). The \emph{bottom} row shows the flattening image
$\mathcal{Q}_{\lambdabar}$, the flat sources $\mathcal{S}_{{\lambdabar}{\rm D}}$ from Eq.~(\ref{flatimage}), and its standard
deviations ${\rm sd}_{O_{\!\lambda}}(\mathcal{S}_{{\lambda}{\rm R}}\mathcal{Q}_{{\lambda}}^{-1})$ that are much flatter (outside
the sources) across the image. Square-root color mapping, except in the \emph{right} panels, which show logarithmic mapping.
} 
\label{srcflatten}
\end{figure*}

The initial backgrounds in Eq.~(\ref{background}) are only the first approximations because they contain substantial residual
contributions from the original structures. It is straightforward to define iterations to improve the backgrounds by decomposing
them and clipping the residual shapes from each single scale. The algorithm described by
Eqs.~(\ref{reassembly})--(\ref{background}) remains the same, with two substitutions,
\begin{equation} 
\{\mathcal{I}_{{\!\lambda}}|\mathcal{B}_{{\lambda}{X}}\}^{i}{\,\leftarrow\,}\mathcal{B}^{\,i-1}_{{\lambda}\{{X}|{Y}\}}, \,\,\, 
{i{\,=\,}1,2,\dots,N_{\rm I}},
\label{iterations}
\end{equation} 
where $N_{\rm I}$ is the number of iterations. Each successive iteration reduces contributions of the residual structures and
improves the backgrounds until corrections in all pixels become small compared to the originals,
\begin{equation} 
\delta\mathcal{B}^{\,i}_{{\lambda}\{{X}|{Y}\}} < 0.003\left(\{\mathcal{I}_{{\!\lambda}}|\mathcal{B}_{{\lambda}{X}}\} + 
10\sigma_{\!\lambda}\right),
\label{convergence}
\end{equation} 
where the additional term helps avoid unnecessary iterations in rare cases when the images contain extremely faint pixels.

The final background-subtracted structural components are computed as
\begin{equation} 
\{\mathcal{S}|\mathcal{F}\}_{{\lambda}} = \{\mathcal{I}_{{\!\lambda}}|\mathcal{B}_{{\lambda}{X}}\} - 
\mathcal{B}_{{\lambda}\{{X}|{Y}\}}.
\label{bgsubtracted}
\end{equation} 
The original images can be recovered by summing the three separated components:
$\mathcal{I}_{{\!\lambda}}{\,=\,}\mathcal{S}_{{\lambda}}{\,+\,}\mathcal{F}_{{\lambda}}{\,+\,}\mathcal{B}_{{\lambda}{Y}}$. The
positive parts of the small-scale background fluctuations and instrumental noise are contained in the component
$\mathcal{S}_{{\lambda}}$, hence the component $\mathcal{F}_{{\lambda}}$ appears fairly smooth (Fig.~\ref{bgderivation}).

\begin{figure*}                                                               
\centering
\centerline{
  \resizebox{0.328\hsize}{!}{\includegraphics{bench.165.obs.pdf}}
  \resizebox{0.328\hsize}{!}{\includegraphics{bench.165.obs.bgs.s350as.pdf}}
  \resizebox{0.328\hsize}{!}{\includegraphics{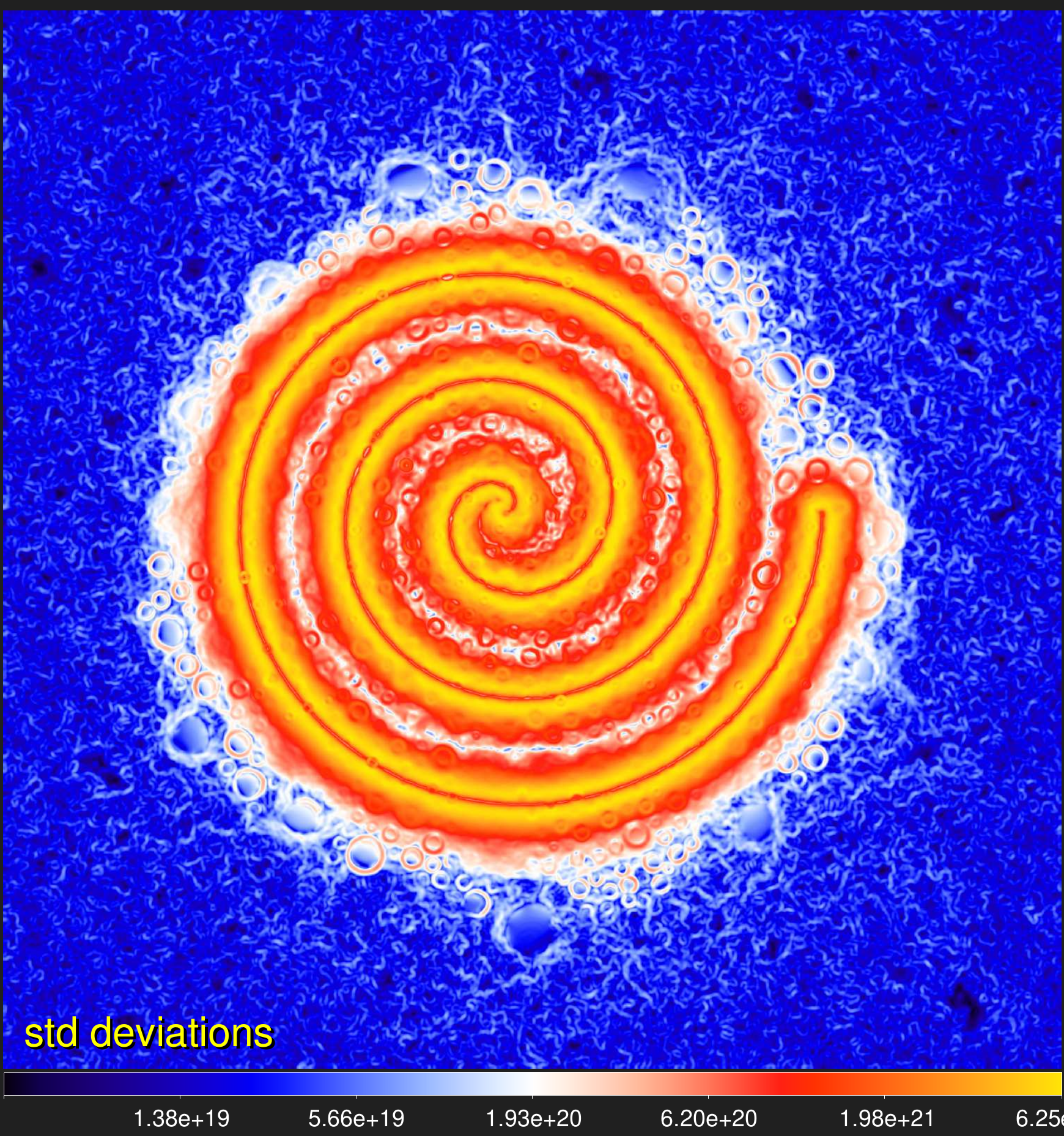}}}
\vspace{0.5mm}
\centerline{
  \resizebox{0.328\hsize}{!}{\includegraphics{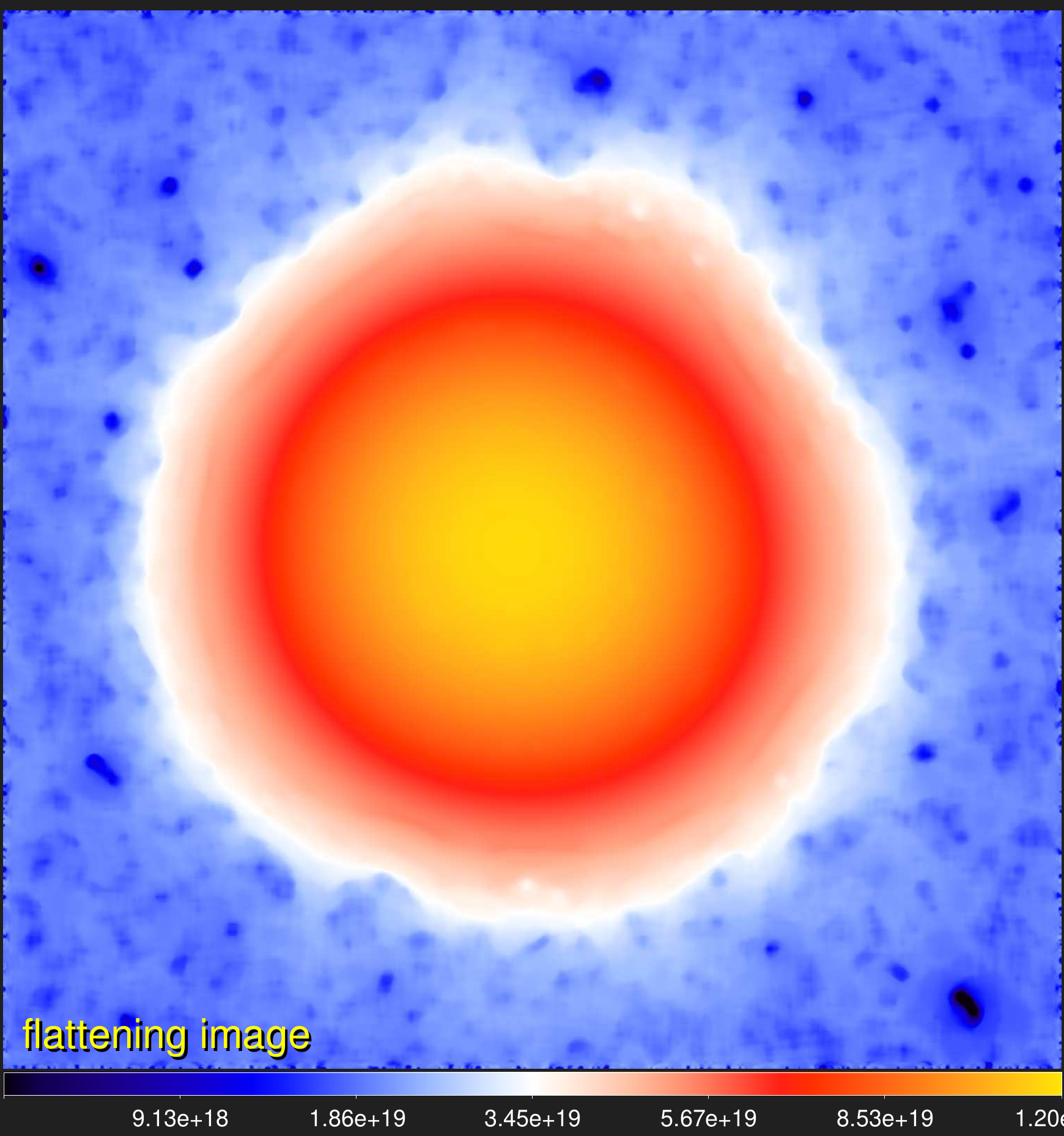}}
  \resizebox{0.328\hsize}{!}{\includegraphics{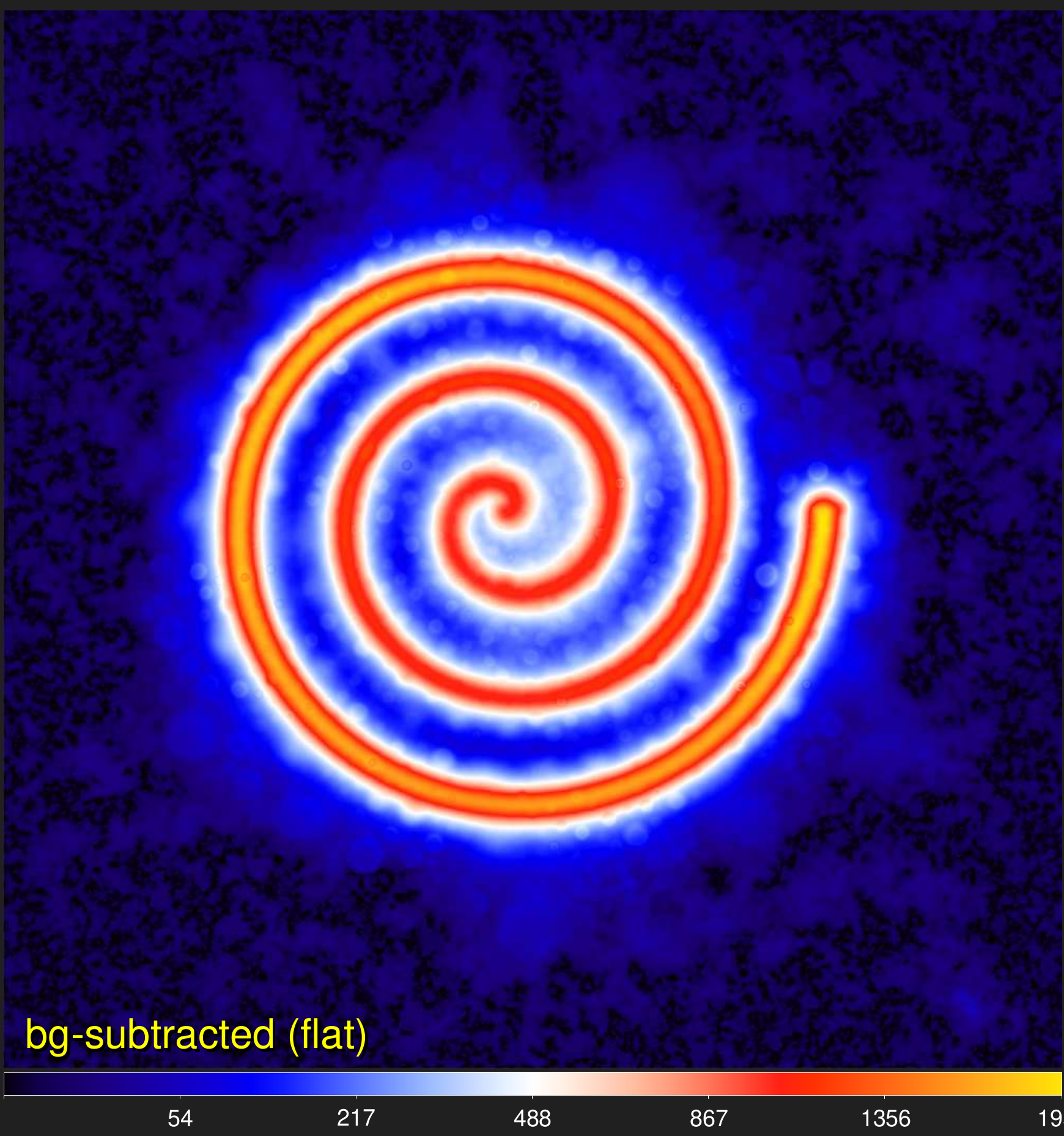}}
  \resizebox{0.328\hsize}{!}{\includegraphics{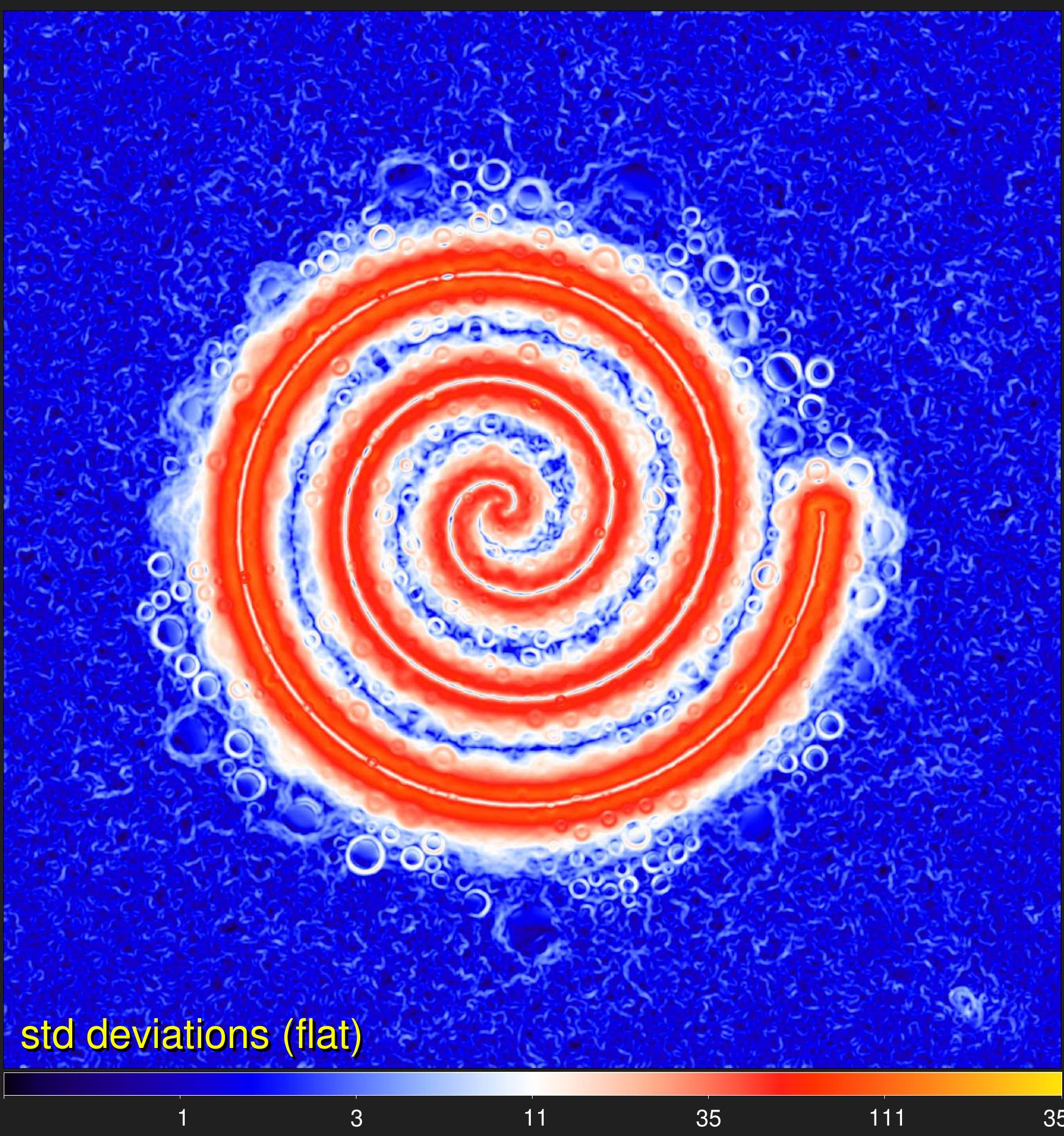}}}
\caption
{ 
Flattening for the component $\mathcal{F}_{{\lambdabar}}$ (Sect.~\ref{flattening}) for $\mathcal{I}_{{\!\lambdabar}} \equiv
\mathcal{D}_{13{\arcsec}}$ from Eq.~(\ref{superdens}). The \emph{top} row shows the original $\mathcal{I}_{\!\lambdabar}$, the
background-subtracted $\mathcal{F}_{{\lambdabar}}$ from Eq.~(\ref{bgsubtracted}), and the standard deviations
$\mathcal{V}_{{\lambdabar}}$ from Eq.~(\ref{stdflatten}). The \emph{bottom} row shows the flattening image
$\mathcal{R}_{\lambdabar}$, the flat filaments $\mathcal{F}_{{\lambdabar}{\rm D}}$ from Eq.~(\ref{flatimage}) and its standard
deviations ${\rm sd}_{O_{\!\lambda}}(\mathcal{F}_{{\lambda}{\rm R}}\mathcal{R}_{{\lambda}}^{-1}),$  which are much flatter (outside
the filament) across the image. Square-root color mapping, except in the \emph{right} panels, which show logarithmic mapping.
} 
\label{filflatten}
\end{figure*}


\subsection{Flattening of the structural components}
\label{flattening}

Observations demonstrate that the levels of the large-scale backgrounds and their smaller-scale fluctuations often differ by orders
of magnitude in various parts of large images. Although the subtraction of the smooth backgrounds
$\mathcal{B}_{{\lambda}\{{X}|{Y}\}}$ greatly simplifies the original images, it does not reduce the strong variations of the
smaller-scale fluctuation levels across $\{\mathcal{S}|\mathcal{F}\}_{{\lambda}}$. As a consequence, many structures detected with
global thresholds in the areas of stronger fluctuations may actually be spurious and unrelated to any real physical objects. On the
other hand, faint real structures in the image areas with the lower levels of fluctuations may escape detection because the global
threshold value is likely to be overestimated for those areas. To produce complete and reliable extractions using constant
thresholds, it is necessary to make small-scale fluctuations uniform over the entire background-subtracted images.

The fluctuation levels are equalized using flattening images $\mathcal{Q}_{{\lambda}}$ and $\mathcal{R}_{{\lambda}}$ that are
derived by \textsl{getsf} from the images $\mathcal{U}_{{\lambda}}$ and $\mathcal{V}_{{\!\lambda}}$ of the standard deviations
computed in the structural components with a circular sliding window of a radius $O_{\lambda}$,
\begin{equation} 
\{\mathcal{U}|\mathcal{V}\}_{{\lambda}} = {\rm sd}_{O_{\!\lambda}\!}\left(\{\mathcal{S}|\mathcal{F}\}_{{\lambda}{\rm R}}\right),
\label{stdflatten}
\end{equation} 
where $\mathcal{S}_{{\lambda}{\rm R}}$ and $\mathcal{F}_{{\lambda}{\rm R}}$ are the regularized images
$\mathcal{S}_{{\lambda}}$ and $\mathcal{F}_{{\lambda}}$, obtained using a smoother version of their backgrounds that is median-filtered
using a sliding window of a radius $2 O_{\lambda}$ and convolved with a Gaussian kernel of a half-maximum size $O_{\lambda}$,
\begin{equation} 
\{\mathcal{S}|\mathcal{F}\}_{{\lambda}{\rm R}} = \{\mathcal{I}_{{\!\lambda}}|\mathcal{B}_{{\lambda}{X}}\} - 
\mathcal{G}_{O_{\lambda}\!}{\,*\,}\mathrm{mf}_{2 O_{\lambda}\!}\left( \mathcal{B}_{{\lambda}\{{X}|{Y}\}}\right).
\label{stdregular}
\end{equation} 
This is done to improve the quality of $\{\mathcal{U}|\mathcal{V}\}_{{\lambda}}$ for further processing because the structural
components from by Eq.~(\ref{bgsubtracted}) are positively defined and have large areas of zero pixels. The regularized components
in Eq.~(\ref{stdregular}) acquire small-scale fluctuations resembling the background and noise fluctuations of the original images.


\subsubsection{Decomposition of the standard deviations}
\label{decompstd}

The advantages of the spatial decomposition (Appendix~\ref{decomposition}) apply also to the standard deviations
$\{\mathcal{U}|\mathcal{V}\}_{{\lambda}}$. The \textsl{getsf} method produces the single-scales
$\{\mathcal{U}|\mathcal{V}\}_{{\lambda}{j}}$ and employs the same iterative algorithm (Appendix~\ref{decomposition}) to determine
the single-scale standard deviation $\sigma_{{\!\lambda}{j}}$ and its total value $\sigma_{{\!\lambda}}$. This is done using the
same procedure as was applied to $\mathcal{I}_{{\!\lambda}}$ in Sect.~\ref{decompos}.


\subsubsection{Removal of the structural features}
\label{clipstd}

The $\{\mathcal{U}|\mathcal{V}\}_{{\lambda}}$ images sample local fluctuations and intensity gradients, revealing all sources and
filaments present in $\mathcal{I}_{{\!\lambda}}$ (Figs.~\ref{srcflatten}, \ref{filflatten}). To produce the corresponding
flattening images, it is necessary to remove all such features from $\{\mathcal{U}|\mathcal{V}\}_{{\lambda}}$, hence to determine
their $\{{X}|{Y}\}_{\lambda}$-scale backgrounds. Deriving the latter, \textsl{getsf} creates single-scale slices
$\{\mathcal{U}|\mathcal{V}\}_{{\lambda}{j}{l}}$, in a complete analogy with $I_{{\lambda}{j}{l}}$ in Sect.~\ref{clipping}, and
clips from them all source- and filament-like shapes according to their definitions in Eq.~(\ref{defsrcfil}). The reconstructed
backgrounds $\mathcal{Q}_{{\lambda}{X}}$ and $\mathcal{R}_{{\lambda}{Y}}$ are computed using the iterative algorithm described in
Sect.~\ref{iterateback}, with the largest spatial scale set to $2.5\{{X}|{Y}\}_{\lambda}$.

When the background iterations converge, numerous sharp craters remain in the derived backgrounds
$\{\mathcal{Q}|\mathcal{R}\}_{{\lambda}{\{{X}|{Y}\}}}$ that could create spurious structures if the images were used to flatten
the structural components. To avoid this, the final flattening images $\mathcal{Q}_{{\lambda}}$ and $\mathcal{R}_{{\lambda}}$
(Figs.~\ref{srcflatten}, \ref{filflatten}) are obtained by median filtering the background in circular sliding windows of radii $2
O_{\lambda}$ and $5 O_{\lambda}$, respectively,
\begin{equation} 
\{\mathcal{Q}|\mathcal{R}\}_{{\lambda}} = \mathrm{mf}_{{\{2|5\}}{O_{\lambda}\!}}\left(
\{\mathcal{Q}|\mathcal{R}\}_{{\lambda}{\{{X}|{Y}\}}}\right).
\label{flatfact}
\end{equation} 
This important step ensures that flattening would never produce spurious structures in the detection images. 


\subsubsection{Flattening of the detection images}
\label{detimages}

The detection images of the separated structural components are used to identify peaks of the sources and skeletons of the
filaments, respectively (Sect.~\ref{extraction}). Both source- and filament-detection images are flattened, that is, divided by the
flattening images,
\begin{equation} 
\{\mathcal{S}|\mathcal{F}\}_{{\lambda}{\rm D}} = 
\frac{\{\mathcal{S}|\mathcal{F}\}_{{\lambda}}}{\{\mathcal{Q}|\mathcal{R}\}_{{\lambda}}}.
\label{flatimage}
\end{equation} 
The standard deviations ${\rm sd}_{O_{\!\lambda}}(\{\mathcal{S}|\mathcal{F}\}_{{\lambda}{\rm
R}}\{\mathcal{Q}|\mathcal{R}\}_{{\lambda}}^{-1})$ in the regularized flattened components demonstrate that the detection images are
remarkably flat outside the structures, as shown in Figs.~\ref{srcflatten} and \ref{filflatten}. This ensures an accurate separation
of significant structures from the fainter background and noise fluctuations during the subsequent extraction of sources and
filaments.


\subsection{Extraction of the structural components}
\label{extraction}

To extract sources and filaments means to detect them and measure their properties. The background subtraction and flattening
algorithms presented in Sects.~\ref{deriveback} and \ref{flattening} radically simplify the originals $\mathcal{I}_{\!\lambda}$,
separating two distinct structural components and creating the independent flat detection images
$\{\mathcal{S}|\mathcal{F}\}_{{\lambda}{\rm D}}$. In contrast to the originals, the flat images are suitable for the detection
techniques that apply a threshold value for the entire image.


\subsubsection{Decomposition of the detection images}
\label{decompflat}

In order to accurately extract various structures that widely range in brightness and size, it is essential to use the benefits
offered by the single-scale spatial decomposition (Appendix~\ref{decomposition}). Following its general approach
(Fig.~\ref{flowchart} and Sects.~\ref{decompos}, \ref{decompstd}), \textsl{getsf} decomposes the detection images
$\mathcal{S}_{{\lambda}{\rm D}}$ and $\mathcal{F}_{{\lambda}{\rm D}}$ into single scales
$\{{\mathcal{S}|\mathcal{F}}\}_{{\lambda}{\rm D}{j}}$ and estimates the corresponding standard deviations $\sigma_{{\!\lambda}{\rm
S}{j}}$ and $\sigma_{{\!\lambda}{\rm F}{j}}$ (Appendix~\ref{decomposition}) that are necessary for separating significant
structures from all other fluctuations. The decomposed components $\mathcal{S}_{{\lambda}{\rm D}{j}}$ and
$\mathcal{F}_{{\lambda}{\rm D}{j}}$ are shown in Figs.~\ref{flatsrccomb} and \ref{flatfilcomb} after they were cleaned and
combined over wavebands.


\subsubsection{Cleaning of the single-scale detection images}
\label{cleaning}

Cleaning is the removal of insignificant background and noise fluctuations from detection images that needs to be done before
combining them over wavebands (Sect.~\ref{combining}). The clean images of the structures are obtained by preserving only the
pixels with values above the cleaning thresholds $\varpi_{{\lambda}{\{\rm S|F\}}{j}}$ and by setting all fainter pixels to
zero,
\begin{equation} 
\{\mathcal{S}|\mathcal{F}\}_{{\lambda}{\rm D}{j}{\rm C}} = \max\left(\{\mathcal{S}|\mathcal{F}\}_{{\lambda}{\rm D}{j}}, 
\varpi_{{\lambda}{\{\rm S|F\}}{j}}\right),
\label{cleanimg}
\end{equation} 
where $\varpi_{{\lambda}{\rm S}{j}}{\,=\,}5\sigma_{{\!\lambda}{\rm S}{j}}$ and $\varpi_{{\lambda}{\rm
F}{j}}{\,=\,}2\sigma_{{\!\lambda}{\rm F}{j}}$. The filament threshold is significantly lower than that for sources because
\textsl{getsf} additionally cleans $\mathcal{F}_{{\lambda}{\rm D}{j}{\rm C}}$ of the residual source-like clusters of connected
pixels according to their definition in Eq.~(\ref{defsrcfil}).

\begin{figure*}                                                               
\centering
\centerline{
  \resizebox{0.328\hsize}{!}{\includegraphics{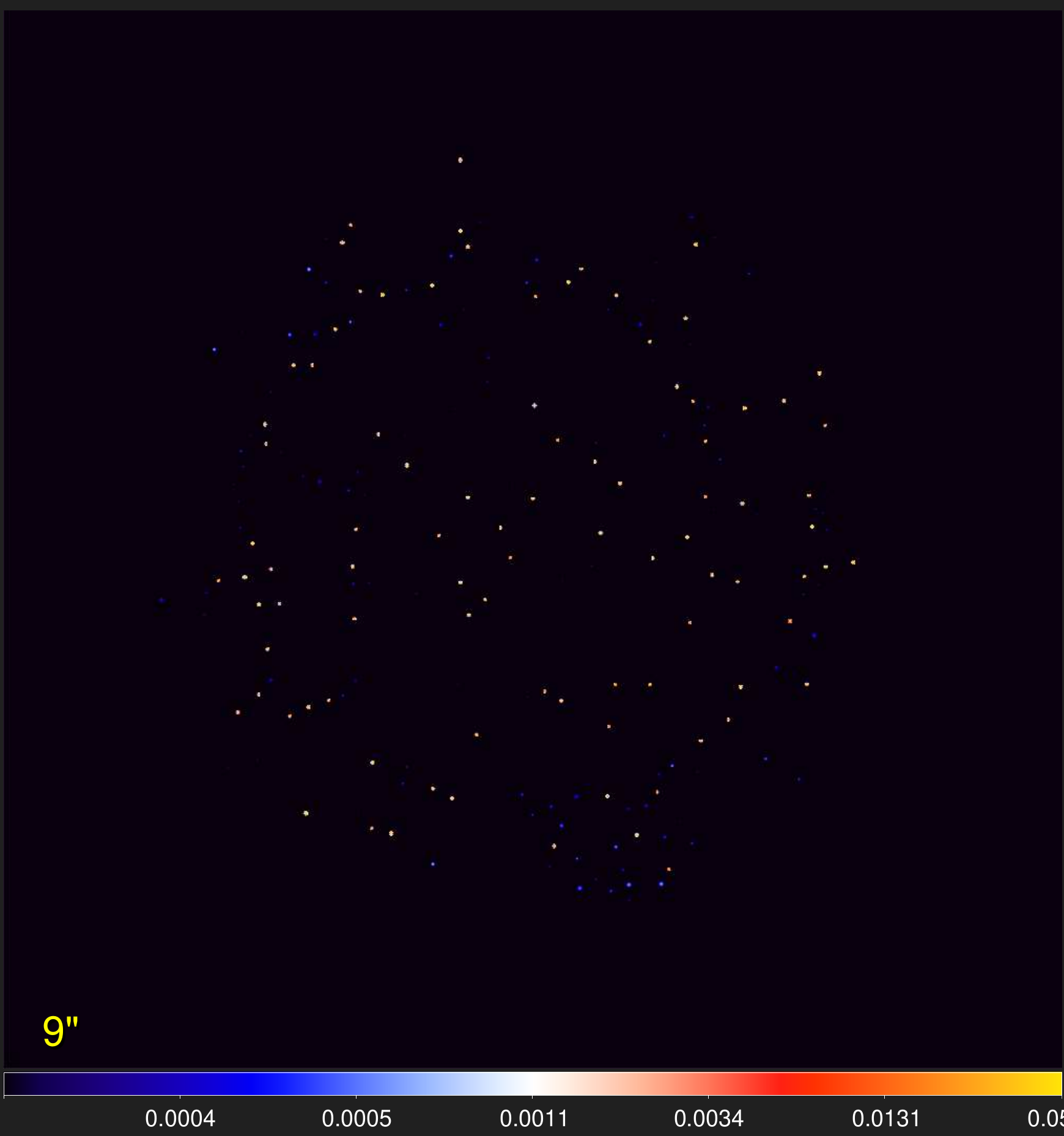}}
  \resizebox{0.328\hsize}{!}{\includegraphics{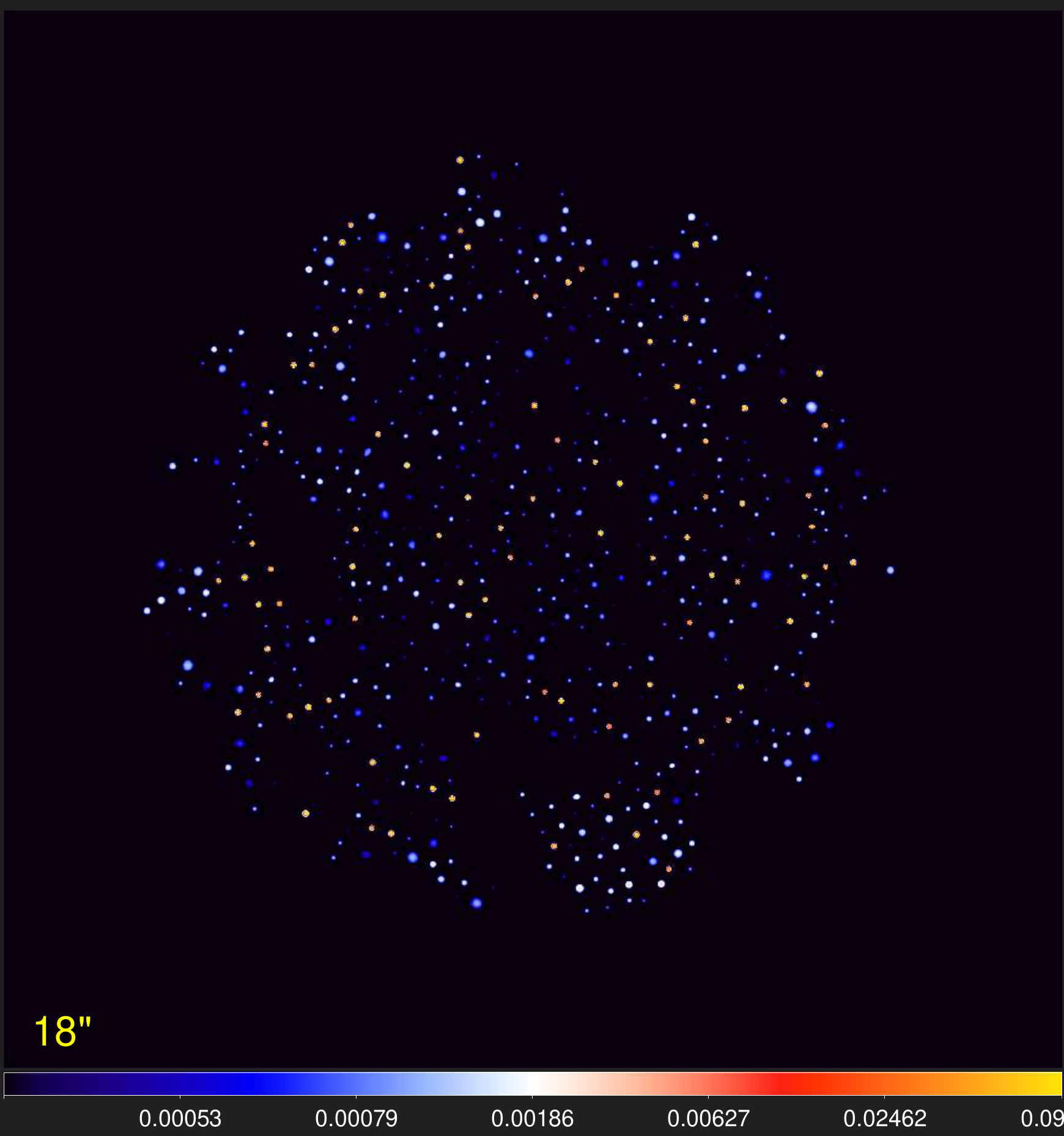}}
  \resizebox{0.328\hsize}{!}{\includegraphics{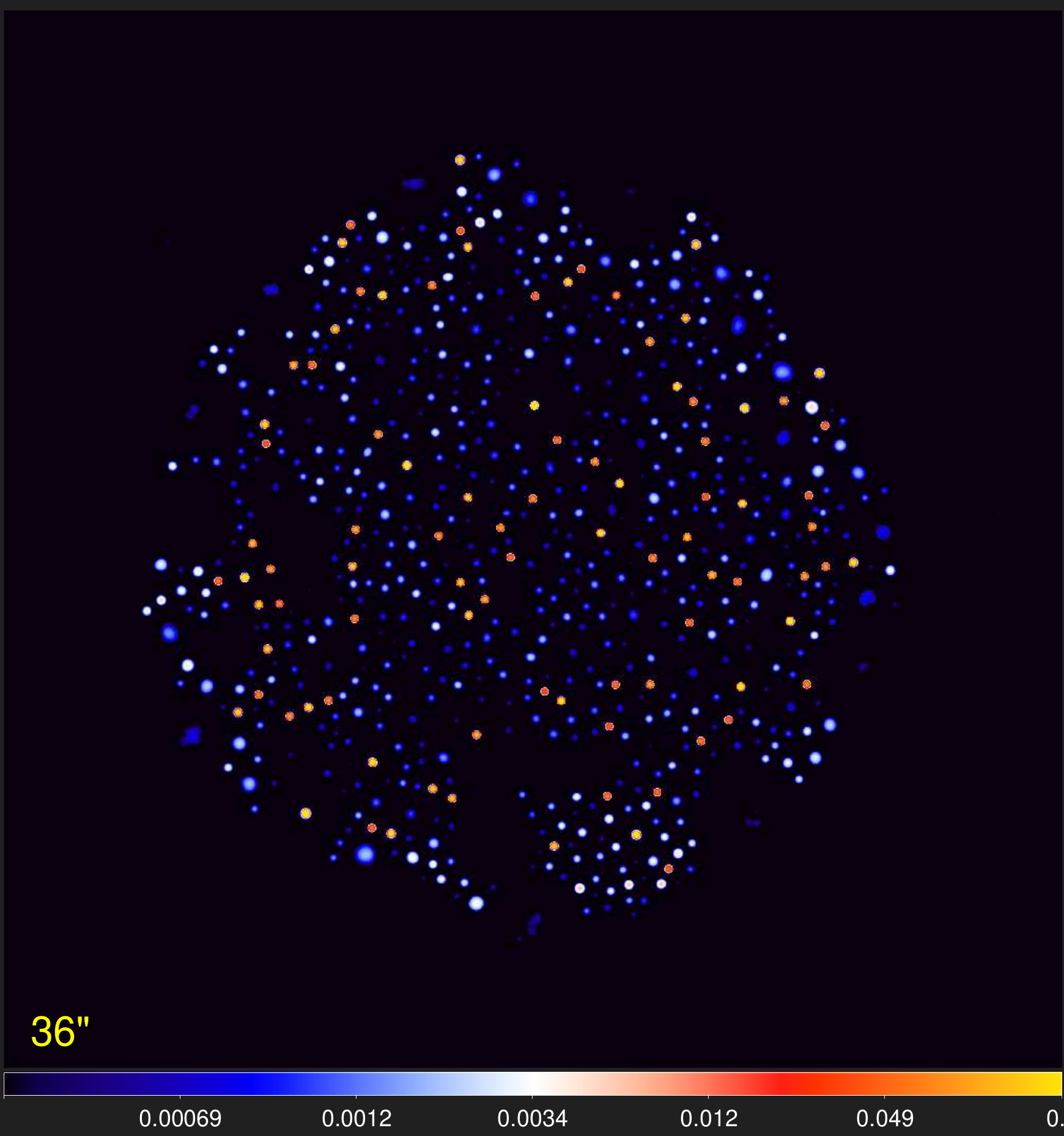}}}
\vspace{0.5mm}
\centerline{
  \resizebox{0.328\hsize}{!}{\includegraphics{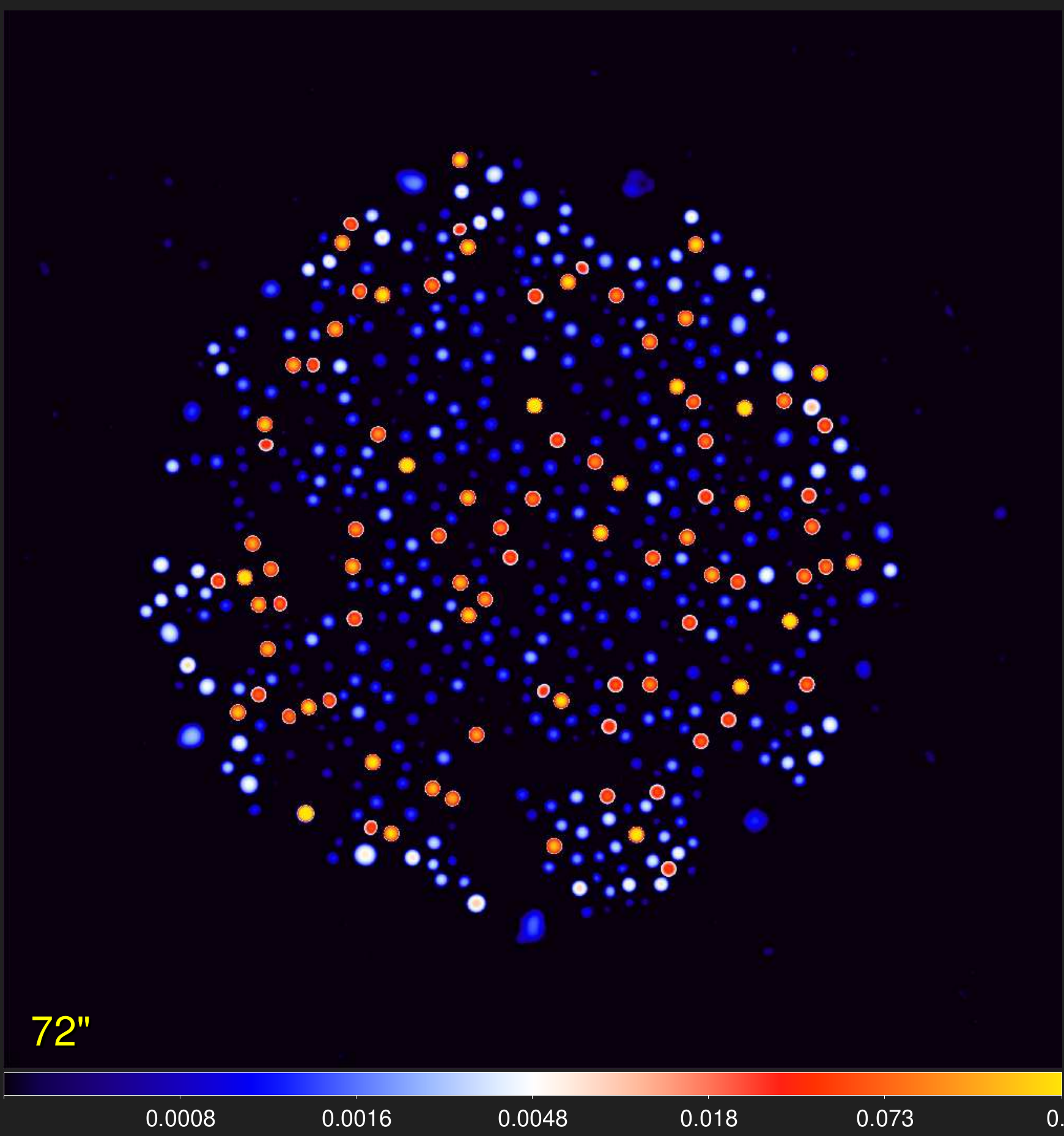}}
  \resizebox{0.328\hsize}{!}{\includegraphics{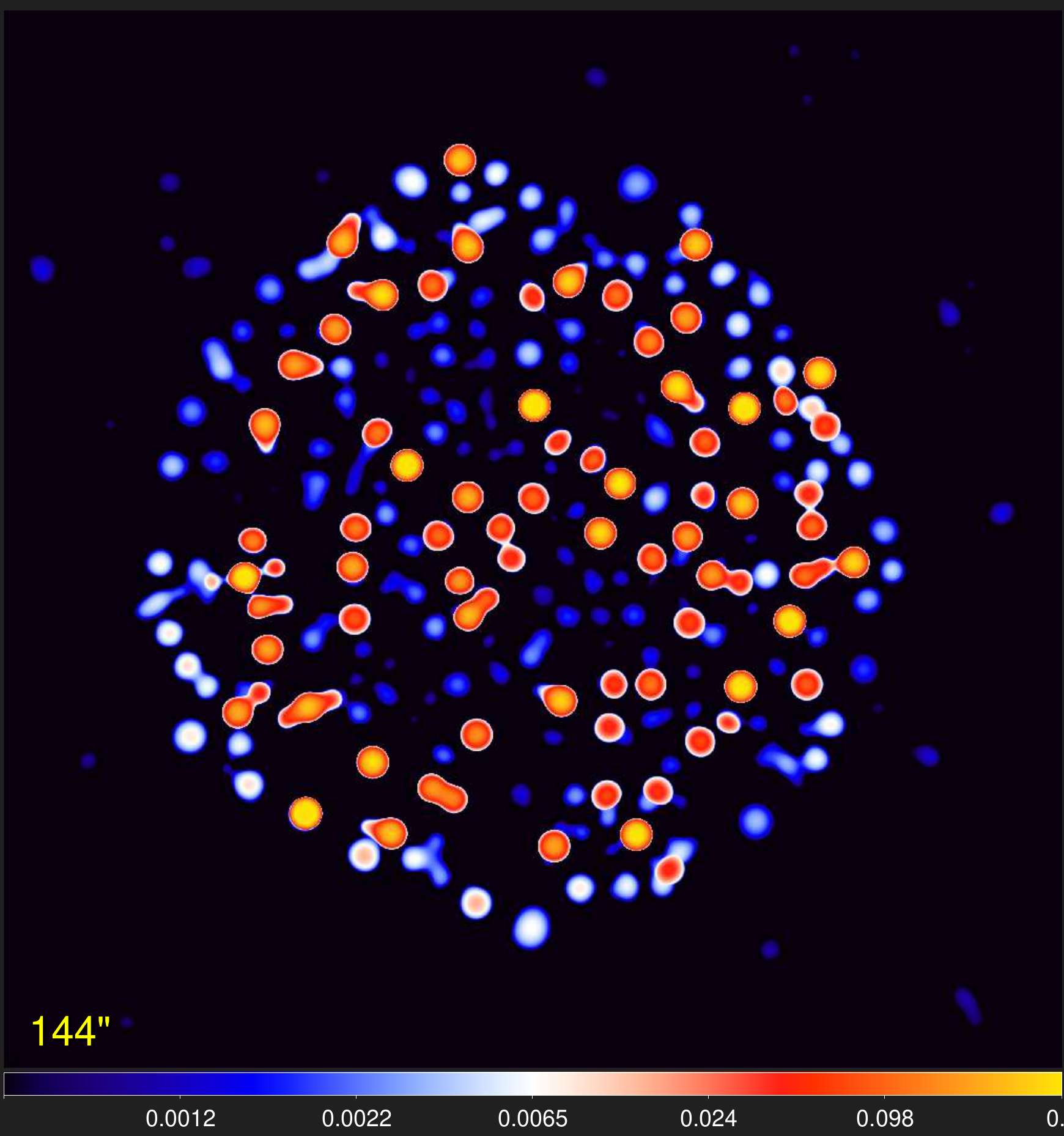}}
  \resizebox{0.328\hsize}{!}{\includegraphics{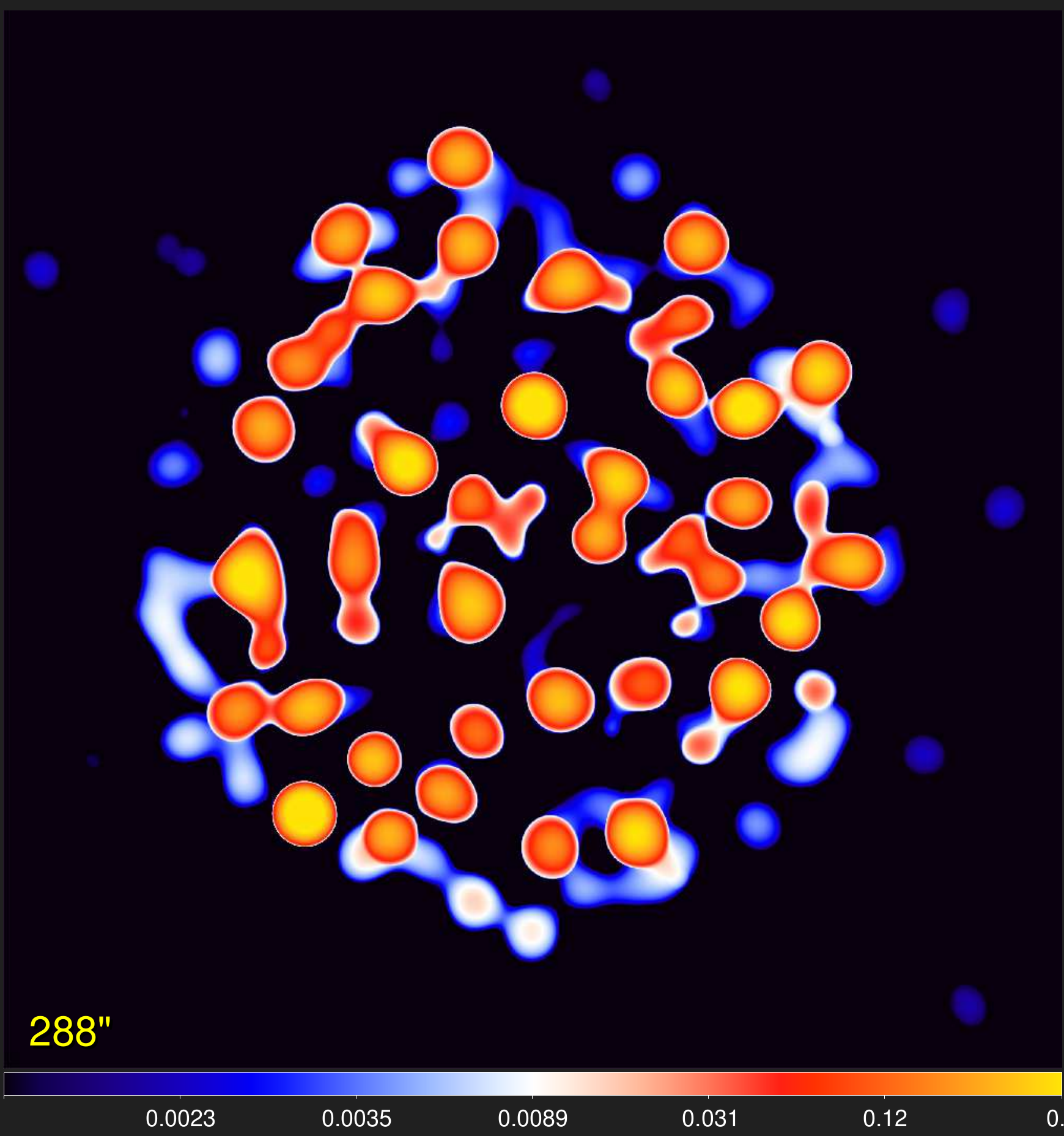}}}
\caption
{ 
Combination of the detection images $\mathcal{S}_{{\lambda}{\rm D}{j}{\rm C}}$ (Sect.~\ref{cleaning}) for the set of images
$\mathcal{I}_{{\!\lambda}}$ containing all \emph{Herschel} wavebands and $\mathcal{I}_{{\!\lambdabar}} \equiv
\mathcal{D}_{13{\arcsec}}$ from Eq.~(\ref{superdens}). The clean $\mathcal{S}_{{\rm D}{j}{\rm C}}$ thresholded above
$\varpi_{{\lambda}{\rm S}{j}}{\,=\,}5\sigma_{{\!\lambda}{\rm S}{j}}$ and combined over all wavebands are shown. Several faint spurious peaks
visible on large scales near edges in the \emph{bottom} row are the background and noise fluctuations that happened to be stronger
than the threshold. They may be discarded during the subsequent detection and measurement steps. Logarithmic color mapping.
} 
\label{flatsrccomb}
\end{figure*}

\begin{figure*}                                                               
\centering
\centerline{
  \resizebox{0.328\hsize}{!}{\includegraphics{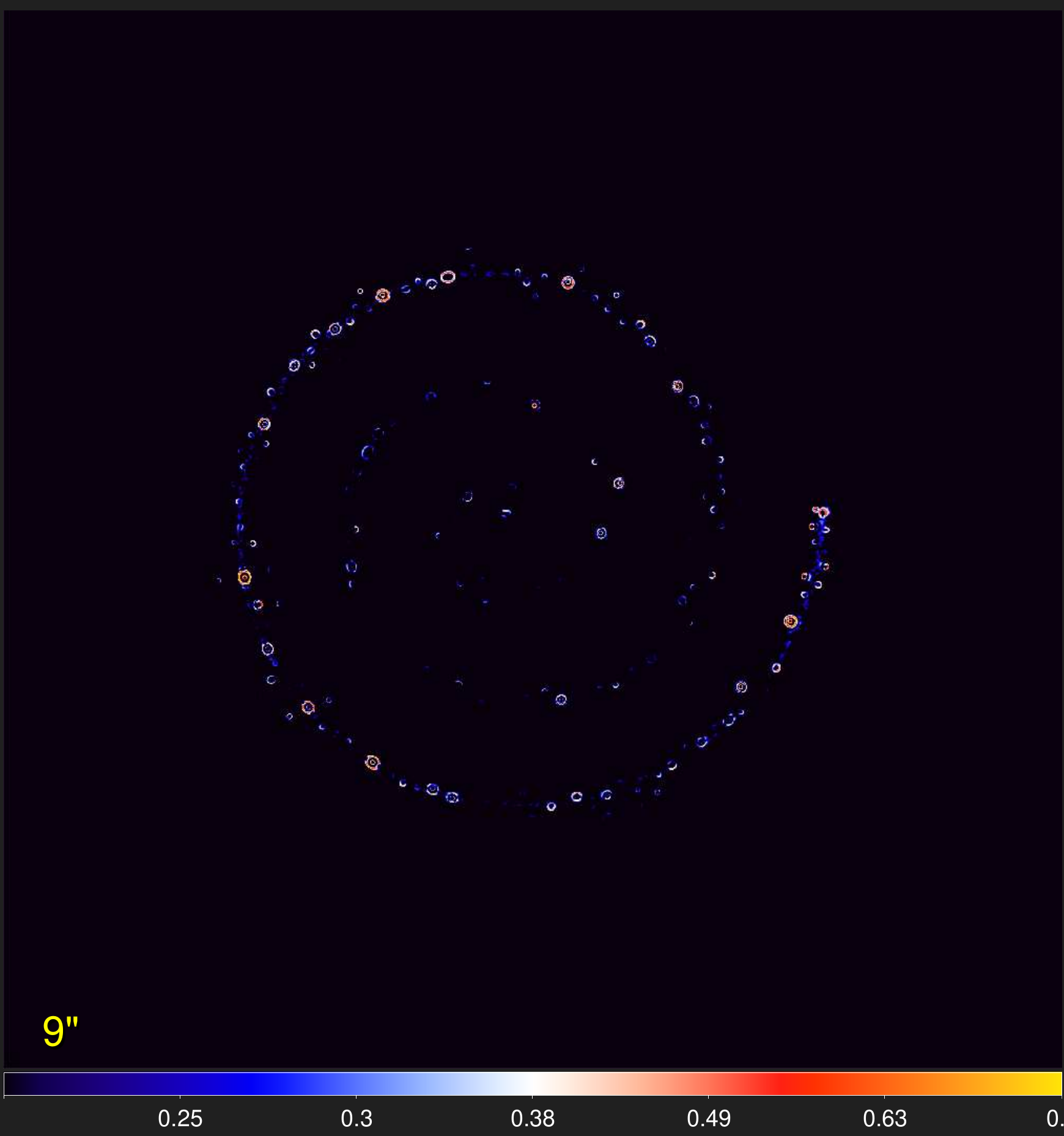}}
  \resizebox{0.328\hsize}{!}{\includegraphics{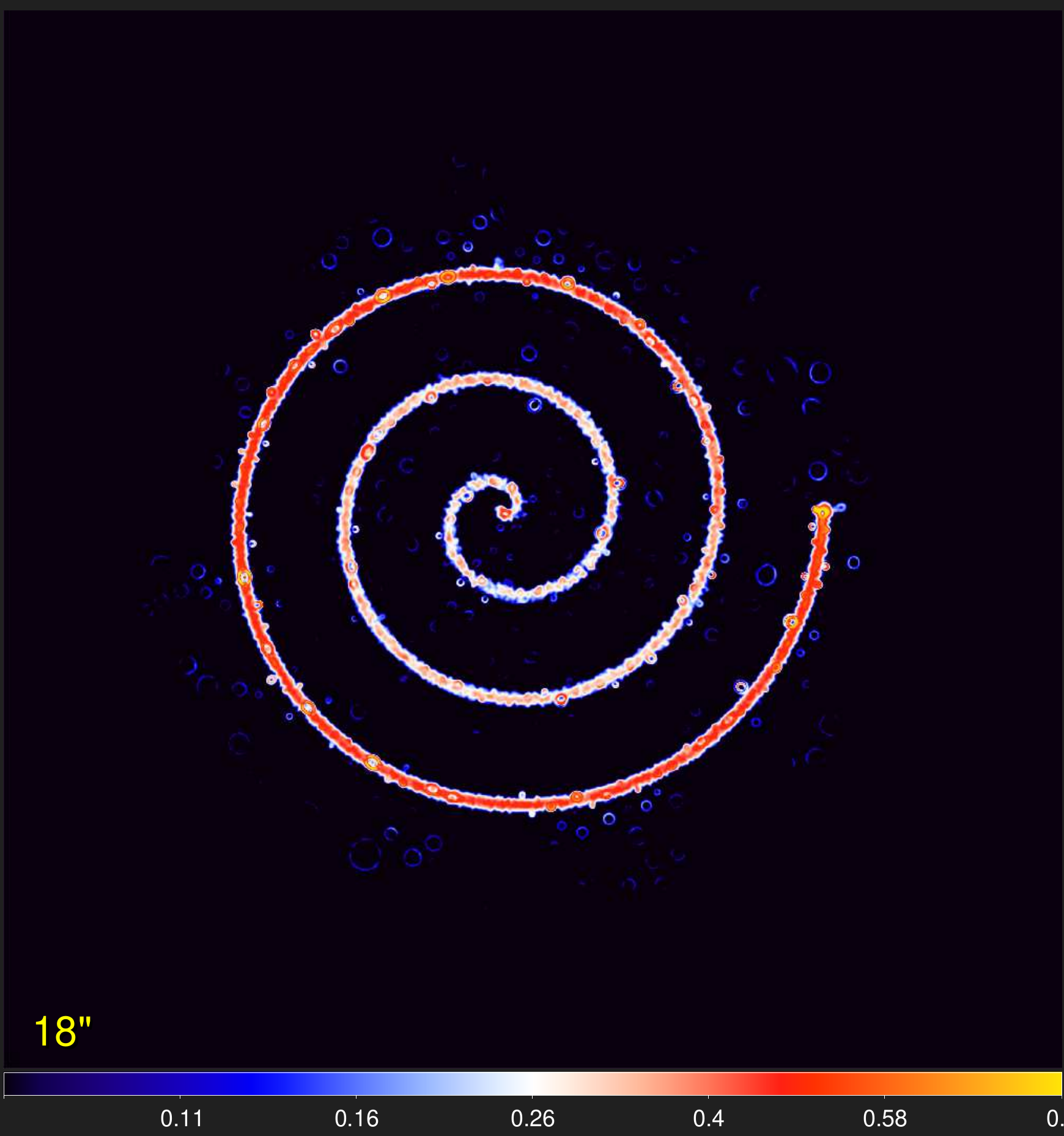}}
  \resizebox{0.328\hsize}{!}{\includegraphics{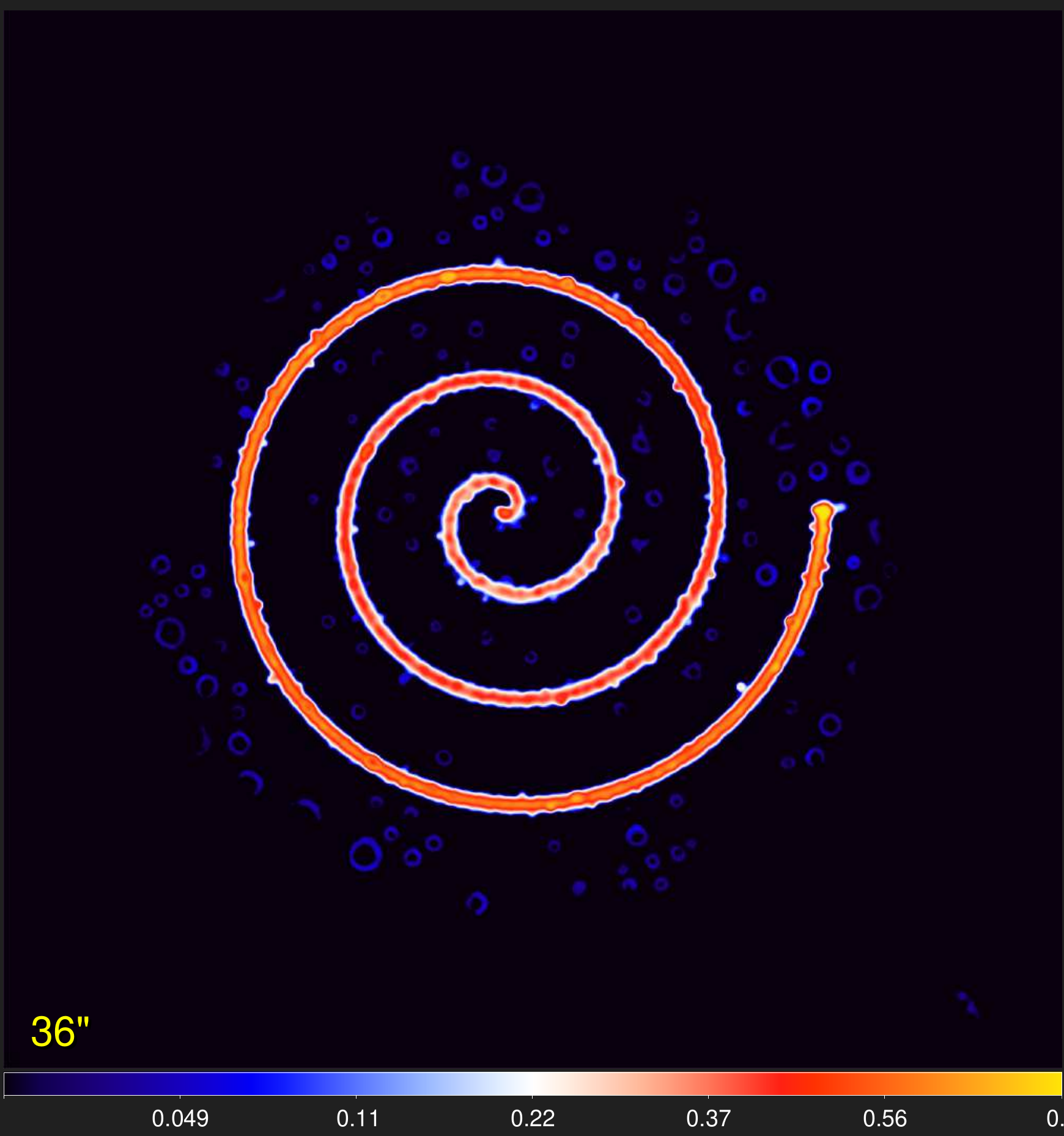}}}
\vspace{0.5mm}
\centerline{
  \resizebox{0.328\hsize}{!}{\includegraphics{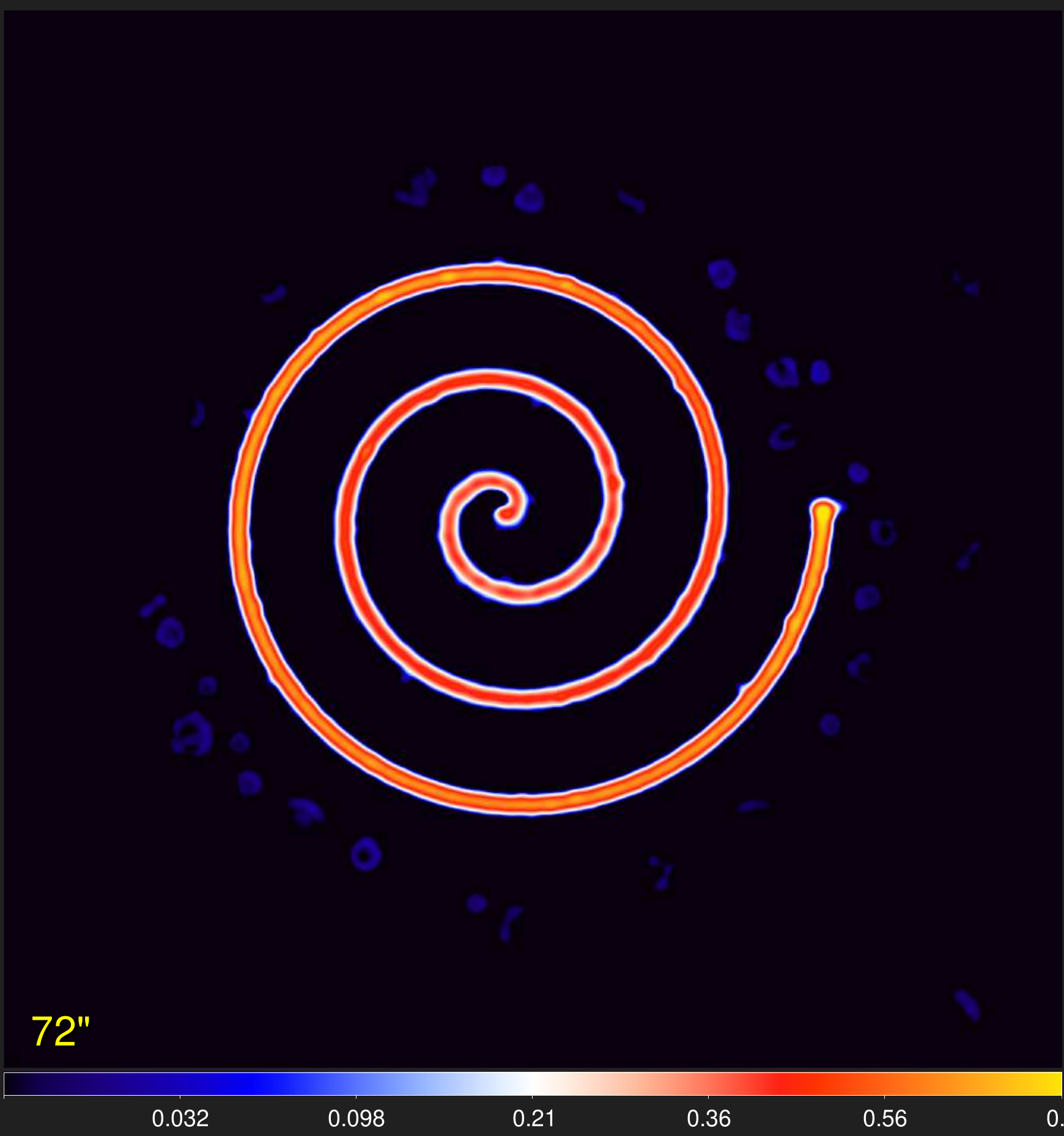}}
  \resizebox{0.328\hsize}{!}{\includegraphics{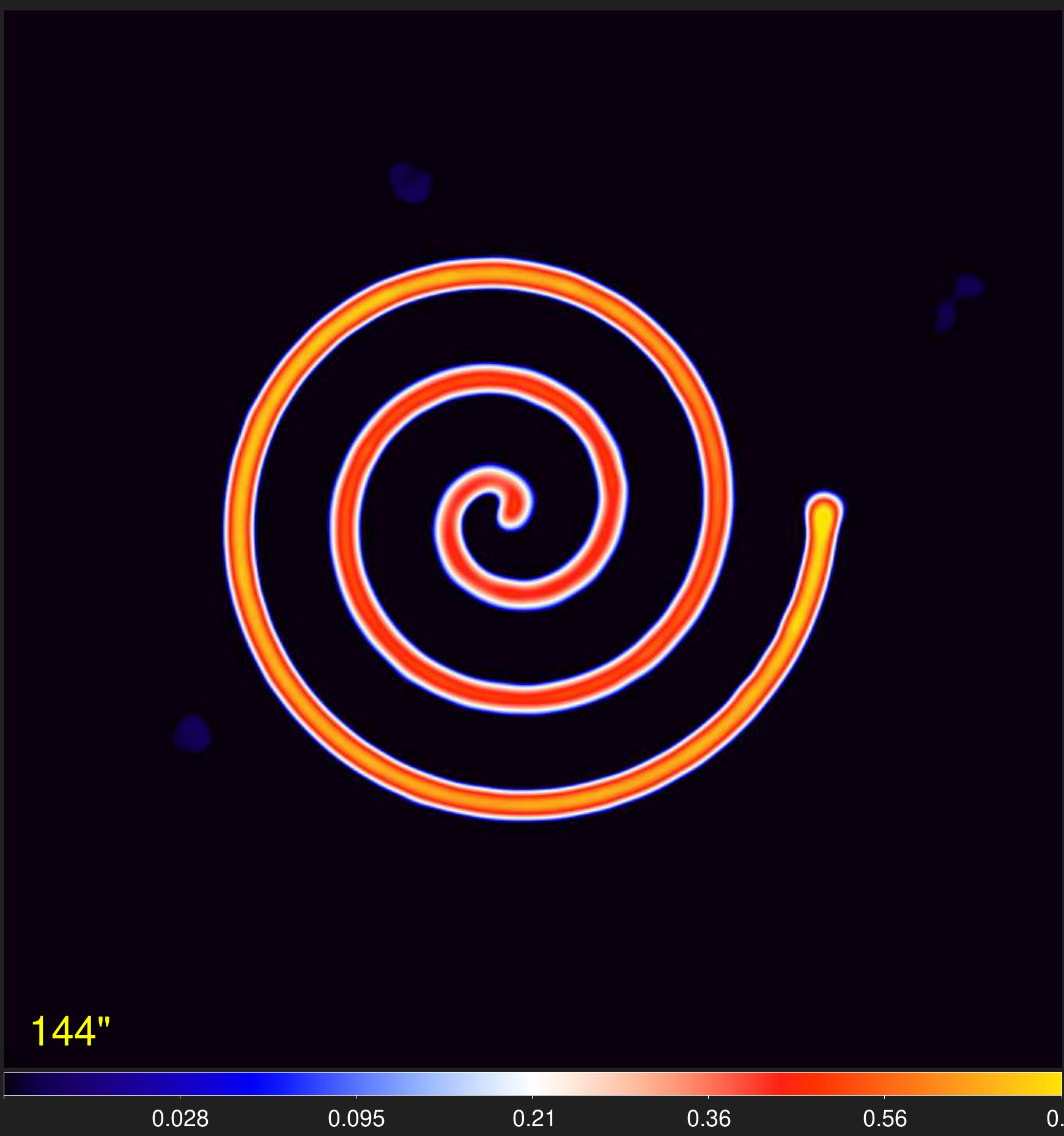}}
  \resizebox{0.328\hsize}{!}{\includegraphics{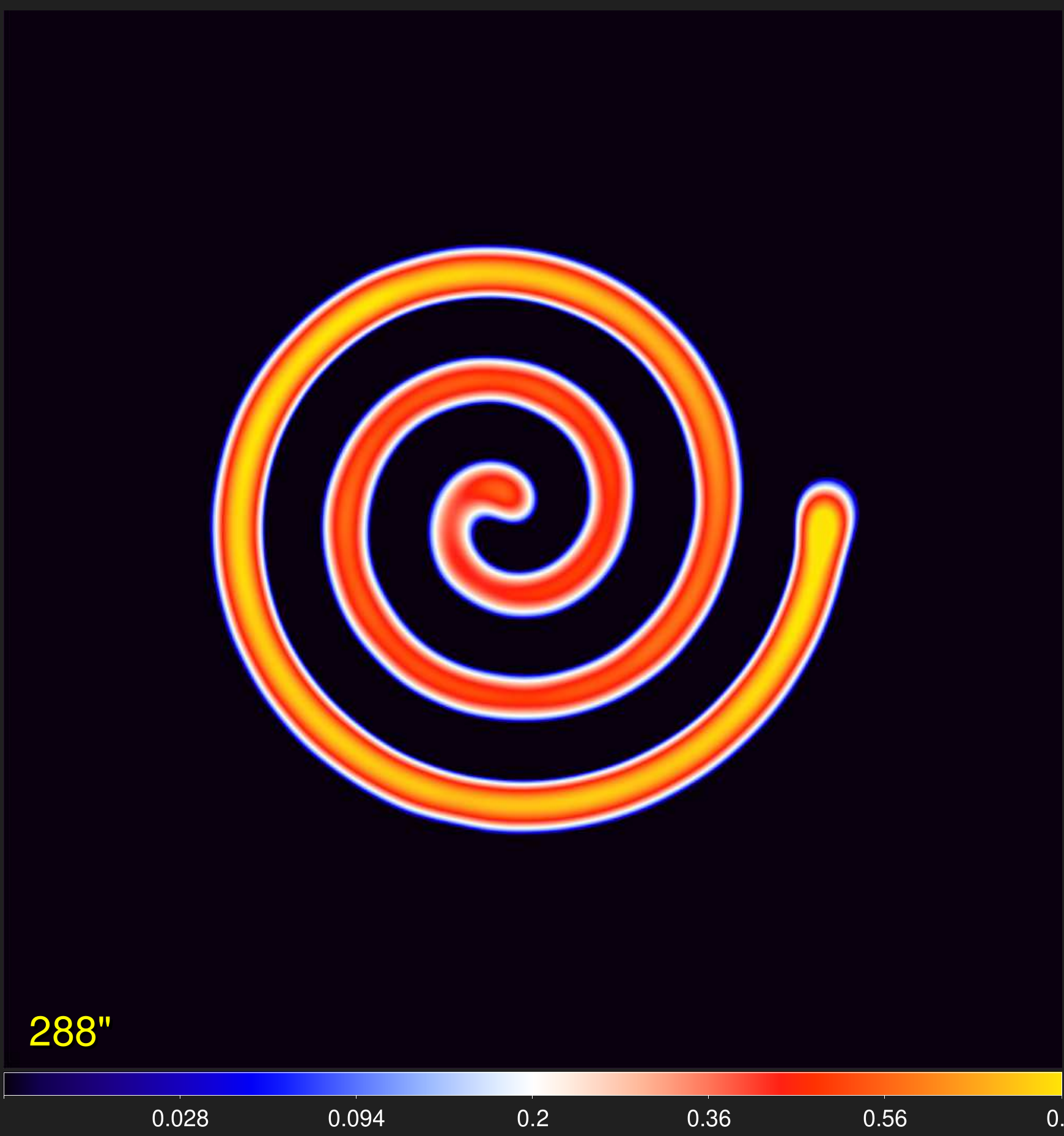}}}
\caption
{ 
Combination of the detection images $\mathcal{F}_{{\lambdabar}{\rm D}{j}{\rm C}}$ (Sect.~\ref{cleaning}) for the set of images
$\mathcal{I}_{{\!\lambda}}$ containing all \emph{Herschel} wavebands and $\mathcal{I}_{{\!\lambdabar}} \equiv
\mathcal{D}_{13{\arcsec}}$ from Eq.~(\ref{superdens}). The clean $\mathcal{F}_{{\rm D}{j}{\rm C}}$ thresholded above
$\varpi_{{\lambda}{\rm F}{j}}{\,=\,}2\sigma_{{\!\lambda}{\rm F}{j}}$ and combined over five wavebands are shown (excluding the noisier $70$ and
$100$\,{${\mu}$m} images). The faint ring-like structures that are visible on some scales are the source residuals originating from the
derived surface densities that have substantial inaccuracies over the sources (cf. Figs.~\ref{addhires} and \ref{bgderivation};
Sect.~\ref{hiresinacc}). Square-root color mapping.
} 
\label{flatfilcomb}
\end{figure*}

The resulting clean images $\{\mathcal{S}|\mathcal{F}\}_{{\lambda}{\rm D}{j}{\rm C}}$ (Figs.~\ref{flatsrccomb} and
\ref{flatfilcomb}) are deemed to have signals only from the sources and filaments, respectively. In practice, some of them may have
several faint spurious peaks and other structures that are discarded during the subsequent detection and measurement steps.

\subsubsection{Combination of the clean single scales over $\lambda$}
\label{combining}

All previous image processing was done independently for each wavelength. It is recommended to always process the input images in
parallel, independent \textsl{getsf} runs to reduce the total extraction time approximately by a factor $N_{\rm W}$, the number of
wavelengths. Now, \textsl{getsf} accumulates the clean single-scale images $\mathcal{S}_{{\lambda}{\rm D}{j}{\rm C}}$ and
$\mathcal{F}_{{\lambda}{\rm D}{j}{\rm C}}$ over the wavebands in order to use the independent information from all images and
enhance the signal of the significant structures. This procedure follows the \textsl{getold} approach (Paper I), with the important
improvement that filaments are handled in the same way as sources. When decomposed images on each scale are combined, differences
in the angular resolutions between the wavebands are much less important because the single-scale images select and enhance the
structures with widths similar to the scale size $S_{\!j}$, not the resolution $O_{\lambda}$.

The clean detection images are normalized before their accumulation over wavelengths to make all cleaning thresholds equal to unity
in all bands. The combination process is described by the following expression:
\begin{equation} 
\{\mathcal{S}|\mathcal{F}\}_{{\rm D}{j}{\rm C}} = N^{-1}_{\{\rm S|F\}} \sum_{\lambda} f_{{\lambda}{j}}
\max\left(\{\mathcal{S}|\mathcal{F}\}_{{\lambda}{\rm D}{j}{\rm C}}, 
\mathcal{Z}_{{\lambda}{\{\rm S|F\}}{j}}\right) \varpi^{-1}_{{\lambda}{\{\rm S|F\}}{j}},
\label{comboimg1}
\end{equation} 
where $N_{\{\rm S|F\}}{\,\le\,}N_{\rm W}$ is the number of the wavebands chosen to be used in the combination,
$\mathcal{Z}_{{\lambda}{\{\rm S|F\}}{j}}$ is the threshold image (equal to $\varpi_{{\!\lambda}{\{\rm S|F\}}{j}}$ in all pixels),
and $f_{{\lambda}{j}}$ is a factor that gradually turns the smallest scales on,
\begin{equation} 
f_{{\lambda}{j}} = \min\left(\left(S_{\!j}\,O^{-1}_{\lambda}\right)^{3},1\right).
\label{turnonfact}
\end{equation} 
This factor ensures that the small-scale noise or artifacts appearing on top of the resolved structures do not produce spurious
detections in the combined images on scales $S_{\!j}{\,<\,}O_{\lambda}$. Sufficiently bright unresolved structures still contribute
to $\{\mathcal{S}|\mathcal{F}\}_{{\rm D}{j}{\rm C}}$ on the smallest scales below $O_{\lambda}$. This super-resolution is useful
to detect blended unresolved peaks. Selected combined images of the two structural components are shown in
Figs.~\ref{flatsrccomb} and \ref{flatfilcomb}.

The normalization to a common threshold in Eq.~(\ref{comboimg1}) is a natural way of maximizing sensitivity of the combined images.
This procedure modifies the original dependence of the source brightness on spatial scales, however, which is analyzed by the
detection algorithm (Sect.~\ref{srcdetection}) to determine the characteristic size for each source. Therefore a second set of
combined images is defined for the component of sources, normalized to the smallest scale in each waveband,
\begin{equation} 
\tilde{\mathcal{S}}_{{\rm D}{j}{\rm C}} = \sum_{\lambda} \frac{w_{{\lambda}}}{\mathcal{S}_{{\lambda}{\rm D}{1}{\rm C}}}\,
\mathcal{S}_{{\lambda}{\rm D}{j}{\rm C}},
\label{comboimg2}
\end{equation} 
where $w_{\lambda}$ is the weight that enhances the contribution of the images with higher angular resolutions,
\begin{equation} 
w_{{\lambda}} = \left(\frac{\bar{O}}{O_{\lambda}}\right)^{7}, \,\,\, \bar{O} = N^{-1}_{\rm W} \sum_{\lambda} O_{\lambda},
\label{weighting}
\end{equation} 
where $\bar{O}$ is the average resolution, and the power of $7$ ensures complete separation of the contributions of different
wavebands in Eq.~(\ref{comboimg2}). After the weighting, the summation of $\mathcal{S}_{{\lambda}{\rm D}{j}{\rm C}}$ preserves the
individual dependence of the peak intensity of each source on spatial scales, which provides an initial estimate of its size during
detection before the actual measurements.


\subsubsection{Detection of sources in the combined images}
\label{srcdetection}

Sources are detected in $\mathcal{S}_{{\rm D}{j}{\rm C}}$ with almost the same algorithm that was used by \textsl{getold}
(Sect.~2.5 of Paper I), which is briefly summarized here for completeness. An inspection of the entire set of single-scale images
$\mathcal{S}_{{\rm D}{j}{\rm C}}$ shows that sources appear on relatively small scales become the brightest on scales roughly
equal to their size and vanish on significantly larger scales (cf. Fig.~\ref{gaussdecomposed}). All detectable sources appear
isolated on small scales and become blended with other nearby sources on larger scales. The \textsl{getsf} source detection scheme
identifies the sources in $\mathcal{S}_{{\rm D}{j}{\rm C}}$ and tracks their evolution from small to large scales, until they
disappear or merge with a nearby brighter source.

To detect sources, \textsl{getsf} slices $\mathcal{S}_{{\rm D}{j}{\rm C}}$ by a number $N_{\rm L}$ of intensity levels
$I_{{j}{l}}$, spaced by $\delta\ln{I_{{j}}}{\,=\,}0.01$, from the image maximum down to the lowest non-zero value. Each slice $l$
cuts through all peaks brighter than $I_{{j}{l}}$, producing a set of partial images,
\begin{equation} 
\mathcal{S}_{{\rm D}{j}{\rm C}{l}} = \max\left(\mathcal{S}_{{\rm D}{j}{\rm C}}, I_{{j}{l}}\right), \,\,\, 
{l{\,=\,}1, 2,\dots, N_{\rm L}}.
\label{levelsd} 
\end{equation} 
The source detection algorithm works on the sequence of partial images, creating and updating source segmentation masks (for each
$j$ and $l$). This is done with the same \textsl{tintfill} algorithm that was applied in Sect.~\ref{clipping} to remove the source-
and filament-like shapes. The resulting single-scale segmentation images of sources sets all pixels belonging to a source to its
number.

The scale $j_{\rm F}$ on which a source $n$ becomes the brightest is referred to as the footprinting scale. It provides an
initial estimate for its half-maximum size $H_{n}{\,=\,}S_{\!j_{\rm F}}$ (cf. Appendix~\ref{decomposition}), which defines the
initial footprint, that is, the entire area of all pixels making non-negligible contributions to the total flux of the source. From a
practical point of view, \textsl{getsf} defines the initial footprint diameter of a circular source as $\phi_{n} H_{n}$, where the
footprint factor $\phi_{n}{\,=\,}3$. For the Gaussian sources (e.g., Fig.~\ref{gaussdecomposed}), these footprints lead to the total
fluxes that are underestimated by only $1.6\%$, well within the usual measurement uncertainties. Having detected the sources,
\textsl{getsf} creates their initial footprints with the diameters $\{A,B\}_{{\rm F}{n}}{\,=\,}\phi_{n} H_{n}$. The footprints
become elliptically shaped in the wavelength-dependent measurements, reflecting the elongation of sources that is obtained from intensity
moments. During the measurement iterations (Sect.~\ref{srcmeasurement}), \textsl{getsf} changes $\phi_{n}$ to expand or shrink the
footprints for those sources that are bright enough and whose intensity distributions indicate that their initial footprints are
not optimal.

\begin{figure*}
\centering
\centerline{
  \resizebox{0.328\hsize}{!}{\includegraphics{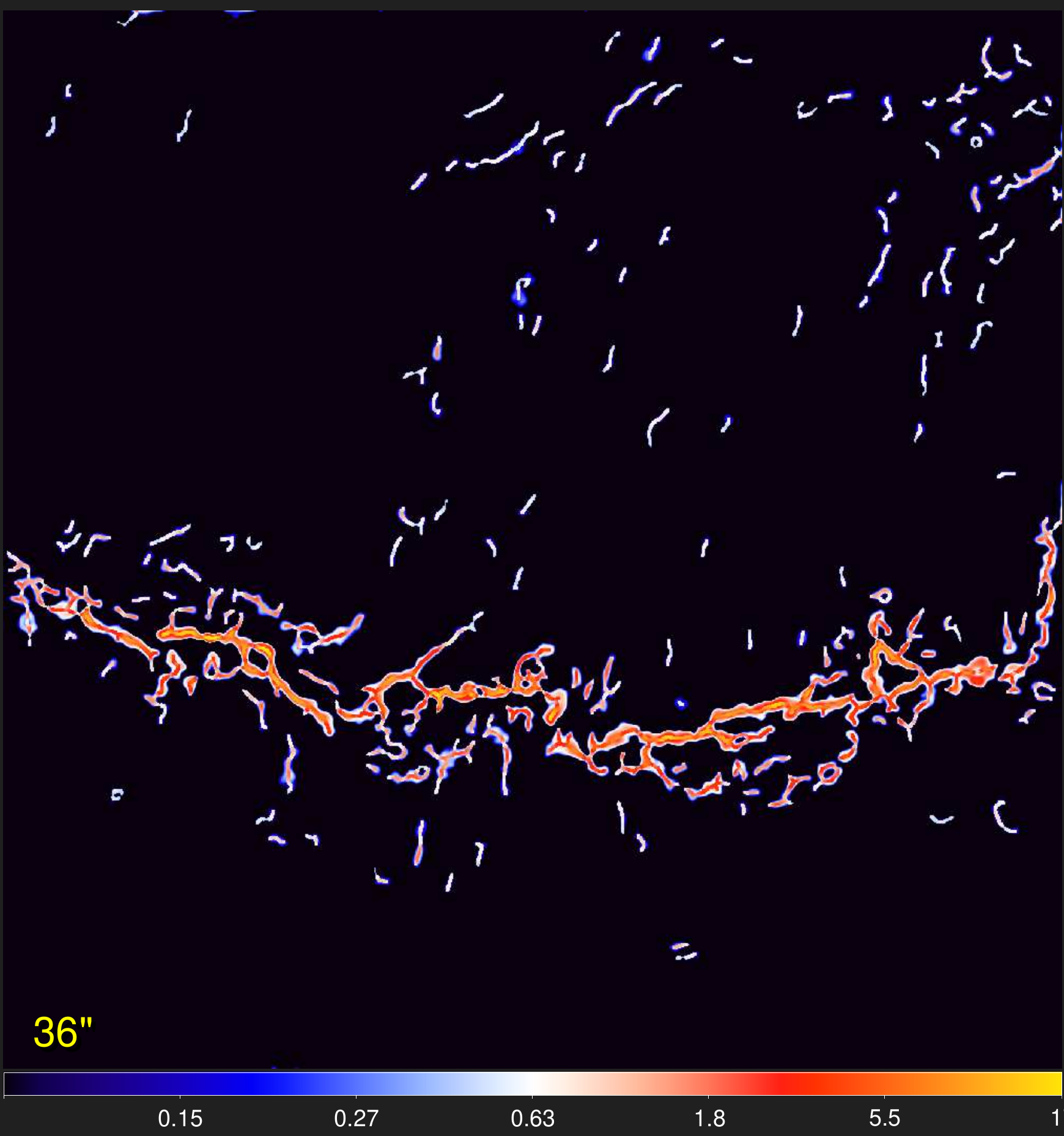}}
  \resizebox{0.328\hsize}{!}{\includegraphics{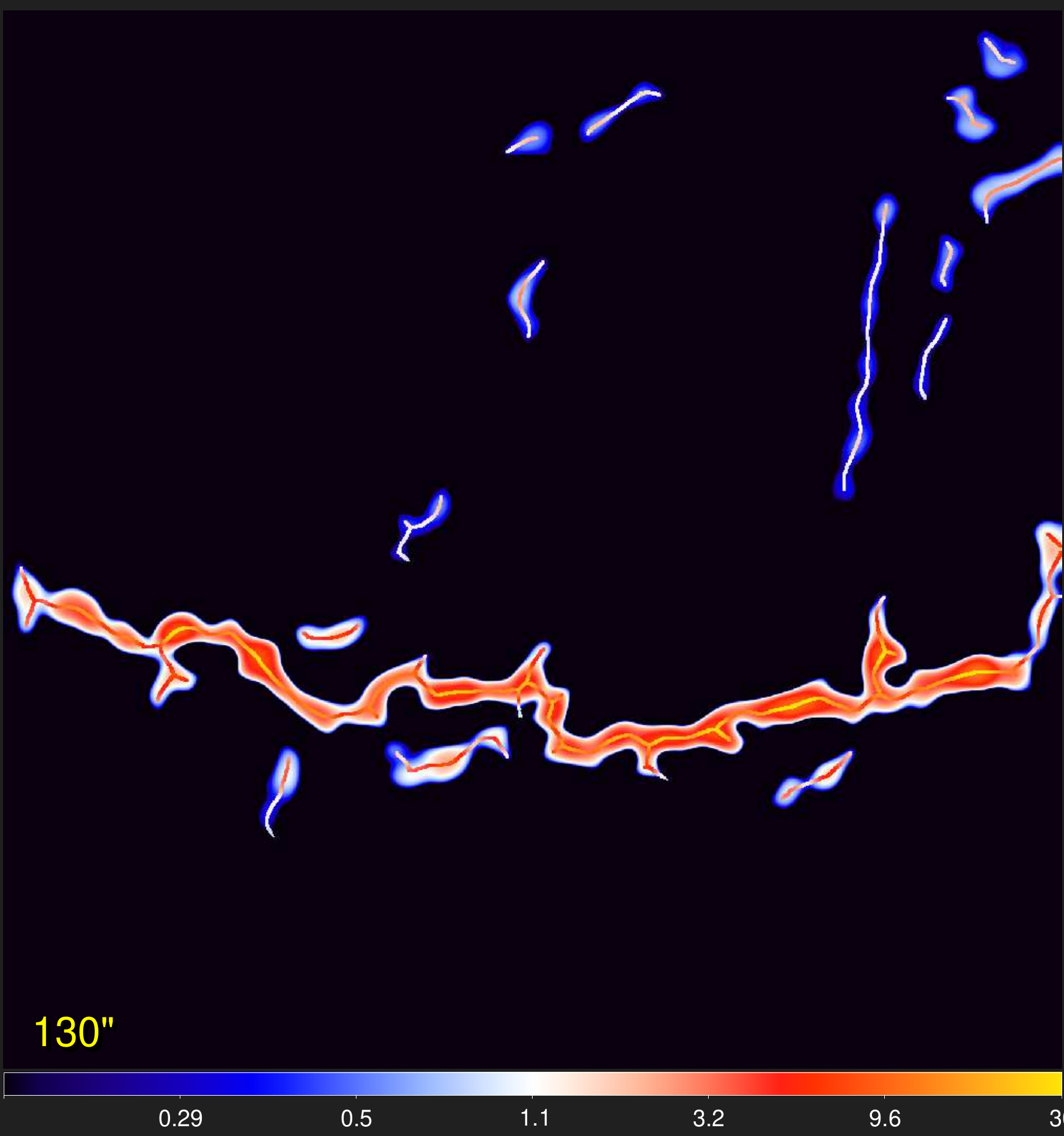}}
  \resizebox{0.328\hsize}{!}{\includegraphics{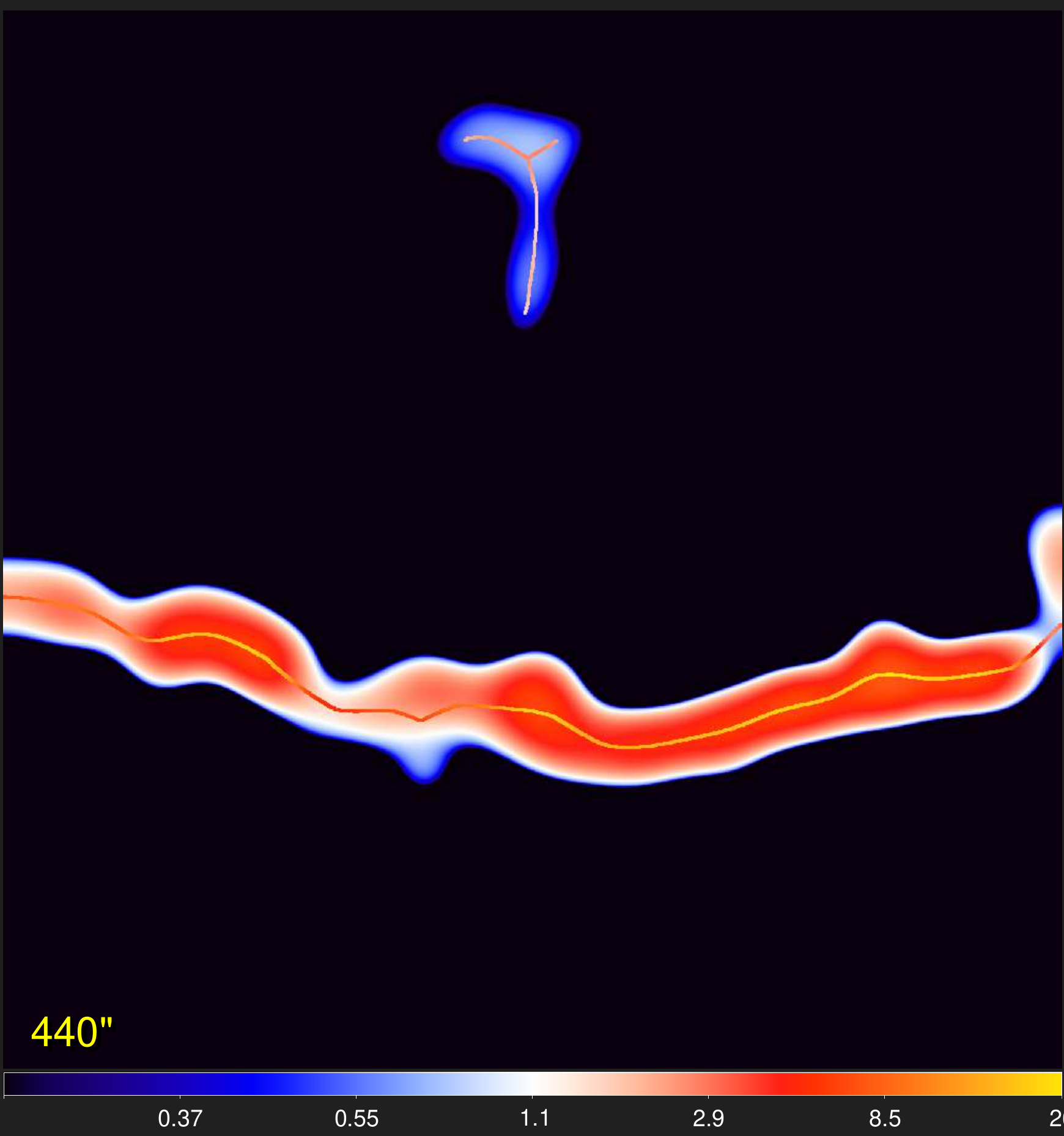}}}
\vspace{0.5mm}
\centerline{
  \resizebox{0.328\hsize}{!}{\includegraphics{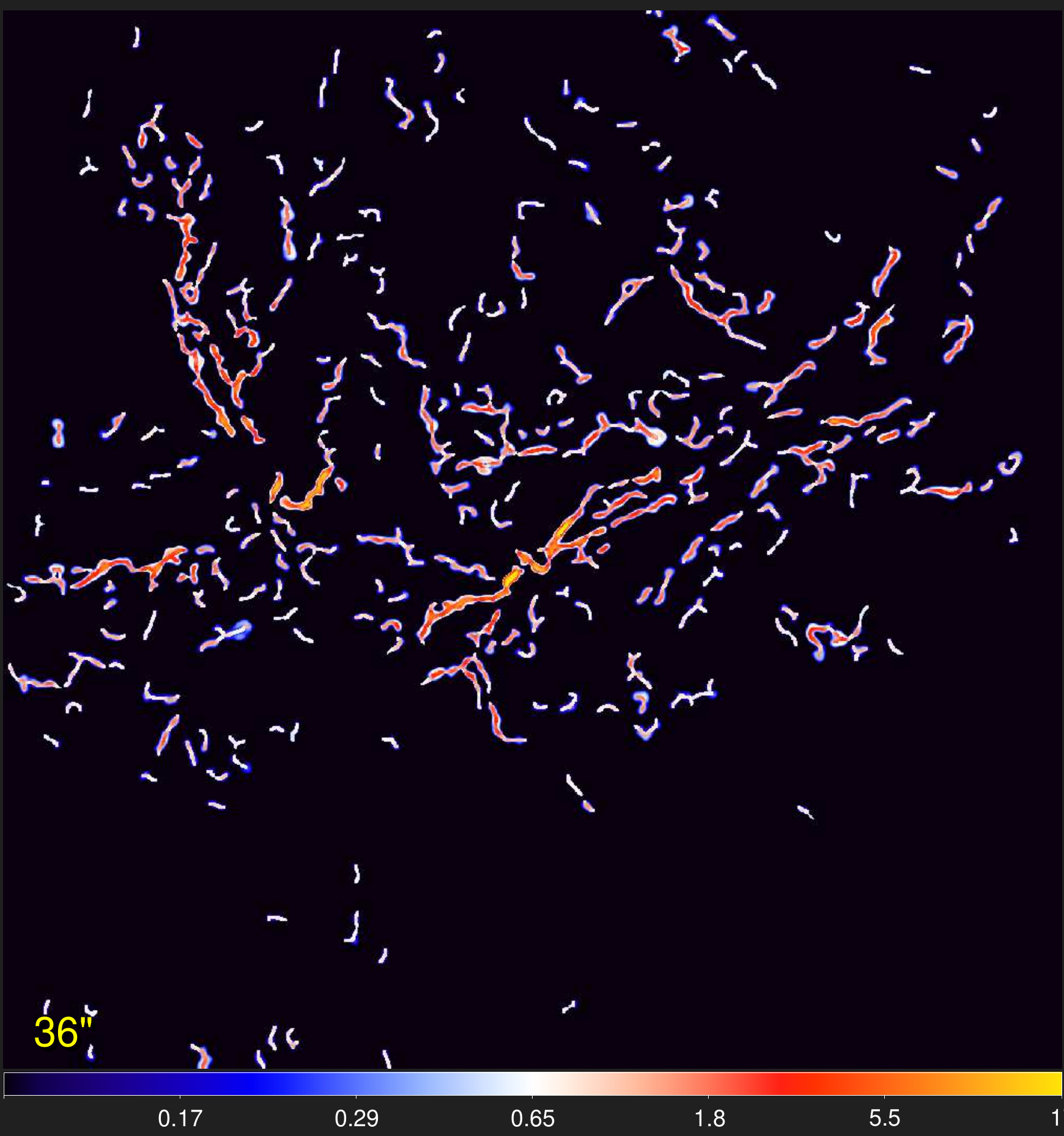}}
  \resizebox{0.328\hsize}{!}{\includegraphics{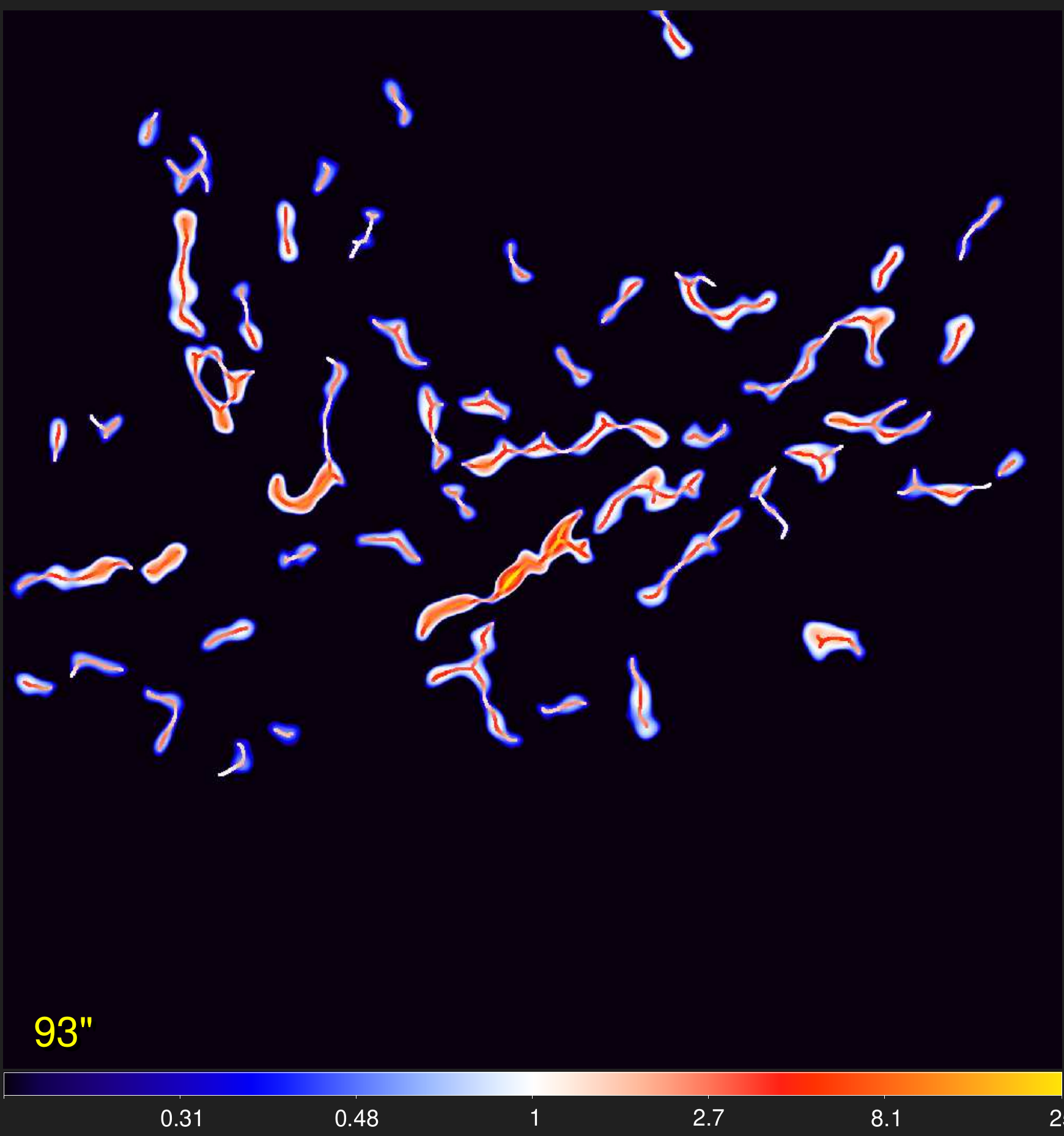}}
  \resizebox{0.328\hsize}{!}{\includegraphics{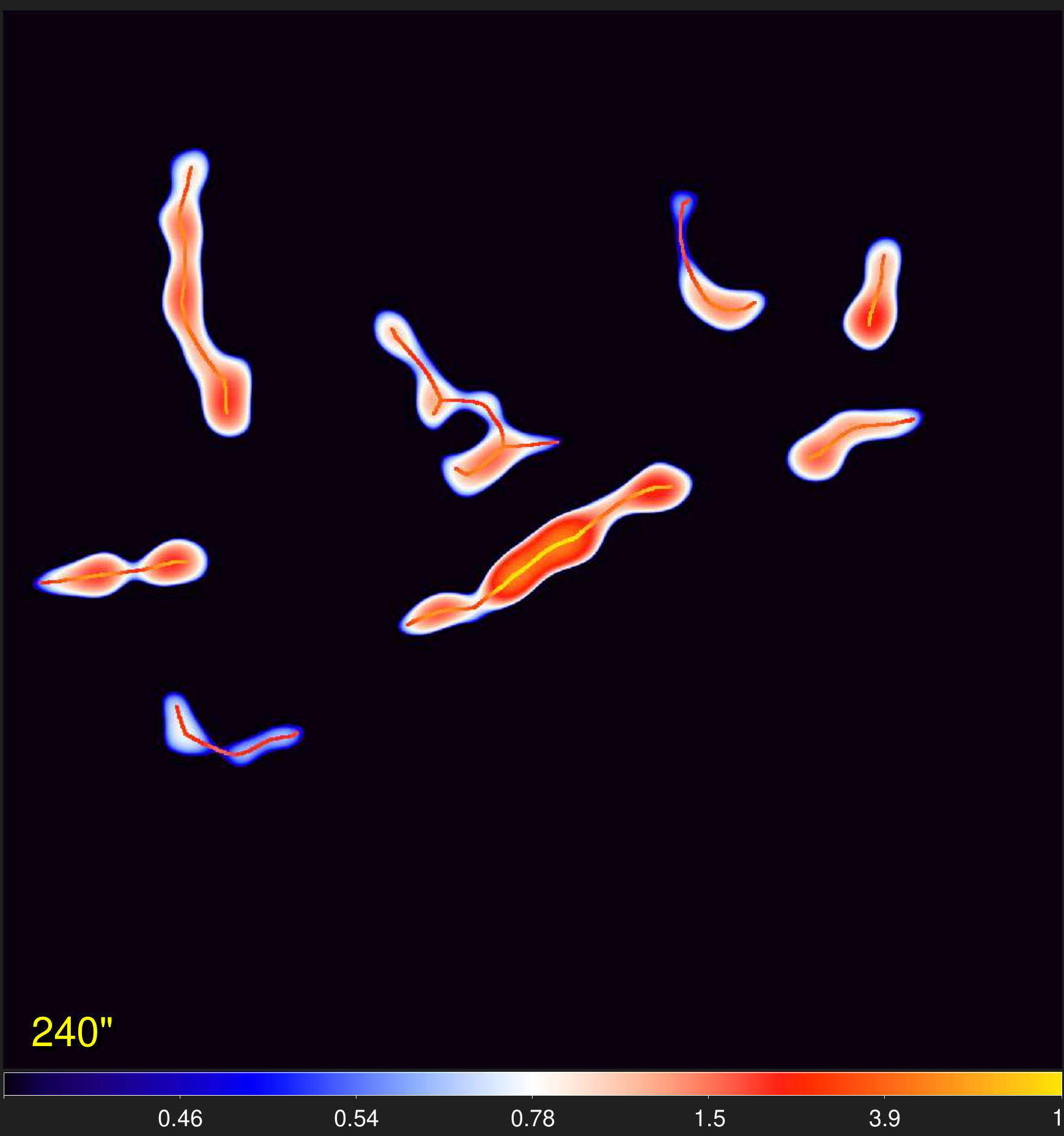}}}
\vspace{0.5mm}
\centerline{
  \resizebox{0.328\hsize}{!}{\includegraphics{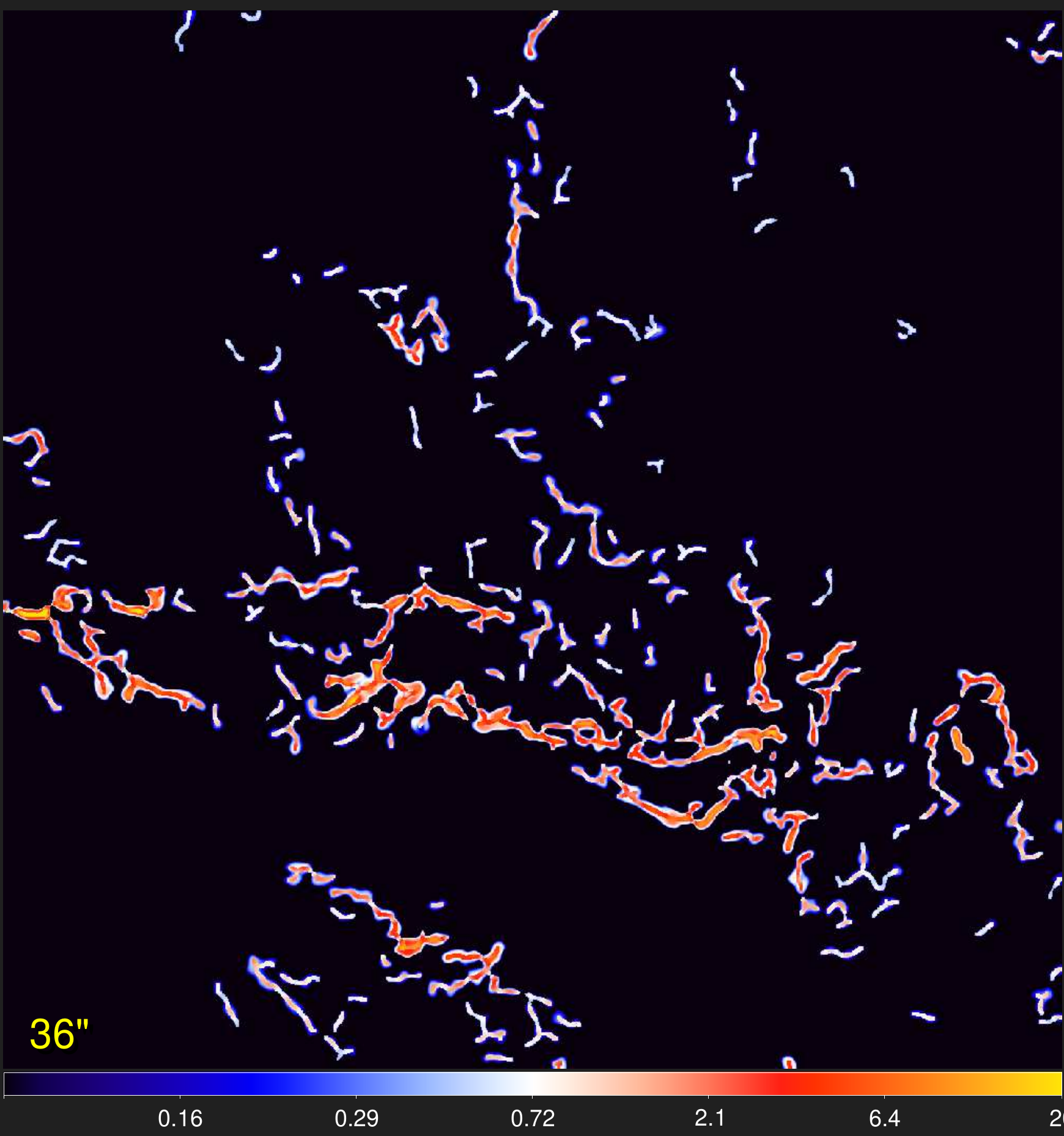}}
  \resizebox{0.328\hsize}{!}{\includegraphics{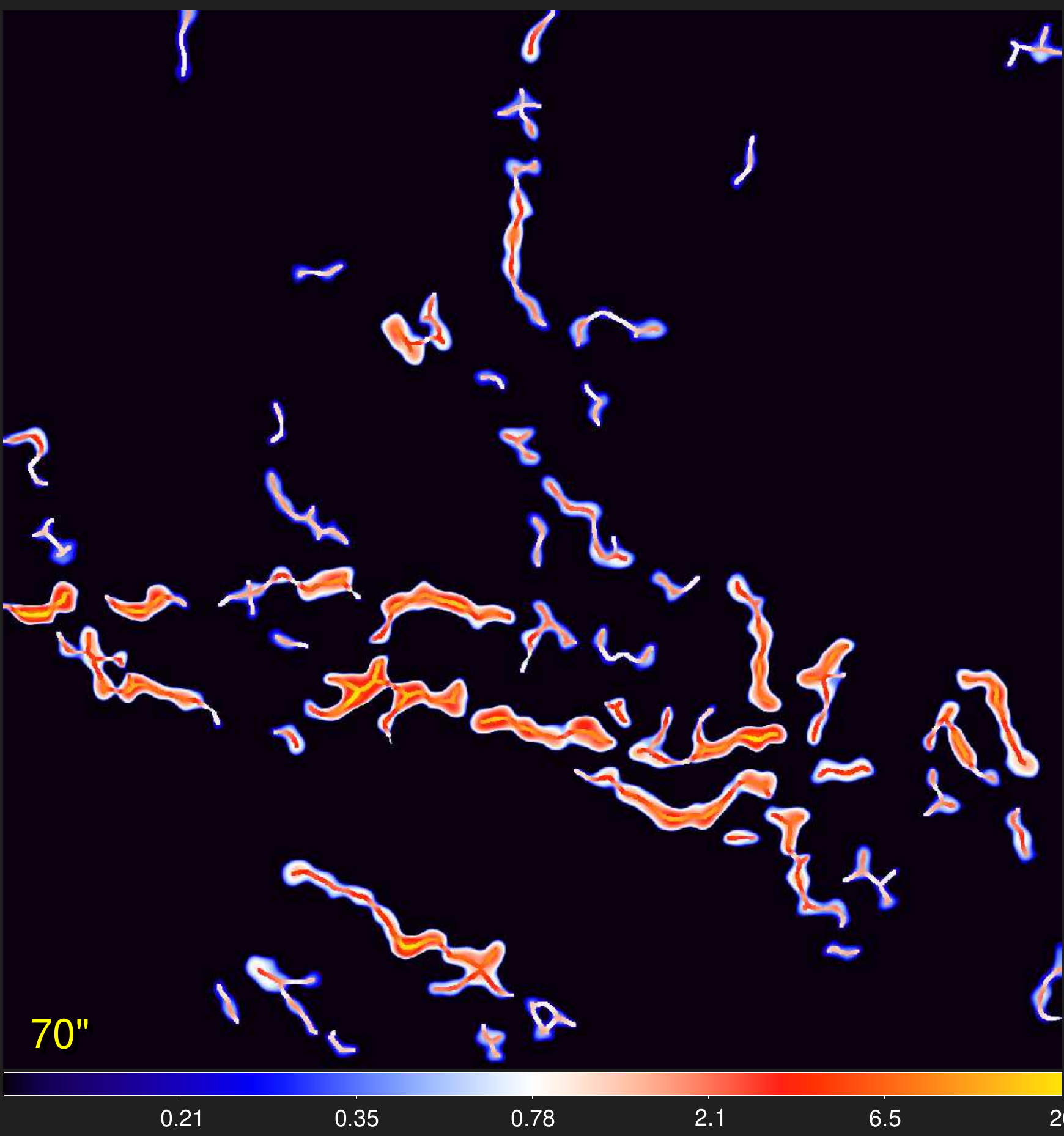}}
  \resizebox{0.328\hsize}{!}{\includegraphics{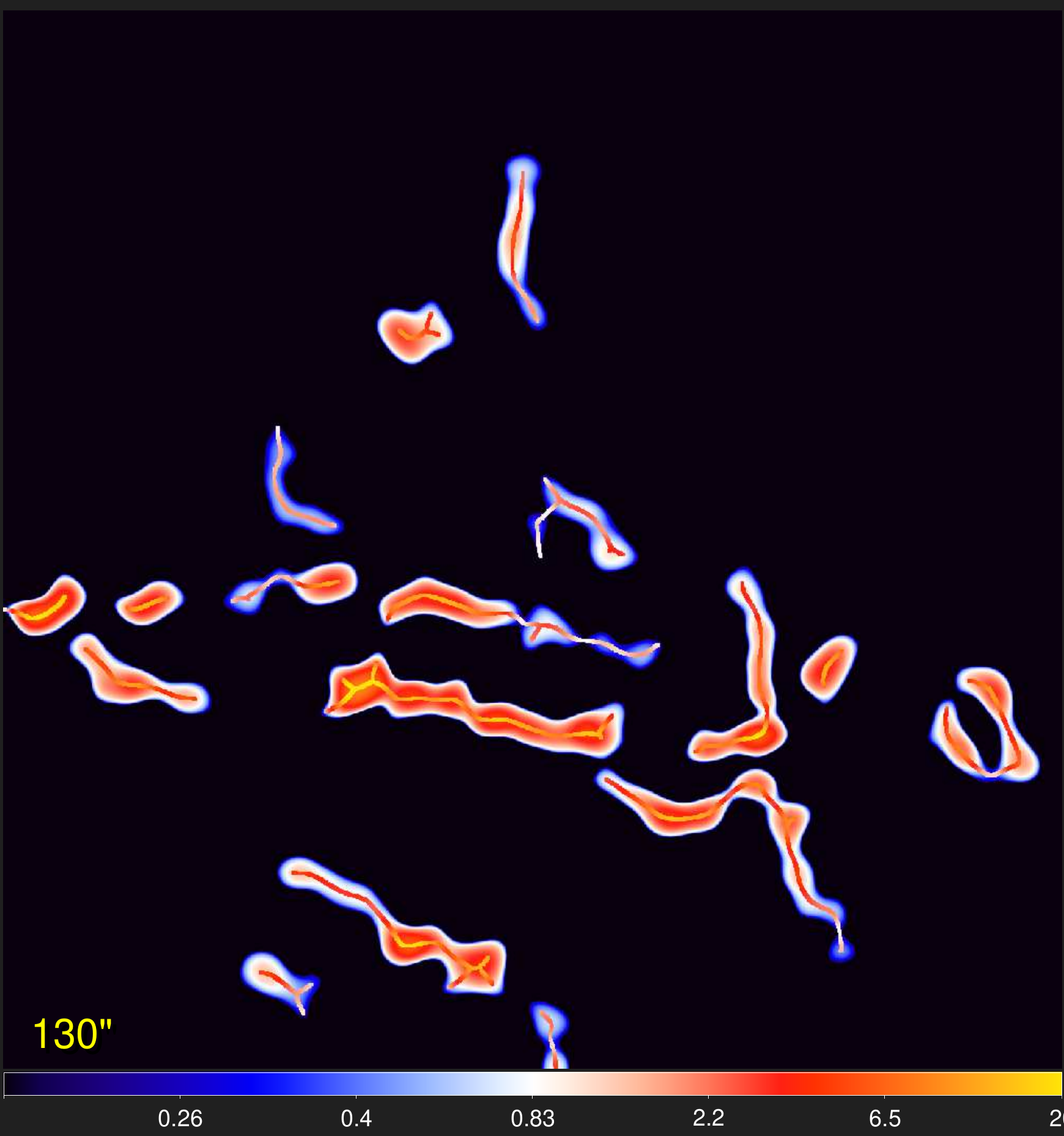}}}
\caption
{ 
Filaments extracted by \textsl{getsf} on selected spatial scales in three star-forming regions: \object{Taurus} (\emph{top}),
\object{Aquila} (\emph{middle}), and \object{IC\,5146} (\emph{bottom}). The flattened components
$\mathcal{F}_{{\!\lambdabar}{\rm D}}$ derived from the \textsl{hires} surface densities $\mathcal{D}_{13{\arcsec}}$ obtained from
Eq.~(\ref{superdens}) using the \emph{Herschel} $160$, $250$, $350$, and $500$\,{${\mu}$m} images are shown. The minimum scales of
$36${\arcsec} (\emph{left} column) correspond to $2.8$ times the angular resolution, whereas the maximum scales (\emph{right}
column) correspond to $0.3$\,pc at the adopted distances of the regions ($140$, $260$, and $460$\,pc, respectively). Intermediate
scales between the two extremes are displayed in the \emph{middle} column. The images were cleaned using the default threshold
$\varpi_{{\lambdabar}{\rm F}{j}}{\,=\,}2\sigma_{{\!\lambdabar}{\rm F}{j}}$. Overlaid on the filaments are their skeletons obtained
from the images using the Hilditch algorithm (Sect.~\ref{fildetection}). The observed filaments are heavily substructured, and
their appearance, detected skeletons, and measured properties depend strongly on spatial scales. Logarithmic color mapping.
} 
\label{ssfilams}
\vspace{20mm}
\end{figure*}


\subsubsection{Detection of filaments in the combined images}
\label{fildetection}

Filaments are detected in $\mathcal{F}_{{\rm D}{j}{\rm C}}$ with a completely new approach. In the \textsl{getold} algorithm
(Sect.~2.4.4 of Paper II), intensity profiles at each pixel of the component of filaments are measured in four directions, and the
pixel is deemed to belong to the crest (marks a skeleton point) if it has the highest value for each of the profiles. In practice,
this simple approach sometimes creates artifacts at the filament end points, where the skeletons sometimes appear forked like a
snake tongue. An important limitation of the \textsl{getold} skeletons is that they trace crests of the images of filaments without
any dependence on the spatial scales.

The \emph{Herschel} observations of nearby star-forming molecular clouds demonstrated that filaments are very complex, multiscale
structures \citep[e.g.,][]{Men'shchikov_etal2010}, unlike the simple case of the relatively round sources, whose intensities
rapidly decrease in all directions from their peaks. Resolved sources are produced by the emission of dense, compact objects and
may be reasonably well characterized by a single value of their half-maximum size (or spatial scale). In contrast to the sources,
detection of filaments is fundamentally a scale-dependent problem, and a single skeleton that may be appropriate for a certain
spatial scale cannot fully describe the complexity of the observed multiscale, profoundly substructured filaments. Resolved
filaments often appear to be composed of thinner filaments on smaller scales, down to the angular resolution, and their widths,
profiles, and crest intensities are quite variable along their skeletons.

The strong dependence of the observed filaments on spatial scales is illustrated in Fig.~\ref{ssfilams}, which shows the surface
densities of the filaments in three well-studied star-forming regions: \object{Taurus}, \object{Aquila}, and \object{IC\,5146}. The
observed images of the regions were downloaded from the \emph{Herschel} Gould Belt Survey \citep{Andre_etal2010}
archive\footnote{\url{http://gouldbelt-herschel.cea.fr/archives}}, and the \textsl{hires} surface densities
$\mathcal{D}_{13{\arcsec}}$ of the regions were computed from Eq.~(\ref{superdens}). Figure~\ref{ssfilams} shows the images of the
spatially decomposed filaments on three selected scales: small, intermediate, and large. The images demonstrate that the observed
filaments are highly substructured in the regions, and their shapes as traced by the skeletons are very different on various
spatial scales. The skeletons, obtained on the small scales, are completely incompatible with the shapes and crests of the
filaments on larger scales. The detected small-scale skeletons are often very curved, meandering back and forth even at the right
angles, which implies a high degree of self-blending and leads to significant inaccuracies in the measured profiles and other
derived properties of filaments. Therefore it is necessary to detect their skeletons on the scales that correspond to the widths of
the structures being studied. Moreover, the small-scale substructures of the larger-scale filaments may even be the key to
understanding the filament properties, the physical processes taking place inside them, and the formation of stars.

Instead of tracing the original image intensity profiles, \textsl{getsf} employs the Hilditch algorithm \citep{Hilditch1969}, which
skeletonizes two-dimensional shapes by erasing their outer pixels until the thinnest representation of the shapes is found. The
original Hilditch algorithm has a deficiency in that the shapes oriented along the two main diagonals become completely erased
during the iterations. To enable its application in \textsl{getsf}, the algorithm has been improved to preserve the diagonal
skeletons.

Through the multiscale decomposition, \textsl{getsf} allows finding crests without any explicit analysis of the filament
intensities. The single-scale images $\mathcal{F}_{{\rm D}{j}{\rm C}}$ not only enhance the structures of the widths
$W{\,\approx\,}S_{\!j}$, but also cause these filtered intensity distributions to become well centered on their zero-level
footprints. The crests of the isolated decomposed filaments always approximate the medial axes of their footprints (cf.
Figs.~\ref{flatfilcomb} and \ref{ssfilams}). If a single-scale filament blends with other filaments (or with itself), there is
always a smaller scale on which it is isolated. This allows determining the scale-dependent skeletons as the medial axes of the
zero-level filament masks.

The single-scale skeletons $\mathcal{K}_{j}$ are created using the Hilditch algorithm, with a width of three pixels to tolerate
one-pixel displacements in the skeleton coordinates between scales. They are further accumulated over a limited range of scales to
produce a set of $N_{\rm K}$ skeletons tracing the filamentary structures of various widths,
\begin{equation} 
\mathcal{K}_{k} = \sum^{J^{+}}_{j=J^{-}} \mathcal{K}_{j}, \,\,\, {k{\,=\,}1, 2,\dots, N_{\rm K}},
\label{partial}
\end{equation} 
where $J^{-}$ and $J^{+}$ are the numbers of the smallest and the largest scales, $S_{\!J^{-}}{=\,}2^{-1/2} S_{\!k}$ and
$S_{\!J^{+}}{=\,}2^{+1/2} S_{\!k}$, in the accumulated skeleton $\mathcal{K}_{k}$. The scale-dependent skeletons $\mathcal{K}_{k}$
sample the following scales:
\begin{equation} 
S_{\!{k}}\!= 2^{1/2} S_{\!{k-1}}, \,\,\, {k{\,=\,}2, 3,\dots, N_{\rm K}},
\label{partrat}
\end{equation} 
where the scale $S_{\!{1}}{=\,}\bar{O}$ is defined by Eq.~(\ref{weighting}) as the average angular resolution over the wavebands
combined in $\mathcal{F}_{{\rm D}{j}{\rm C}}$ (Sect.~\ref{combining}), and $S_{\!{N_{\rm K}}}{\,=\,}4\max_{\lambda}({Y_{\lambda}})$
is the largest spatial scale for the filament detection.

Each pixel of the accumulated skeleton $\mathcal{K}_{k}$ in Eq.~(\ref{partial}) contains information on the filament detection
significance $\xi$, defined as the number of scales between $J^{-}$ and $J^{+}$, on which the single-scale skeleton
$\mathcal{K}_{j}$ contributes to $\mathcal{K}_{k}$ in that pixel. Depending on the filament intensity at the skeleton pixel, the
significance range is $1{\,\le\,}\xi{\,\la\,}\ln 2\,(\ln f)^{-1}$ (${\approx\,}14$, assuming $f{\,\approx\,}1.05$,
Appendix~\ref{decomposition}). The algorithm automatically creates the final one-pixel-wide skeletons by thresholding:
$\mathcal{K}_{{k}{\xi}}{\,=\,}\max\left(\mathcal{K}_{k}, \xi\right)$ with a default $\xi{\,=\,}2,$ and applying the Hilditch
algorithm to the resulting shapes. Segmentation images of the skeletons $\mathcal{K}_{{k}{\xi}}$ are computed using the
\textsl{tintfill} algorithm, which sets all pixels belonging to a filament to its number.


\subsubsection{Measurement of the sources}
\label{srcmeasurement}

The source-measurement algorithm is an improved version of the one employed by \textsl{getold} (Sect.~2.6 of Paper I). Sources
cannot be measured in their component $\mathcal{S}_{{\lambda}{X}}$ (Sect.~\ref{iterateback}) because the subtracted background
$\mathcal{B}_{{\lambda}{X}}$ contains substantial source residuals at low intensity levels (Fig.~\ref{bgderivation}). The
background $\mathcal{B}_{{\lambda}{X}}$ is derived specifically for the most complete and reliable source detection, not for
accurate measurements. The sources are measured by \textsl{getsf} in the original $\mathcal{I}_{{\!\lambda}}$ after subtracting
their backgrounds and deblending them from overlapping sources, which entails iterations. The background determination and
deblending are more accurate for the sources with relatively small footprints. However, in crowded regions with larger areas
of overlapping footprints and strongly fluctuating backgrounds, they may become very inaccurate.

The background $\mathcal{B}_{{\rm F}{\lambda}}$ of each source is determined by a linear interpolation of
$\mathcal{I}_{{\!\lambda}}$ across its footprint. The interpolation along two image axes and two diagonals is based on the adjacent
pixels (not belonging to any source) outside the footprint, as was done by \textsl{getold}. To improve the background estimate in
the presence of overlapping footprints, \textsl{getsf} evaluates the background only along those of the four directions for which
the distances between the outside points being interpolated are within a factor of two of the smallest distance. For each pixel of
the source footprint area, the background value is averaged between the directions used in the interpolation. The background
$\mathcal{B}_{{\rm F}{\lambda}}$ is median filtered using a sliding window with a radius $O_{\lambda}$ , and the
background-subtracted image of a source is then obtained as $\mathcal{I}_{{\!\rm S}{\lambda}}{\,=\,}\mathcal{I}_{{\!\lambda}} -
\mathcal{B}_{{\rm F}{\lambda}}$.

In the measurements, the source coordinates $x_{n},y_{n}$ are known from the detection step and are kept unchanged. For the first
measurement iteration, it uses the initial characteristic size $H_{n}{\,=\,}S_{\!j_{\rm F}}$, provided by the detection algorithm
(Sect.~\ref{srcdetection}). The corresponding initial footprint $\{A,B\}_{{\rm F}{n}}{\,=\,}\phi_{n} H_{n}$ is a good approximation
for only Gaussian sources, when $H_{n}$ is close to the actual widths $\{A,B\}_{{\lambda}{n}}$. However, the initial factor
$\phi_{n}{\,=\,}3$ may strongly underestimate the footprints of the resolved power-law sources and overestimate those of the
resolved flat-topped sources (see below). In the subsequent measurement iterations, the sizes and orientation
$\{A,B,\omega\}_{{\lambda}{n}}$ from the previous iteration are used.

The size derivation algorithm in \textsl{getsf} has become more accurate, hence it requires some clarifications. The half-maximum
sizes were computed by \textsl{getold} using intensity moments (cf. Appendix F in Paper I) that give accurate sizes only for the
sources with Gaussian shapes. In real-life observations, however, some sources are markedly non-Gaussian and their intensity
moments give either over- or underestimated sizes, corresponding to the levels well below or above the half-maximum intensity.

\begin{figure*}                                                               
\centering
\centerline{
  \resizebox{0.328\hsize}{!}{\includegraphics{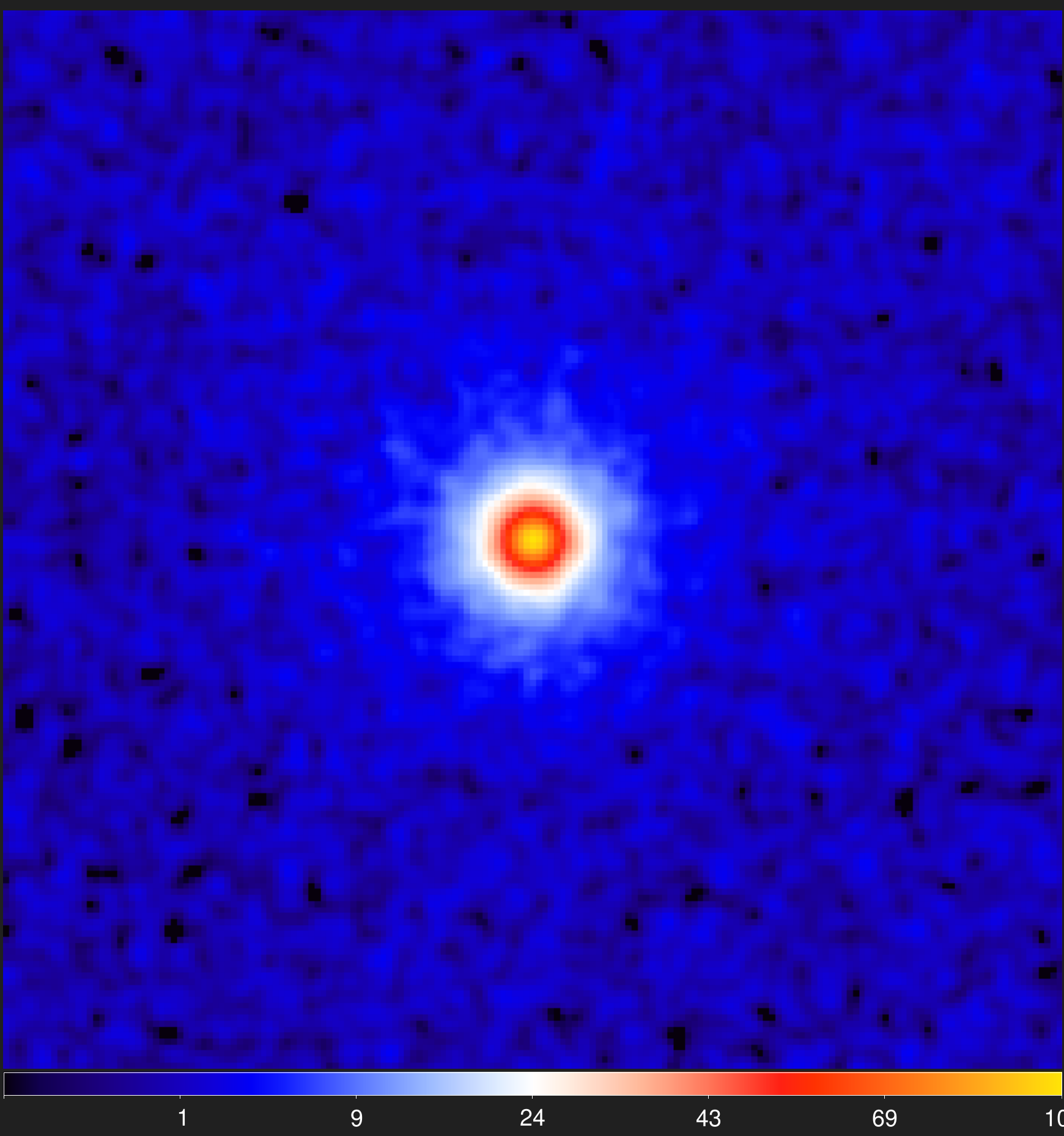}}
  \resizebox{0.328\hsize}{!}{\includegraphics{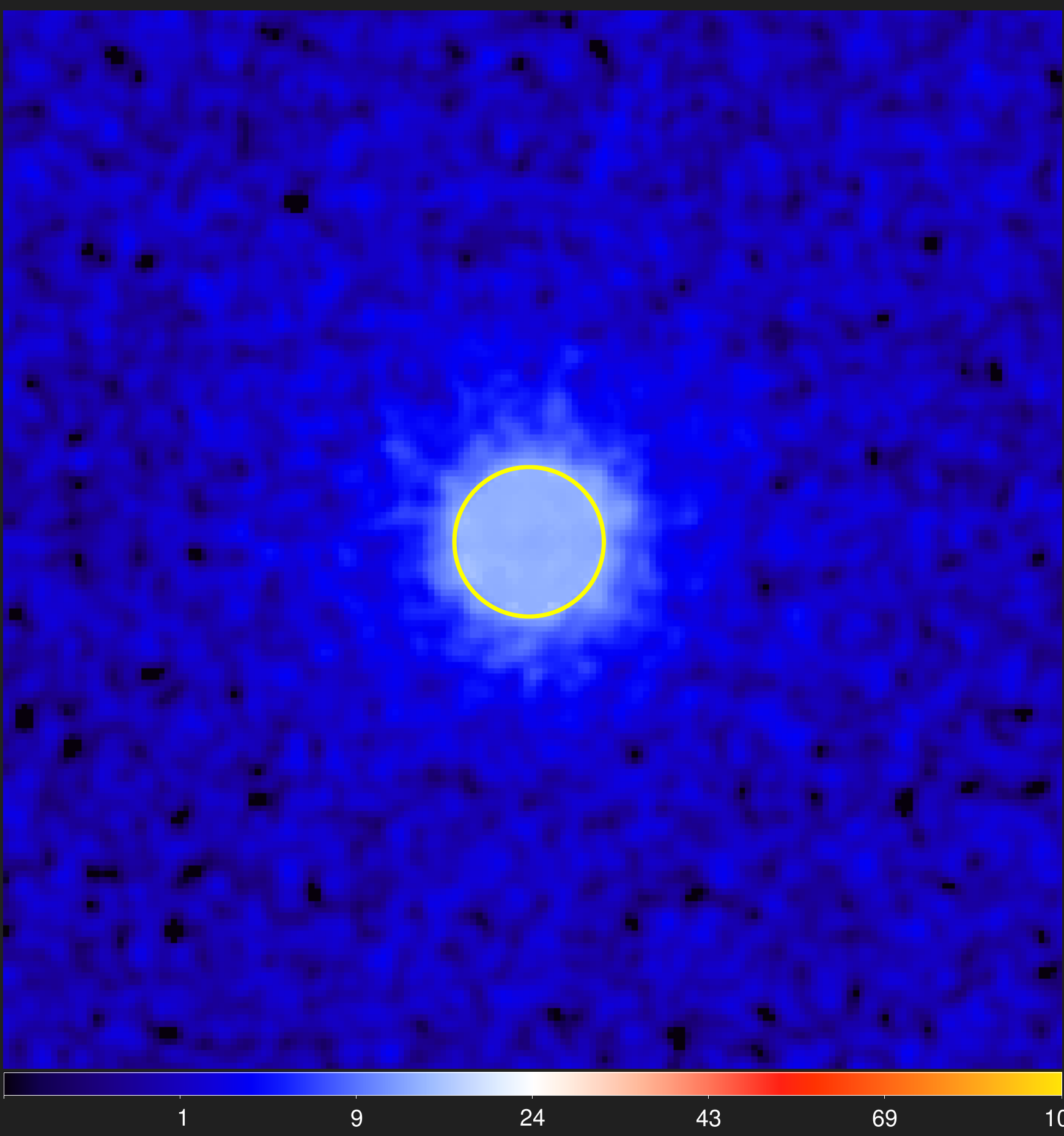}}
  \resizebox{0.328\hsize}{!}{\includegraphics{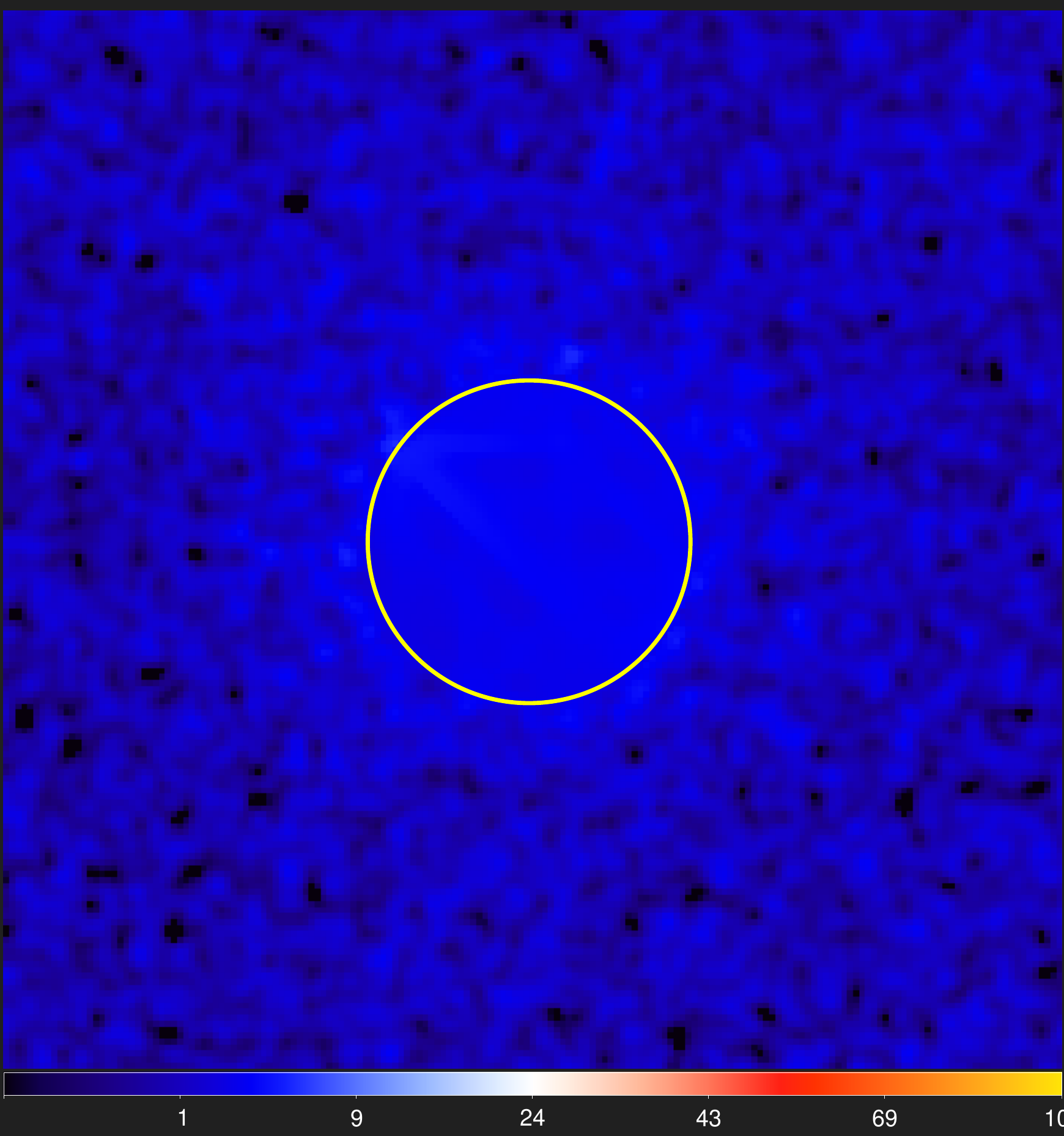}}}
\caption
{ 
Footprint expansion, illustrated in an image with $3${\arcsec} resolution of a source with a peak intensity of $100$, half-maximum size
of $10${\arcsec}, and S/N of $100$ (\emph{left}). The source has an intensity profile defined by Eq.~(\ref{moffatfun}) with
$\Theta{\,=\,}5${\arcsec} and $\zeta{\,=\,}1$, transforming into a power law $I{\,\propto\,}\theta^{-2}$ for
${\theta{\,\gg\,}\Theta}$ and filling up the entire image, its faint outer areas ($I{\,\sim\,}0.2$) are largely lost within the
noise. The initial footprint factor $\phi_{n}{\,=\,}3$ (Sect.~\ref{srcdetection}) is too small for these power-law sources, hence
background subtraction leaves a relatively bright pedestal containing a large amount of the source emission (\emph{middle}). The
footprint expansion algorithm (Sect.~\ref{srcmeasurement}) enlarges $\phi_{n}$ by a factor of $2.2$ (\emph{right}), which lowers
the source background by a factor of $5,$ and as a result, increases the source flux by a factor of $2.7$. The improved flux is
still below the true value by a factor of $1.9$ because the actual footprint is ${}\text{about three}$ times larger. Square-root color mapping.
} 
\label{footexpansion}
\end{figure*}

The inaccuracies of the moment sizes become very large for the resolved sources with power-law intensity distributions. The
simulated image of such a source shown in Fig.~\ref{footexpansion} has a half-maximum size of $10${\arcsec}. However, according to
the intensity moments (over the entire image), the model source has a diameter of $76${\arcsec}. It is easy to see that this value
corresponds to a level that is lower by an order of magnitude than the half-maximum intensity. The source size depends on the
adopted footprint. Within the two footprints shown in the middle and right panels of Fig.~\ref{footexpansion}, the moment sizes are
$10.2$ and $22.5${\arcsec}. The source flux is also underestimated by correspondingly large factors of $5.2$ and $1.8$.

\begin{figure*}                                                               
\centering
\centerline{
  \resizebox{0.328\hsize}{!}{\includegraphics{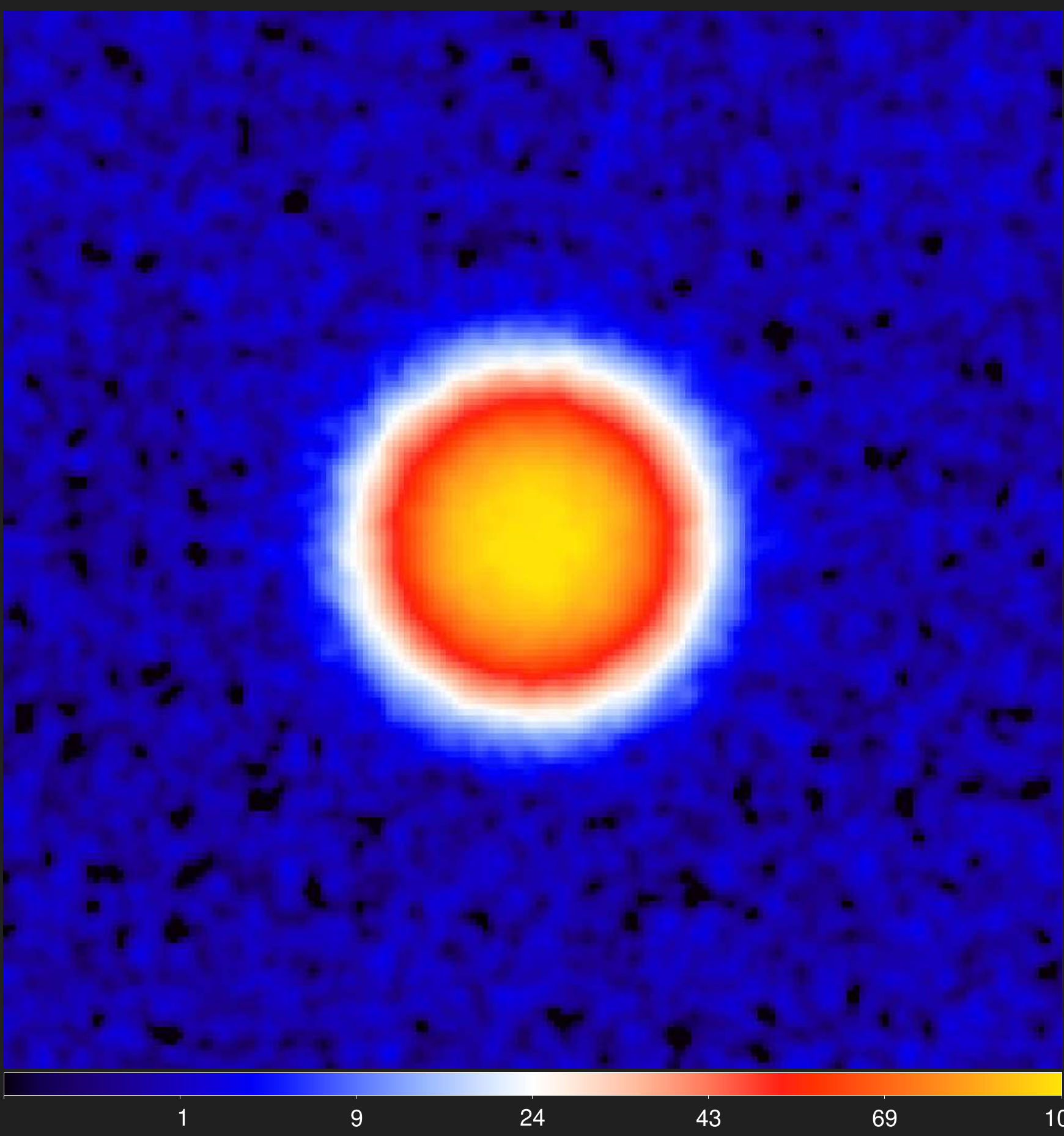}}
  \resizebox{0.328\hsize}{!}{\includegraphics{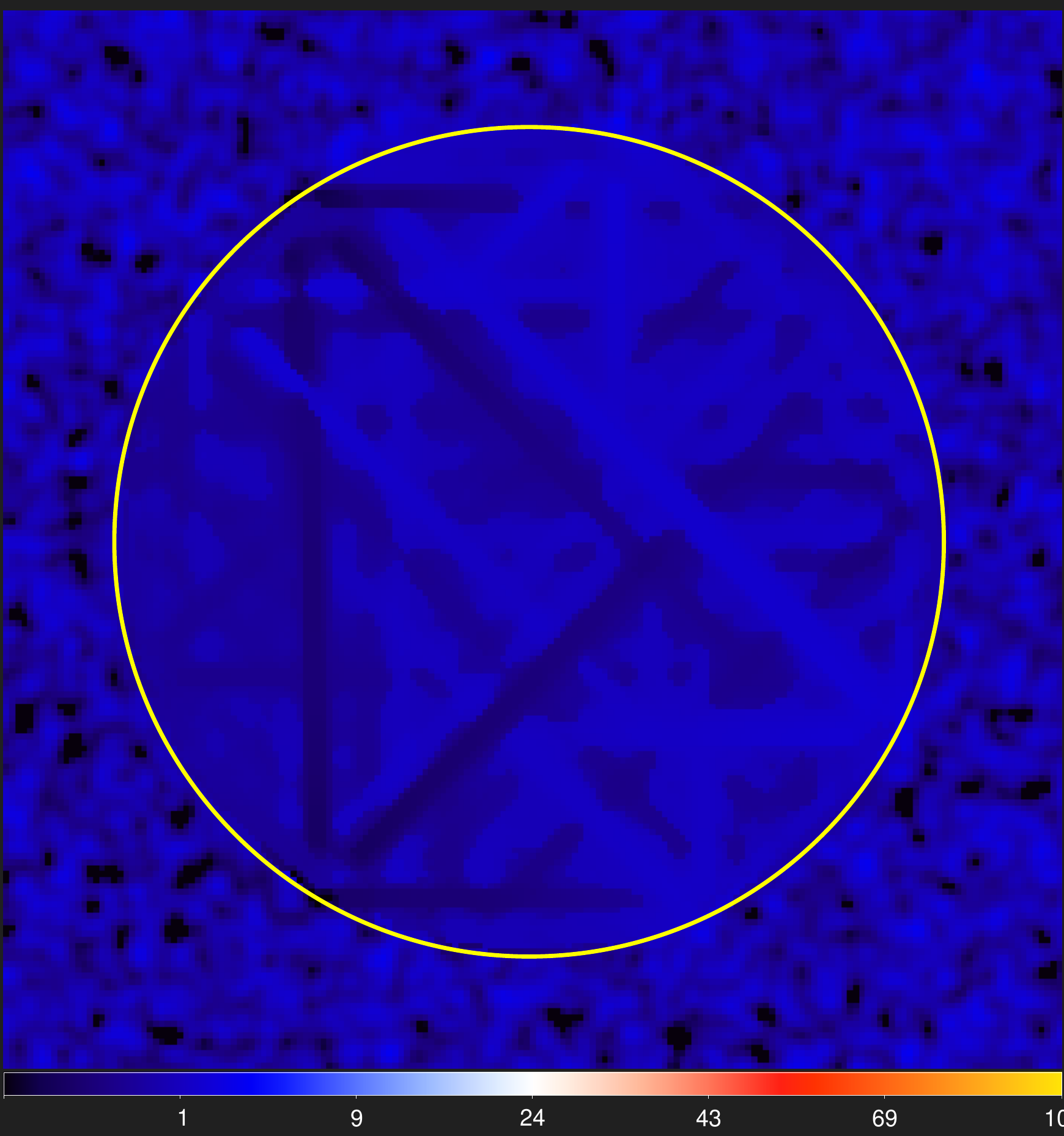}}
  \resizebox{0.328\hsize}{!}{\includegraphics{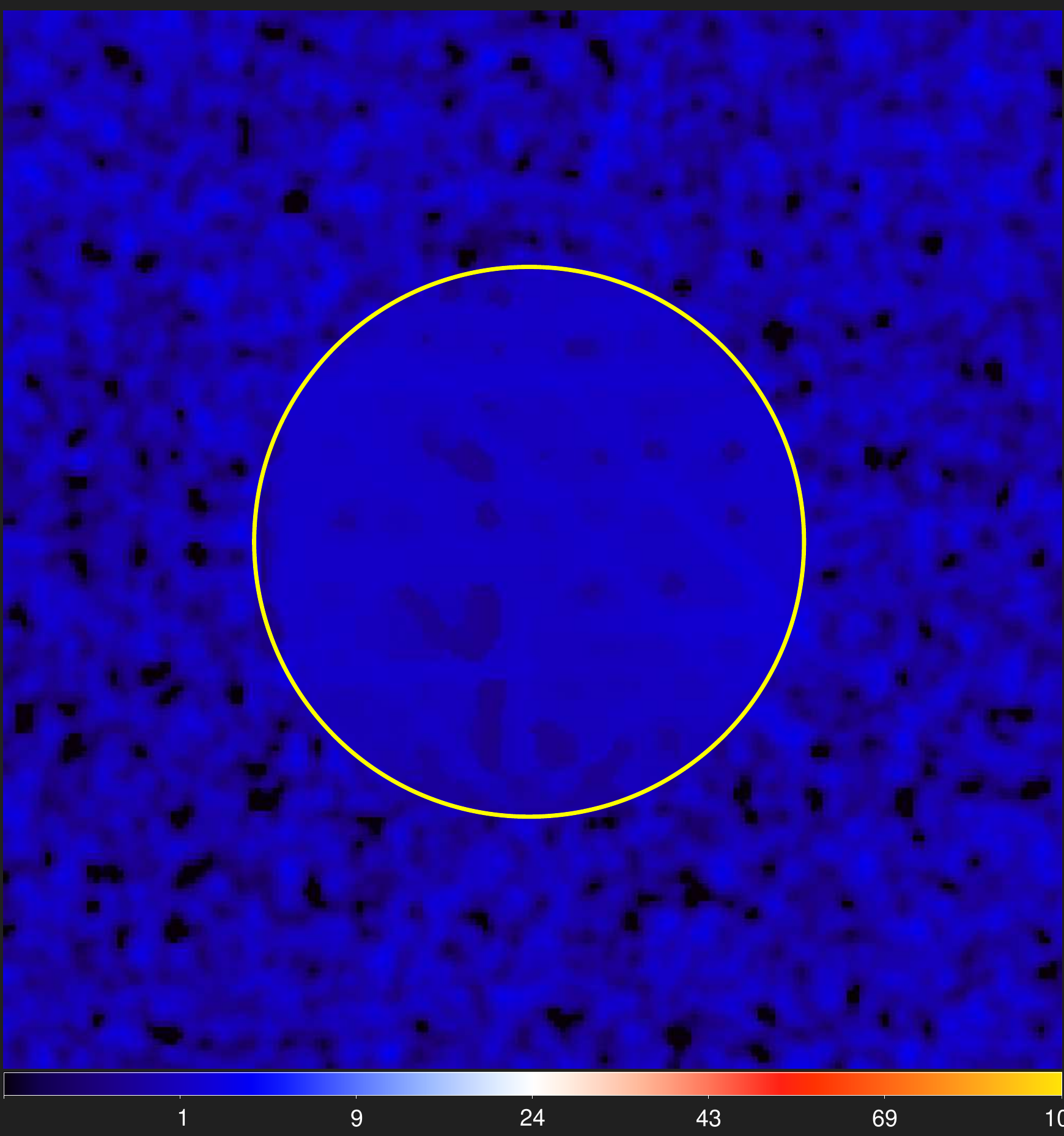}}}
\caption
{ 
Footprint shrinkage, illustrated in an image with $3${\arcsec} resolution of a flat-topped source with a peak intensity of $100$,
half-maximum size of $49${\arcsec}, and S/N of $100$ (\emph{left}), modeled as a $50${\arcsec} cylinder convolved with a
$10${\arcsec} Gaussian kernel. The initial footprint factor $\phi_{n}{\,=\,}3$ (Sect.~\ref{srcdetection}) is too large for the
flat-topped source (\emph{middle}), whose actual footprint relates to the FWHM value by a factor $\phi_{n}{\,=\,}1.5$. The
footprint shrinkage algorithm (Sect.~\ref{srcmeasurement}) reduces $\phi_{n}$ by a factor of $1.5$ (\emph{right}), which shrinks the
footprint and confines it to the pixels belonging to the source alone. This footprint adjustment improves the accuracy of
background interpolation and flux measurement on complex backgrounds. Square-root color mapping.
} 
\label{footshrinkage}
\end{figure*}

Large inaccuracies of the half-maximum sizes also occur for the resolved starless cores that tend to have flat-topped shapes at
short wavelengths ($\lambda{\,\la\,}250$\,${\mu}$m, cf. Fig.~\ref{simcores}), where the emission of their low-temperature interiors
fades away. A simulated image of such a source is shown in Fig.~\ref{footshrinkage}, with the model half-maximum size of
$49${\arcsec}. However, the intensity moments (over the entire image) indicate that its diameter is $31${\arcsec}, which
corresponds to a level by a factor of $2$ above the half-maximum intensity. In the simple example in Fig.~\ref{footshrinkage}, the
source size and flux do not depend on the footprint size because the intensity profile in its outer parts is steep and the
background is flat (zero).

The above examples demonstrate that the intensity moments do not provide accurate estimates of the half-maximum source sizes in the
general case of arbitrary non-Gaussian intensity profiles. Therefore \textsl{getsf} determines accurate half-maximum sizes by the
direct Gaussian interpolation of the source intensity distribution at its half-maximum and averaging the resulting distances from
the source peak, thereby estimating an average radius $h_{{\lambda}{n}}$. The source elongation $E_{{\rm
M}{\lambda}{n}}{\,=\,}A_{{\rm M}{\lambda}{n}}/B_{{\rm M}{\lambda}{n}}$ and position angle $\omega_{{\rm M}{\lambda}{n}}$ are
computed independently from the intensity moments above the $10$\% level of the peak, excluding the low-intensity pixels that may
be affected by the noise and background fluctuations. The major and minor half-maximum axes of the source are then computed from
\begin{equation} 
B_{{\lambda}{n}} = e^{-0.05(E_{{\lambda}{n}}-1)} h_{{\lambda}{n}} E^{-1/2}_{{\rm M}{\lambda}{n}}, \,\,\, 
A_{{\lambda}{n}} = B_{{\lambda}{n}} E_{{\rm M}{\lambda}{n}},
\label{sourcesizes}
\end{equation} 
where the (empirical) exponential factor converts the average radius $h_{{\lambda}{n}}$ into the equivalent-area radius
$(A_{{\lambda}{n}} B_{{\lambda}{n}})^{1/2}$ of an ellipse. The FWHM ellipse $\{A,B,\omega\}_{{\lambda}{n}}$ from
Eq.~(\ref{sourcesizes}) is guaranteed to correspond to the source half-maximum intensity, in contrast to the ellipse estimated from
the intensity moments. The moment sizes $\{A,B,\omega\}_{{\rm M}{\lambda}{n}}$ are also computed by \textsl{getsf} because they
contain independent information that can be useful for the analysis of the extracted sources.

During the measurement iterations (Sect.~2.6 of Paper I), \textsl{getsf} employs a footprint expansion and shrinkage algorithm to
correct the footprint areas of those sources that need such adjustments. It is based on a simple observation that when a footprint
area is too small, the source background contains a pedestal of the residual intensity distribution of the source
(Fig.~\ref{footexpansion}); when the source pedestal does not exist or is negative (Fig.~\ref{footshrinkage}), the footprint may be
accurate or too large. The analysis is made in the regularized component $\mathcal{S}_{{\lambda}{\rm R}}$ from
Eq.~(\ref{stdregular}) without contribution from the complex background and filaments.

The presence of the background pedestal is indicated by the positive difference between the background values below the source and
those in an external annulus just outside the source,
\begin{equation} 
1.1 B_{{\Phi}{\lambda}{n}} > B_{{\Psi}{\lambda}{n}} + D_{{\Psi}{\lambda}{n}},
\label{footexpand}
\end{equation} 
where $B_{{\Phi}{\lambda}{n}}$ is the median value within the footprint and $B_{{\Psi}{\lambda}{n}}$ and $D_{{\Psi}{\lambda}{n}}$
are the mean and the standard deviation inside the annulus. When the condition of Eq.~(\ref{footexpand}) is fulfilled and the
source is not too elongated ($A_{{\lambda}{n}}{\,<\,}1.3 B_{{\lambda}{n}}$) and bright enough ($\Xi_{{\lambda}{n}}{\,>\,}50$ and
$\Omega_{{\lambda}{n}}{\,>\,}15$, see Eq.~(\ref{monoquant})), \textsl{getsf} increases the factor $\phi_{n}$ by $5\%$ before proceeding
to the next measurement iteration. The footprint expansion terminates when the residual background pedestals
(Fig.~\ref{footexpansion}) are reduced as much as possible and the condition in Eq.~(\ref{footexpand}) becomes false. As a final
adjustment, the footprint is expanded once more by the factor $0.9{\,+\,}0.1(\phi_{n}/3)$ to reduce the residual pedestal.

The footprints of the sources that do not need any expansion are attempted to be reduced in size. It is important to confine the
footprints to the most local area occupied by the sources because oversized footprints may strongly decrease the accuracy of
background subtraction and flux measurement for sources on complex (filamentary) backgrounds and in crowded areas. The need to
shrink a source footprint is indicated by a negative difference between the background values below the source and in an external
annulus just outside the source,
\begin{equation} 
1.1 B_{{\Phi}{\lambda}{n}} < B_{{\Psi}{\lambda}{n}} + D_{{\Psi}{\lambda}{n}},
\label{footshrink}
\end{equation} 
where the quantities are the same as in Eq.~(\ref{footexpand}). When the condition of Eq.~(\ref{footshrink}) is fulfilled,
\textsl{getsf} decreases the factor $\phi_{n}$ by $2\%$ before proceeding to the subsequent measurement iterations. The footprint
shrinkage is completed when the condition in Eq.~(\ref{footshrink}) becomes false, that is, when the reduced footprint causes a small residual background pedestal. In a final adjustment, the footprint is expanded by a factor of $1.1$ to
eliminate the pedestal created in the process (Fig.~\ref{footshrinkage}).

Extensive testing has shown that the simple footprint expansion and shrinkage algorithm performs well for most sources in
complicated environments and backgrounds in both benchmarks and real-life observations. Despite the footprint expansion, total
fluxes of the power-law sources may still remain underestimated by large factors because the faint outer areas of these sources,
with an unknown full extent, vanish into the fluctuating backgrounds and noise.

After computing the background-subtracted images $\mathcal{I}_{{\rm S}{\lambda}}$, \textsl{getsf} deblends overlapping sources,
calculating the peak intensities $F_{{\rm P}{\lambda}{n}}$ and the total fluxes $F_{{\rm T}{\lambda}{n}}$ for each source $n$. The
iterative deblending algorithm employs the Gaussian shapes $G_{{\lambda}{n}}(x,y)$ defined by the source ellipse
$\{A,B,\omega\}_{{\lambda}{n}}$ and peak intensity $F_{{\rm P}{\lambda}{n}}$. The intensity $I_{{\rm S}{\lambda}}(x,y)$ is split
between the source $n$ and all overlapping sources $n^{\prime}$ according to a fraction of the shape intensities,
\begin{equation} 
I_{{\lambda}{n}}(x,y) = \frac{G_{{\lambda}{n}}(x-x_{n},y-y_{n})}{\sum\limits_{n^{\prime}} 
G_{{\lambda}{n^{\prime}}}(x-x_{n^{\prime}},y-y_{n^{\prime}})}\,I_{{\rm S}{\lambda}}(x,y),
\label{deblending}
\end{equation} 
where the summation is done over all surrounding sources whose footprints cover the pixel $(x,y)$. The iterative deblending of the
peak intensities starts with the original image values $I_{{\rm S}{\lambda}}(x_{n},y_{n})$ of each source and proceeds with the
splitting of the pixel values until $I_{{\lambda}{n}}(x_{n},y_{n})$ converges to the deblended peak intensity $F_{{\rm
P}{\lambda}{n}}$. After obtaining $F_{{\rm P}{\lambda}{n}}$ for all sources, \textsl{getsf} computes the deblended intensities
$I_{{\lambda}{n}}(x,y)$ of all pixels within their footprints, estimates the ellipses $\{A,B,\omega\}_{{\lambda}{n}}$ and
$\{A,B,\omega\}_{{\rm M}{\lambda}{n}}$, and integrates the total fluxes $F_{{\rm T}{\lambda}{n}}$. It also computes an independent
flux estimate $F_{{\rm G}{\lambda}{n}}$ by integrating $G_{{\lambda}{n}}(x,y)$, which may only be accurate when a source shape
resembles the two-dimensional Gaussian.

Uncertainties of the peak intensities $F_{{\rm P}{\lambda}{n}}$ are estimated by \textsl{getsf} as the standard deviations
$\sigma_{{\rm P}{\lambda}{n}}$, evaluated in the original image $\mathcal{I}_{{\lambda}}$, in an elliptical annulus around each
source $n$ just outside its footprint. In heavily crowded fields, no local source-free annulus can be found near the sources, in
which case the uncertainties are estimated from the more distant source-free pixels. The uncertainties $\sigma_{{\rm
T}{\lambda}{n}}$ of the total fluxes $F_{{\rm T}{\lambda}{n}}$ are computed with the same assumptions as in \textsl{getold}
(Sect.~2.6 of Paper I),
\begin{equation} 
\sigma_{{\rm T}{\lambda}{n}}{\,=\,}\sigma_{{\rm P}{\lambda}{n}} \frac{(A_{{\rm F}{\lambda}{n}} 
B_{{\rm F}{\lambda}{n}})^{1/2}}{\phi_{n} O_{\lambda}},
\label{totfluxerr}
\end{equation} 
where $A_{{\rm F}{\lambda}{n}}$ and $B_{{\rm F}{\lambda}{n}}$ are the major and minor axes of the source footprints.

It is convenient to define the detection significance $\Xi_{{\lambda}{n}}$ and the signal-to-noise ratios (S/Ns) $\Omega_{{\lambda}{n}}$
and $\Psi_{\!{\lambda}{n}}$, describing the detection and measurement properties of each extracted source,
\begin{equation} 
\Xi_{{\lambda}{n}}{\,=\,}\frac{S_{\!{\lambda}{\rm D}{j_{\rm F}}{n}}}{\sigma_{\!{\lambda}{\rm S}{j_{\rm F}}}}, \,\,\,
\Omega_{{\lambda}{n}}{\,=\,}\frac{F_{{\rm P}{\lambda}{n}}}{\sigma_{{\rm P}{\lambda}{n}}}, \,\,\,
\Psi_{\!{\lambda}{n}}{\,=\,}\frac{F_{{\rm T}{\lambda}{n}}}{\sigma_{{\rm T}{\lambda}{n}}}, \,\,\,
\label{monoquant}
\end{equation} 
where $j_{\rm F}$ is the footprinting scale (Sect.~\ref{srcdetection}) and $S_{\!{\lambda}{\rm D}{j_{\rm F}}{n}}$ is the intensity
at the source position in $\mathcal{S}_{{\lambda}{\rm D}{j_{\rm F}}}$ (Sect.~\ref{decompflat}). The above quantities can be
combined together to characterize the overall ``goodness'' of a source,
\begin{equation} 
\Gamma_{{\lambda}{n}}{\,=\,} \frac{\Xi_{{\lambda}{n}}}{5} \frac{\left(\Omega_{{\lambda}{n}}
\Psi_{{\lambda}{n}}\right)^{1/2}}{2} \frac{B_{{\lambda}{n}}}{A_{{\lambda}{n}}},
\label{goodness}
\end{equation} 
normalized such that all acceptable sources in the extraction catalogs have $\Gamma_{{\lambda}{n}}{\,\ga\,}1$. The sources with
$\Gamma_{{\lambda}{n}}{\,\la\,}1$ may have quite unreliable measurements in waveband $\lambda$. The corresponding global quantities
$\Xi_{n}$ and $\Gamma_{n}$ describe the source detection significance and goodness, respectively, in all wavebands,
\begin{equation} 
\Xi_{n}{\,=\,}\left(\sum_{\lambda} \Xi_{{\lambda}{n}}^2\right)^{1/2}, \,\,\,
\Gamma_{n}{\,=\,}\left(\sum_{\lambda} \Gamma_{{\lambda}{n}}^2\right)^{1/2}\!.
\label{globalquant}
\end{equation} 

The \textsl{getsf} source extraction catalogs contain detailed headers, documenting the extraction parameters and explaining the
tabulated quantities. Each data line presents the source number $n$, coordinates $x_{n}, y_{n}$ (in pixels), world coordinates
$\alpha_{n}, \delta_{n}$ \citep[computed with the \textsl{xy2sky} utility,][]{Mink2002}, global flag $f_{n}$, significance
$\Xi_{n}$, and goodness $\Gamma_{n}$,
\begin{displaymath} 
n\: x_{n}\: y_{n}\: \alpha_{n}\: \delta_{n}\: f_{n}\: \Xi_{n}\: \Gamma_{n,}\: 
\label{catalogline1}
\end{displaymath} 
followed (in the same line) by the measured quantities in each of the $N_{\rm W}$ wavebands,
\begin{displaymath} 
\left( f_{{\lambda}{n}}\: \Xi_{{\lambda}{n}}\: \Gamma_{{\lambda}{n}}\: F_{{\rm P}{\lambda}{n}}\: \sigma_{{\rm P}{\lambda}{n}}\: 
F_{{\rm T}{\lambda}{n}}\: \sigma_{{\rm T}{\lambda}{n}}\: A_{{\lambda}{n}}\: B_{{\lambda}{n}}\: A_{{\rm M}{\lambda}{n}}\: 
B_{{\rm M}{\lambda}{n}}\: \omega_{{\lambda}{n}} \right)_{N_{\rm W}}\!,
\label{catalogline2}
\end{displaymath} 
where $f_{{\lambda}{n}}$ is a wavelength-dependent flag. In addition to this information, an expanded version of the catalog adds (to
the same line) the Gaussian flux $F_{{\rm G}{\lambda}{n}}$, characteristic size $S_{\!j_{\rm F}}$, footprint factor $\phi_{n}$, and
footprint axes $A_{{\rm F}{\lambda}{n}}$, $B_{{\rm F}{\lambda}{n}}$. For surface density images, the $F_{{\rm G}{\lambda}{n}}$ 
column is replaced with source mass $M_{{\lambdabar}{n}}$.

It is necessary to emphasize that sources from extraction catalogs must always be carefully selected (for each waveband separately)
to ensure that only sufficiently good and accurately measurable sources are used in further analysis. This is especially important
for the multiwavelength extraction catalogs, where sources can be prominent in one waveband and completely undetectable or not
measurable in another. The \textsl{getsf} catalogs provide various quantities to enable the evaluation and selection of
only acceptable sources and recommended the following selection criteria:
\begin{eqnarray} 
\left.\begin{aligned}
&{\Xi_{{\lambda}{n}} > 1} \,\land\, {\Gamma_{{\lambda}{n}} > 1} \,\land\, {\Omega_{{\lambda}{n}} > 2} \,\land\, 
{\Psi_{\!{\lambda}{n}} > 2} \,\land\, \\
&{A_{{\lambda}{n}} < 2 B_{{\lambda}{n}}} \,\land\, A_{{\rm F}{\lambda}{n}} > 1.15 A_{{\lambda}{n}}.
\end{aligned}\right.
\label{acceptable} 
\end{eqnarray} 
These empirical conditions, based on numerous test results obtained in various benchmarks (Pouteau et al., in prep.; Men'shchikov
2021, submitted), and verified in applications to a variety of observed images (e.g., Sect.~\ref{applications}), ensure that the
selected subset of sources is reliable (not contaminated by significant numbers of spurious sources) and that selected sources have
acceptably accurate measurements.


\subsubsection{Measurements of the filaments}
\label{filmeasurement}

Filaments are measured in their background-subtracted $\mathcal{F}_{{\lambda}{Y}}$, derived in Sect.~\ref{iterateback}. When the
maximum size $Y_{{\lambda}}$ of the filaments of interest is estimated sufficiently accurately (Sect.~\ref{maxsizes}), their
background $\mathcal{B}_{{\lambda}{Y}}$ does not reveal any filamentary residuals. Nevertheless, the background may well have
substantial inaccuracies, especially when the filaments are wide and blended (Fig.~\ref{bgderivation}). Observed filaments are the
two-dimensional projections of the complex three-dimensional structures, which are much more difficult to disentangle, deblend,
measure, and analyze than sources with their well-defined round shapes and compact footprints.

Sources can be represented by their peak intensity and half-maximum size, but filaments are extremely complicated in their shapes
and widths, often interconnected with each other and with various nearby branches, and have variable intensity along their crests.
It is quite clear that blending of the structures is a major source of large inaccuracies in the measured quantities of general
interest (widths, fluxes, masses, profiles) and in other properties, derived from the measurements.

Another difficulty in understanding filaments (distinct physical structures) is that the filament length cannot be determined
objectively. In most cases, it is quite unclear where a physical filament starts, where it ends, and which branches of the
three-dimensional filamentary network belong to that filament. Fortunately, the global properties of the entire filaments (even if
the latter could be clearly defined) are not as important for studying star formation as the local properties of their relatively
short segments that develop appropriate physical conditions for the formation of prestellar cores.

The approach that is adopted in \textsl{getsf} is to simplify the very complex problem by separating all branches of the skeleton
network, converting the latter into the simple, non-branching skeletons. The set of non-branching skeletons is derived during the
segmentation of the skeletons $\mathcal{K}_{k{\xi}}$, the last step of the filament detection process (Sect.~\ref{fildetection}).
The simplified skeletons enable an easy selection and better measurements of only the well-behaving, preferably isolated (not
blended), and relatively straight parts of the filaments. No attempt is made by \textsl{getsf} to deblend filaments because a
general algorithm for accurately deblending them is not available.

The segmentation image of all skeletons is scanned to trace each skeleton $n$ and find coordinates of all its pixels; to smooth the
skeletons, the integer coordinates of their pixels are averaged within a seven-pixel window. The resulting high-resolution
coordinates $x_{n}(i), y_{n}(i)$ of each skeleton point $i$ are cataloged, together with the local position angles
$\vartheta_{n}(i)$ of the skeleton direction and $\alpha_{n}(i), \beta_{n}(i)$ of the left and right normals. A normal is called
left ($\alpha$) or right ($\beta$) depending on which side it is located from the first skeleton point to the last. With an adopted
distance to the observed region, \textsl{getsf} converts the angular units of the pixels into parsecs and measures each filament as
a function of the length $l$ along its skeleton and the distance $r$ along its normals. If the distance is unknown or unspecified,
a default distance of $100$ pc is used; the measurements can be scaled to another distance after the extraction.

The observed filaments usually meander, hence their skeleton normals diverge from each other on one side and intersect with each
other on the other side. In the absence of deblending, more accurate measurements for them are usually obtained from the one-sided
quantities that correspond to the side on which the filament is the least affected by blending with itself and with other nearby
structures. The filament surface density (or intensity) profiles and their full half-maximum widths are cataloged as the one-sided
quantities $D_{\{{\alpha|\beta}\}{n}}(l, r)$ and $W_{\{{\alpha|\beta}\}{n}}(l)$ and as the average quantities $D_{n}(l, r)$ and
$W_{n}(l)$ between the two sides. Also cataloged are the corresponding average profiles $D_{\{{\alpha|\beta}\}{n}}(r)$ and
$D_{n}(r)$ along the skeleton with their standard deviations $\varsigma_{\{{\alpha|\beta}\}{n}}(r)$ and $\varsigma_{n}(r)$, as well
as the median widths $W_{n}$ and the slopes $\gamma(r)$ of the filament profiles.

Although the total length $L_{n}$ of a skeleton and mass $M_{n}$ of a filament may not always be objective and physically
meaningful quantities (see the discussion above), \textsl{getsf} derives the mass by direct integration of
$\mathcal{F}_{{\lambda}{Y}}$ within a filament footprint, assuming that the image is obtained from surface densities,
\begin{equation} 
M_{\{{\alpha|\beta}\}{n}} = 2\,\mu m_{\rm H} 
\!\iint\limits_{\,\Upsilon_{\{{\alpha|\beta}\}{n}}}\!\mathcal{F}_{{\lambda}{Y}{n}}(x, y)\,{\rm d}x {\rm d}y,
\label{filmass}
\end{equation} 
where $M_{\{{\alpha|\beta}\}{n}}$ are the one-sided mass estimates, from which the average mass $M_{n}$ between the two sides is
obtained. The one-sided footprints $\Upsilon_{\{{\alpha|\beta}\}{n}}$ used in Eq.~(\ref{filmass}) are defined as the areas between
the skeleton and the maximum extent of the filament on either side. In practice, a filament footprint $\Upsilon_{{n}}$ is the set
of all pixels whose shortest distances from the skeleton are smaller than the filament normals.

When the filament mass $M_{\{{\alpha|\beta}\}{n}}$ and length $L_{n}$ are known, the one-sided estimates of the average linear
density\footnote{In some publications, the filament linear density is also referred to as the mass per unit length.} of the entire
filament are readily obtained,
\begin{equation} 
\bar{\Lambda}_{\{{\alpha|\beta}\}{n}} = M_{\{{\alpha|\beta}\}{n}\,} L^{-1}_{n},
\label{lineardens1}
\end{equation} 
together with the average linear density $\bar{\Lambda}_{{n}}$ between the two sides. The linear density of filaments is also
computed by \textsl{getsf} as a function of the coordinate $l$ along their skeletons,
\begin{equation} 
\Lambda_{\{{\alpha|\beta}\}{n}}(l) = 2\,\mu m_{\rm H} 
\!\!\!\int\limits^{R_{\{{\alpha|\beta}\}{n}}(l)}_{0} \!\mathcal{F}_{{\lambda}{Y}{n}}(l, r)\,{\rm d}r,
\label{lineardens2}
\end{equation} 
where the integration limits $R_{\{{\alpha|\beta}\}{n}}(l)$ along the left and right normals are chosen at zero surface density
values or at a radial distance of the profile minimum at which the filament becomes blended with another structure. The median
one-sided linear densities $\Lambda_{\{{\alpha|\beta}\}{n}}$ for the entire length $L_{n}$ of the filament and its average linear
density $\Lambda_{{n}}$ are also computed and cataloged. The linear density values from Eq.~(\ref{lineardens1}) and
Eq.~(\ref{lineardens2}) are expected to be similar to each other for the well-behaved filaments.


\section{Applications to observed regions}
\label{applications}

The multiscale, multiwavelength source- and filament-extraction method presented in Sect.~\ref{getsf} was very extensively tested
using ${\sim\,}40$ images that were observed with different instruments and both ground-based and orbital telescopes during the
past two decades. Multiwaveband observations of star-forming regions obtained in the \emph{Herschel} Gould Belt Survey
\citep{Andre_etal2010} and HOBYS \citep{Motte_etal2010} key projects, as well as the most recent interferometric images observed in
the \emph{ALMA}-IMF program (Motte et al., in prep.), played an important role in validating \textsl{getsf.}

The new extraction method has demonstrated very good results in \emph{ALMA} benchmarks (Pouteau et al., in prep.) on images,
created from a magnetohydrodynamic (MHD) simulation of a star-forming region \citep{NtormousiHennebelle2019} that was populated
with model cores and processed by the \textsl{casa} task \textsl{simobs} \citep{McMullin_etal2007} to resemble the real \emph{ALMA}
observations (Louvet et al., in prep.). Furthermore, \textsl{getsf} has been applied to source extraction in $15$ regions of the
\emph{ALMA}-IMF program and $12$ infrared dark clouds of the ASHES survey \citep{Sanhueza_etal2019,Li_etal2020}. However, the most
significant and definitive testing and validation of extraction tools is achieved with simulated benchmarks for which everything is
fully known about their components. The second paper (Men'shchikov 2021, submitted) presents a quantitative analysis of
\textsl{getsf} extractions using several variants of the new benchmark (Sect.~\ref{skybench}) and old benchmark (Papers I and III).

Sections \ref{xray}--\ref{alma} illustrate the performance of \textsl{getsf} on nine images obtained with different telescopes:
\emph{XMM-Newton}, the Galaxy Evolution Explorer (\emph{GALEX}), \emph{Hubble}, \emph{Spitzer}, \emph{Herschel}, the Atacama
Pathfinder Experiment (\emph{APEX}), the James Clerk Maxwell Telescope (\emph{JCMT}), and \emph{ALMA} from the X-ray domain to the
millimeter wavelengths. These examples are presented to demonstrate that the method is applicable to a wide variety of observed
images, visualizing the effects of the separation of structural components and flattening of detection images. Scientific analyses
and discussions of these results, as well as their comparisons with previous studies, are beyond the scope of this paper. This can
be accomplished using the corresponding extraction catalogs that are available on the \textsl{getsf}
website\footnote{\url{http://irfu.cea.fr/Pisp/alexander.menshchikov/\#intro}}.


\subsection{Supernova remnant \object{RXJ\,1713.7-3946}}
\label{xray}

\begin{figure*}                                                               
\centering
\centerline{
  \resizebox{0.328\hsize}{!}{\includegraphics{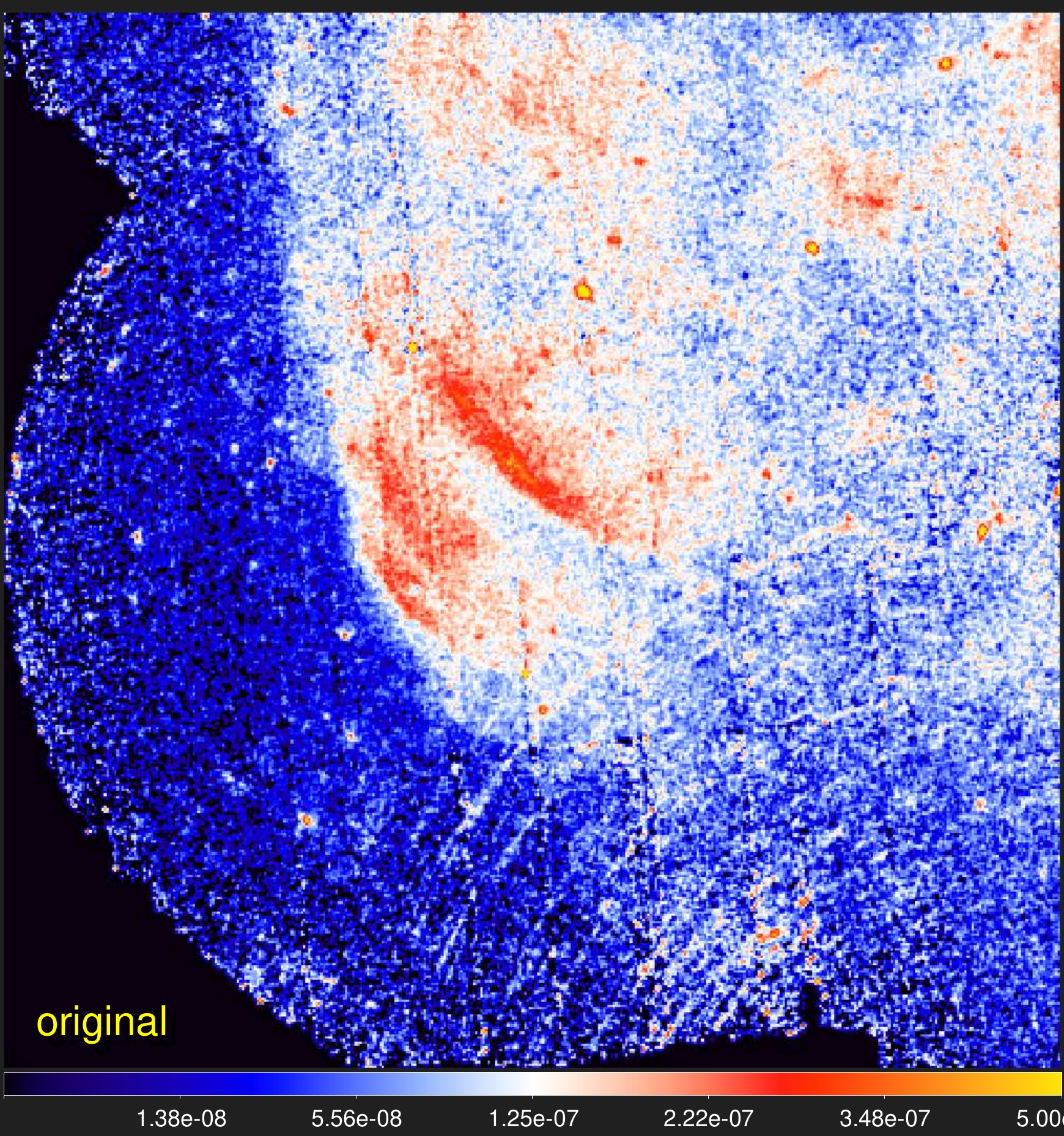}}
  \resizebox{0.328\hsize}{!}{\includegraphics{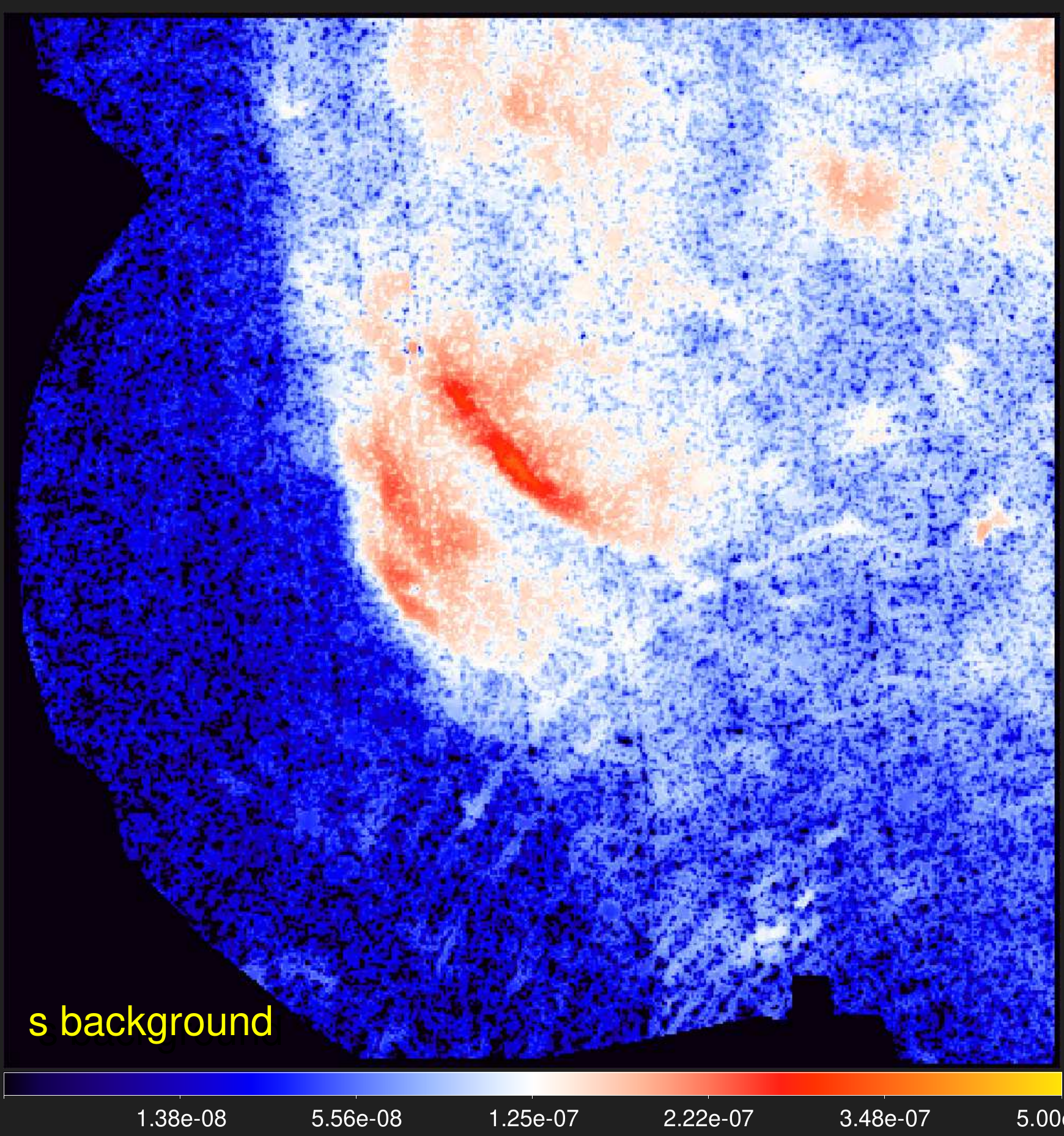}}
  \resizebox{0.328\hsize}{!}{\includegraphics{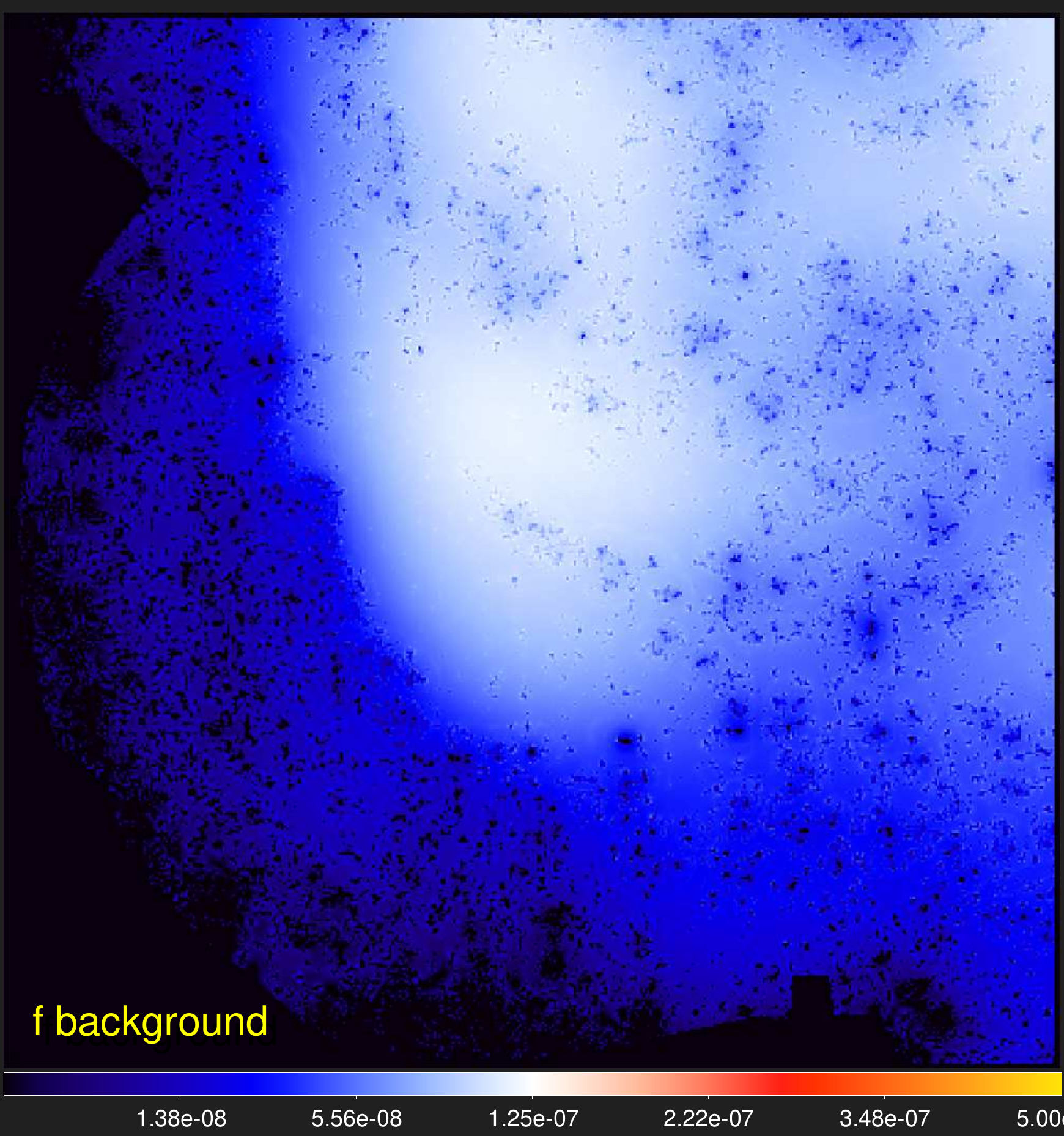}}}
\vspace{0.5mm}
\centerline{
  \resizebox{0.328\hsize}{!}{\includegraphics{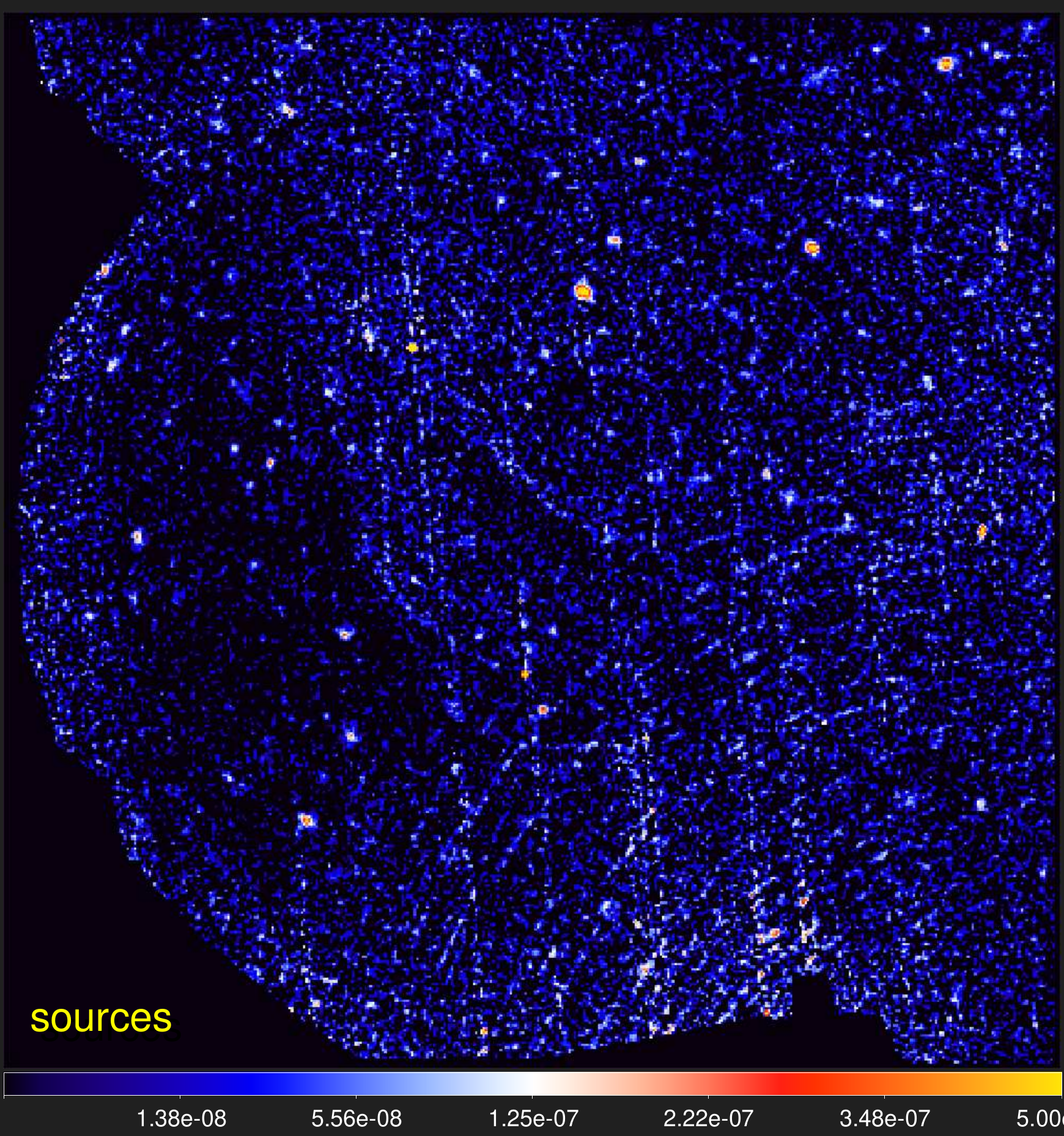}}
  \resizebox{0.328\hsize}{!}{\includegraphics{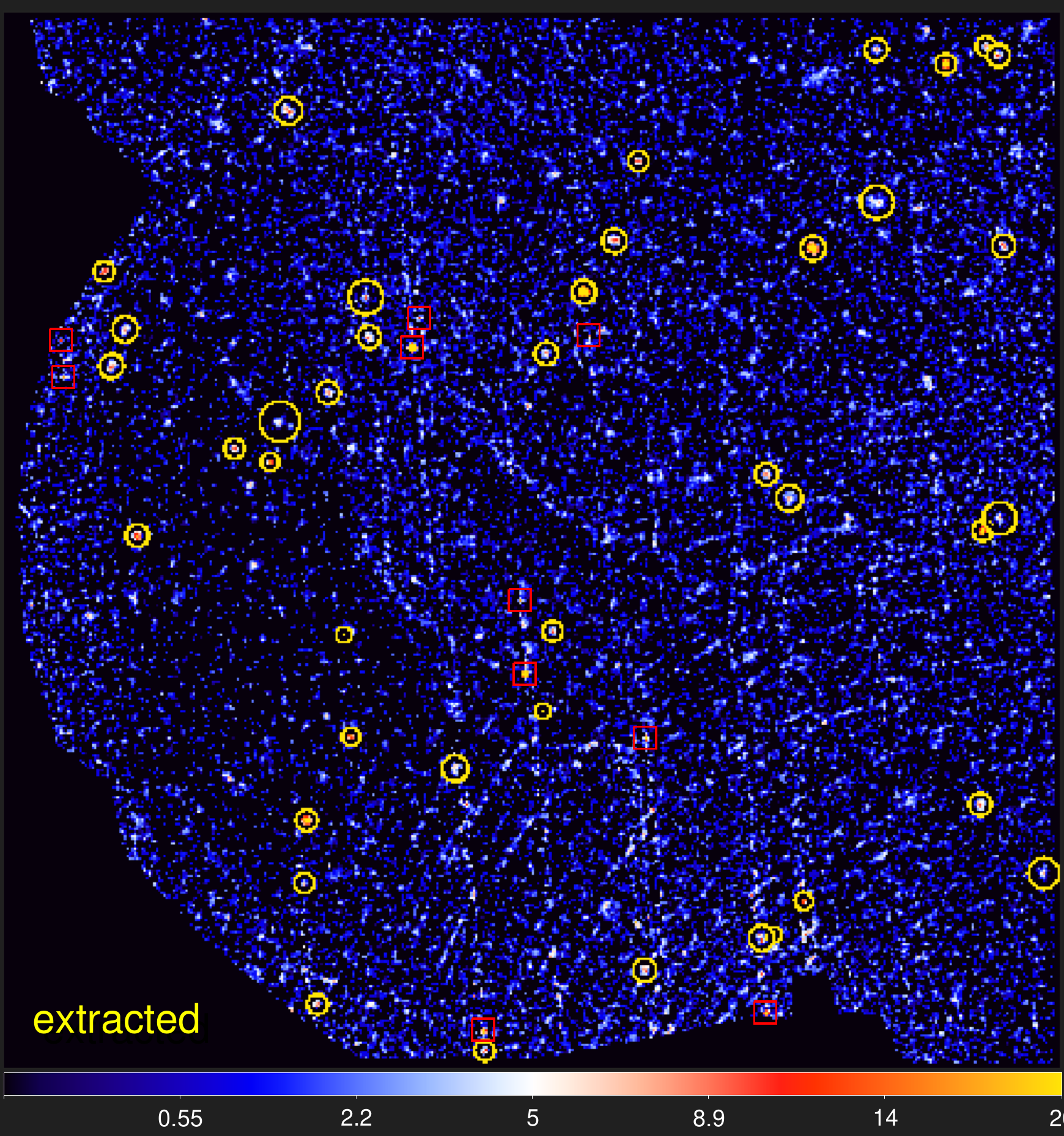}}
  \resizebox{0.328\hsize}{!}{\includegraphics{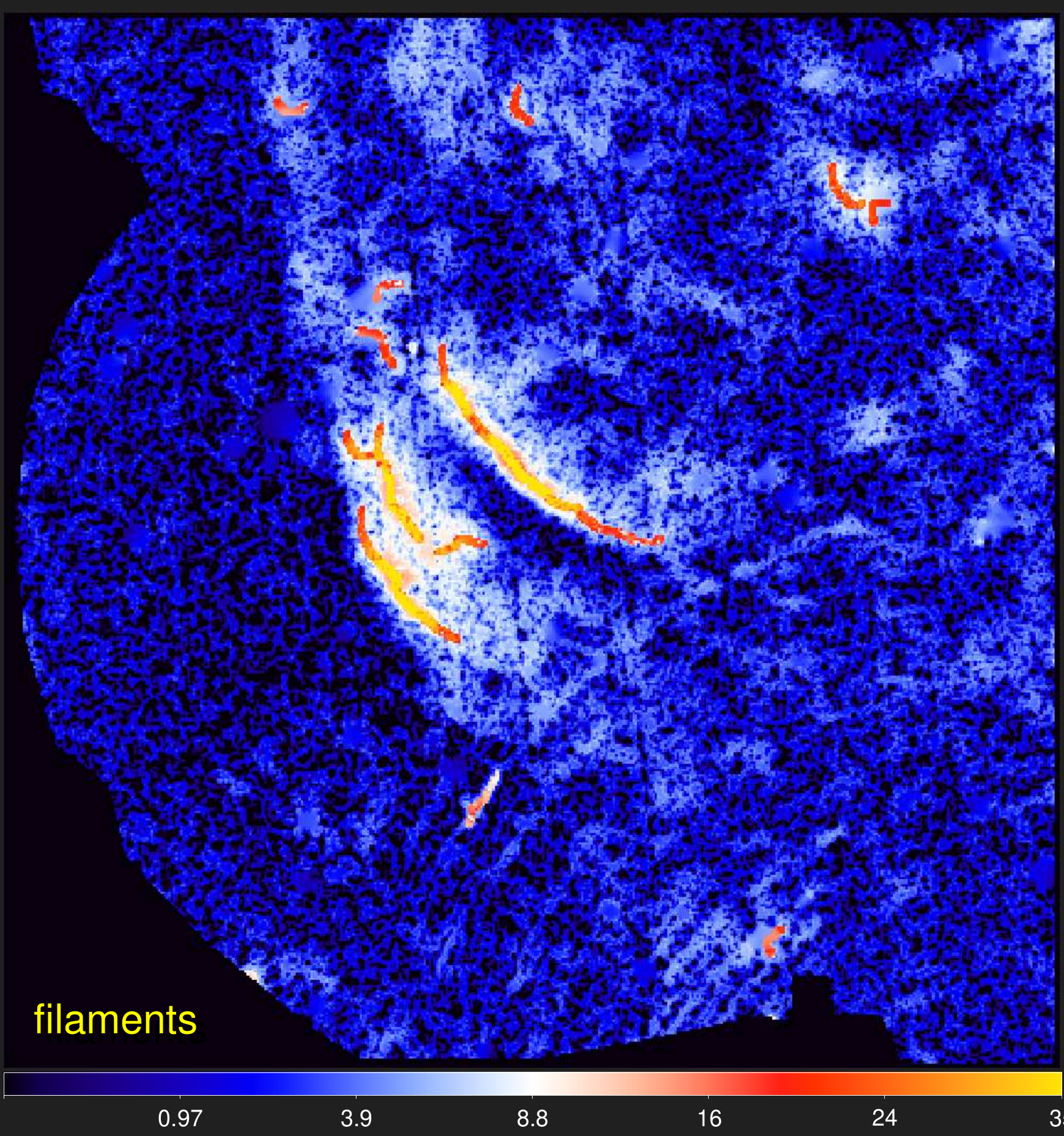}}}
\vspace{0.5mm}
\centerline{
  \resizebox{0.328\hsize}{!}{\includegraphics{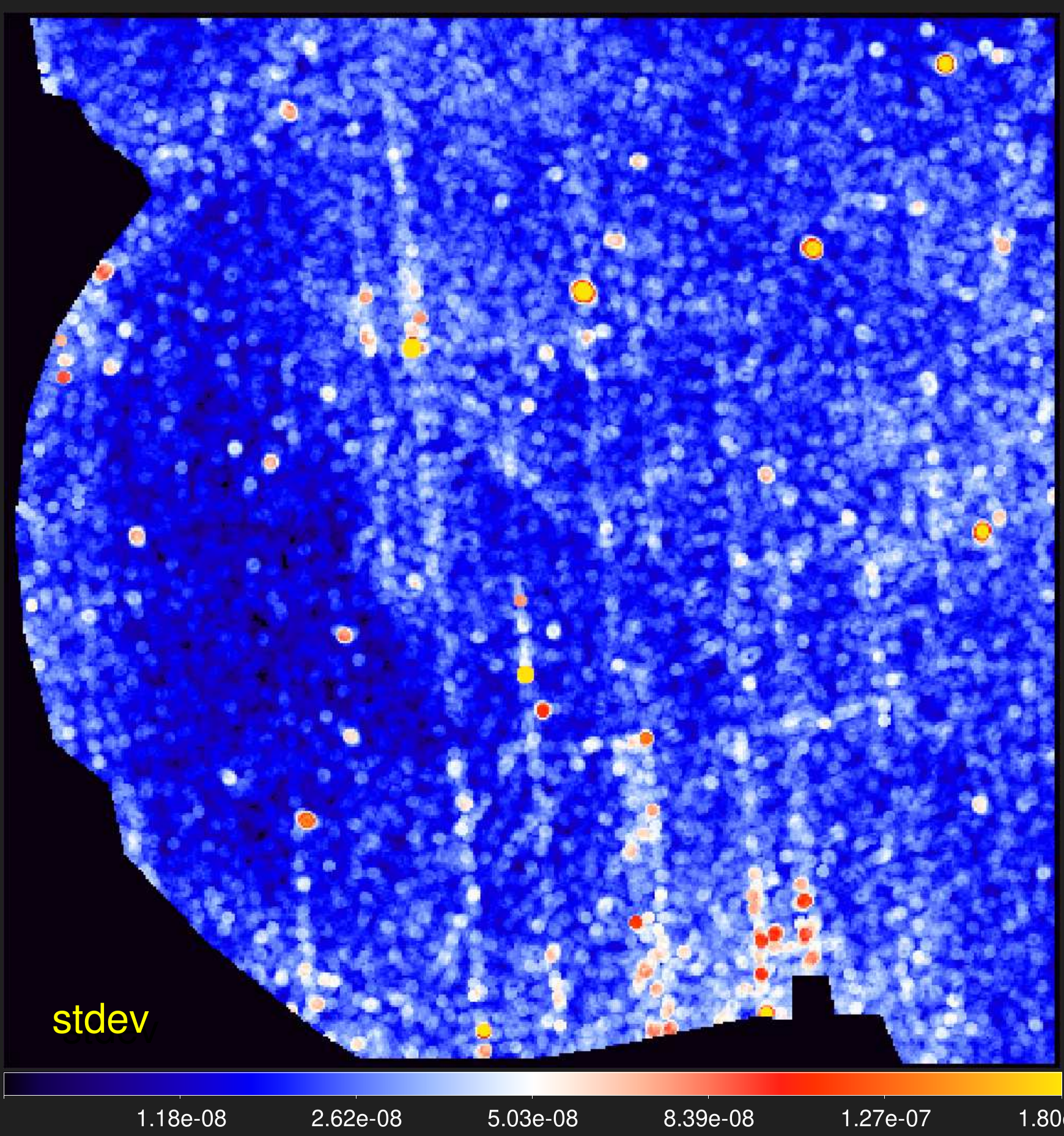}}
  \resizebox{0.328\hsize}{!}{\includegraphics{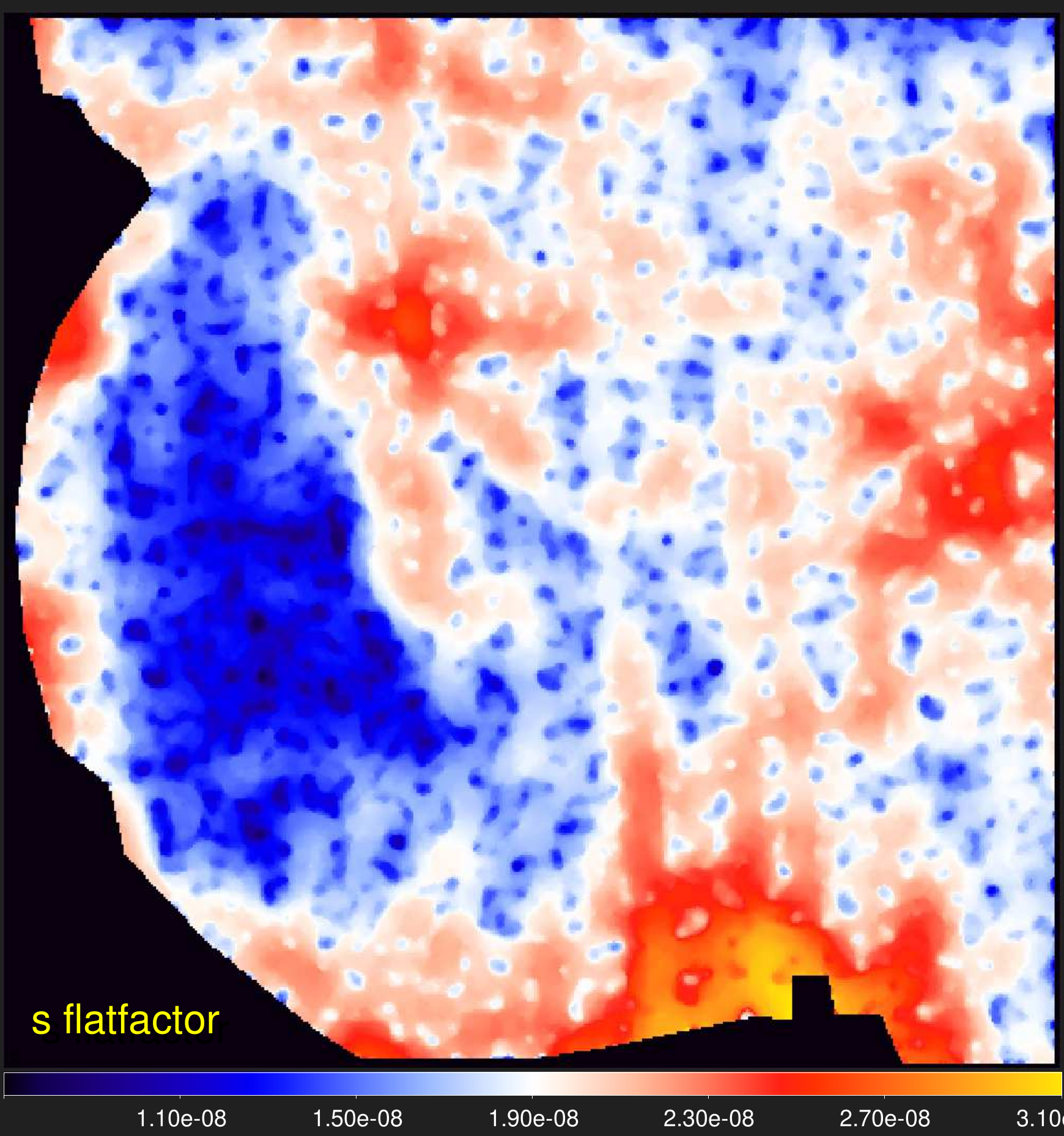}}
  \resizebox{0.328\hsize}{!}{\includegraphics{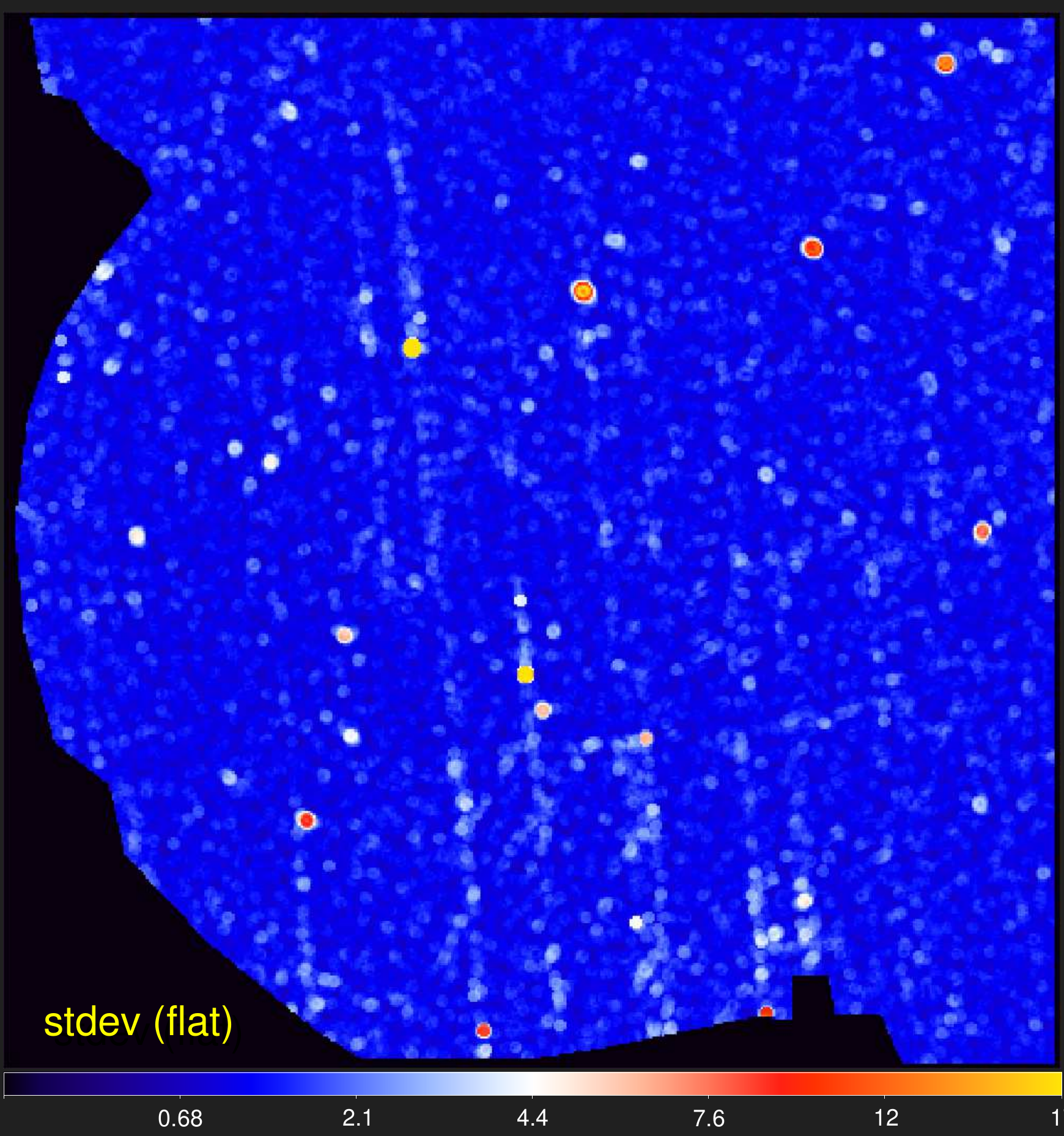}}}
\caption
{ 
Application of \textsl{getsf} to the \emph{XMM-Newton} $\lambda{\,\approx\,}0.0024$\,{${\mu}$m} image ($7${\arcsec\!} resolution)
of the supernova remnant \object{RX\,J1713.7-3946}, adopting $\{X|Y\}_{\lambda}{\,=\,}\{15,25\}${\arcsec}. The \emph{top} row shows
the original image $\mathcal{I}_{\!\lambda}$ and the backgrounds $\mathcal{B}_{{\lambda}{\{X|Y\}}}$ of sources and filaments. The
\emph{middle} row shows the component $\mathcal{S}_{{\lambda}}$, the footprint ellipses of $41$ acceptably good sources on
$\mathcal{S}_{{\lambda}{\rm D}}$ (red squares mark the spurious peaks), and the component $\mathcal{F}_{{\lambda}{\rm D}}$ with
$13$ non-branching skeletons $\mathcal{K}_{{k}{2}}$ corresponding to the scales $S_{\!k}{\,\approx\,}40${\arcsec}. The
\emph{bottom} row shows the standard deviations $\mathcal{U}_{\lambda}$ in the regularized component $\mathcal{S}_{{\lambda}{\rm
R}}$, the flattening image $\mathcal{Q}_{\lambda}$, and the standard deviations in the flattened component
$\mathcal{S}_{{\lambda}{\rm R}}\mathcal{Q}_{\lambda}^{-1}$. Intensities (in photons\,cm$^{-2}$\,s$^{-1}$) are limited in range
with square-root color mapping, except for $\mathcal{Q}_{\lambda}$, which is shown with linear mapping.
} 
\label{snremnant}
\vspace{20mm}
\end{figure*}

\object{RXJ\,1713.7-3946} was observed with \emph{XMM-Newton} (EPIC camera) in the X-ray waveband ($0.6{-}6$\,{keV}) centered at
$0.0024$\,{${\mu}$m}. The $0.5{\degr}{\times\,}0.5${\degr} image in Fig.~\ref{snremnant} is a mosaic of multiple
observations\footnote{\url{http://nxsa.esac.esa.int/nxsa-web/}}, first presented in \cite{Acero_etal2017}. With an average angular
resolution of $7${\arcsec}, it reveals the southeast segment of the supernova remnant shell that may have been created by the
explosion of the historical supernova \object{SN\,393}, whose center of explosion is located beyond the upper right image corner.
For this source and filament extraction with \textsl{getsf}, maximum sizes $\{X|Y\}_{\lambda}{\,=\,}\{15,25\}${\arcsec} were
adopted (Sect.~\ref{maxsizes}).

The observed X-ray image (Fig.~\ref{snremnant}) has relatively low counts of the detected photons per pixel and high levels of
Poisson noise. The image is contaminated by linear artifacts and several spurious single-pixel spikes. The latter may appear in
these images when just one or several photons are detected at an edge of the mapped area.

The image features several elongated shock fronts created by the expanding supernova shell, and a number of faint and bright point
sources, all of them well isolated. The \textsl{getsf} extraction greatly simplified the image by separating the components of
sources $\mathcal{S}_{{\lambda}}$, filaments $\mathcal{F}_{{\lambda}}$, and their backgrounds $\mathcal{B}_{{\lambda}{\{X|Y\}}}$.
The small-scale fluctuation levels across the observed image are only within a factor of two, therefore the improvement caused by
the flattening is not clearly discernible in $\mathcal{S}_{{\lambda}}$. However, the images of standard deviations show that the
flat source detection image $\mathcal{S}_{{\lambda}{\rm D}}$ has uniform fluctuations over the entire image, which is beneficial
for source detection.

The extraction catalog contains measurements of $41$ sources, all of them selected as acceptably good by Eq.~(\ref{acceptable}).
Although the spurious one-pixel spikes were not removed before the extraction, \textsl{getsf} identified them as such (red squares
in Fig.~\ref{snremnant}) and eliminated them from the catalog. Despite the faintness of the observed X-ray image and the Poisson
noise, the three prominent shocks of the supernova shell become clearly visible and are extracted in the filament component.



\subsection{Star-forming galaxy \object{NGC\,6744}}
\label{galex}

\begin{figure*}                                                               
\centering
\centerline{
  \resizebox{0.328\hsize}{!}{\includegraphics{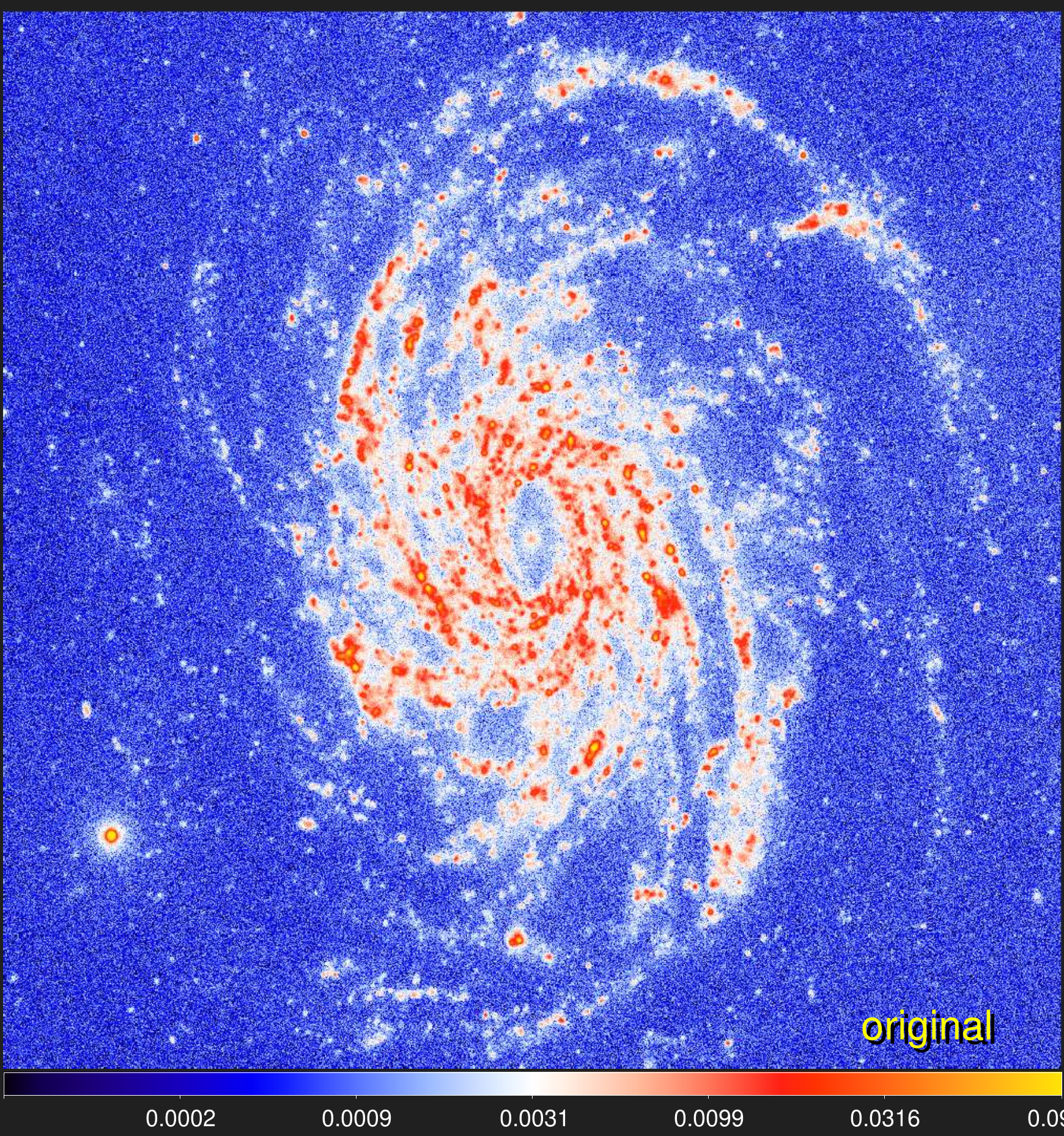}}
  \resizebox{0.328\hsize}{!}{\includegraphics{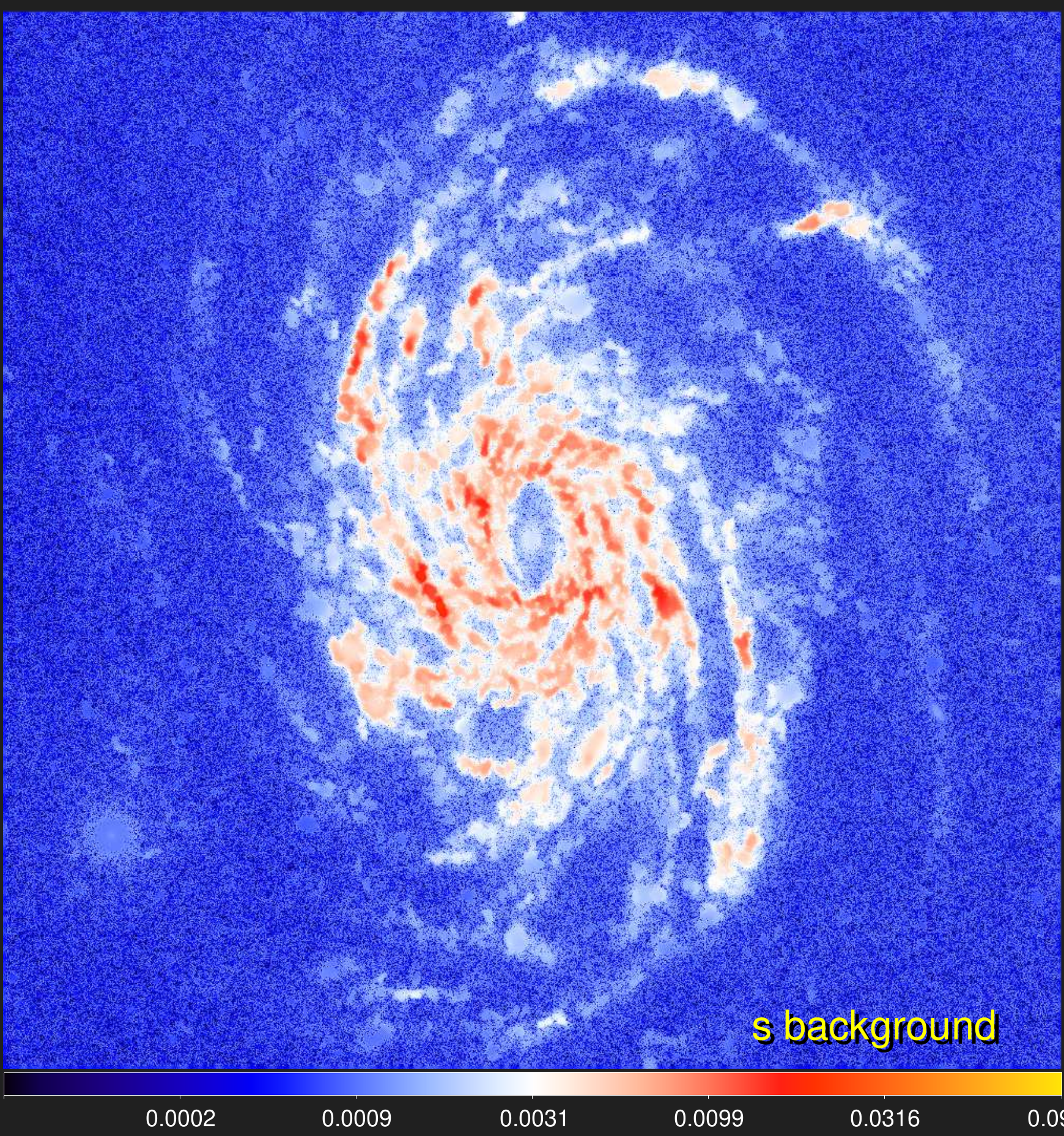}}
  \resizebox{0.328\hsize}{!}{\includegraphics{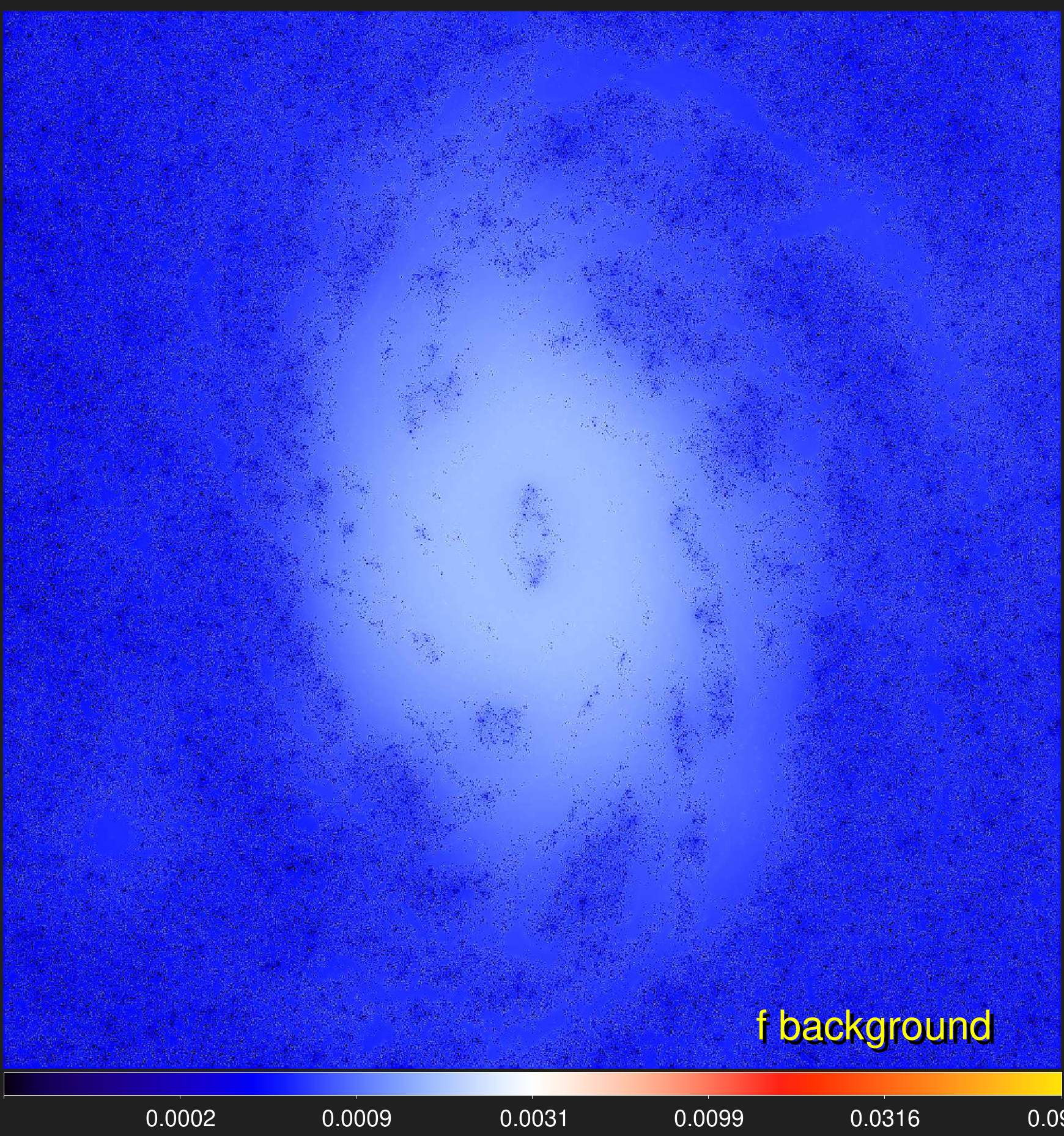}}}
\vspace{0.5mm}
\centerline{
  \resizebox{0.328\hsize}{!}{\includegraphics{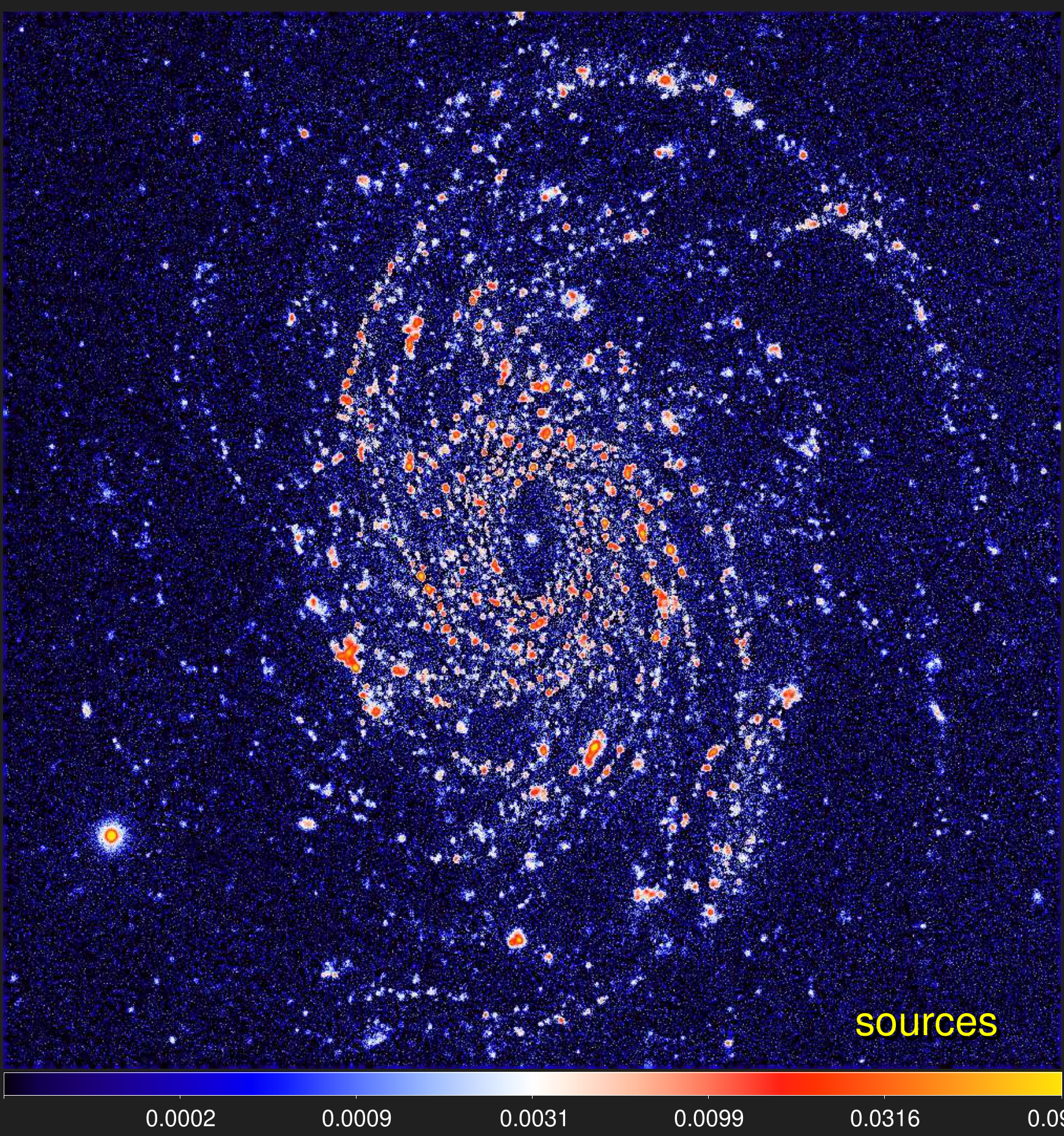}}
  \resizebox{0.328\hsize}{!}{\includegraphics{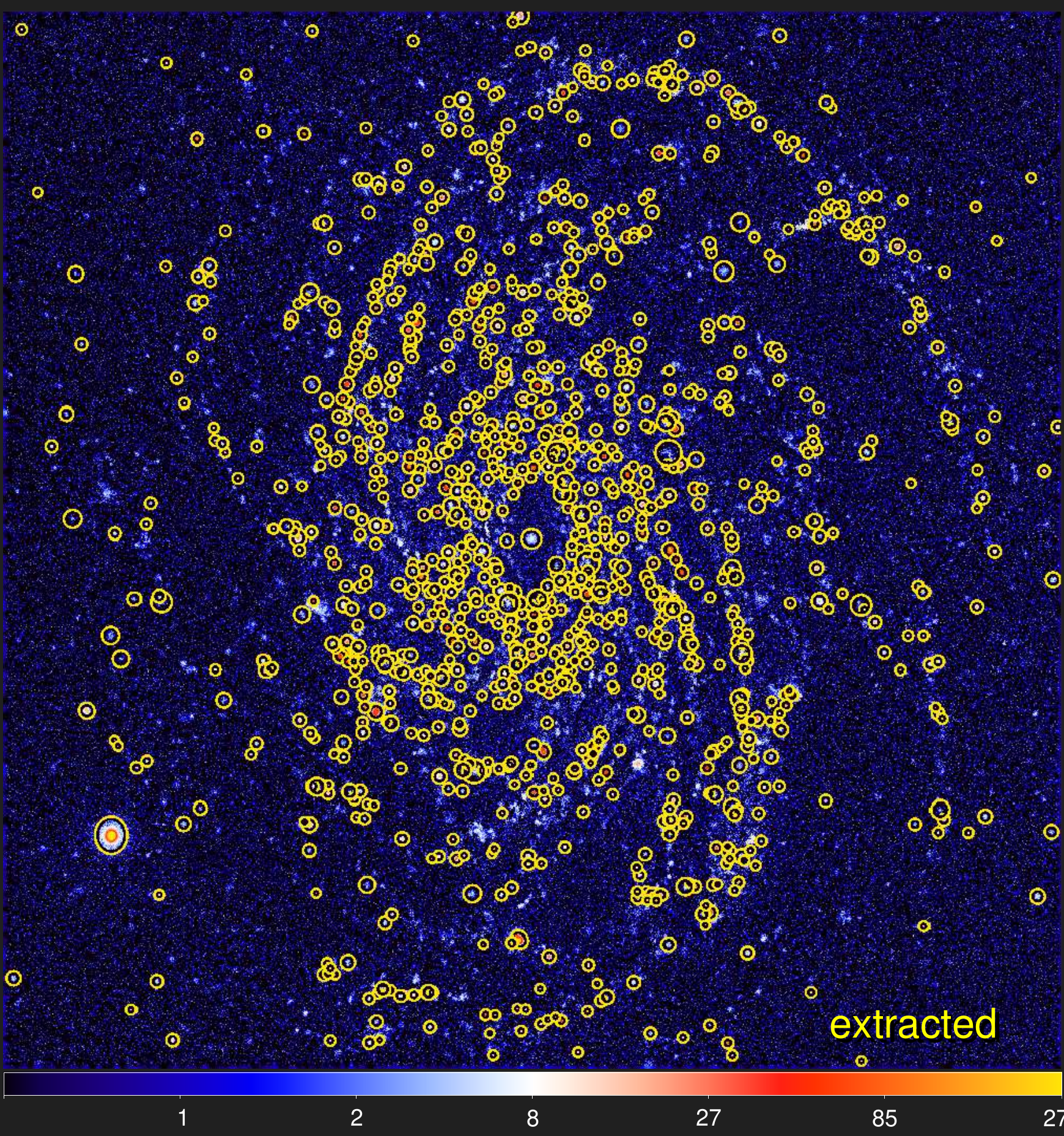}}
  \resizebox{0.328\hsize}{!}{\includegraphics{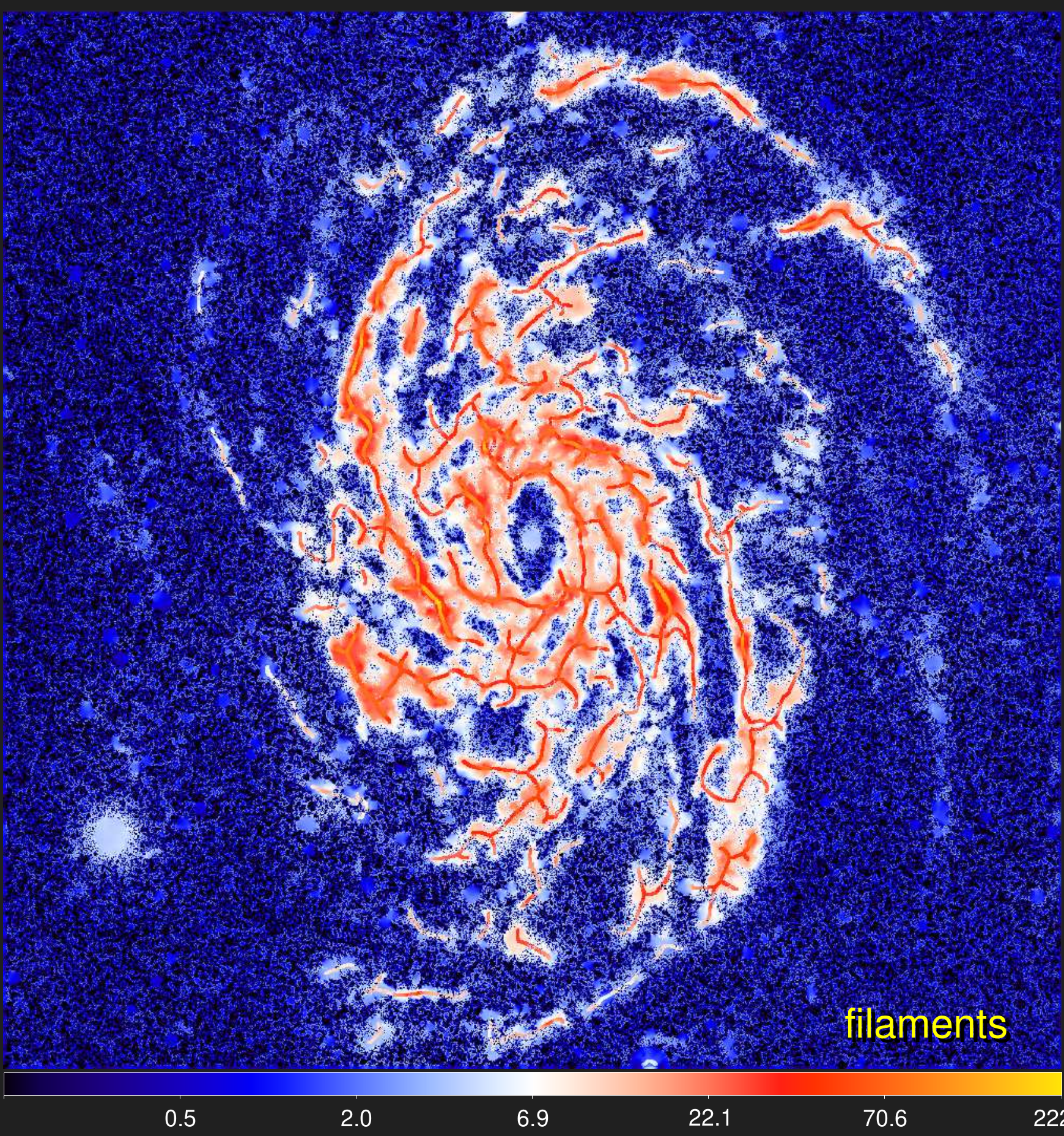}}}
\vspace{0.5mm}
\centerline{
  \resizebox{0.328\hsize}{!}{\includegraphics{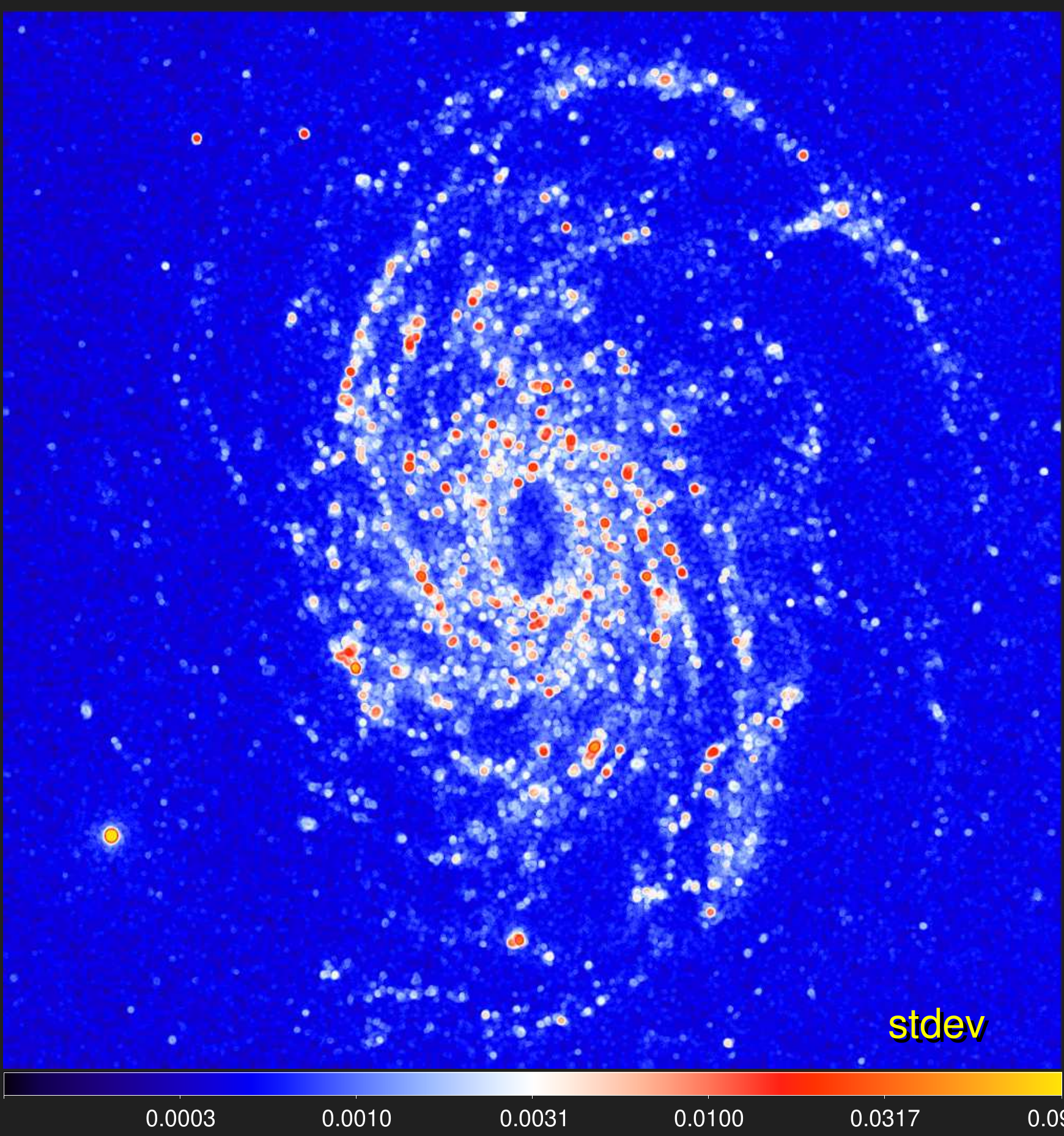}}
  \resizebox{0.328\hsize}{!}{\includegraphics{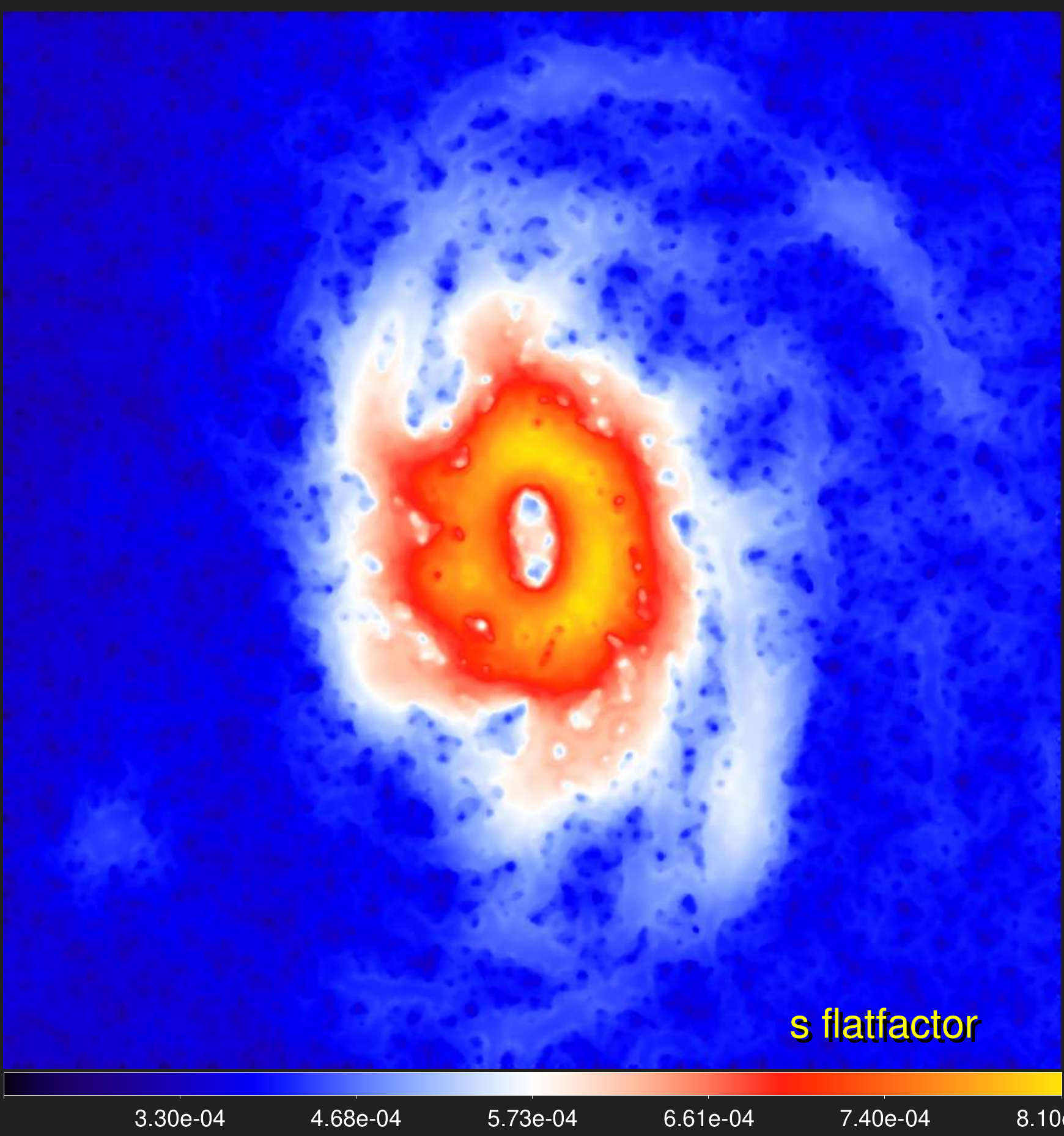}}
  \resizebox{0.328\hsize}{!}{\includegraphics{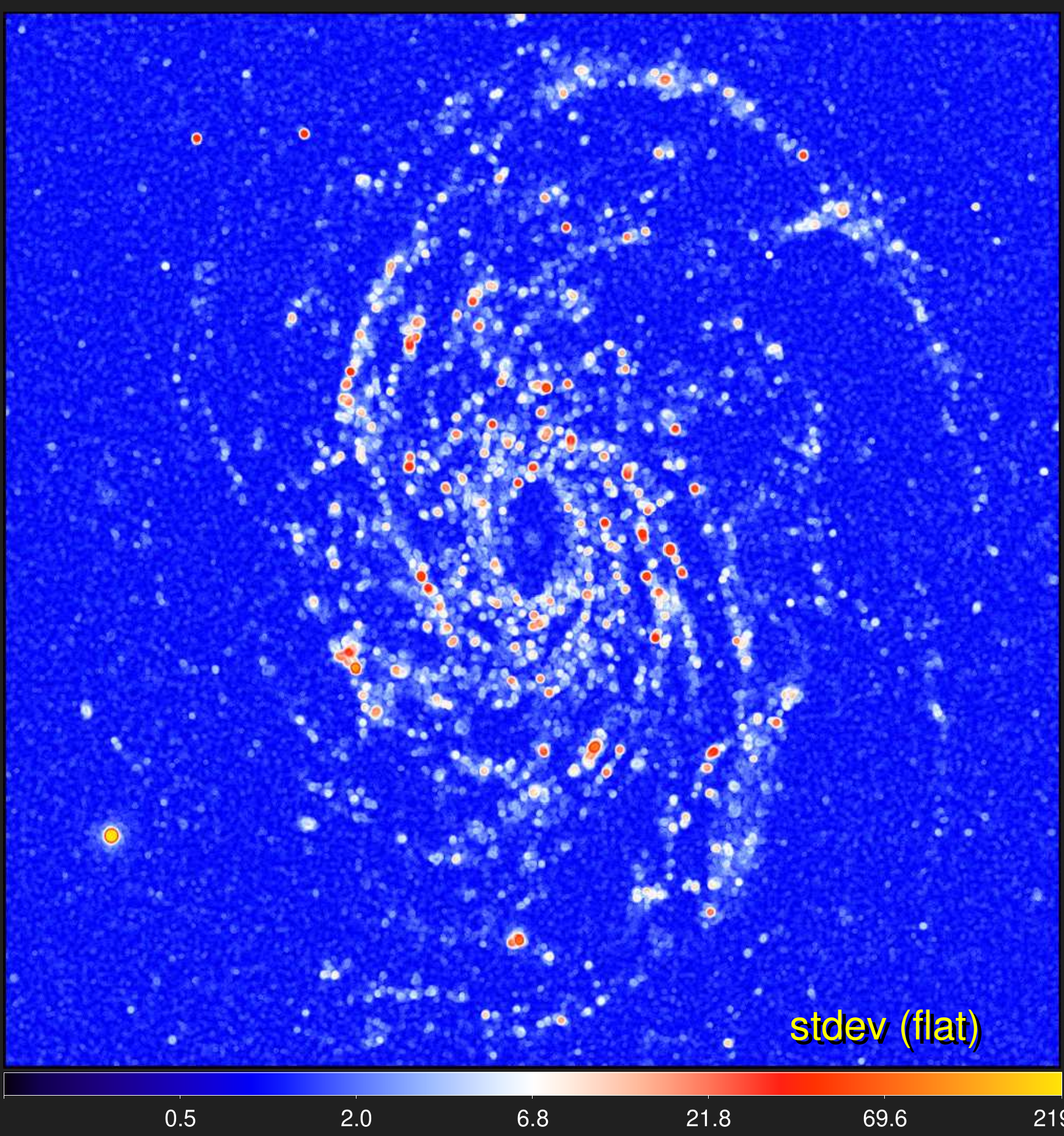}}}
\caption
{ 
Application of \textsl{getsf} to the \emph{GALEX} $\lambda{\,=\,}0.15$\,{${\mu}$m} image ($4${\arcsec\!} resolution) of the spiral
galaxy \object{NGC\,6744}, adopting $\{X|Y\}_{\lambda}{\,=\,}20${\arcsec}. The \emph{top} row shows the original image
$\mathcal{I}_{\!\lambda}$ and the backgrounds $\mathcal{B}_{{\lambda}{\{X|Y\}}}$ of sources and filaments. The \emph{middle} row
shows the component $\mathcal{S}_{{\lambda}}$, the footprint ellipses of $1130$ acceptably good sources on
$\mathcal{S}_{{\lambda}{\rm D}}$, and the component $\mathcal{F}_{{\lambda}{\rm D}}$ with $147$ skeletons $\mathcal{K}_{{k}{2}}$
corresponding to the scales $S_{\!k}{\,\approx\,}30${\arcsec}. The \emph{bottom} row shows the standard deviations
$\mathcal{U}_{\lambda}$ in the regularized component $\mathcal{S}_{{\lambda}{\rm R}}$, the flattening image
$\mathcal{Q}_{\lambda}$, and the standard deviations in the flattened component $\mathcal{S}_{{\lambda}{\rm
R}}\mathcal{Q}_{\lambda}^{-1}$. Some skeletons may only appear to have branches because they were widened for this presentation.
Intensities (in counts\,s$^{-1}$) are limited in range with logarithmic color mapping, except for $\mathcal{Q}_{\lambda}$, which
is shown with squared mapping.
} 
\label{galaxy}
\vspace{20mm}
\end{figure*}

\object{NGC\,6744} was observed with \emph{GALEX} in a far-ultraviolet (FUV) waveband ($1350{-}1750$\,\AA) centered at
$0.15$\,{${\mu}$m} \citep{Lee_etal2011}. The $0.4{\degr}{\times\,}0.4${\degr}
image\footnote{\url{https://archive.stsci.edu/missions-and-data/galex/}} in Fig.~\ref{galaxy} with an angular resolution of
$4${\arcsec\!} shows the spiral galaxy, which is considered to be similar to our own Galaxy. Despite noisiness of the FUV image, it
displays the spiral arms with many hundreds of unresolved emission sources. These are the regions of ongoing star formation, heated
by the embedded young massive stars. For this source and filament extraction, maximum sizes $\{X|Y\}_{\lambda}{\,=\,}20${\arcsec}
were adopted (Sect.~\ref{maxsizes}).

Separation of the structural components by \textsl{getsf} provided independent images of sources $\mathcal{S}_{{\lambda}}$,
filaments $\mathcal{F}_{{\lambda}}$, and their backgrounds $\mathcal{B}_{{\lambda}{\{X|Y\}}}$ (Fig.~\ref{galaxy}). Fluctuation
levels in the observed image vary within a factor of two, largely in the central, brighter part of the galaxy. Flattening of the
$\mathcal{S}_{{\lambda}}$ component effectively equalized the fluctuations across the detection image $\mathcal{S}_{{\lambda}{\rm
D}}$, improving the extraction results. 

The source catalog contains measurements of $1169$ sources, $1130$ of which are selected as acceptably good by
Eq.~(\ref{acceptable}). Most of the sources likely correspond to the star-forming regions along the galactic spiral arms; many of
them overlap with each other, hence they required deblending for accurate measurements of their fluxes. The filaments extracted in
the $\mathcal{F}_{{\lambda}}$ component represent the spiral arms and their branches. The $147$ skeletons trace the simple,
non-branching segments of the filamentary network (Sect.~\ref{fildetection}).



\subsection{Supernova remnant \object{NGC\,6960}}
\label{hubble}

\begin{figure*}                                                               
\centering
\centerline{
  \resizebox{0.328\hsize}{!}{\includegraphics{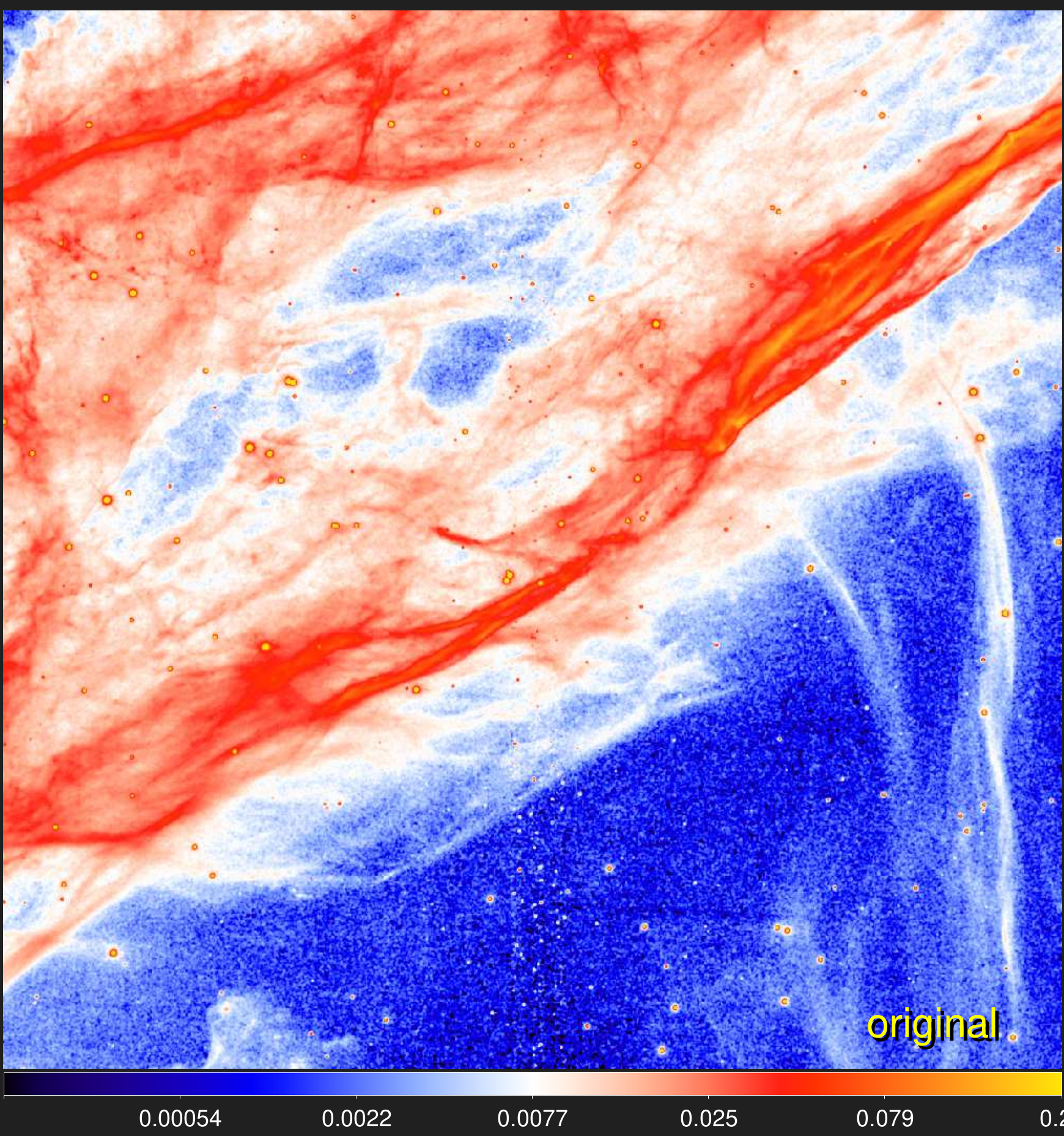}}
  \resizebox{0.328\hsize}{!}{\includegraphics{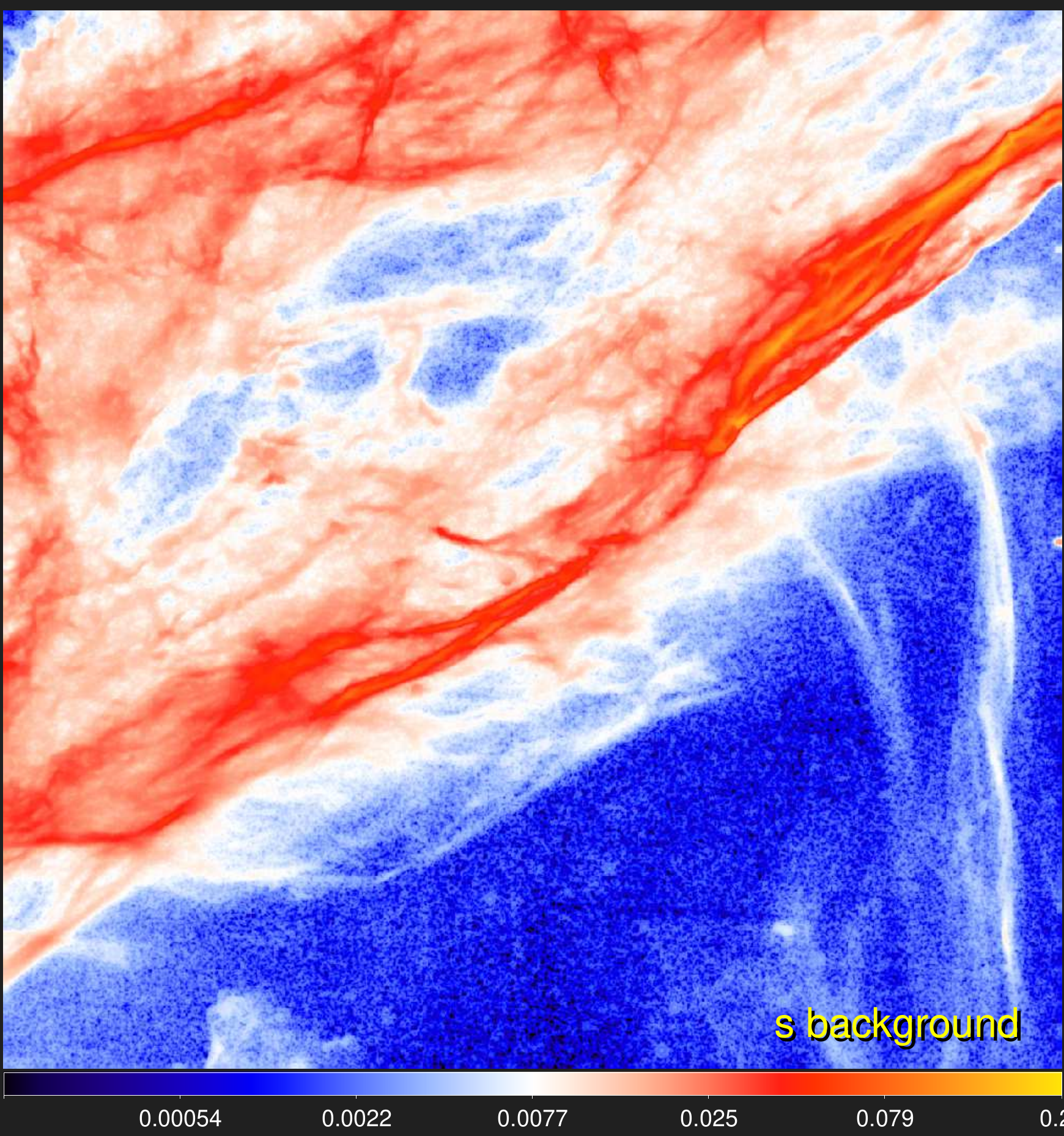}}
  \resizebox{0.328\hsize}{!}{\includegraphics{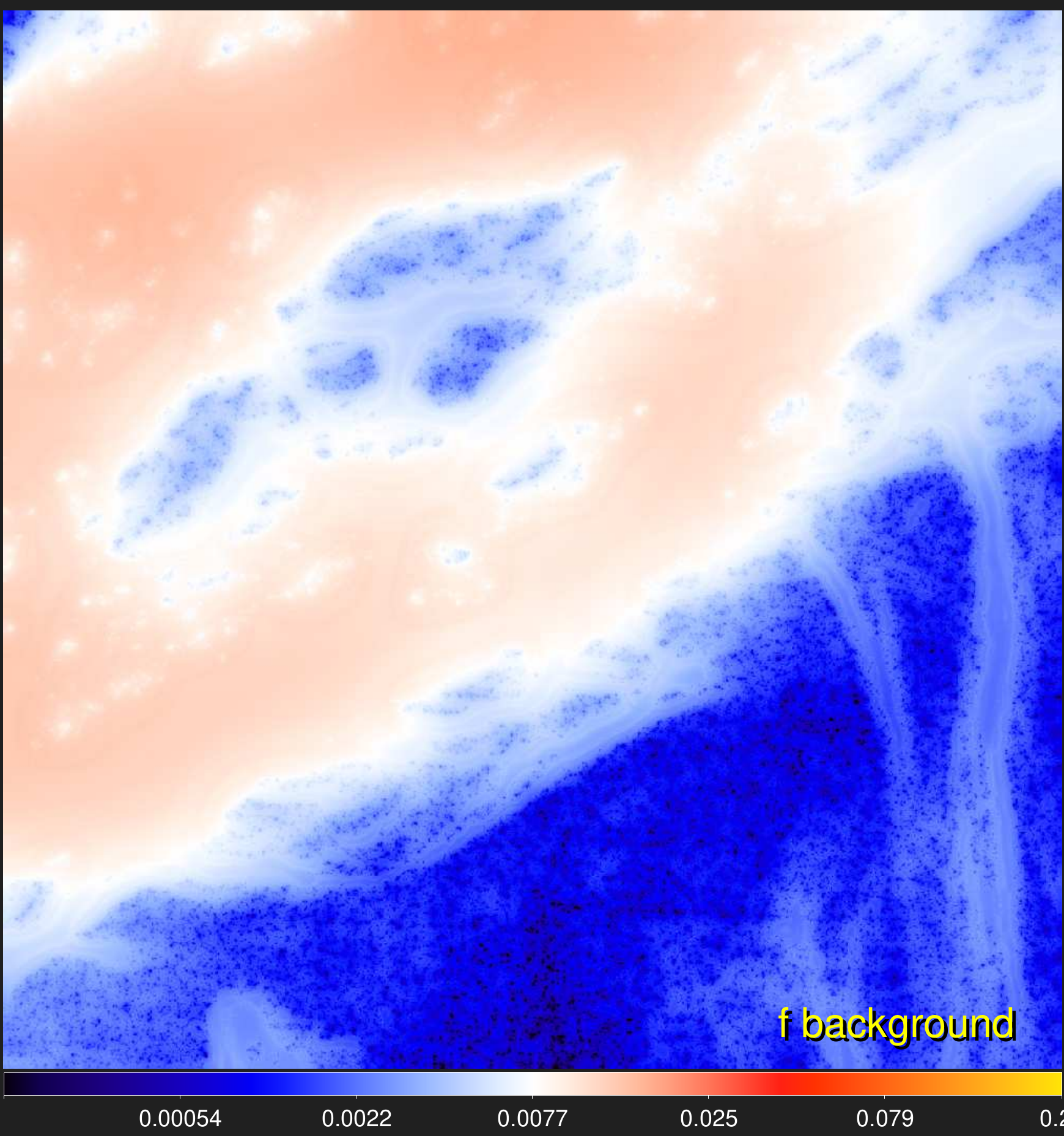}}}
\vspace{0.5mm}
\centerline{
  \resizebox{0.328\hsize}{!}{\includegraphics{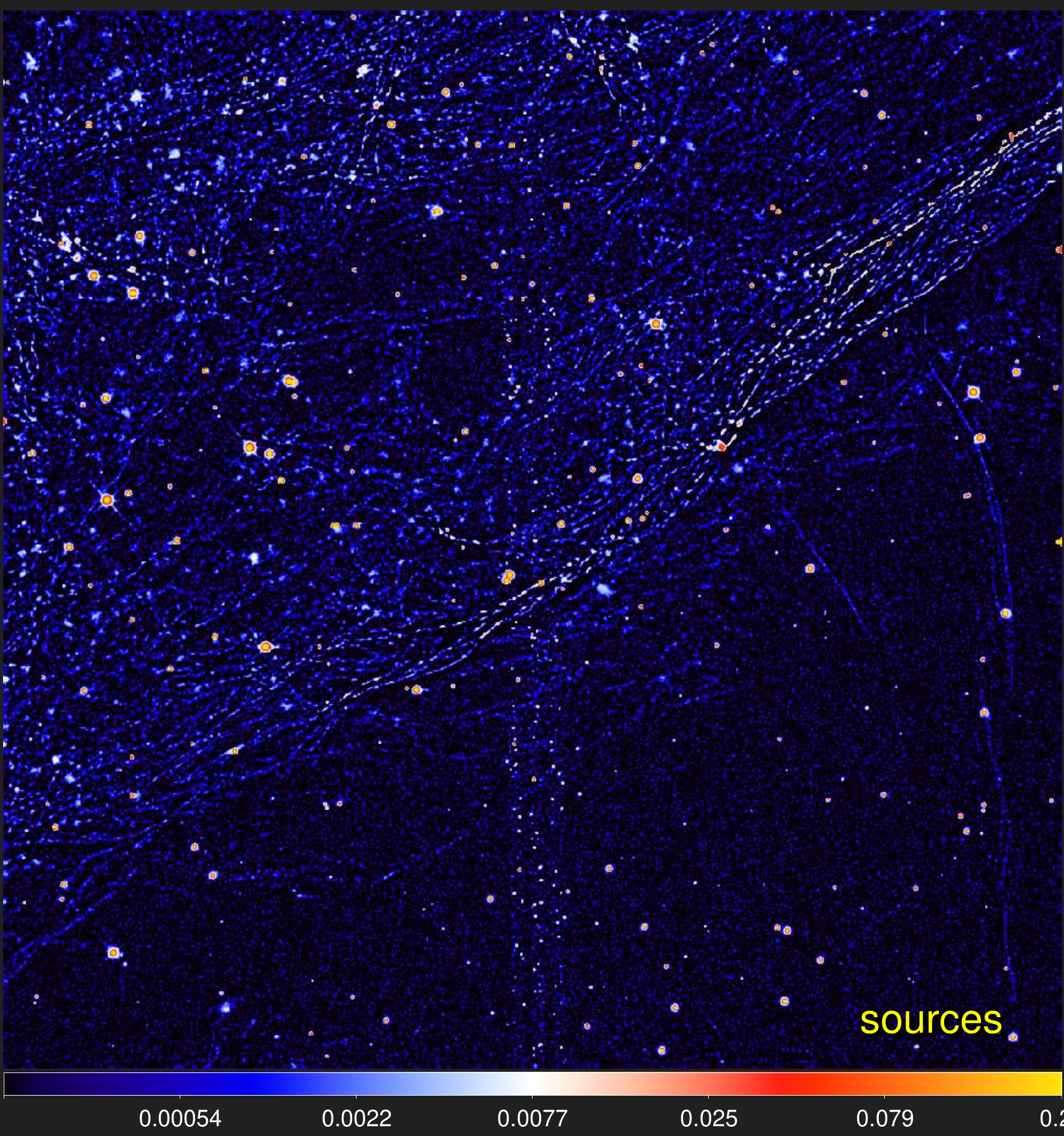}}
  \resizebox{0.328\hsize}{!}{\includegraphics{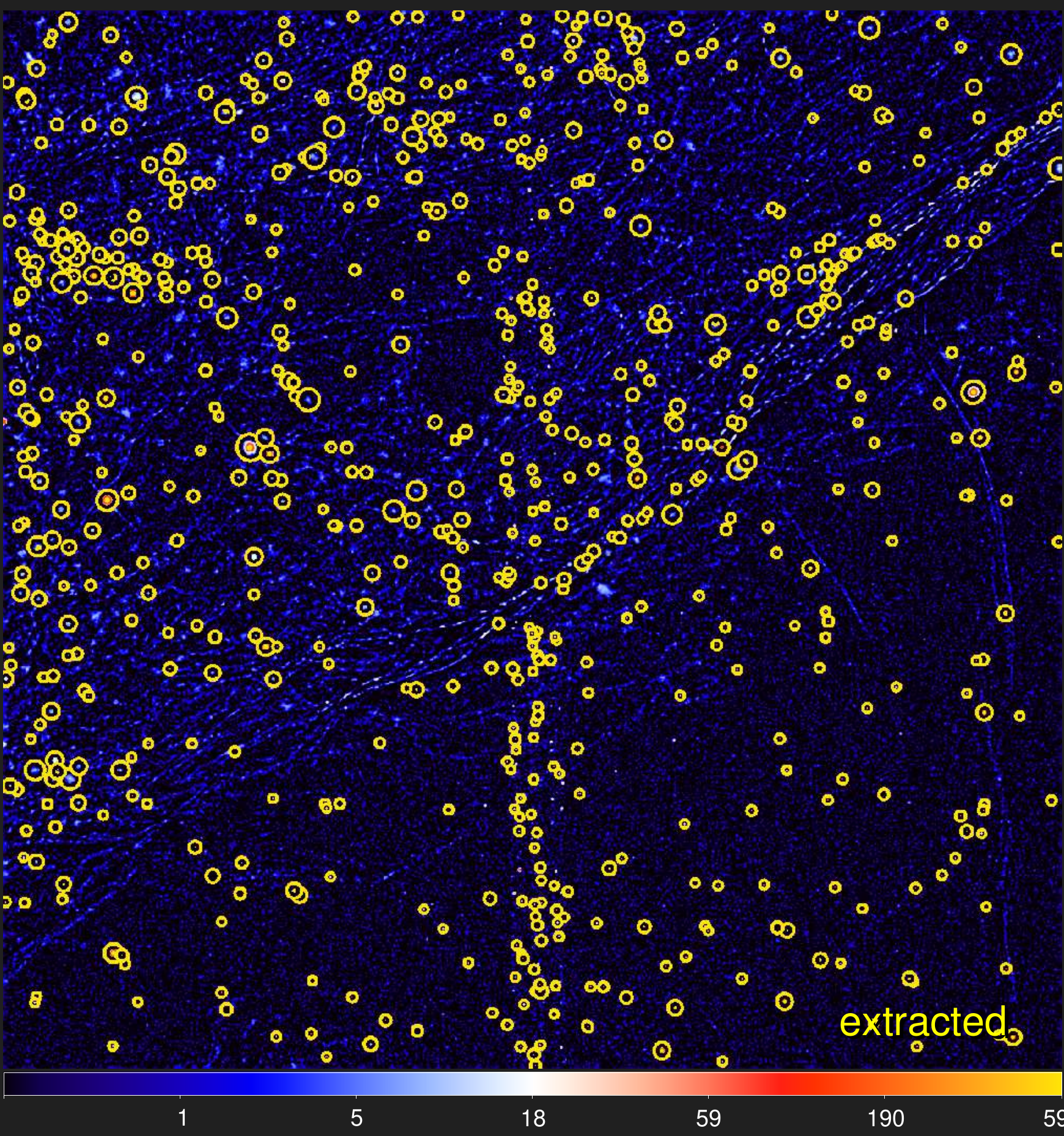}}
  \resizebox{0.328\hsize}{!}{\includegraphics{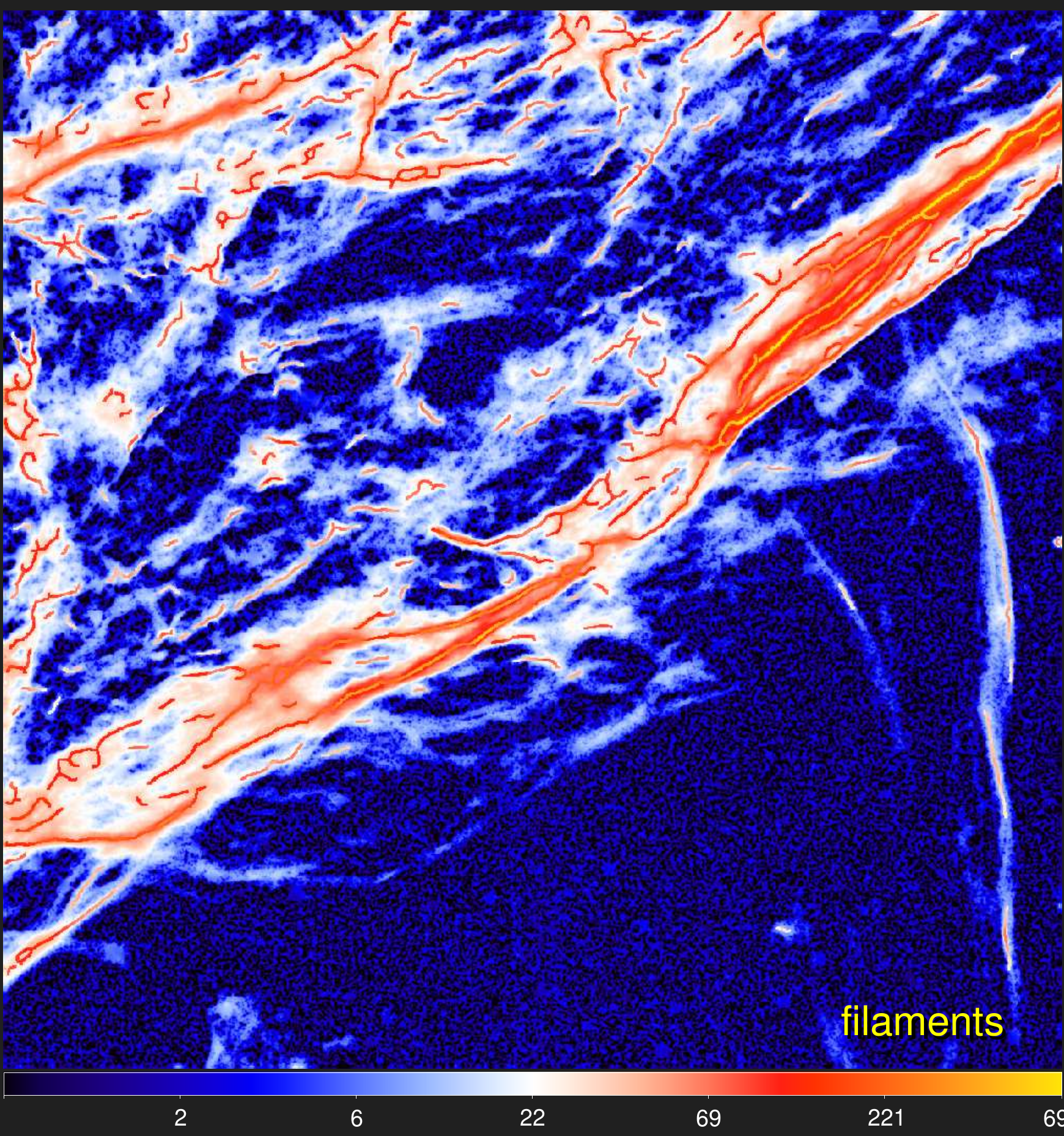}}}
\vspace{0.5mm}
\centerline{
  \resizebox{0.328\hsize}{!}{\includegraphics{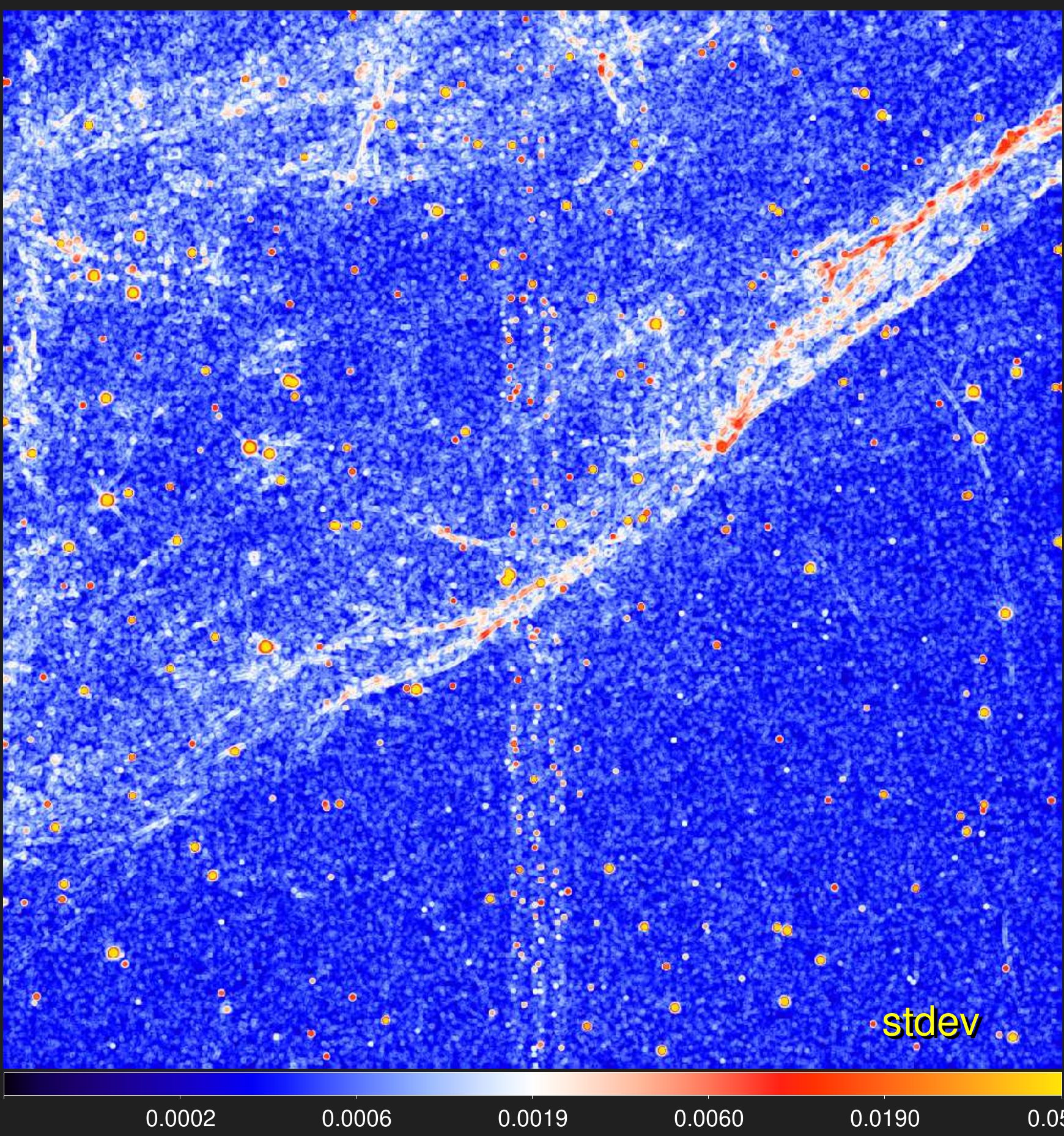}}
  \resizebox{0.328\hsize}{!}{\includegraphics{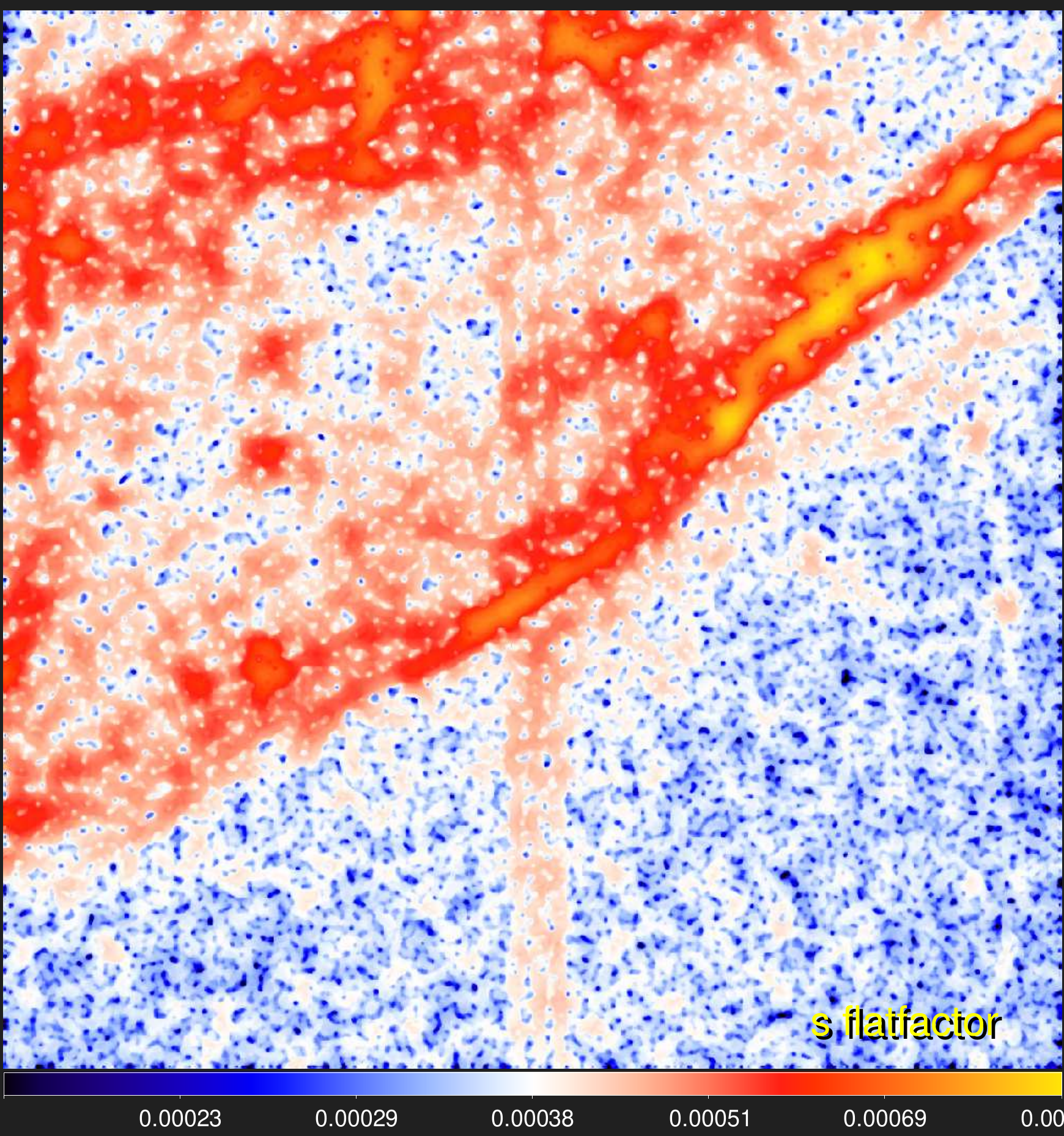}}
  \resizebox{0.328\hsize}{!}{\includegraphics{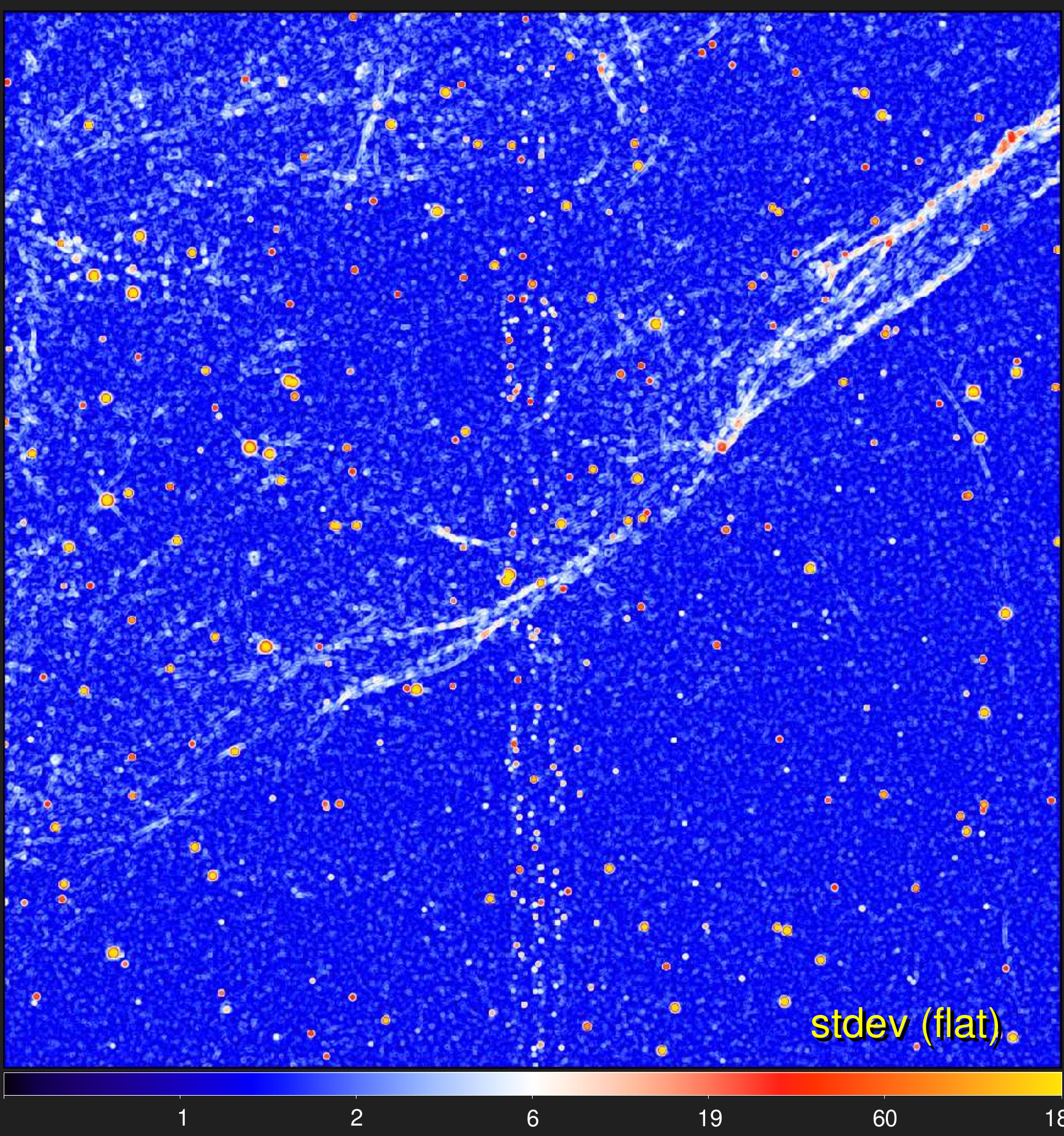}}}
\caption
{ 
Application of \textsl{getsf} to the \emph{Hubble} $\lambda{\,=\,}0.6$\,{${\mu}$m} image ($0.2${\arcsec\!} resolution) of the
supernova remnant \object{NGC\,6960}, adopting $\{X|Y\}_{\lambda}{\,=\,}\{0.5,2\}${\arcsec}. The \emph{top} row shows the original
image $\mathcal{I}_{\!\lambda}$ and the backgrounds $\mathcal{B}_{{\lambda}{\{X|Y\}}}$ of sources and filaments. The \emph{middle}
row shows the component $\mathcal{S}_{{\lambda}}$, the footprint ellipses of $690$ acceptably good sources on
$\mathcal{S}_{{\lambda}{\rm D}}$, and the component $\mathcal{F}_{{\lambda}{\rm D}}$ with $100$ skeletons $\mathcal{K}_{{k}{2}}$
corresponding to the scales $S_{\!k}{\,\approx\,}1${\arcsec}. The \emph{bottom} row shows the standard deviations
$\mathcal{U}_{\lambda}$ in the regularized component $\mathcal{S}_{{\lambda}{\rm R}}$, the flattening image
$\mathcal{Q}_{\lambda}$, and the standard deviations in the flattened component $\mathcal{S}_{{\lambda}{\rm
R}}\mathcal{Q}_{\lambda}^{-1}$. Some skeletons may only appear to have branches because they were widened for this presentation.
Intensities (in electrons\,s$^{-1}$) are limited in range with logarithmic color mapping, except for $\mathcal{Q}_{\lambda}$, which
is shown with square-root mapping.
} 
\label{veil}
\vspace{20mm}
\end{figure*}

\object{NGC\,6960} was observed with \emph{Hubble} in five UVIS wavebands ($0.5{-}0.8$\,nm) centered at $0.6$\,{${\mu}$m}, within
the frame of the \emph{Hubble} Heritage project \citep[][PI: Z.\,Levay]{Mack_etal2015}. The small
$73{\arcsec}{\times\,}73${\arcsec} image\footnote{\url{https://archive.stsci.edu/prepds/heritage/veil/}} in Fig.~\ref{veil} with an
angular resolution of $0.2${\arcsec} represents a small fragment of the \object{Veil Nebula}, which is a segment of the
\object{Cygnus Loop}, the large expanding shell of a supernova remnant \citep[][]{Fesen_etal2018}.
For this source and filament extraction with \textsl{getsf}, maximum sizes $\{X|Y\}_{\lambda}{\,=\,}\{0.5,2\}${\arcsec} were
adopted (Sect.~\ref{maxsizes}).

The observed image (Fig.~\ref{veil}) is dominated by impressive fine filamentary structure of the nebula, seen in emission of a
number of atomic lines. Many unresolved
intensity peaks of sources are less prominent on this bright backdrop. The structural
components were separated by \textsl{getsf} in the independent images of sources $\mathcal{S}_{{\lambda}}$, filaments
$\mathcal{F}_{{\lambda}}$, and backgrounds $\mathcal{B}_{{\lambda}{\{X|Y\}}}$; together with the flattening of detection images,
this greatly facilitates their extraction and analysis. 

The source catalog contains measurements of $786$ sources, $690$ of which are selected as acceptably good by
Eq.~(\ref{acceptable}). The strings of sources that run up through the middle of the image are the spurious peaks created by the
linear artifacts. The spurious spikes were not cut out of the image before this extraction to illustrate that they need to be
removed to avoid contamination of the source catalogs. The finely structured filamentary network of the nebula that is extracted by
\textsl{getsf} in the $\mathcal{F}_{{\lambda}}$ component comprises $100$ skeletons representing its simple, non-branching segments
(Sect.~\ref{fildetection}).



\subsection{Star-forming cloud \object{L\,1688}}
\label{spitzer}

\begin{figure*}                                                               
\centering
\centerline{
  \resizebox{0.328\hsize}{!}{\includegraphics{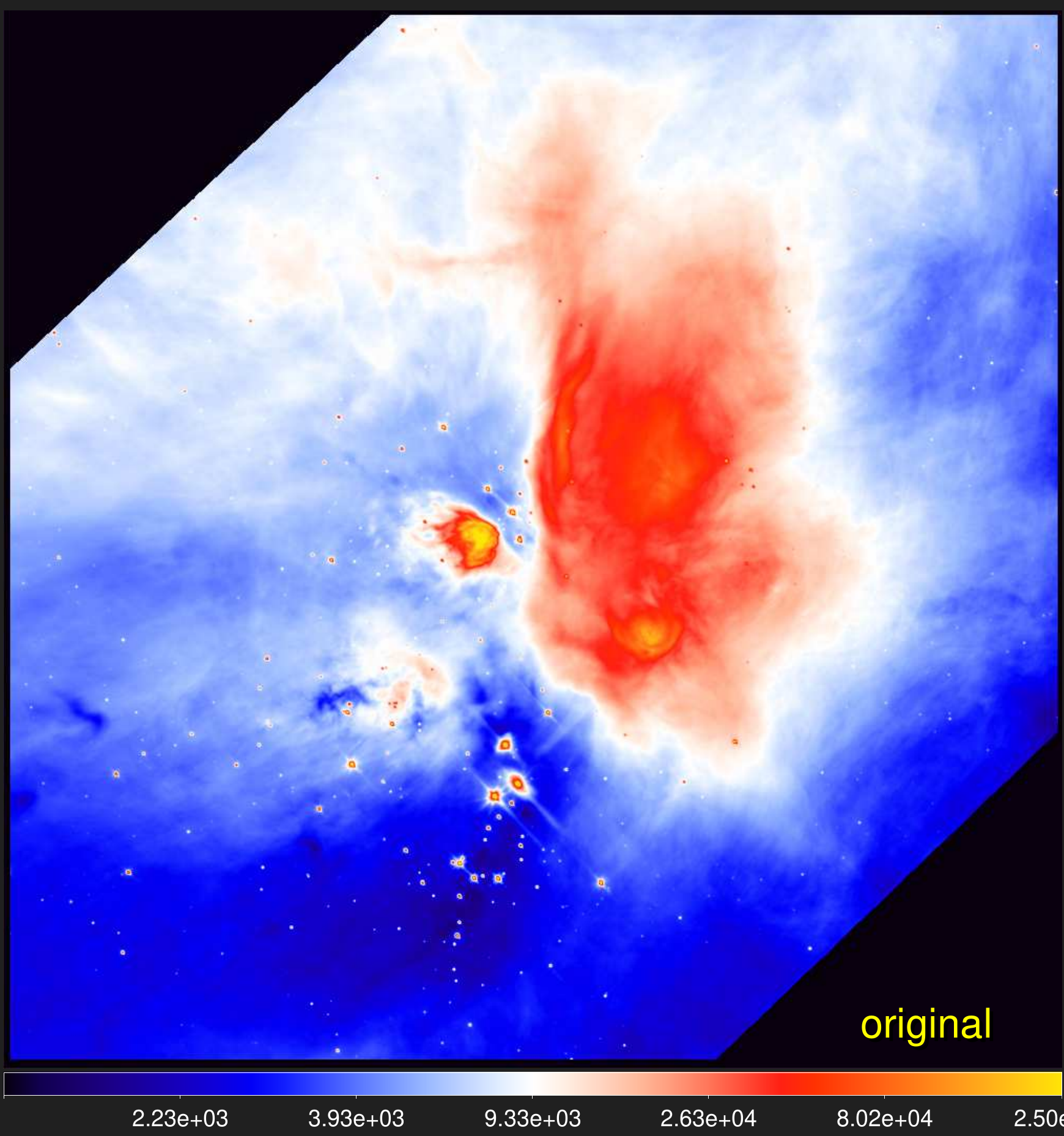}}
  \resizebox{0.328\hsize}{!}{\includegraphics{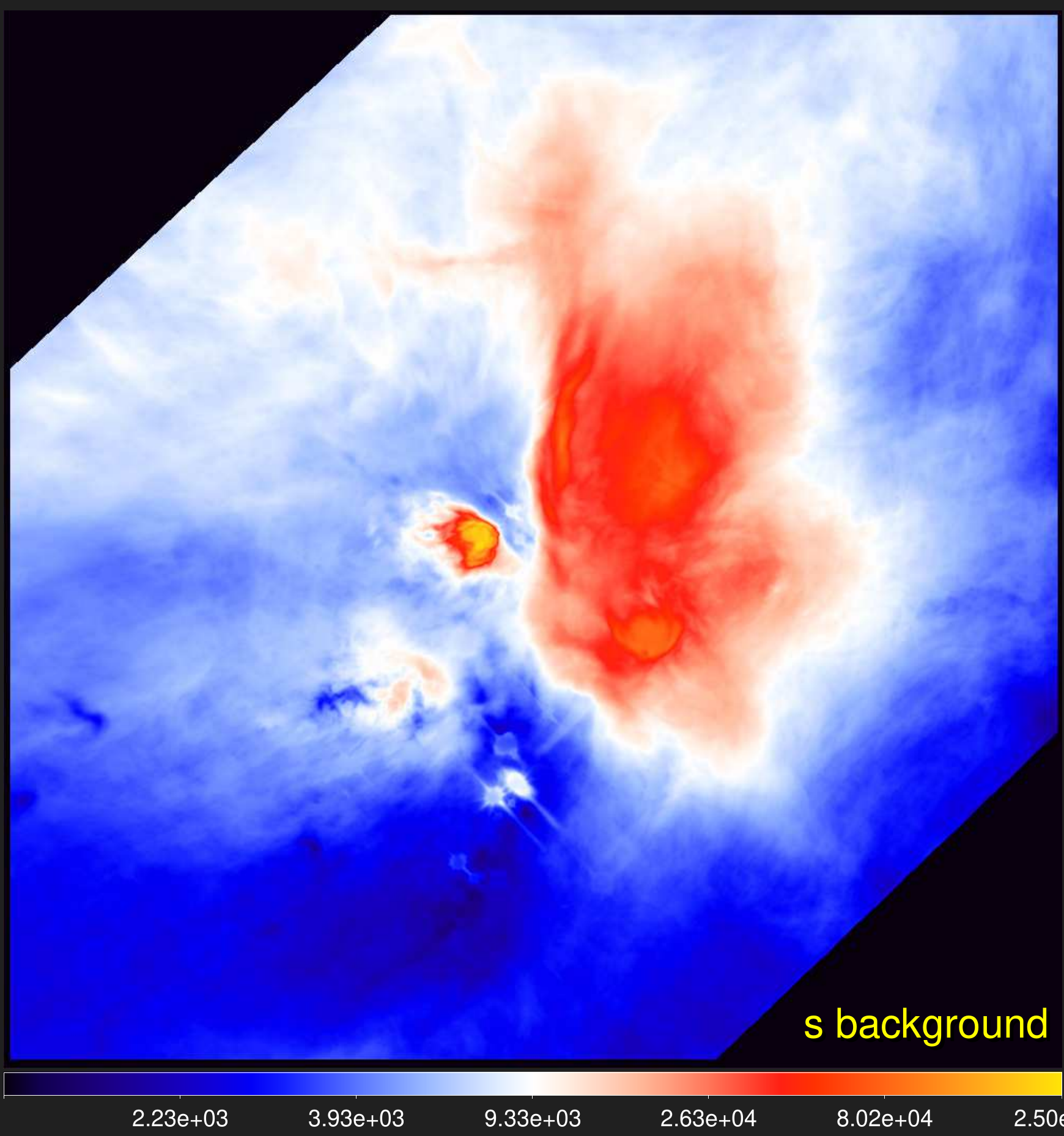}}
  \resizebox{0.328\hsize}{!}{\includegraphics{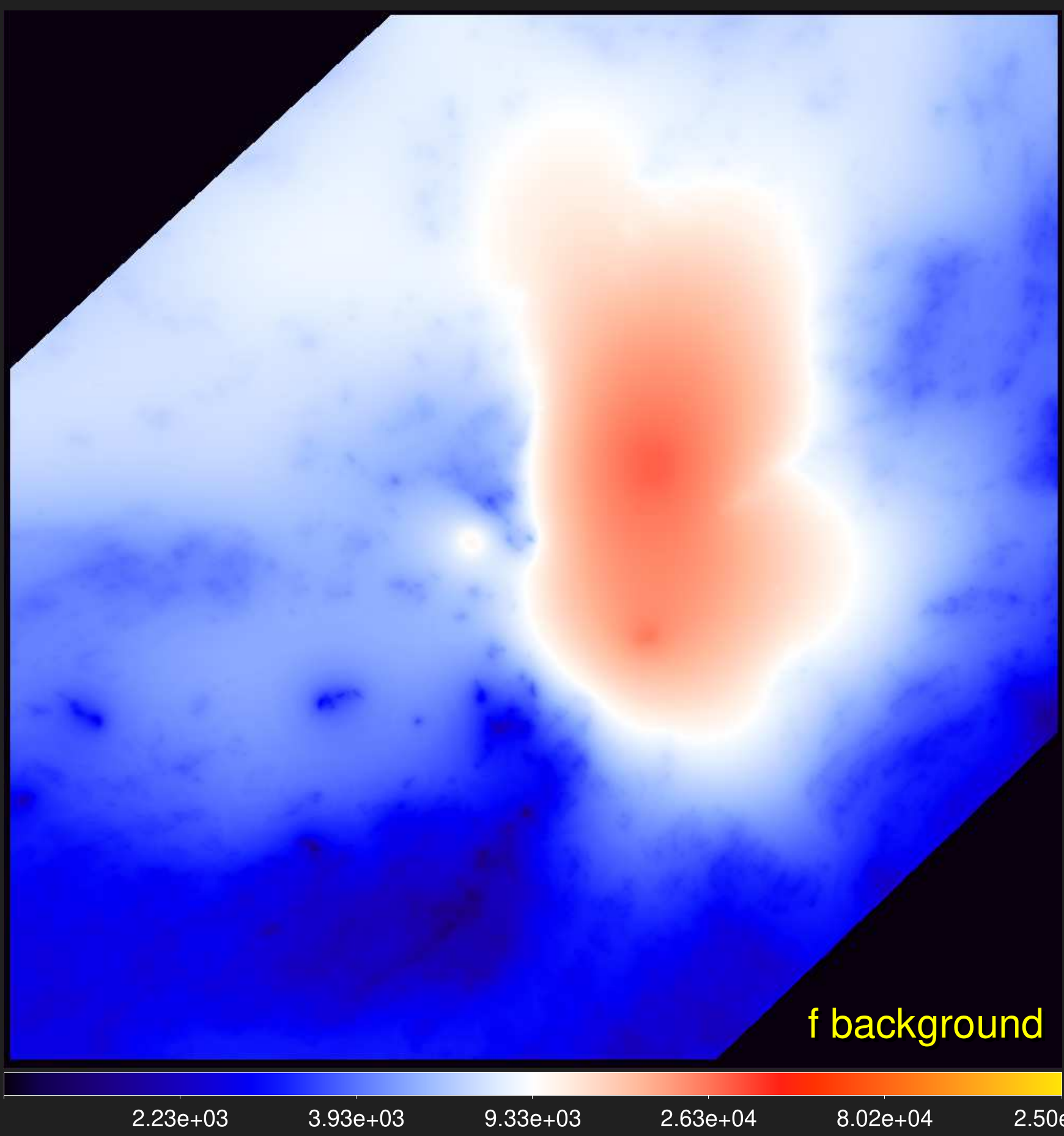}}}
\vspace{0.5mm}
\centerline{
  \resizebox{0.328\hsize}{!}{\includegraphics{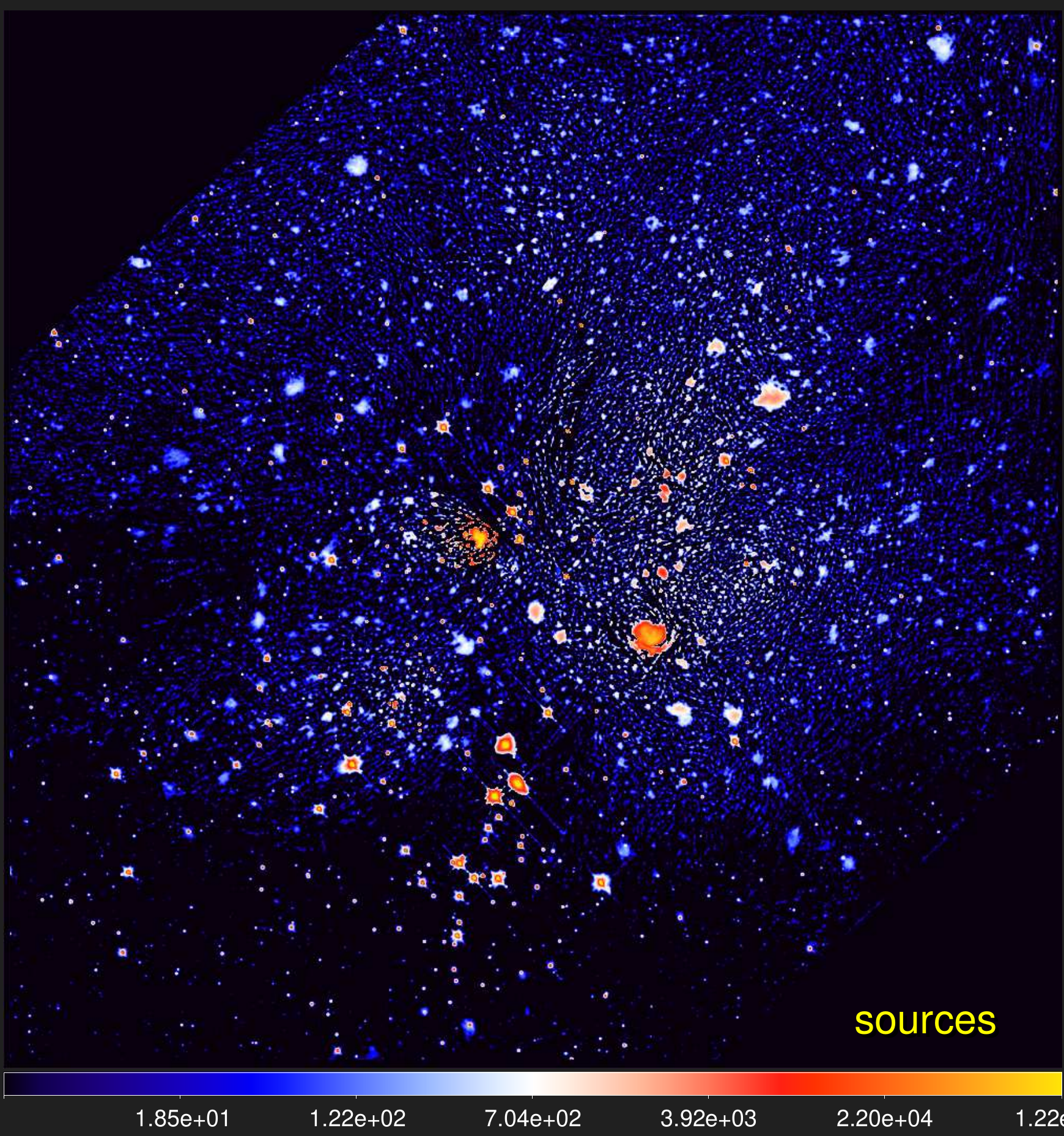}}
  \resizebox{0.328\hsize}{!}{\includegraphics{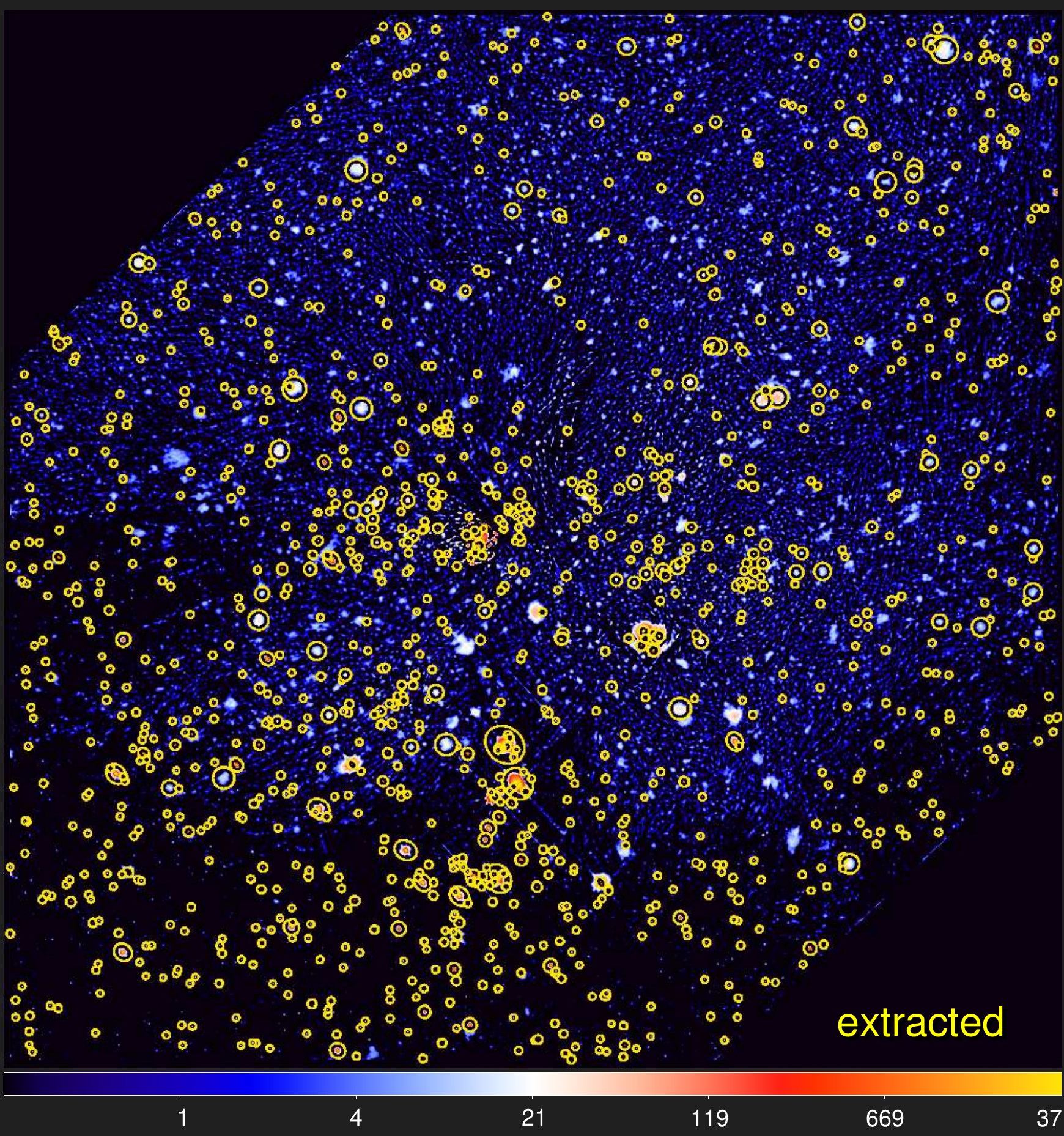}}
  \resizebox{0.328\hsize}{!}{\includegraphics{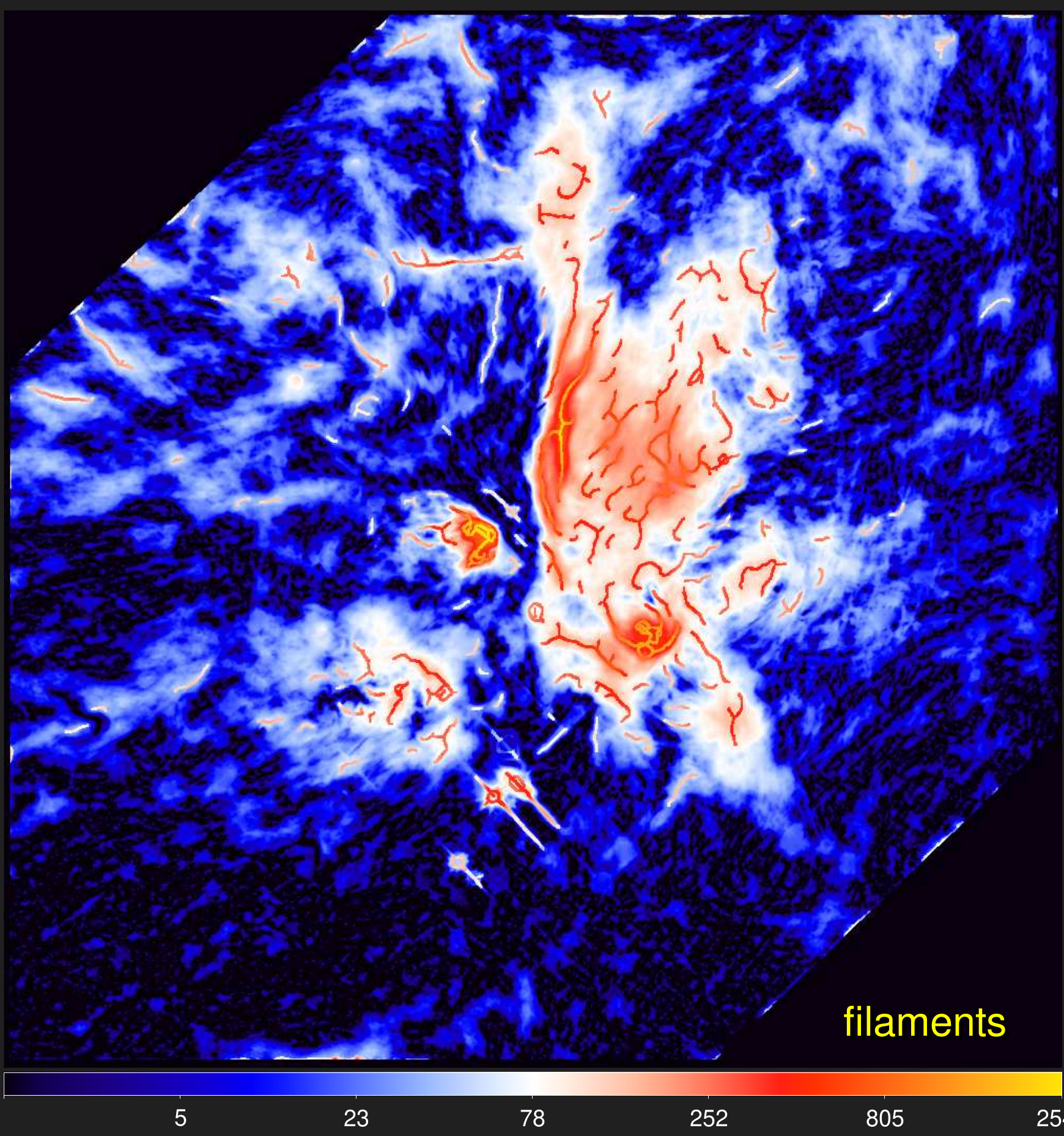}}}
\vspace{0.5mm}
\centerline{
  \resizebox{0.328\hsize}{!}{\includegraphics{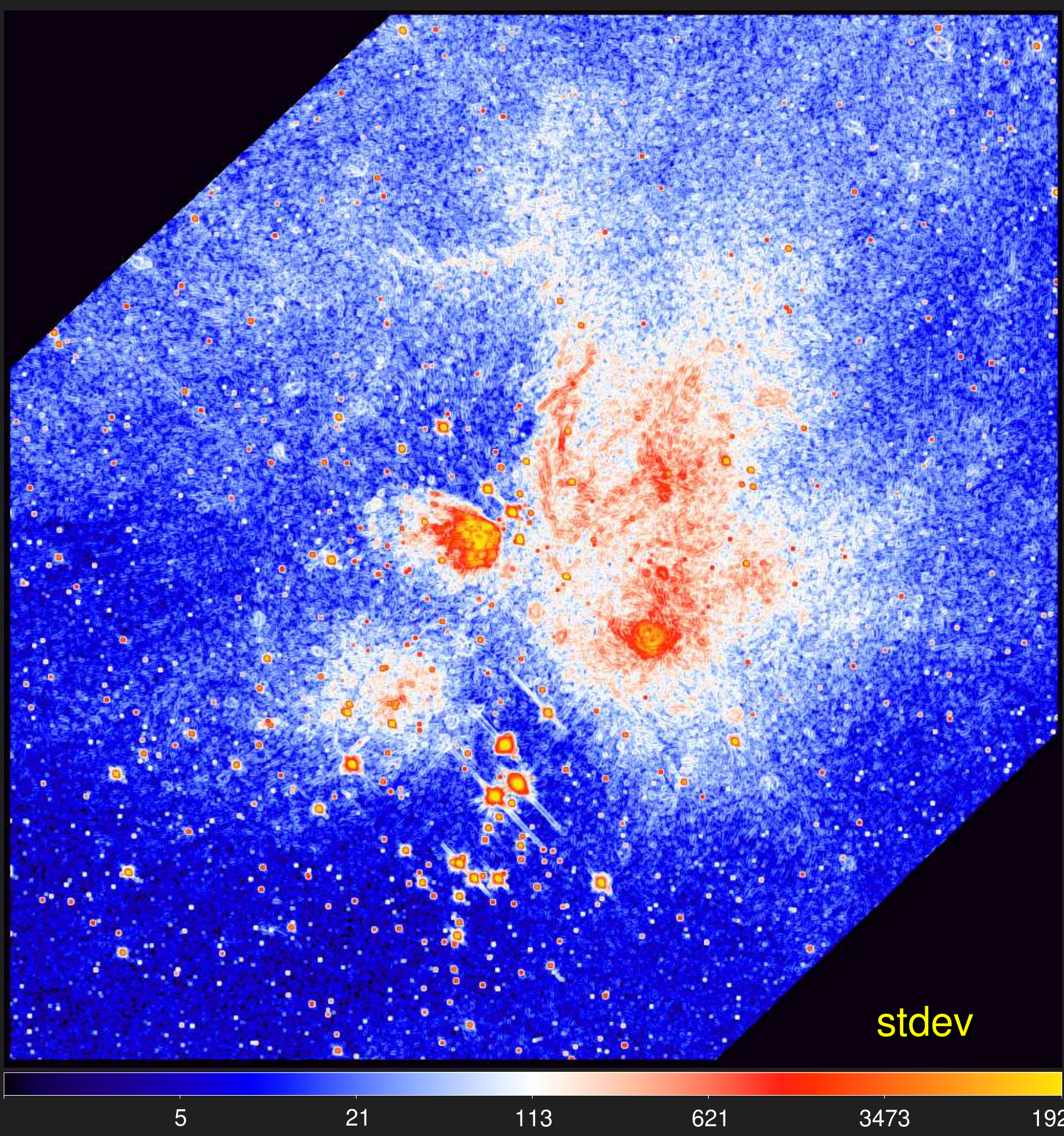}}
  \resizebox{0.328\hsize}{!}{\includegraphics{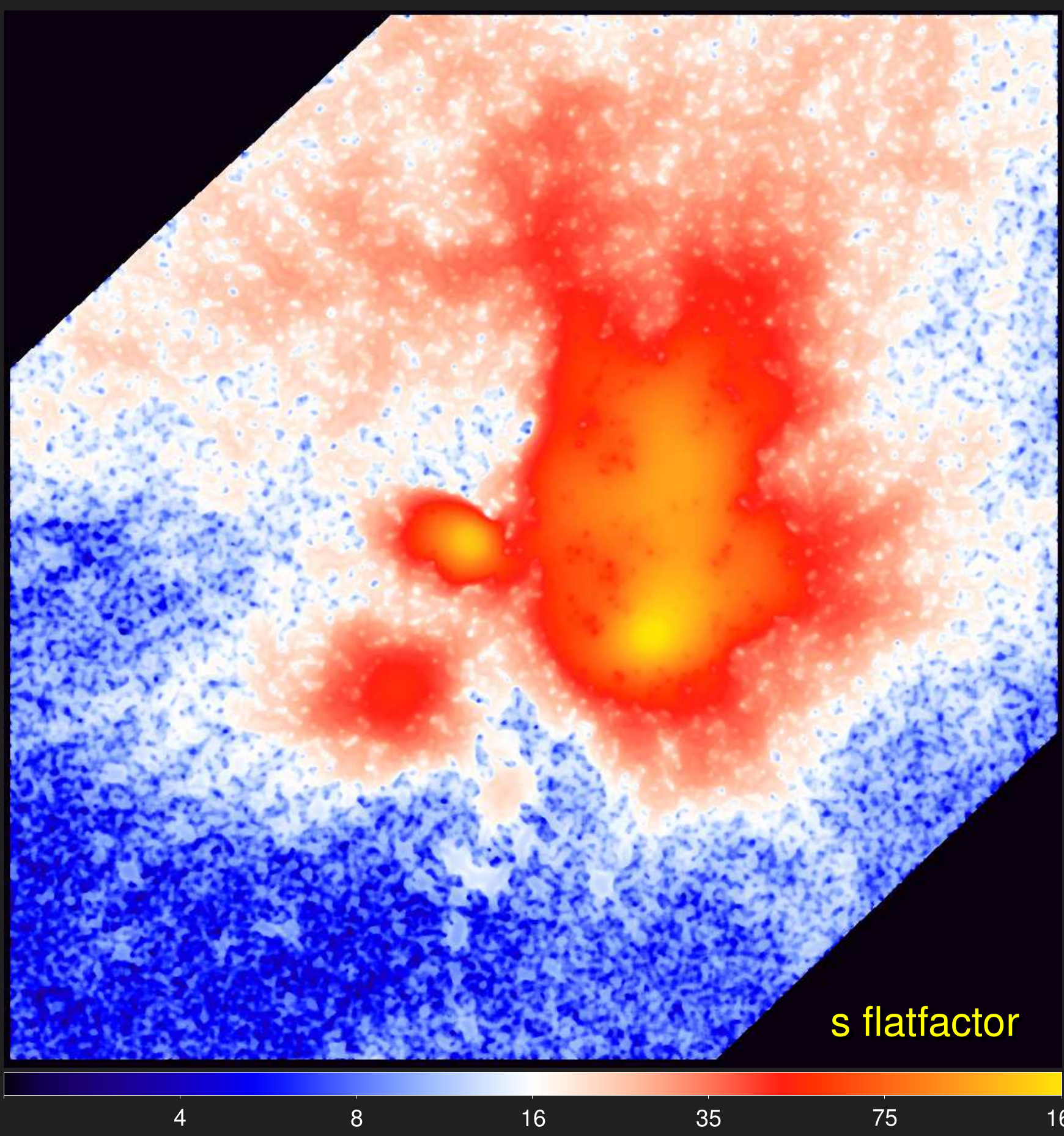}}
  \resizebox{0.328\hsize}{!}{\includegraphics{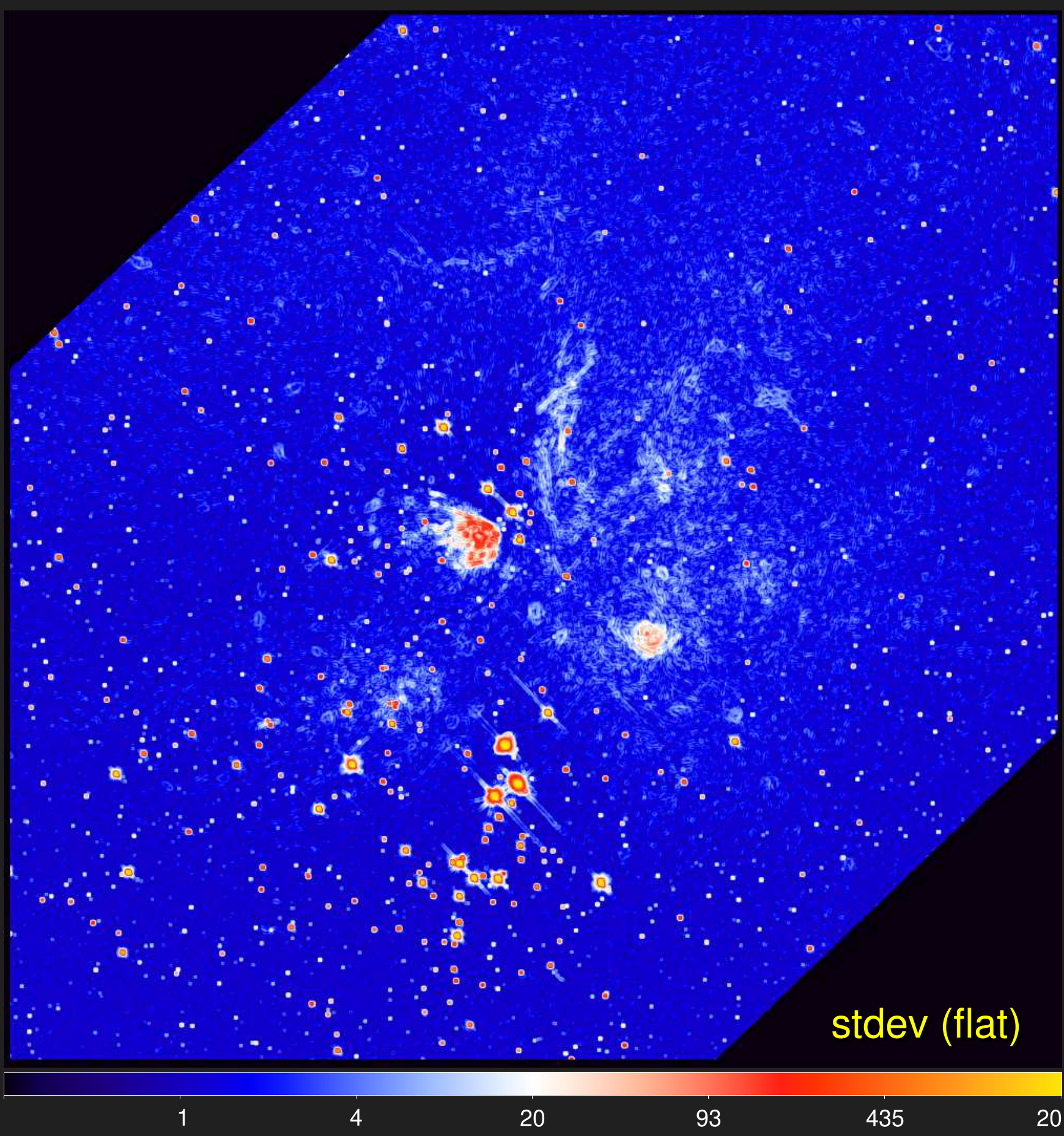}}}
\caption
{ 
Application of \textsl{getsf} to the \emph{Spitzer} $\lambda{\,=\,}8$\,{${\mu}$m} image ($6${\arcsec\!} resolution) of the
\object{L\,1688} star-forming cloud, adopting $\{X|Y\}_{\lambda}{\,=\,}30${\arcsec}. The \emph{top} row shows the original image
$\mathcal{I}_{\!\lambda}$ and the backgrounds $\mathcal{B}_{{\lambda}{\{X|Y\}}}$ of sources and filaments. The \emph{middle} row
shows the component $\mathcal{S}_{{\lambda}}$, the footprint ellipses of $1162$ acceptably good sources on
$\mathcal{S}_{{\lambda}{\rm D}}$, and the component $\mathcal{F}_{{\lambda}{\rm D}}$ with $286$ skeletons $\mathcal{K}_{{k}{2}}$
corresponding to the scales $S_{\!k}{\,\approx\,}30${\arcsec}. The \emph{bottom} row shows the standard deviations
$\mathcal{U}_{\lambda}$ in the regularized component $\mathcal{S}_{{\lambda}{\rm R}}$, the flattening image
$\mathcal{Q}_{\lambda}$, and the standard deviations in the flattened component $\mathcal{S}_{{\lambda}{\rm
R}}\mathcal{Q}_{\lambda}^{-1}$. Some skeletons may only appear to have branches because they were widened for this presentation.
Intensities (in MJy\,sr$^{-1}$) are limited in range, with logarithmic color mapping.
} 
\label{ophiuchus}
\vspace{20mm}
\end{figure*}

\object{L\,1688} was observed with \emph{Spitzer} in the IRAC $8$\,{${\mu}$m} waveband \citep{Evans_etal2009}. The
$1{\degr}{\times\,}1${\degr} image\footnote{\url{https://sha.ipac.caltech.edu/applications/Spitzer/SHA/}} in Fig.~\ref{ophiuchus}
with an angular resolution of $6${\arcsec} shows a complex intensity distribution in this well-known star-forming region, with the
background varying by almost two orders of magnitude and many sources situated in both faint and bright background areas. For this
source and filament extraction with \textsl{getsf}, maximum sizes $\{X|Y\}_{\lambda}{\,=\,}30${\arcsec} were adopted
(Sect.~\ref{maxsizes}).

The clean separation of the components of sources $\mathcal{S}_{{\lambda}}$ and filaments $\mathcal{F}_{{\lambda}}$ from their
backgrounds $\mathcal{B}_{{\lambda}{\{X|Y\}}}$ provided by \textsl{getsf} (Fig.~\ref{ophiuchus}) represents an obvious
improvement over the results obtained with \textsl{getimages} (Fig.~6 in Paper III). The old method of background derivation was
indiscriminate with respect to the shapes of the components, hence the background-subtracted image also contained some filamentary
structures on small scales. In contrast to \textsl{getimages,}  which produced a single background, \textsl{getsf} derived and
subtracted individual backgrounds for $\mathcal{S}_{{\lambda}}$ and $\mathcal{F}_{{\lambda}}$. The component
$\mathcal{S}_{{\lambda}}$ of sources (Fig.~\ref{ophiuchus}) is completely free of the elongated structures. The standard deviations
$\mathcal{U}_{{\lambda}}$ reveal that the small-scale background fluctuation levels vary by roughly three orders of magnitude
across the image. If not equalized, the fluctuations would be extracted as numerous spurious sources and contaminate the source
catalog. The very effective flattening of the detection image $\mathcal{S}_{{\lambda}{\rm D}}$ leads to a much more reliable
extraction.

Several bright unresolved sources in the lower part of the observed image have very wide power-law wings and cross-like artifacts
that are induced by the complex PSF of the \emph{Spitzer} IRAC camera at $8$\,{${\mu}$m}. Their intensity profiles are markedly
non-Gaussian, and for a proper measurement of their integrated fluxes, \textsl{getsf} expanded their footprints by factors
${\sim\,}4$ to $15$ using the footprint expansion algorithm (Fig.~\ref{footexpansion}). The cross-like artifacts from the PSF were
interpreted by \textsl{getsf} as filaments and were moved to the filament component, thereby improving $\mathcal{S}_{{\lambda}{\rm
D}}$ for source detection. In addition to the cross shape, the complex PSF has ${\sim\,}20$ faint peaks that surround the main
beam. They were extracted as several spurious sources, surrounding the brightest peaks; they must be eliminated in a
post-extraction analysis.

The source catalog gives measurements of $1474$ sources, $1162$ of which are selected as acceptably good using
Eq.~(\ref{acceptable}). The filament component produced by \textsl{getimages} (Fig.~7 in Paper III) contains only the brightest
parts of the filaments, their fainter intensities are missing. In contrast, \textsl{getsf} determines the intensity distributions
down to very low intensity levels, with $286$ skeletons tracing the simple, non-branching segments of the filaments
(Sect.~\ref{fildetection}).


\subsection{Embedded starless core \object{L\,1689B}}
\label{herschel}

\begin{figure*}                                                               
\centering
\centerline{
  \resizebox{0.328\hsize}{!}{\includegraphics{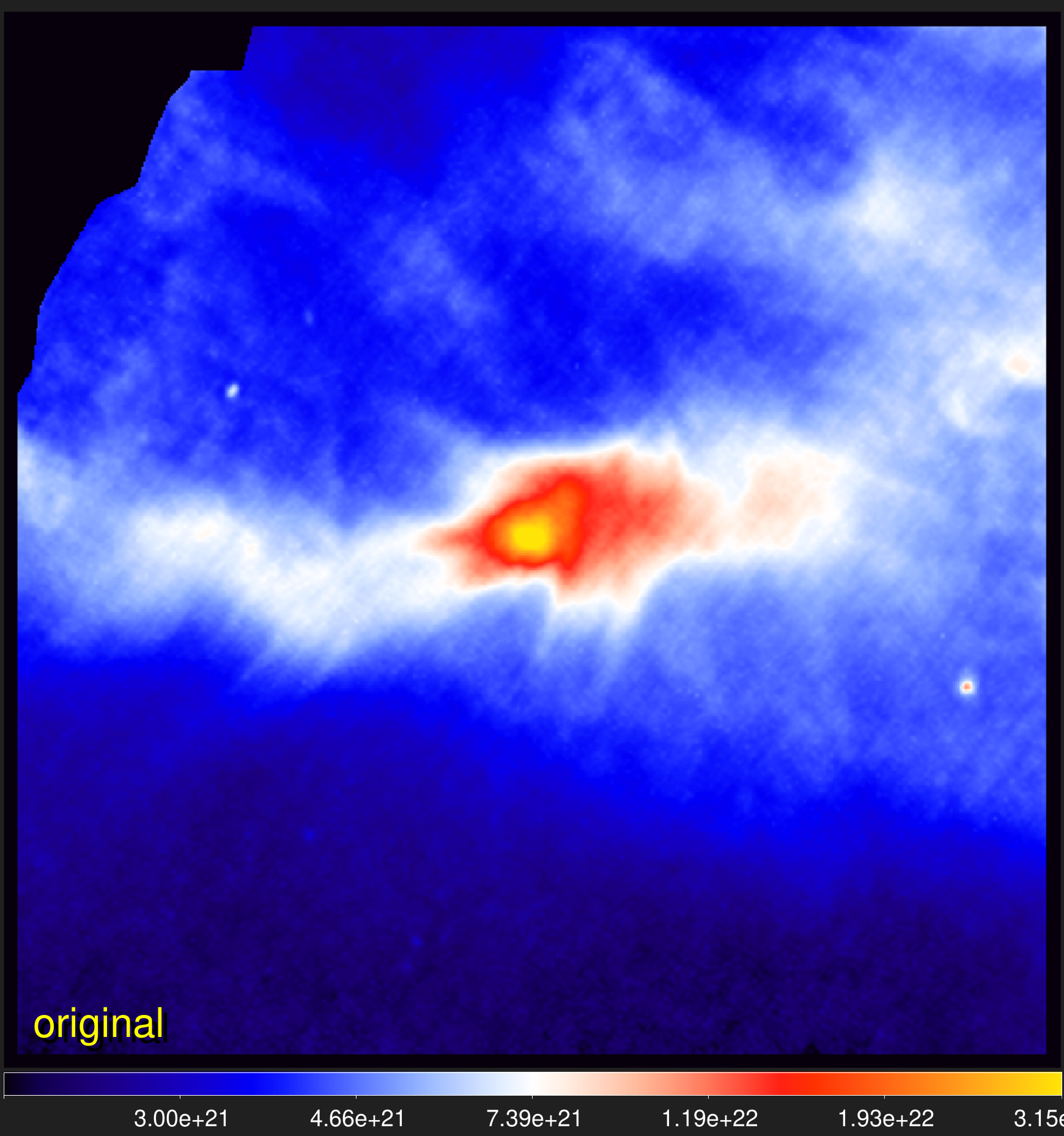}}
  \resizebox{0.328\hsize}{!}{\includegraphics{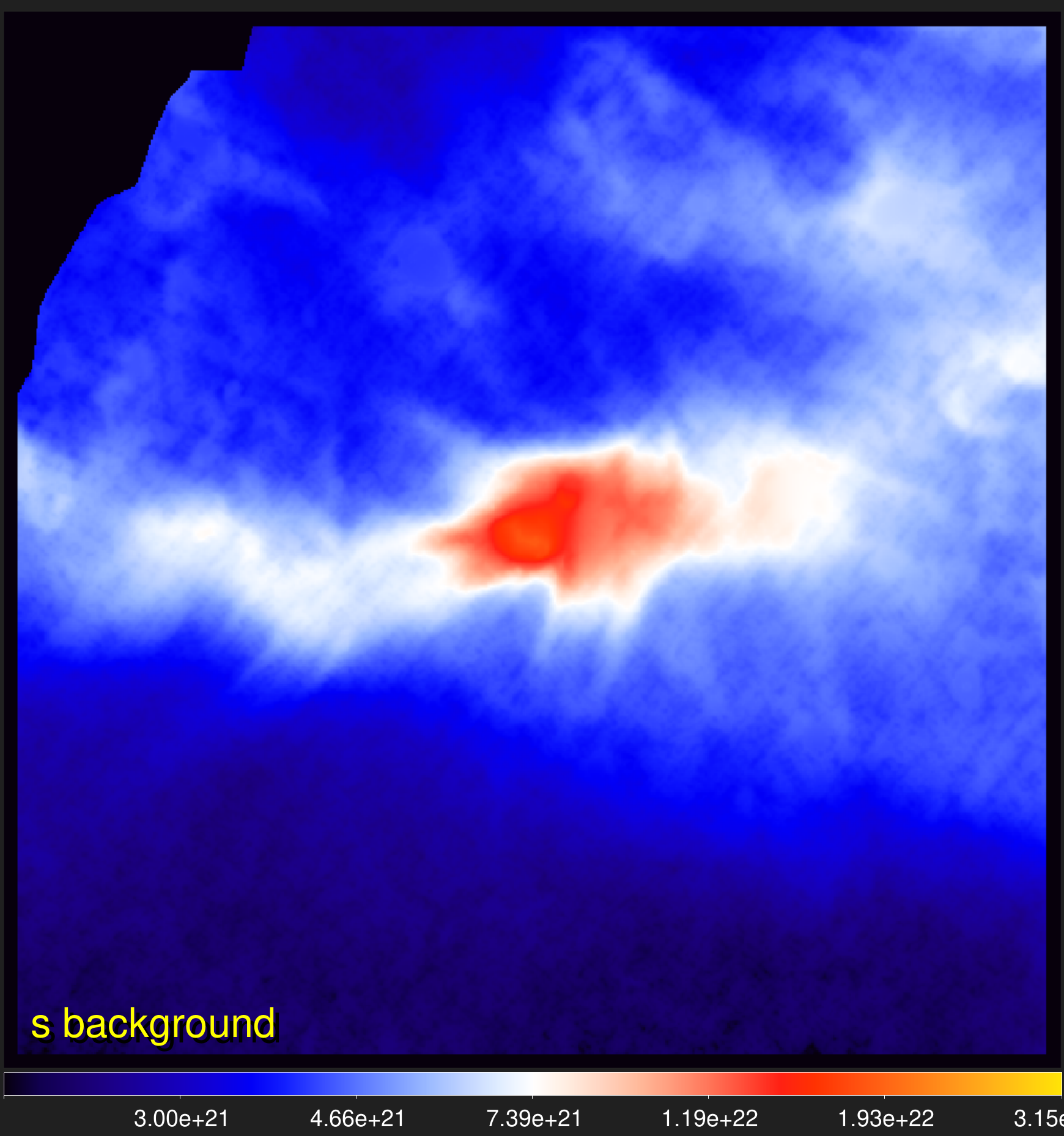}}
  \resizebox{0.328\hsize}{!}{\includegraphics{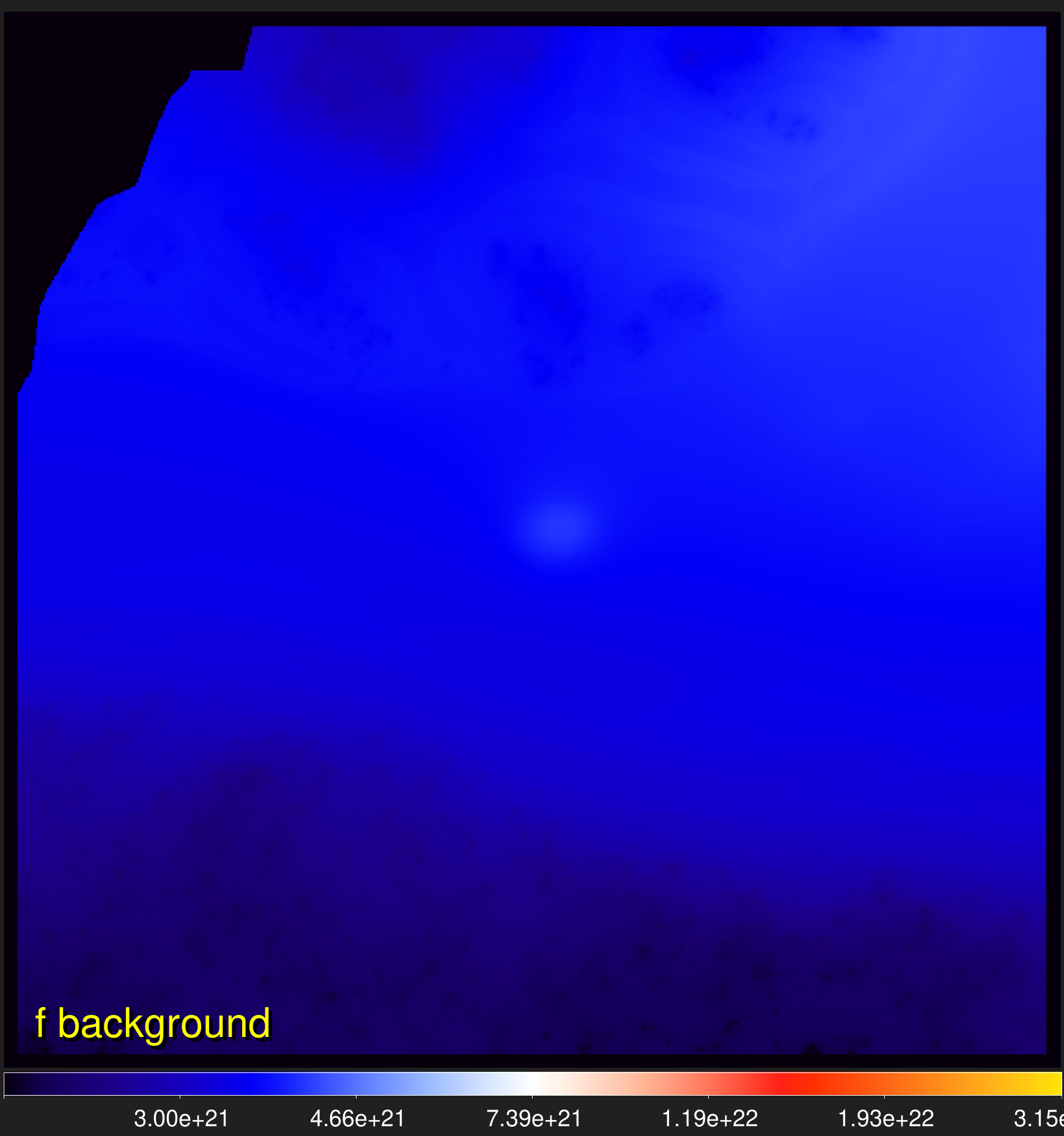}}}
\vspace{0.5mm}
\centerline{
  \resizebox{0.328\hsize}{!}{\includegraphics{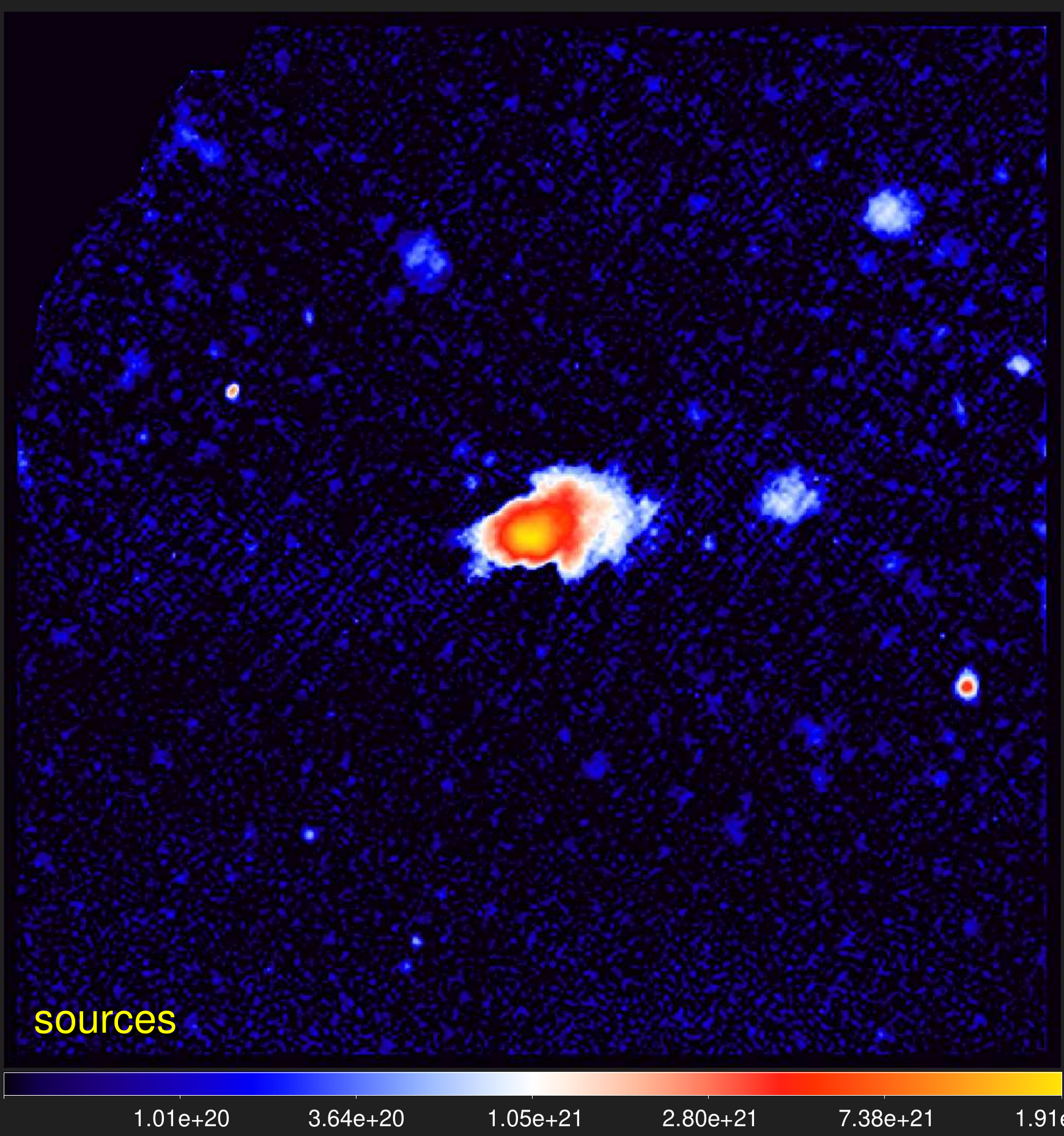}}
  \resizebox{0.328\hsize}{!}{\includegraphics{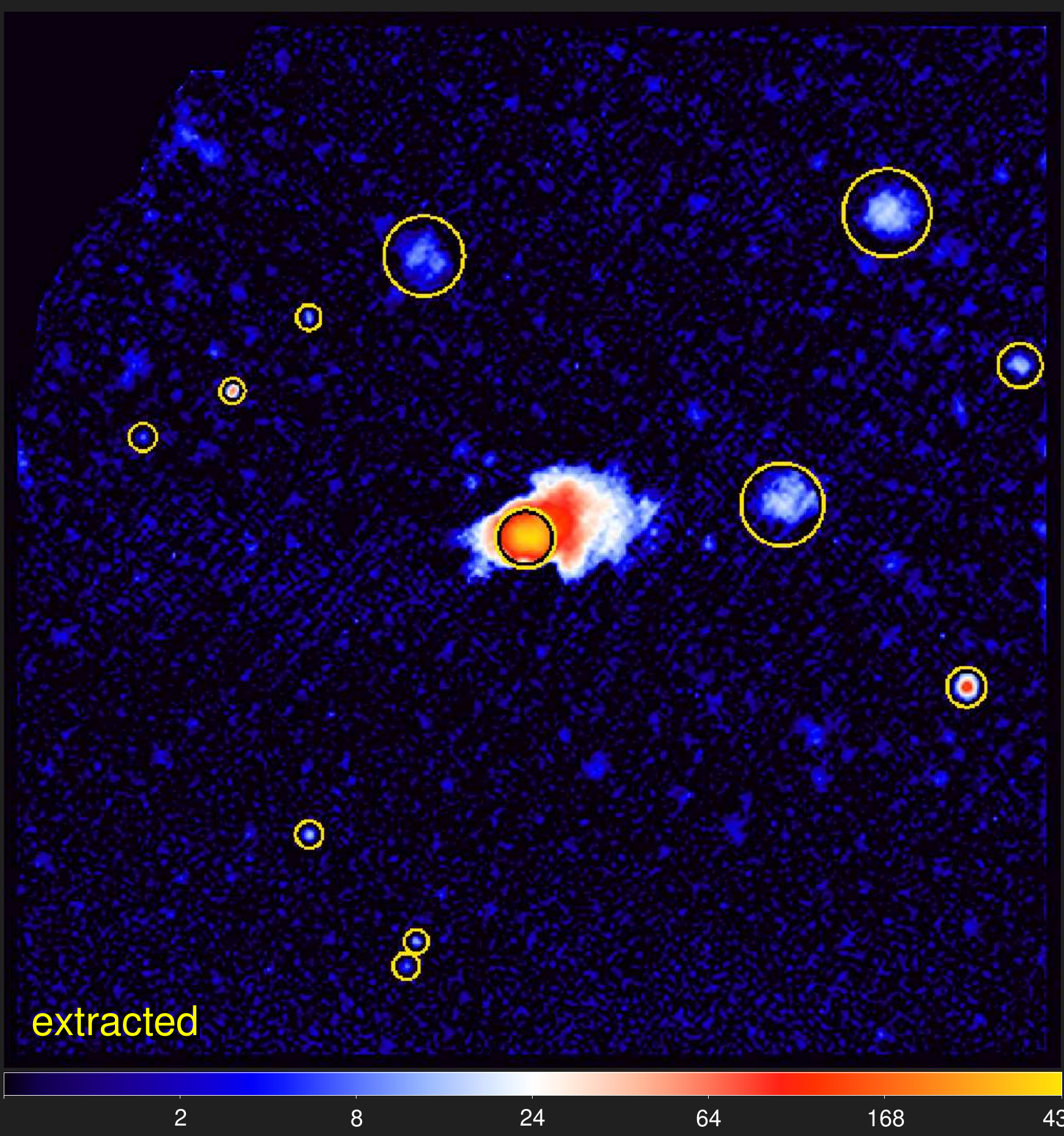}}
  \resizebox{0.328\hsize}{!}{\includegraphics{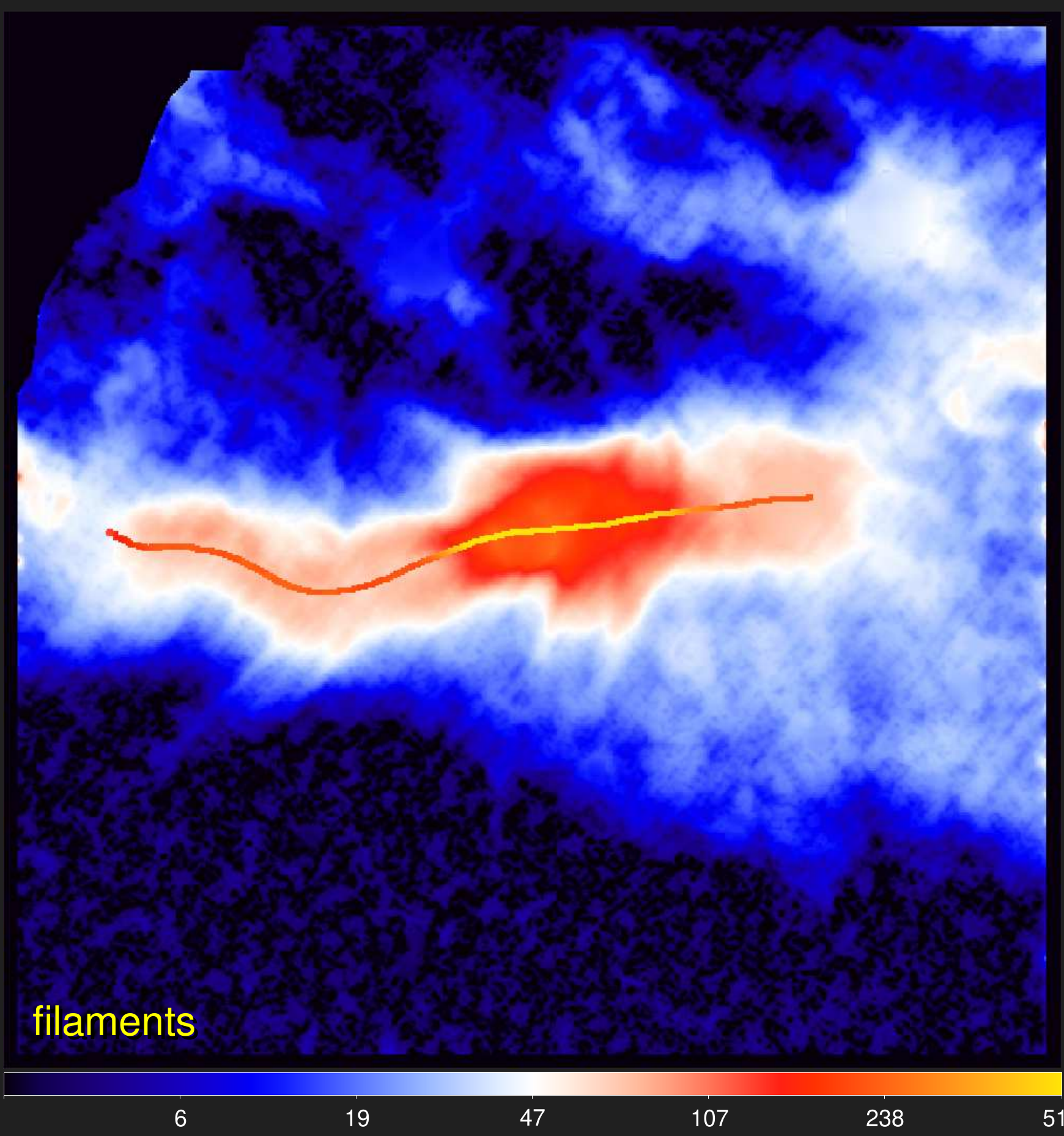}}}
\vspace{0.5mm}
\centerline{
  \resizebox{0.328\hsize}{!}{\includegraphics{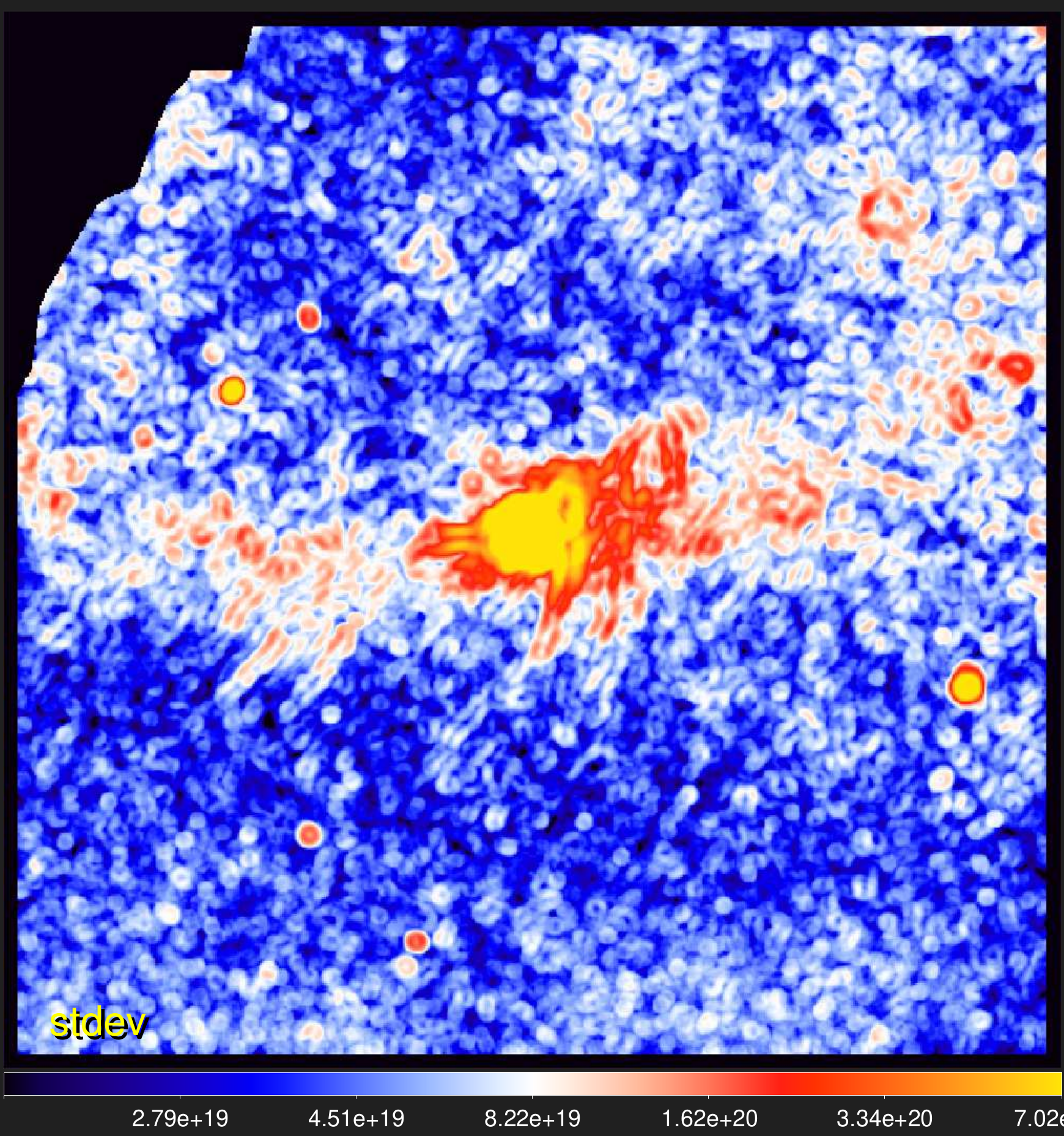}}
  \resizebox{0.328\hsize}{!}{\includegraphics{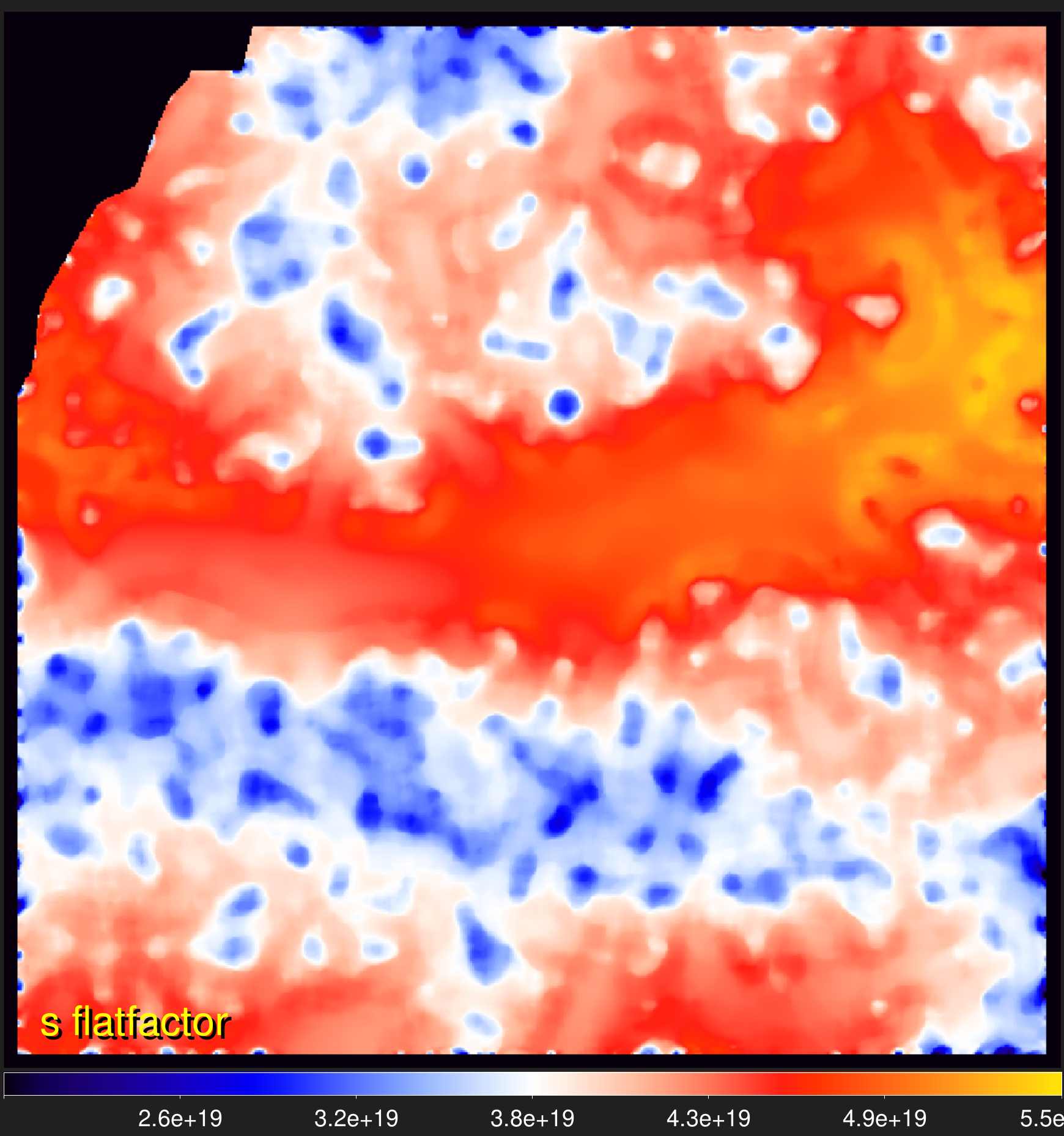}}
  \resizebox{0.328\hsize}{!}{\includegraphics{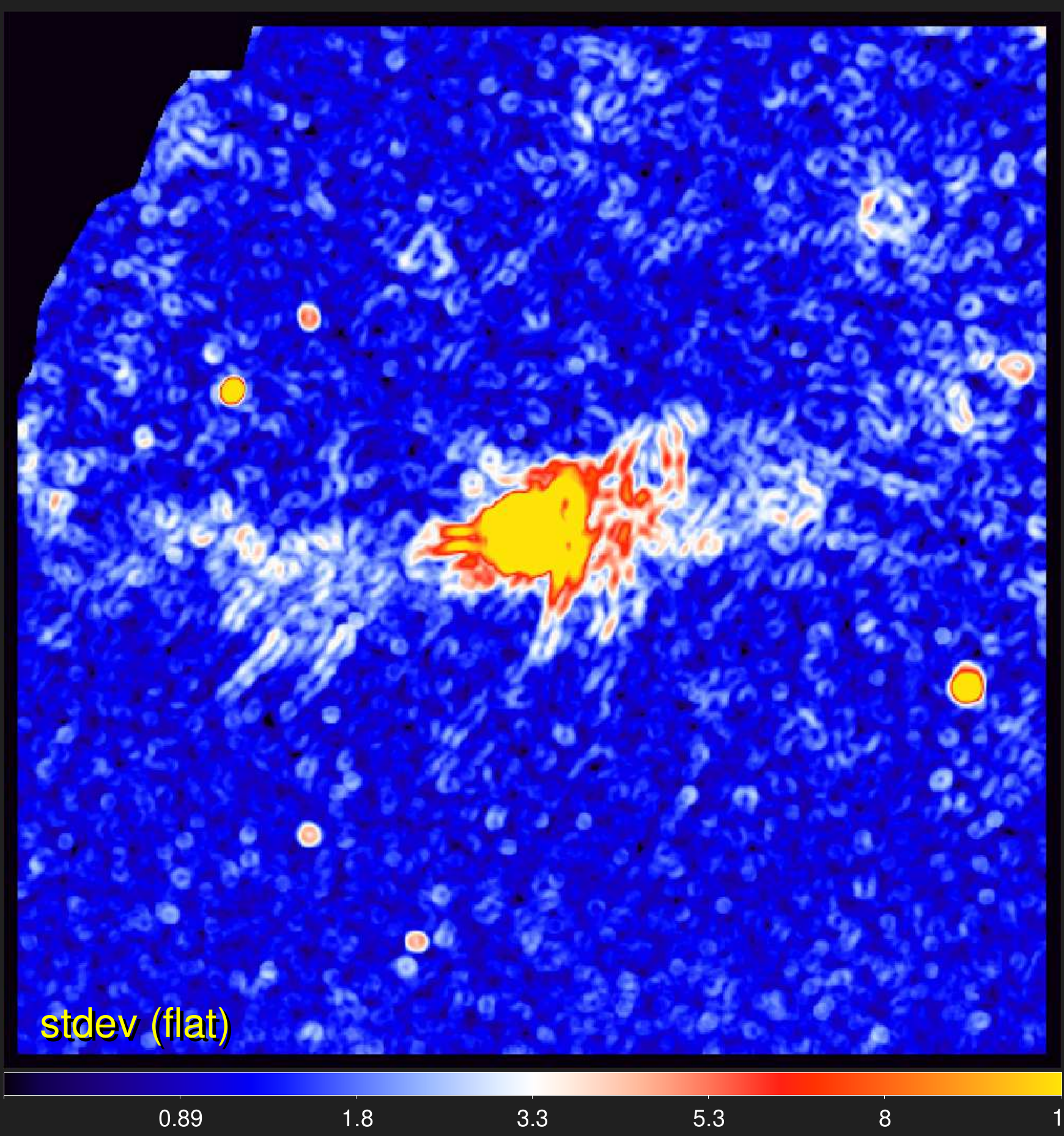}}}
\caption
{ 
Application of \textsl{getsf} to the \emph{Herschel} surface density ($13.5${\arcsec\!} resolution) of the starless core
\object{L\,1689B}, embedded in a filament, adopting $\{X|Y\}_{\lambda}{\,=\,}\{90,180\}${\arcsec}. The \emph{top} row shows the
original \textsl{hires} image $\mathcal{D}_{13{\arcsec}}$ obtained from Eq.~(\ref{superdens}) and the backgrounds
$\mathcal{B}_{{\lambda}{\{X|Y\}}}$ of sources and filaments. The \emph{middle} row shows the component $\mathcal{S}_{{\lambda}}$,
the footprint ellipses of $12$ acceptably good sources on $\mathcal{S}_{{\lambda}{\rm D}}$, and the component
$\mathcal{F}_{{\lambda}{\rm D}}$ with one skeleton $\mathcal{K}_{{k}{2}}$ corresponding to the scales
$S_{\!k}{\,\approx\,}200${\arcsec}. The \emph{bottom} row shows the standard deviations $\mathcal{U}_{\lambda}$ in the regularized
component $\mathcal{S}_{{\lambda}{\rm R}}$, the flattening image $\mathcal{Q}_{\lambda}$, and the standard deviations in the
flattened component $\mathcal{S}_{{\lambda}{\rm R}}\mathcal{Q}_{\lambda}^{-1}$. Surface densities (in $N_{{\rm H}_{2}}$cm$^{-2}$)
are limited in range with logarithmic color mapping, except for $\mathcal{Q}_{\lambda}$, which is shown with linear mapping.
} 
\label{l1689b}
\vspace{20mm}
\end{figure*}

\object{L\,1689B}, one of the nearest well-resolved starless cores (a distance of $140$\,pc) embedded in a resolved filament, was
observed with \emph{Herschel} in five PACS and SPIRE wavebands \citep{Ladjelate_etal2020}. The $160{-}500$\,{${\mu}$m}
images\footnote{\url{http://gouldbelt-herschel.cea.fr/archives}} and Eq.~(\ref{superdens}) were used to compute a
$1.1{\degr}{\times\,}1.1${\degr} surface density image $\mathcal{D}_{13{\arcsec}}$ in Fig.~\ref{l1689b} with a resolution of
$13.5${\arcsec} to illustrate the new extraction method on a single image. In addition to the reduction of the number of images,
the use of surface densities allows \textsl{getsf} to catalog physical parameters of the core and filament. For this extraction,
maximum sizes $\{X|Y\}_{\lambdabar}{\,=\,}\{90,180\}${\arcsec} were adopted (Sect.~\ref{maxsizes}).

The $\mathcal{D}_{13{\arcsec}}$ image in Fig.~\ref{l1689b} presents \object{L\,1689B} in the wide filamentary structure near the
edge of a diffuse cloud, all components are blended. The filament surface density is a factor of ${\sim\,}5$ below the peak
surface density $N_{{\rm H}_{2}}{\,=\,}3.7{\,\times\,}10^{22}$\,cm$^{-2}$, whereas at the values, lower by just a factor of $2$, a
round shape of the source becomes distorted by its complex environment. Separation of the components by \textsl{getsf} greatly
simplifies the image, isolating the structures in their individual images $\mathcal{S}_{{\lambdabar}}$,
$\mathcal{F}_{{\lambdabar}}$, and $\mathcal{B}_{{\lambdabar}{\{X|Y\}}}$ (Fig.~\ref{l1689b}). Subsequent flattening of the
small-scale fluctuation levels allowed a reliable identification of the filament and sources in both low- and high-background areas
of the observed image. The extraction catalog contains measurements of $20$ sources, $12$ of which are selected as acceptably good
by Eq.~(\ref{acceptable}). The single skeleton was obtained on spatial scales of ${\sim\,}200${\arcsec}, corresponding to the
maximum width $Y_{\lambdabar}$.

The main physical parameters of the starless core \object{L\,1689B}, $M{\,=\,}0.6\,M_{\sun}$ and $N_{{\rm
H}_{2}}{\,=\,}10^{22}$\,cm$^{-2}$, are underestimated because of the inaccuracies (Appendix~\ref{hiresinacc}) of the standard
surface density derivation approach (Sect.~\ref{hiresimages}). The errors and correction factors can be found using the benchmark
models from Sect.~\ref{simulcores}. A model of the critical Bonnor-Ebert sphere with $T_{\rm BE}{\,=\,}14$\,K,
$M{\,=\,}1\,M_{\sun}$, and $N_{{\rm H}_{2}}{\,=\,}2.5{\,\times\,}10^{22}$\,cm$^{-2}$ has an FWHM size of $57${\arcsec}, almost the
same as the size $A{\,=\,}58${\arcsec} of \object{L\,1689B}, measured by \textsl{getsf}. However, in the derived
$\mathcal{D}_{13{\arcsec}}$ image, the same model has $M{\,=\,}0.66\,M_{\sun}$ and $N_{{\rm H}_{2}}{\,=\,}10^{22}$\,cm$^{-2}$,
implying correction factors of $1.5$ and $2.5$ for the mass and peak surface density, correspondingly. After correction, the
measured mass of \object{L\,1689B} becomes $M{\,\approx\,}0.9\,M_{\sun}$; masses of the other sources in the image are lower by (at
least) a factor of ${\sim\,}10$ . The filament measurements (Sect.~\ref{filmeasurement}) give its median value
$N_{0}{\,=\,}3.8{\,\times\,}10^{21}$\,cm$^{-2}$, length $L{\,=\,}0.8$\,pc, half-maximum width $W{\,=\,}0.14$\,pc ($205${\arcsec}),
mass $M{\,=\,}15\,M_{\sun}$, and linear density $\Lambda{\,=\,}14\,M_{\sun}\,{\rm pc}^{-1}$; the values are little affected by the
fitting inaccuracies.

\subsection{Star-forming cloud \object{NGC\,6334}}
\label{apex}

\begin{figure*}                                                               
\centering
\centerline{
  \resizebox{0.328\hsize}{!}{\includegraphics{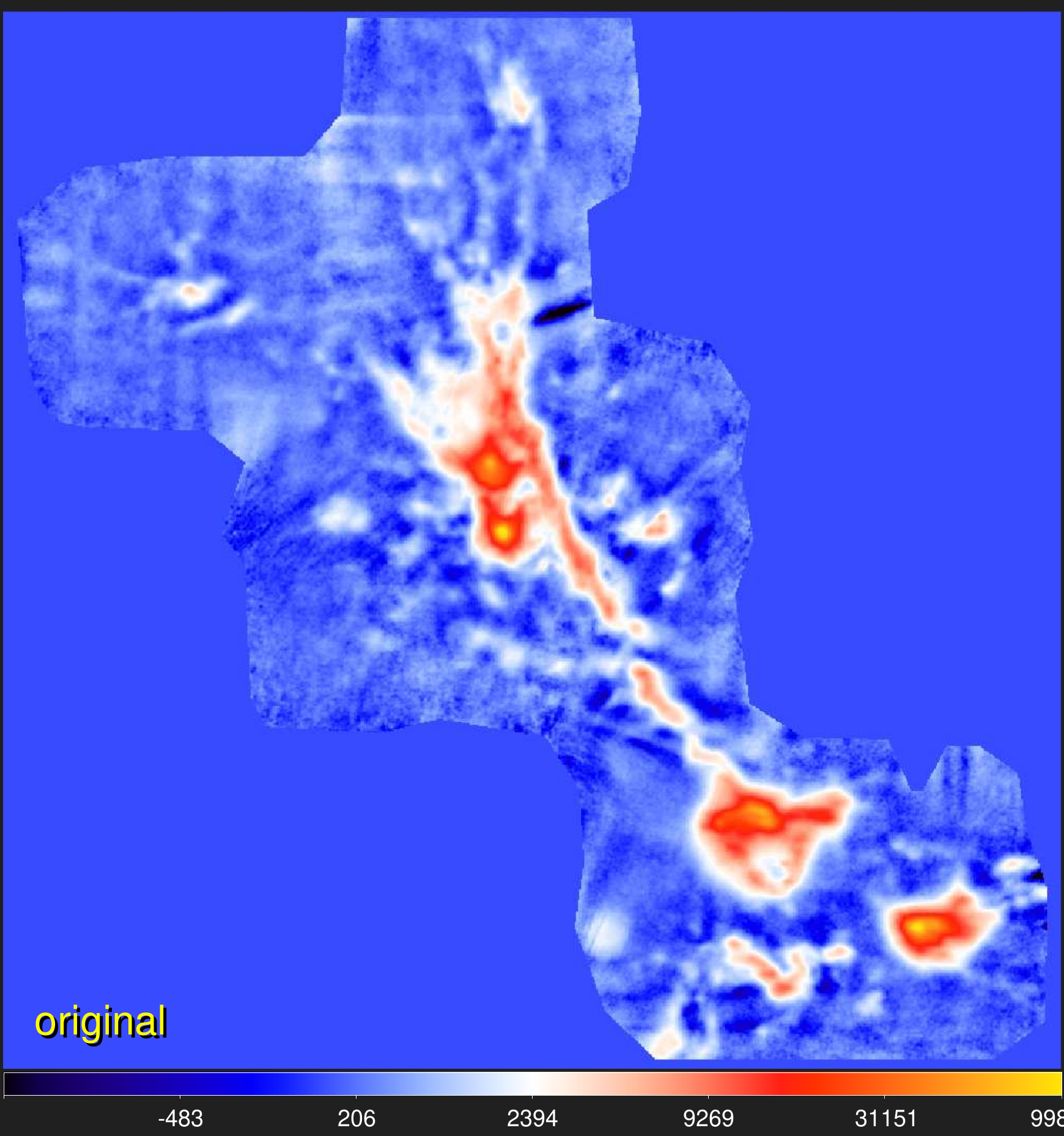}}
  \resizebox{0.328\hsize}{!}{\includegraphics{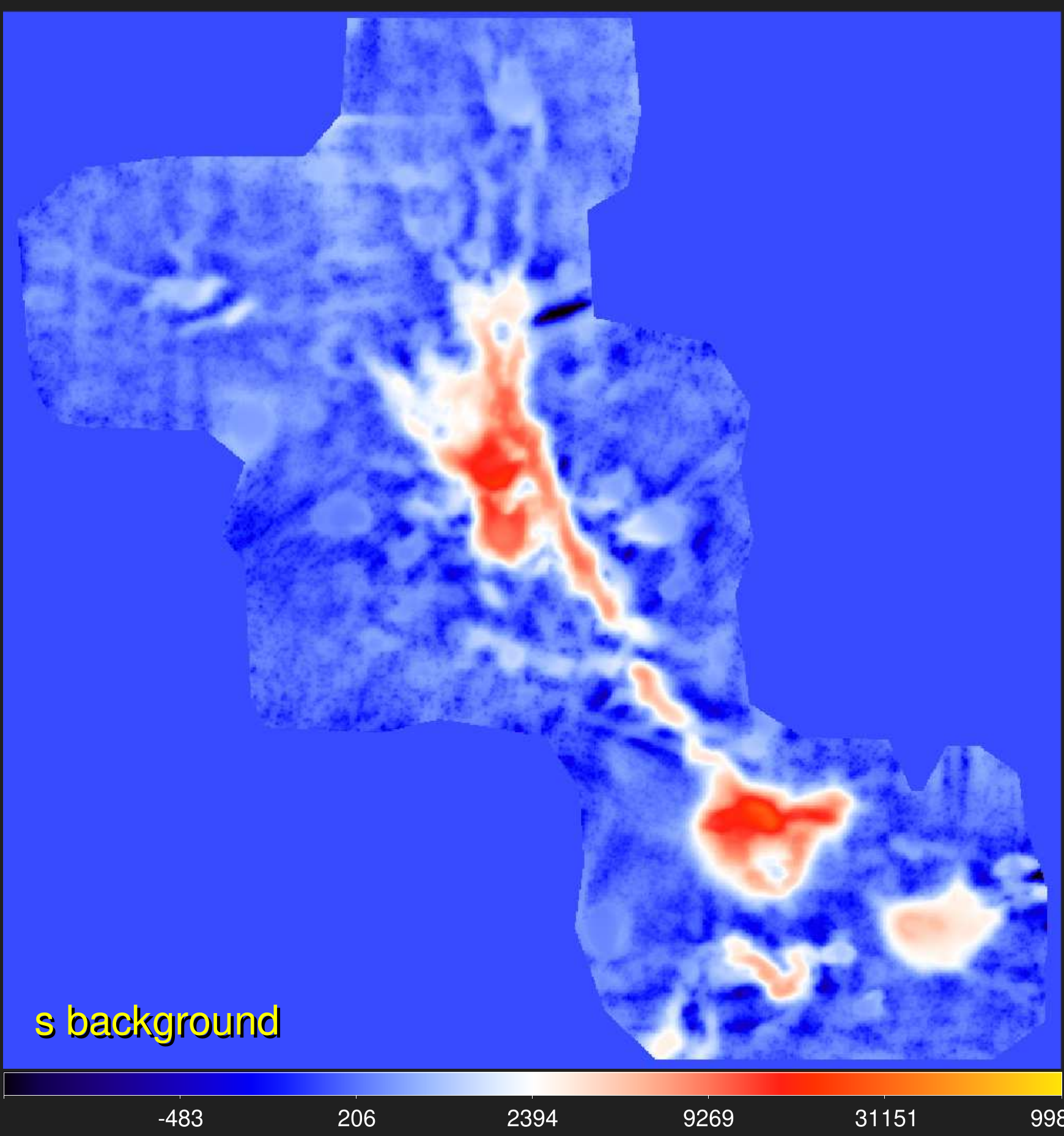}}
  \resizebox{0.328\hsize}{!}{\includegraphics{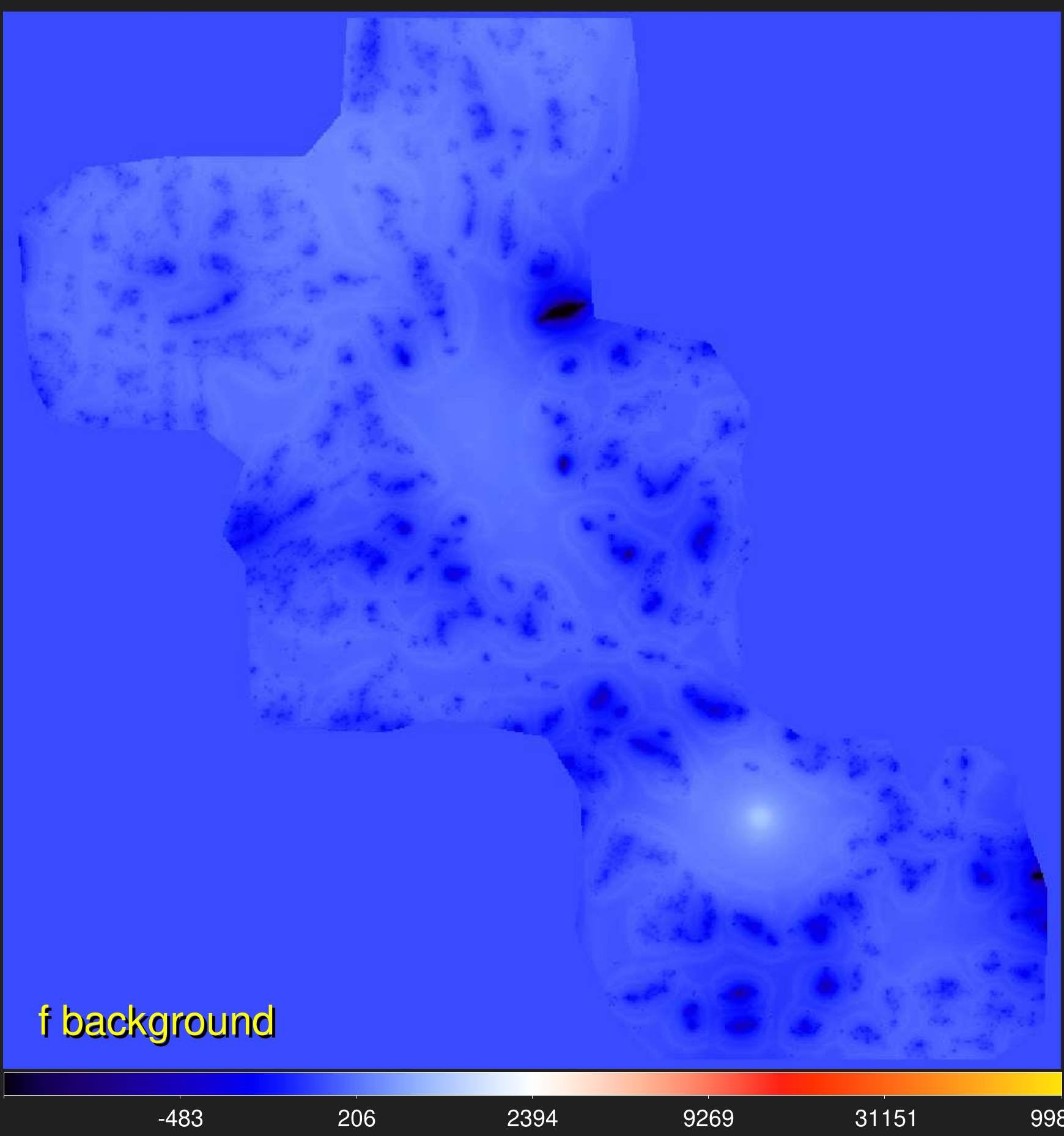}}}
\vspace{0.5mm}
\centerline{
  \resizebox{0.328\hsize}{!}{\includegraphics{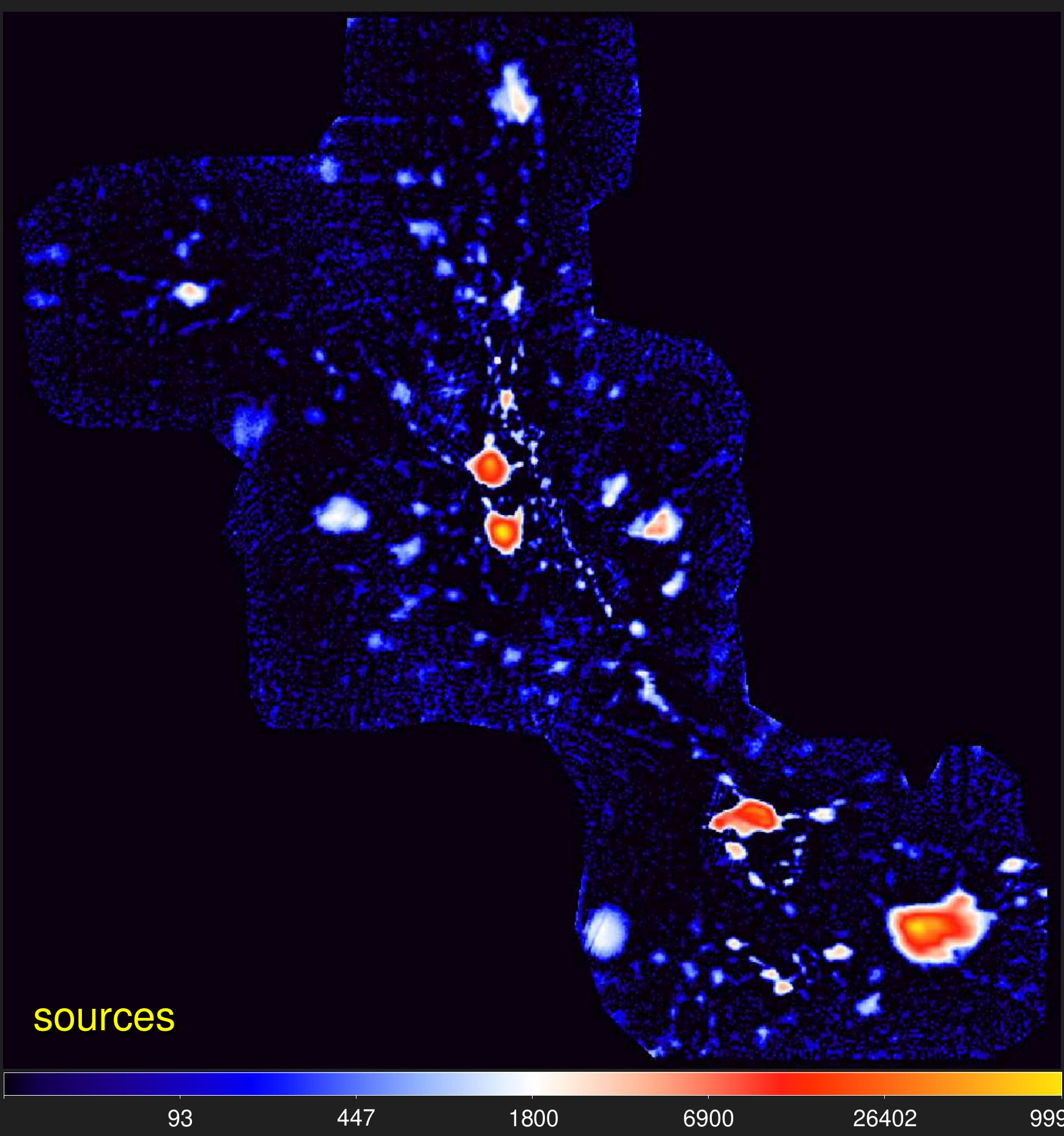}}
  \resizebox{0.328\hsize}{!}{\includegraphics{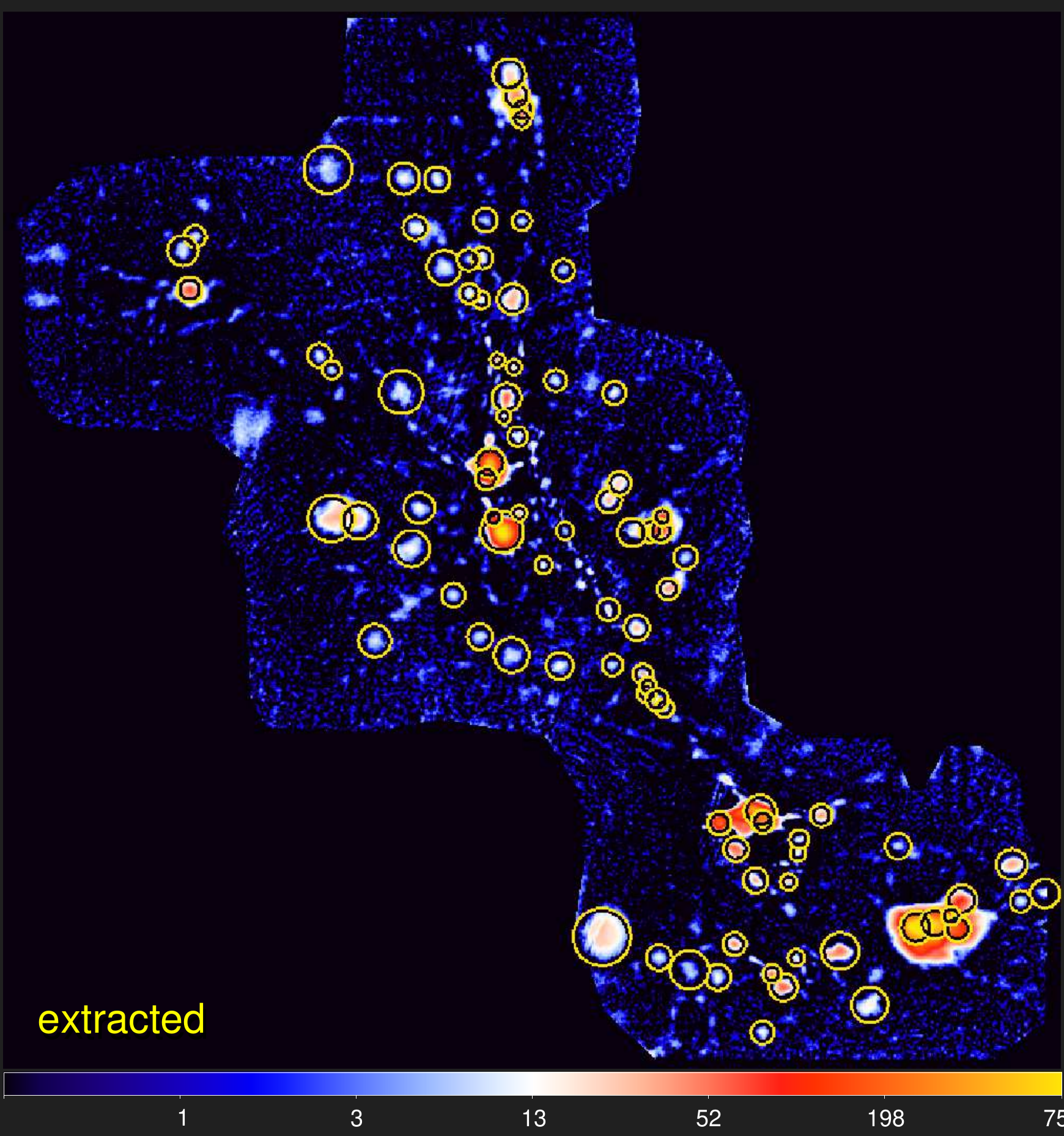}}
  \resizebox{0.328\hsize}{!}{\includegraphics{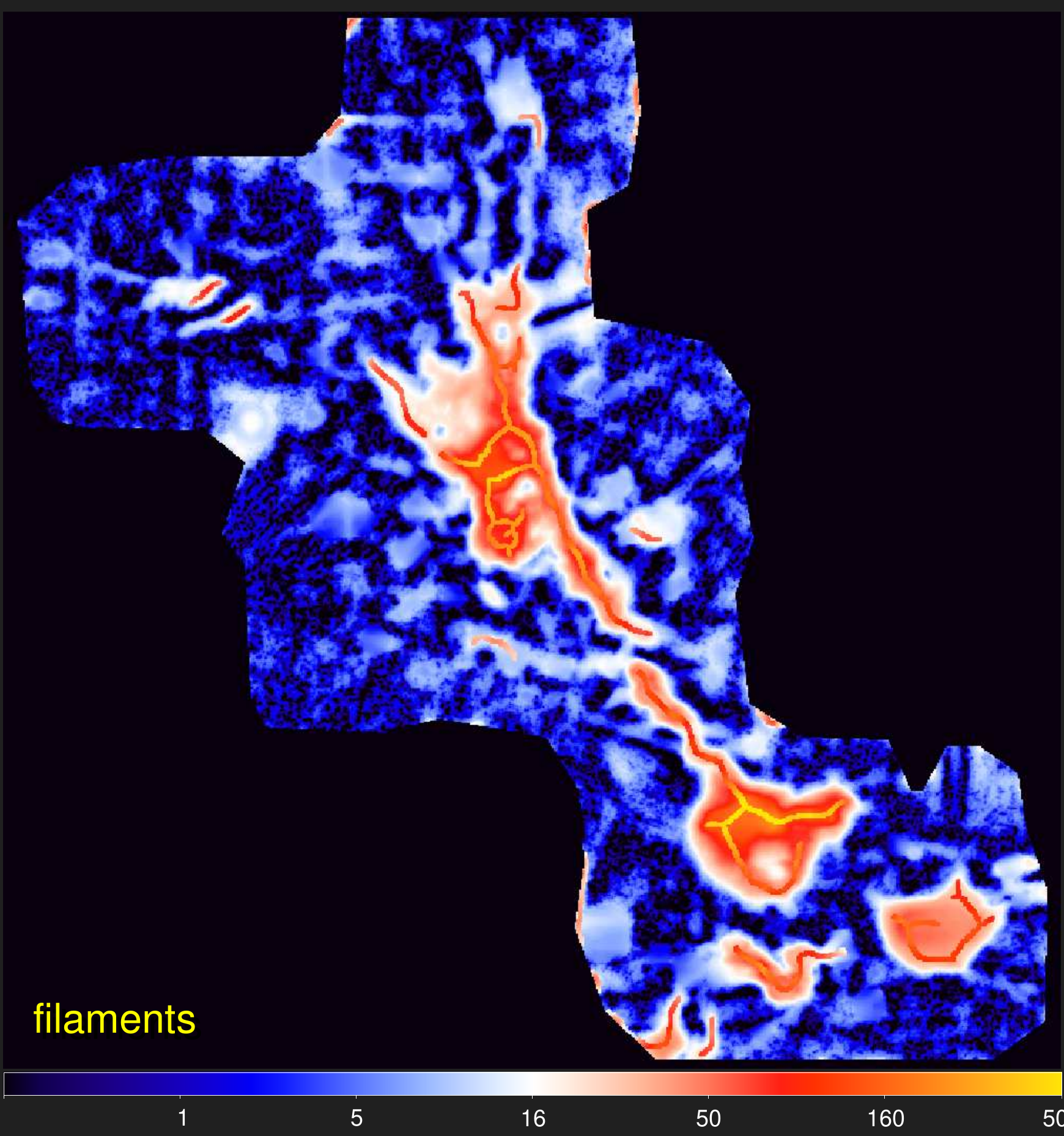}}}
\vspace{0.5mm}
\centerline{
  \resizebox{0.328\hsize}{!}{\includegraphics{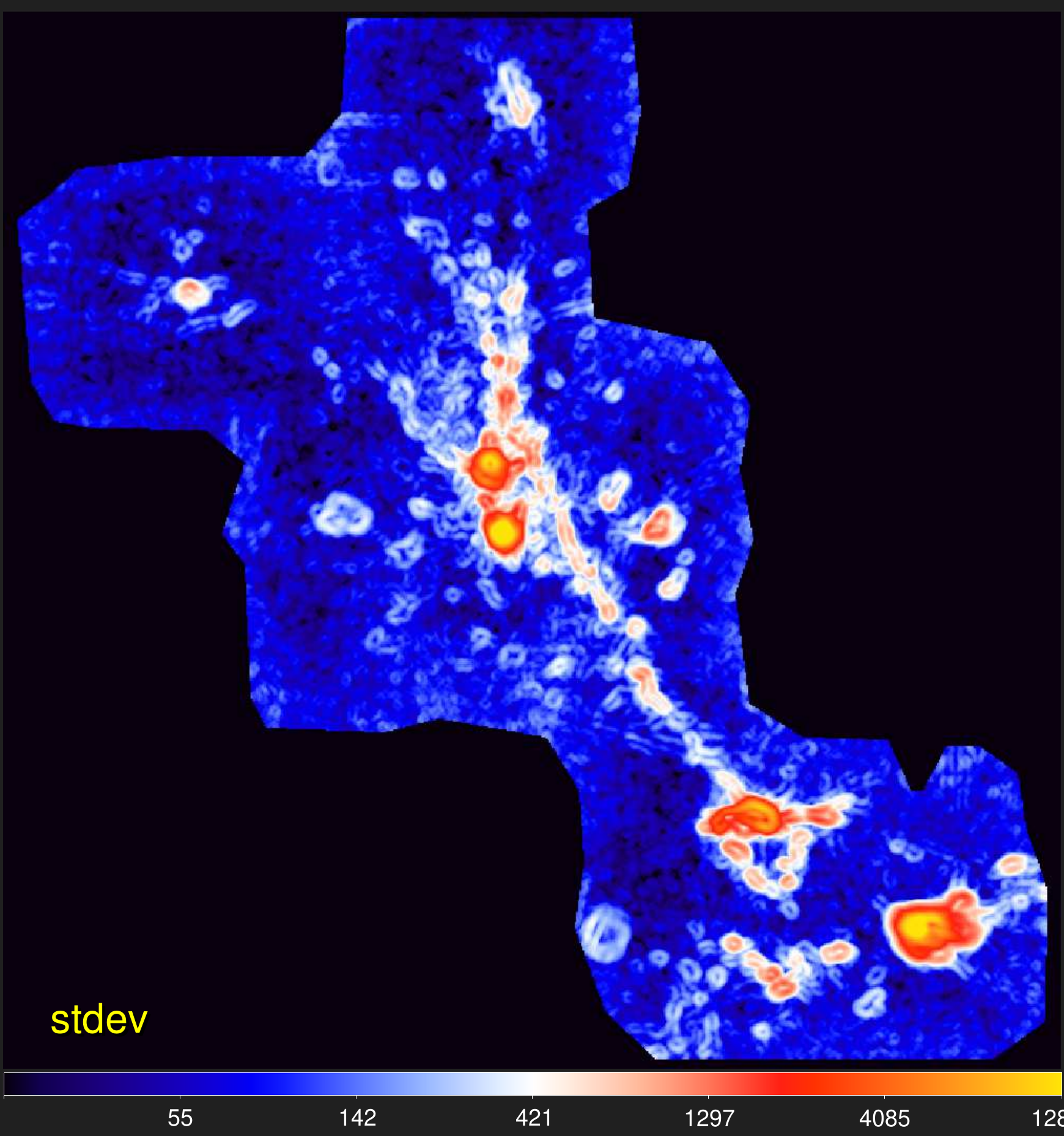}}
  \resizebox{0.328\hsize}{!}{\includegraphics{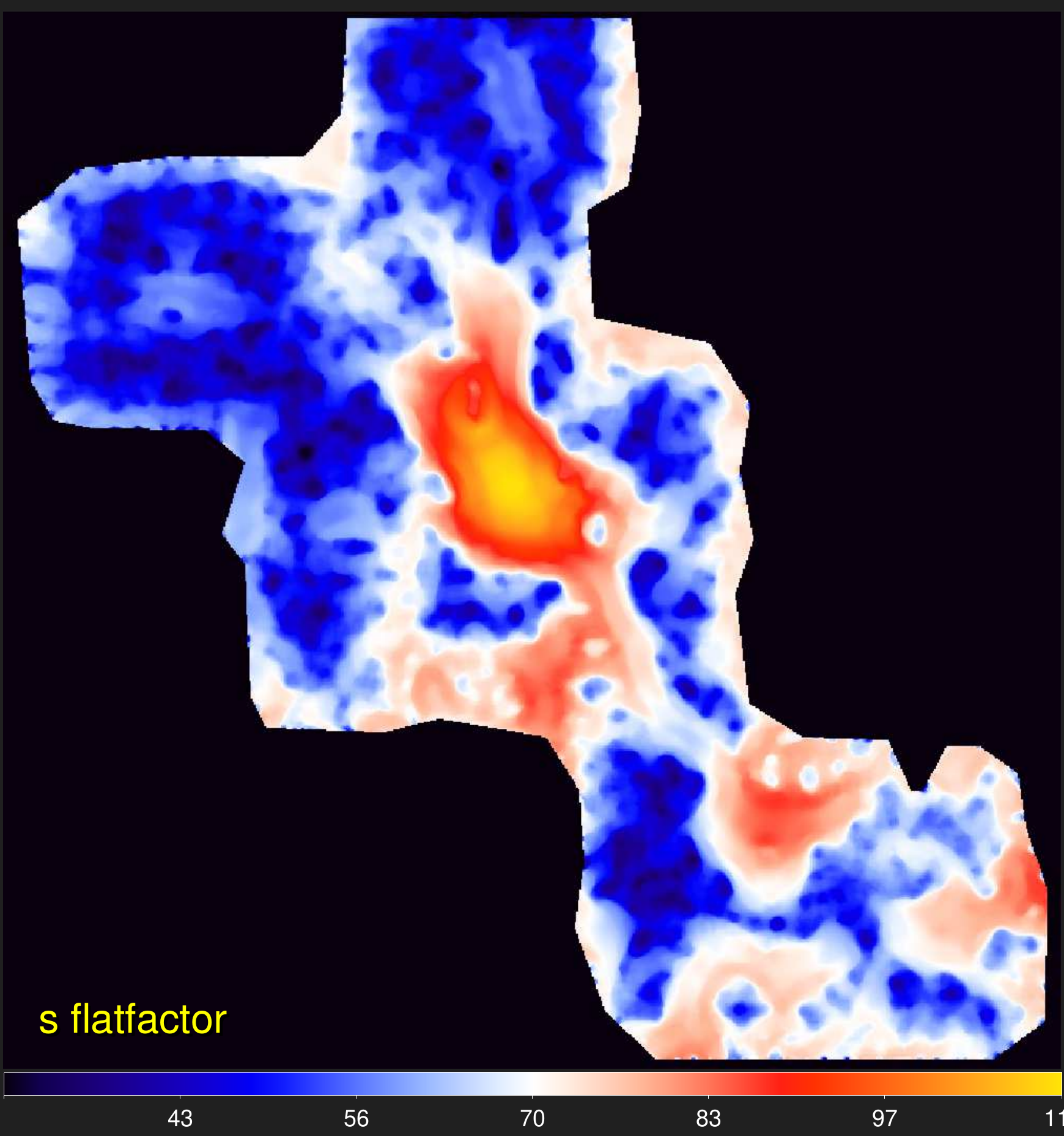}}
  \resizebox{0.328\hsize}{!}{\includegraphics{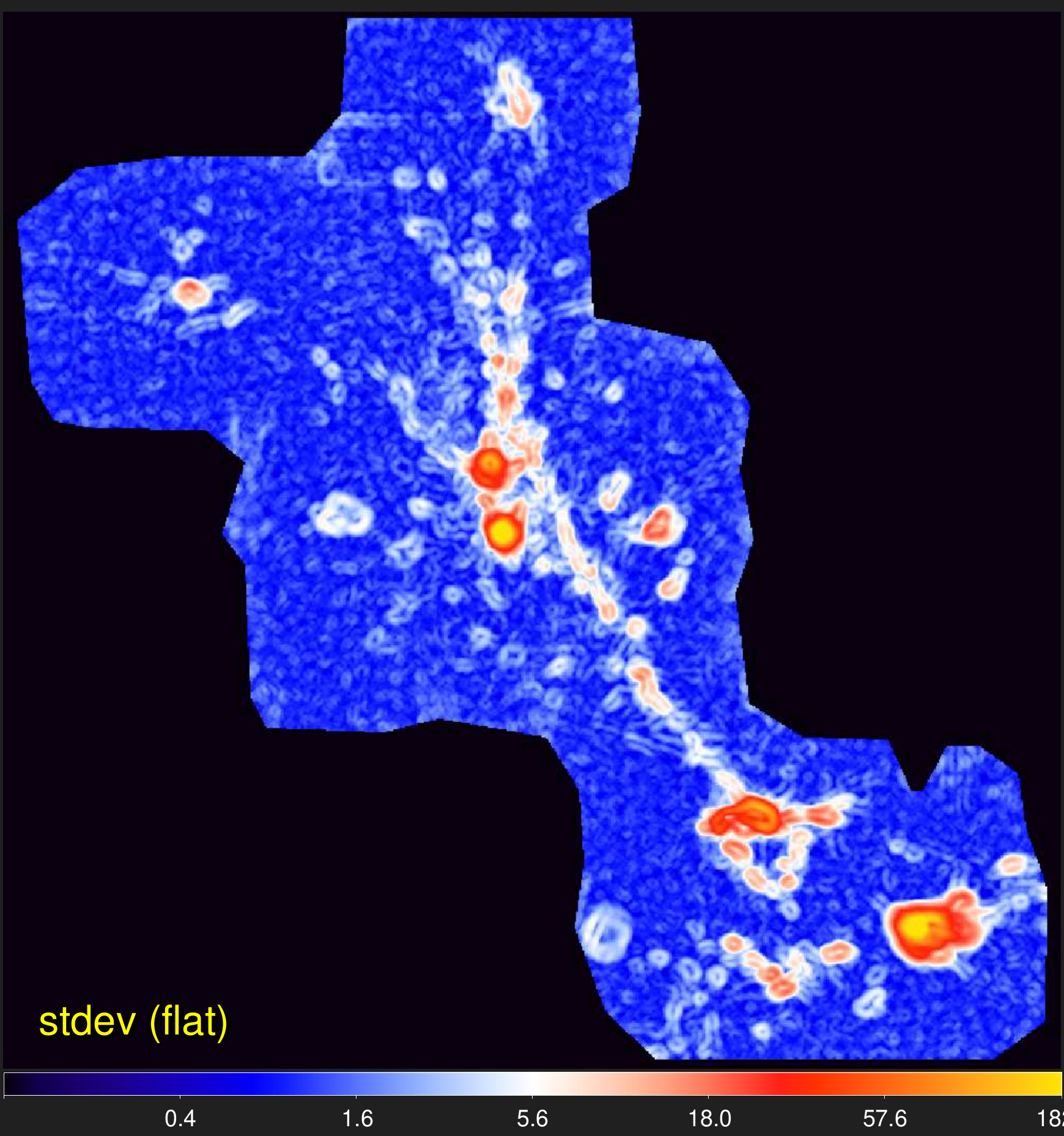}}}
\caption
{ 
Application of \textsl{getsf} to the \emph{APEX} $\lambda{\,=\,}350$\,{${\mu}$m} image ($8${\arcsec\!} resolution) of the
\object{NGC\,6334} star-forming cloud, adopting $\{X|Y\}_{\lambda}{\,=\,}30${\arcsec}. The \emph{top} row shows the original image
$\mathcal{I}_{\!\lambda}$ and the backgrounds $\mathcal{B}_{{\lambda}{\{X|Y\}}}$ of sources and filaments. The \emph{middle} row
shows the component $\mathcal{S}_{{\lambda}}$, the footprint ellipses of $91$ acceptably good sources on
$\mathcal{S}_{{\lambda}{\rm D}}$, and the component $\mathcal{F}_{{\lambda}{\rm D}}$ with $26$ skeletons $\mathcal{K}_{{k}{2}}$
corresponding to the scales $S_{\!k}{\,\approx\,}30${\arcsec}. The \emph{bottom} row shows the standard deviations
$\mathcal{U}_{\lambda}$ in the regularized component $\mathcal{S}_{{\lambda}{\rm R}}$, the flattening image
$\mathcal{Q}_{\lambda}$, and the standard deviations in the flattened component $\mathcal{S}_{{\lambda}{\rm
R}}\mathcal{Q}_{\lambda}^{-1}$. Some skeletons may only appear to have branches because they were widened for this presentation.
Intensities (in MJy\,sr$^{-1}$) are limited in range with logarithmic color mapping, except for $\mathcal{Q}_{\lambda}$, which is
shown with linear mapping.
} 
\label{ngc6334}
\vspace{20mm}
\end{figure*}

\object{NGC\,6334} was observed with \emph{APEX} at $350$\,$\mu$m, equipped with the ArT{\'e}MiS camera \citep{Andre_etal2016}. The
$0.5{\degr}{\times\,}0.5${\degr} image\footnote{\url{http://cdsarc.unistra.fr/viz-bin/cat/J/A+A/592/A54}} in Fig.~\ref{ngc6334}
with an angular resolution of $8${\arcsec} represents an improvement by a factor of $3$ with respect to the \emph{Herschel} images
at $350$\,$\mu$m. Subtraction of the correlated sky noise resulted in an image without signals on spatial scales above
$120${\arcsec} \citep{Andre_etal2016}. Therefore the large-scale background and the zero level of the image are not known, and the
structures in the image are smaller than the largest scale. Fortunately, these observational problems are entirely unimportant for
\textsl{getsf}. For this source and filament extraction, maximum sizes $\{X|Y\}_{\lambda}{\,=\,}30${\arcsec} were adopted
(Sect.~\ref{maxsizes}).

The observed image of \object{NGC\,6334} displays complex blended structures of various shapes and intensities
(Fig.~\ref{ngc6334}), including substantial numbers of negative areas and artifacts from the data reduction and map-making
algorithms. The separated $\mathcal{S}_{{\lambda}}$ component shows all source-like peaks very clearly, even those that are hardly
visible in the original image, because \textsl{getsf} is able to distinguish sources from the elongated filamentary shapes.
Many of the sources overlap each other, therefore they require deblending for accurate measurements. The background
$\mathcal{B}_{{\lambda}{Y}}$ of filaments is fairly low, hence its subtraction enhanced the visibility of filaments in
$\mathcal{F}_{{\lambda}}$ only little. Nonuniform small-scale fluctuations in $\mathcal{S}_{{\lambda}}$ were effectively equalized
in the detection image $\mathcal{S}_{{\lambda}{\rm D}}$ by the flattening algorithm.

The source catalog contains measurements of $124$ sources, $91$ of which are selected as acceptably good by Eq.~(\ref{acceptable}).
In the component of filaments, \textsl{getsf} identified $26$ skeletons, tracing the simple, non-branching segments of the
filaments (Sect.~\ref{fildetection}) on spatial scales of ${\sim\,}30${\arcsec}, corresponding to the maximum width $Y_{\lambda}$.



\subsection{Star-forming cloud \object{Orion\,A}}
\label{scuba2}

\begin{figure*}                                                               
\centering
\centerline{
  \resizebox{0.328\hsize}{!}{\includegraphics{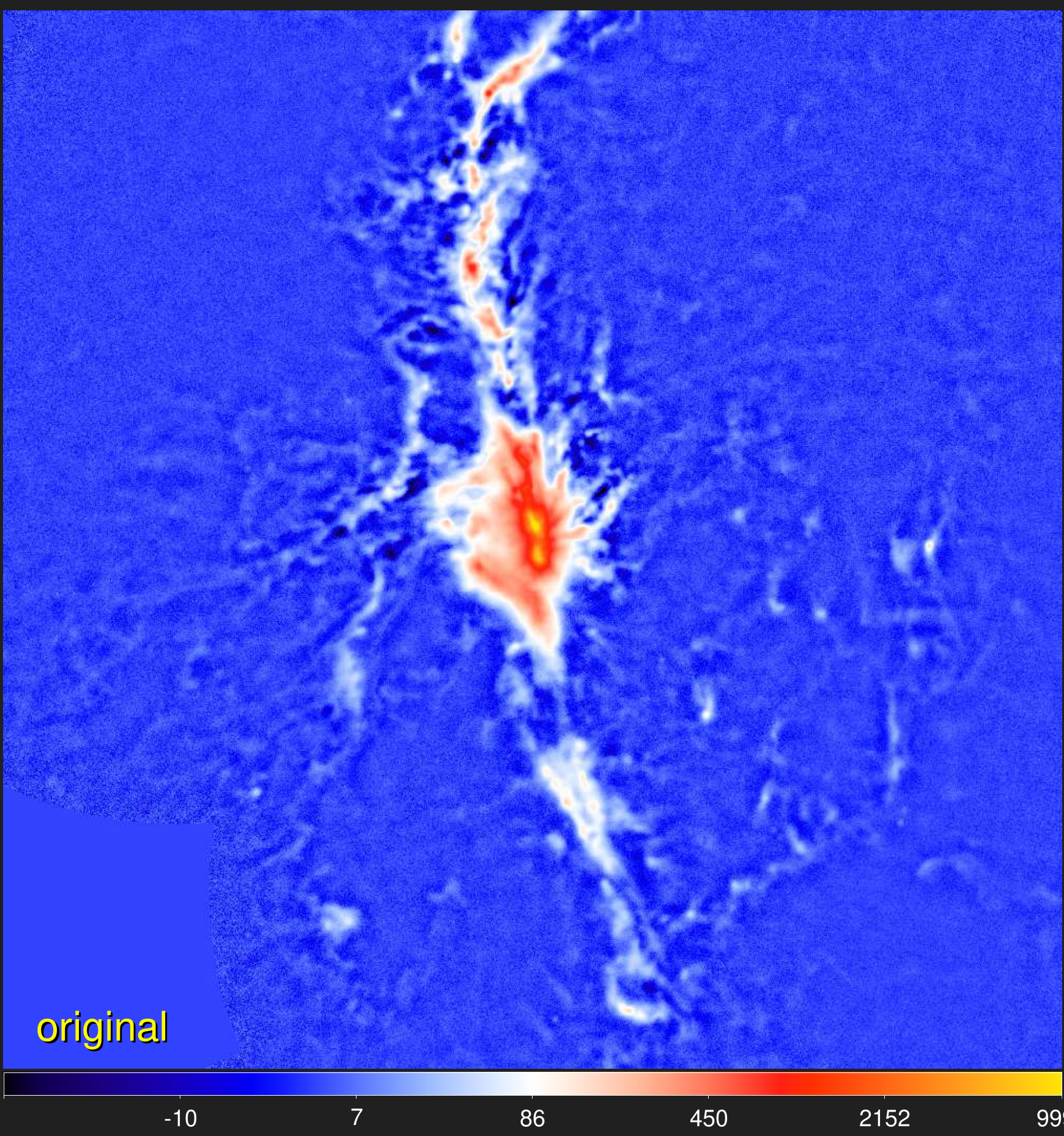}}
  \resizebox{0.328\hsize}{!}{\includegraphics{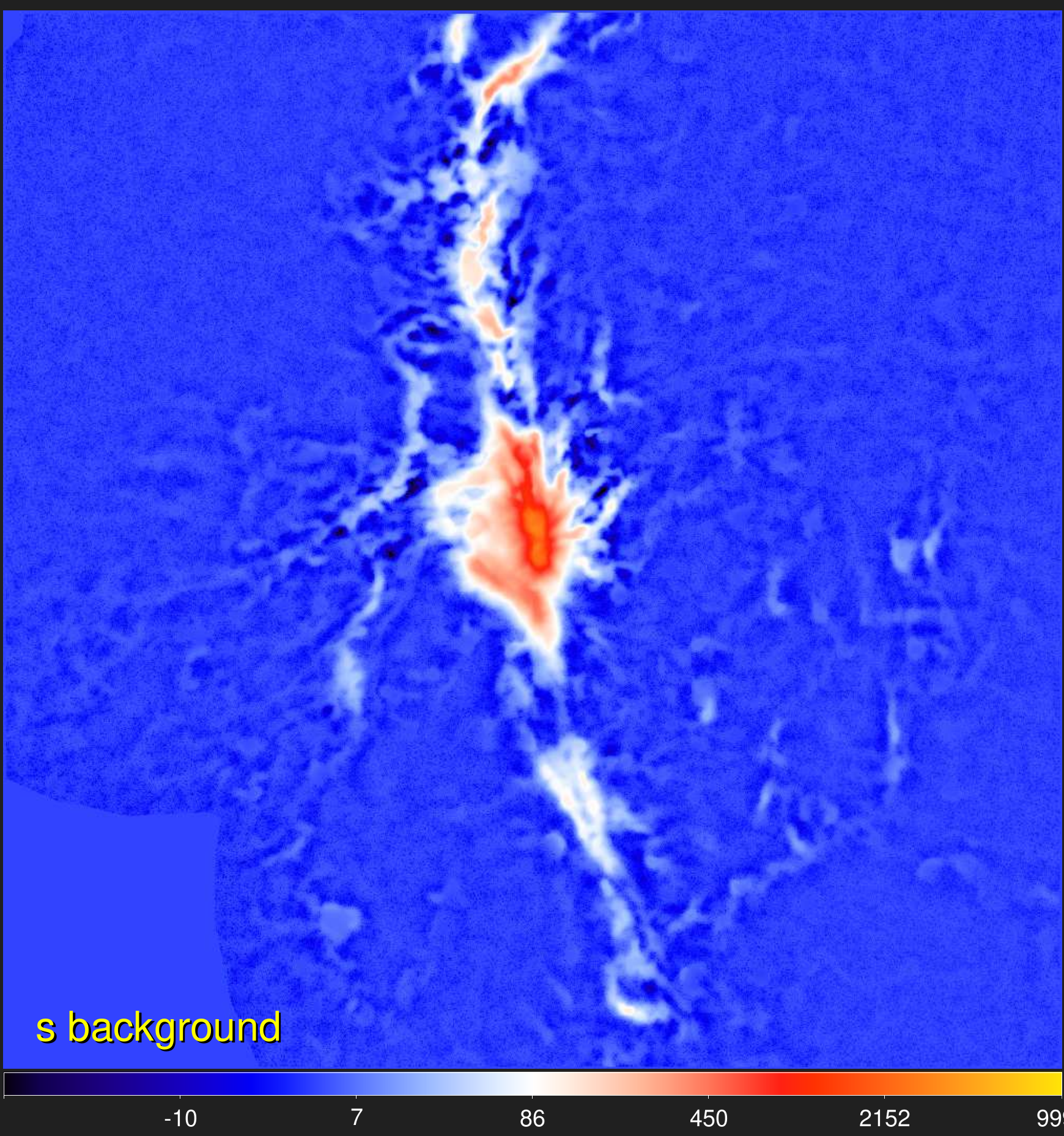}}
  \resizebox{0.328\hsize}{!}{\includegraphics{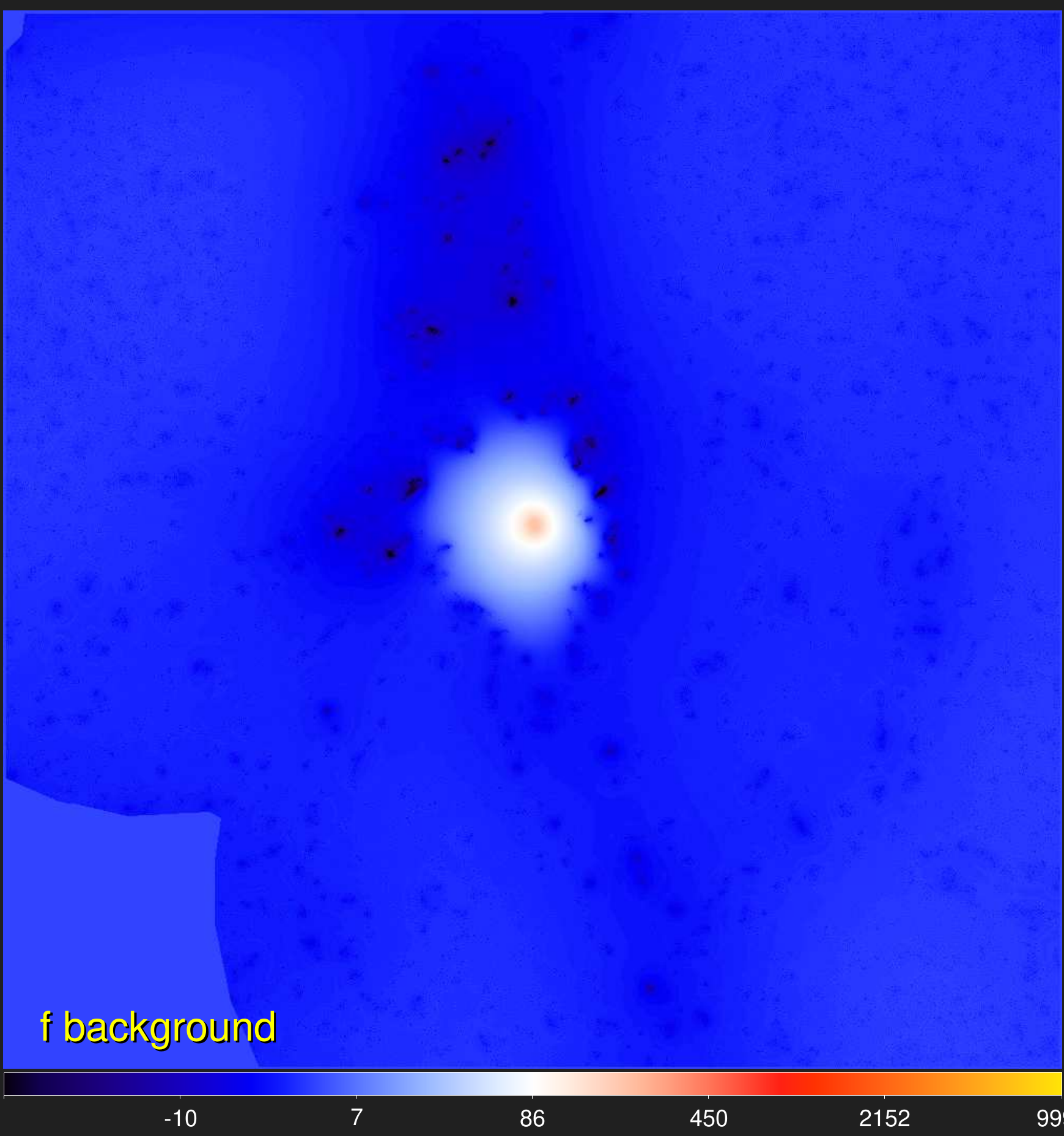}}}
\vspace{0.5mm}
\centerline{
  \resizebox{0.328\hsize}{!}{\includegraphics{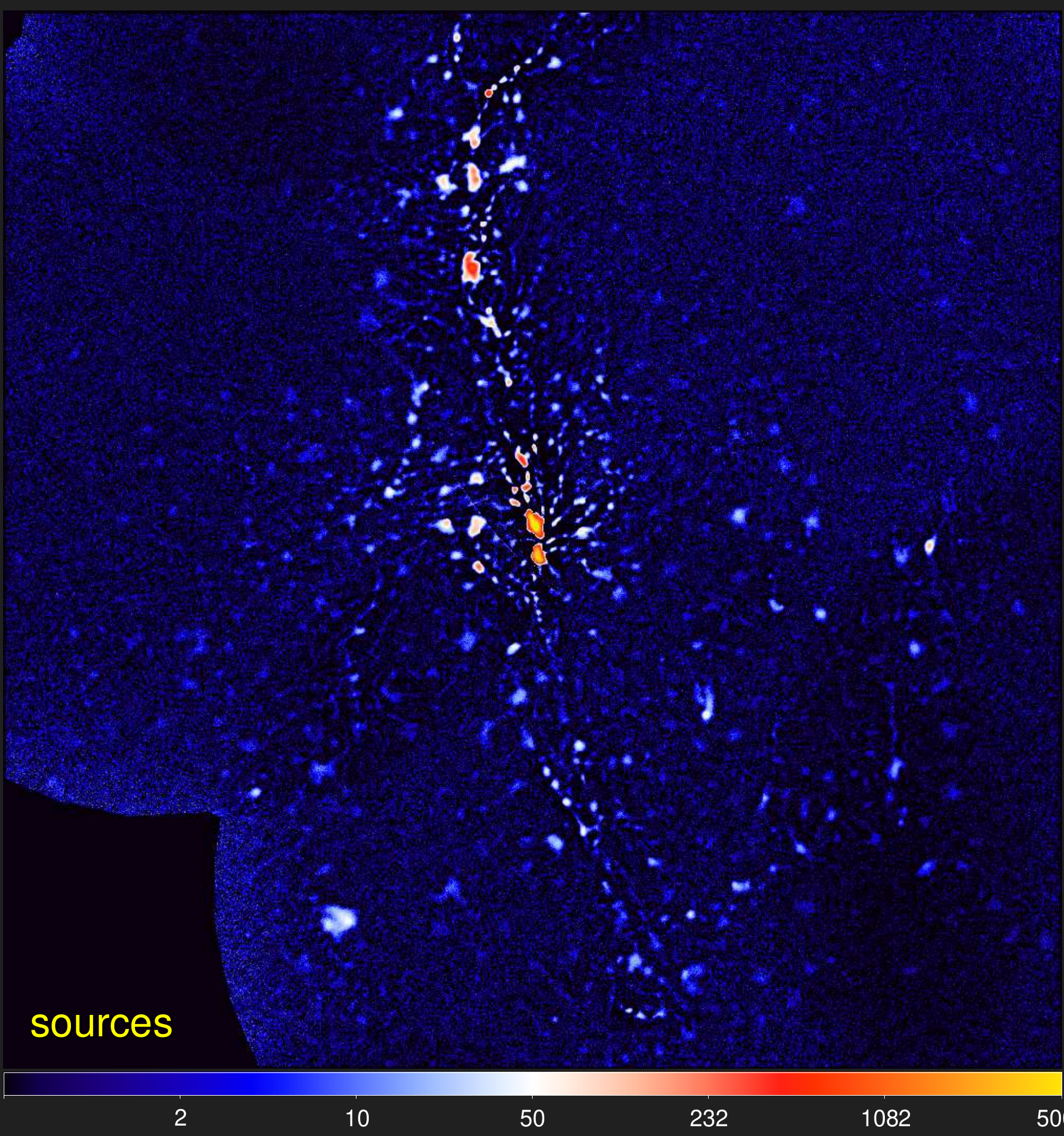}}
  \resizebox{0.328\hsize}{!}{\includegraphics{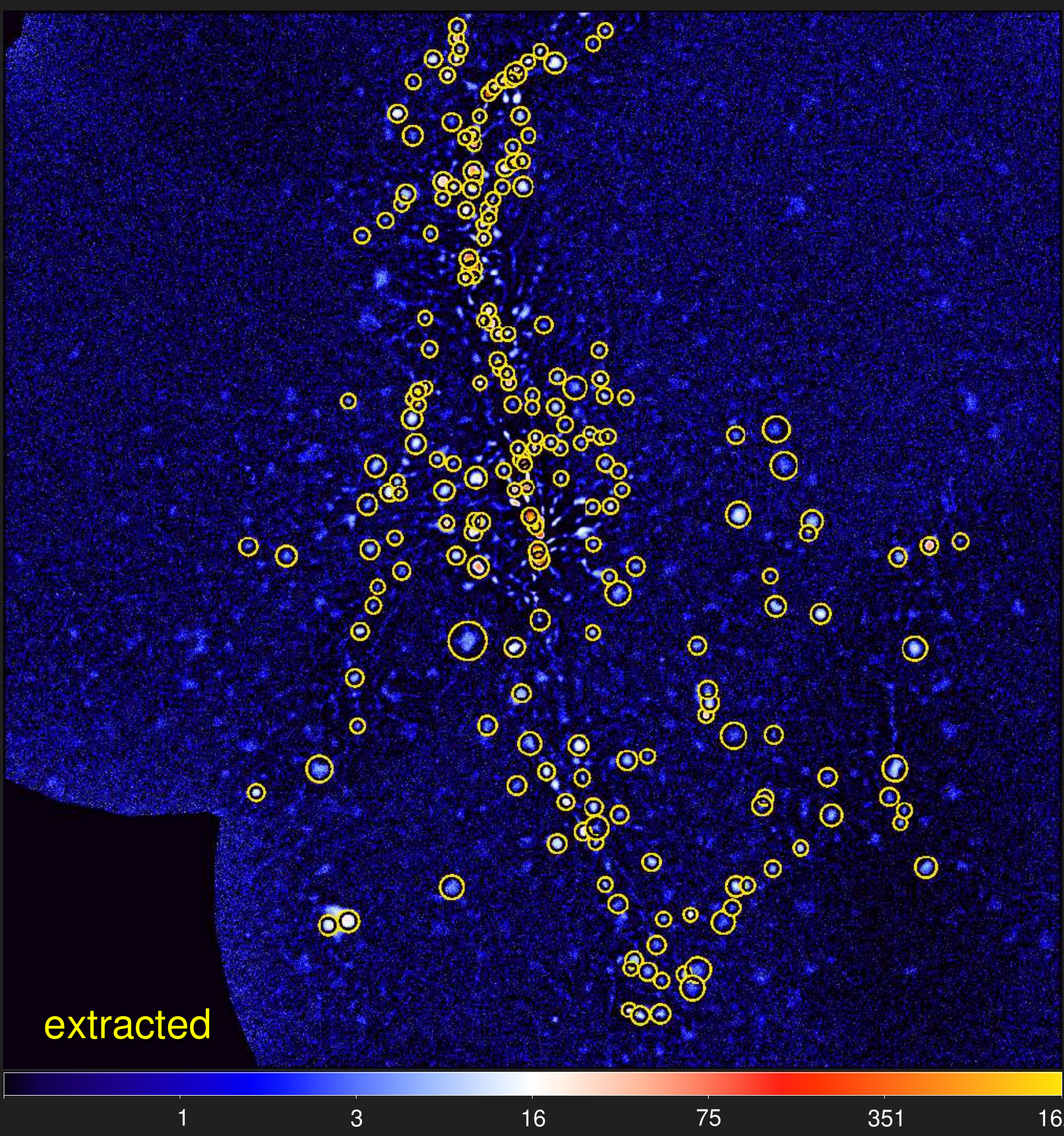}}
  \resizebox{0.328\hsize}{!}{\includegraphics{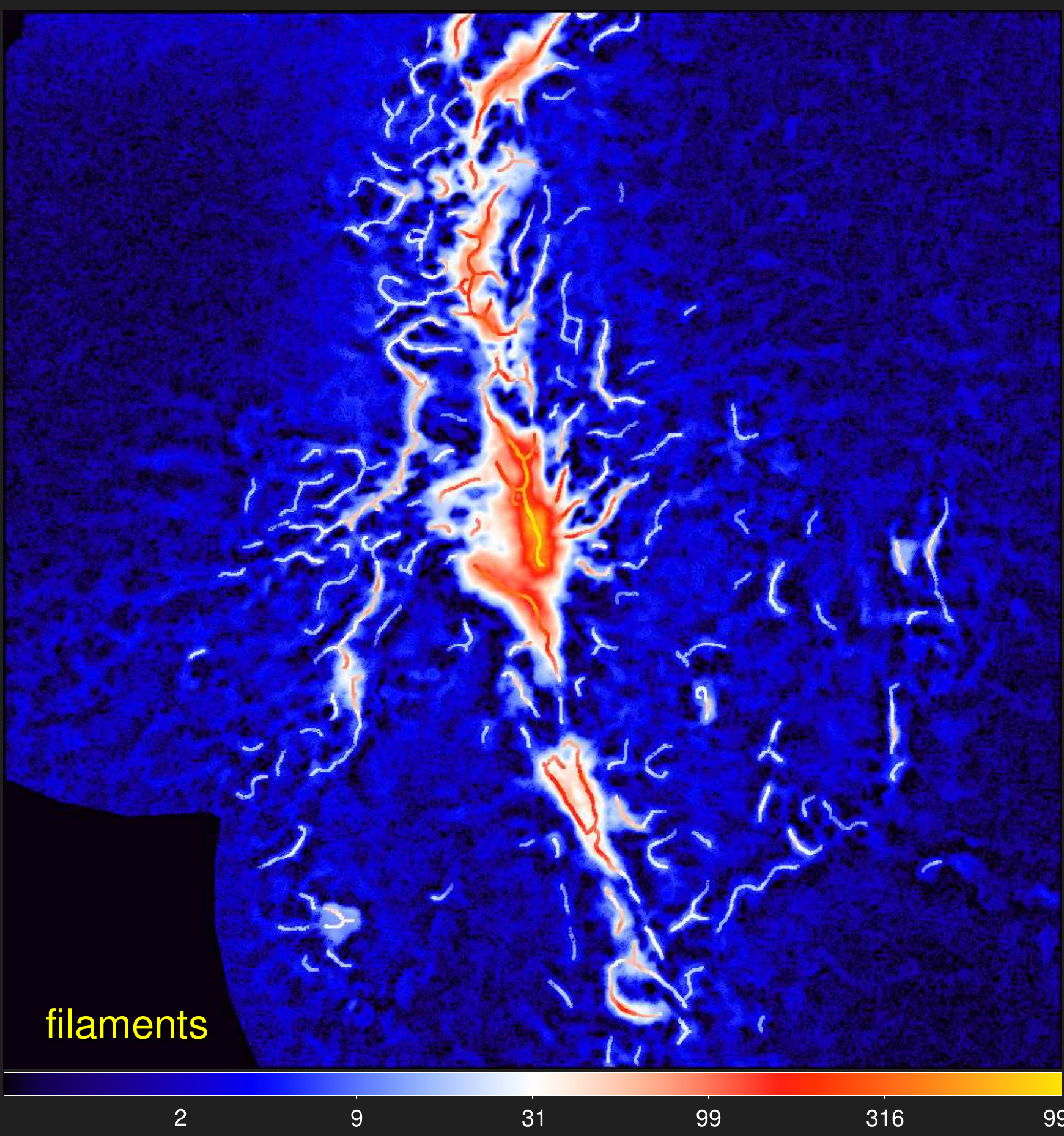}}}
\vspace{0.5mm}
\centerline{
  \resizebox{0.328\hsize}{!}{\includegraphics{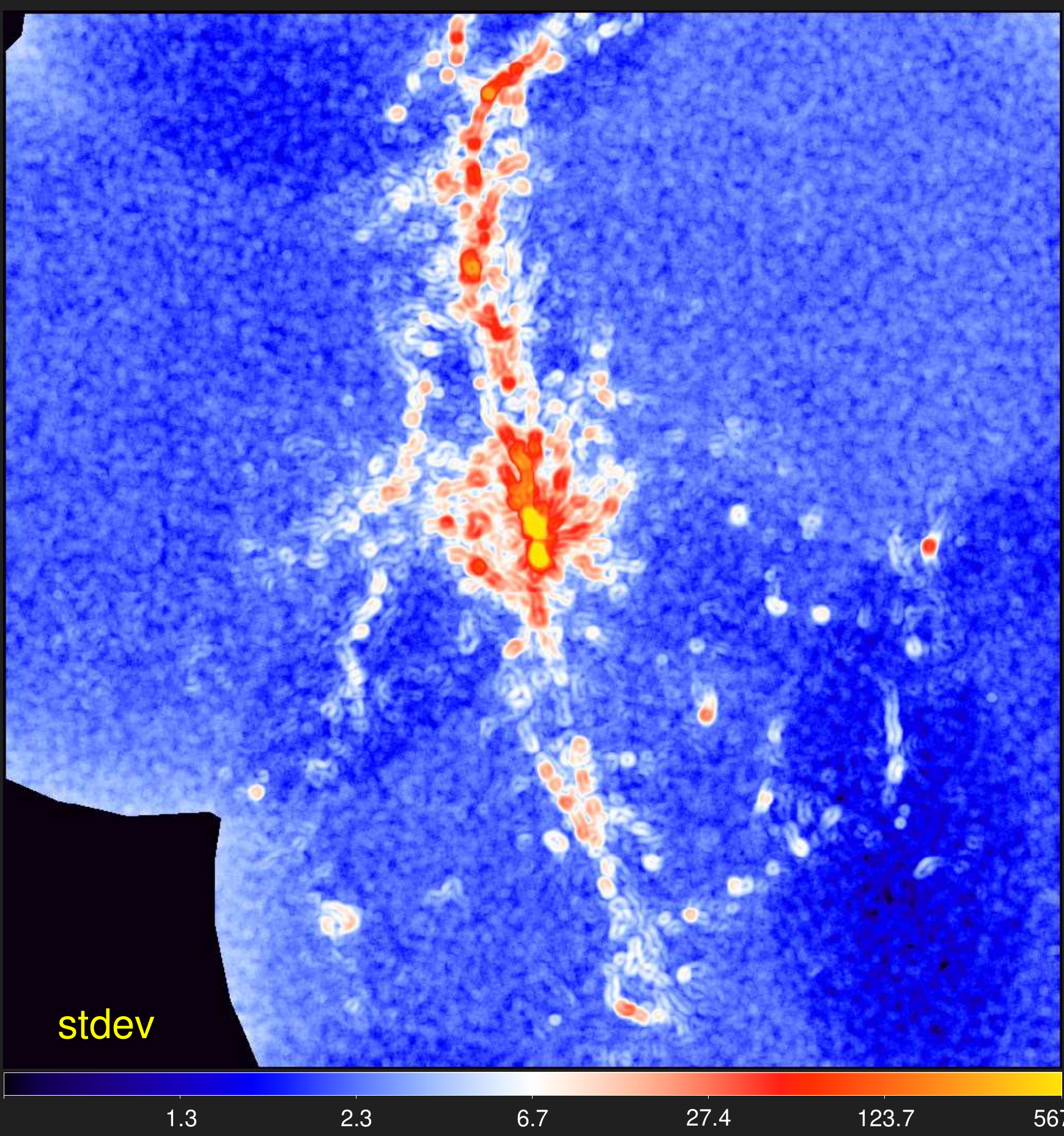}}
  \resizebox{0.328\hsize}{!}{\includegraphics{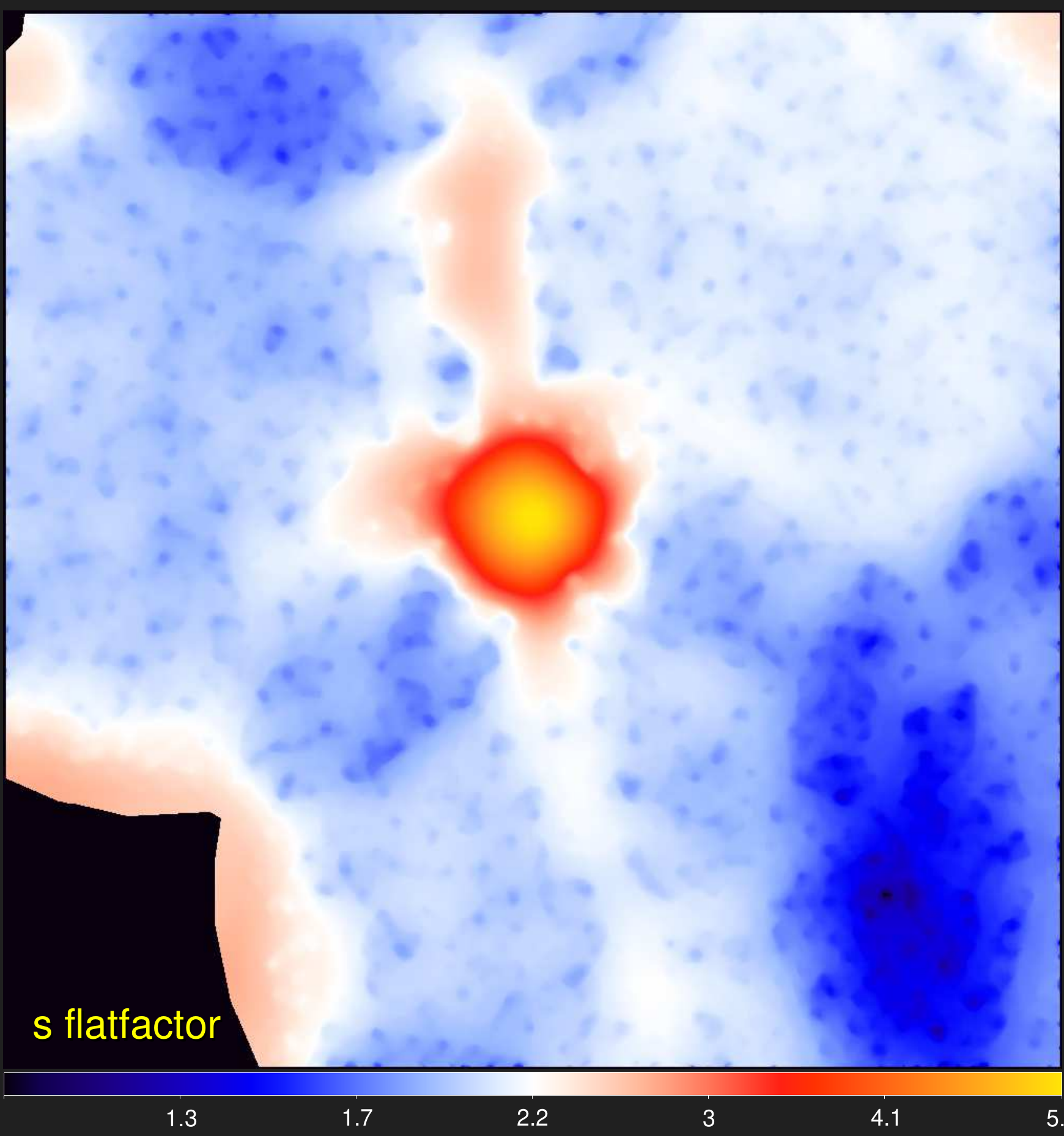}}
  \resizebox{0.328\hsize}{!}{\includegraphics{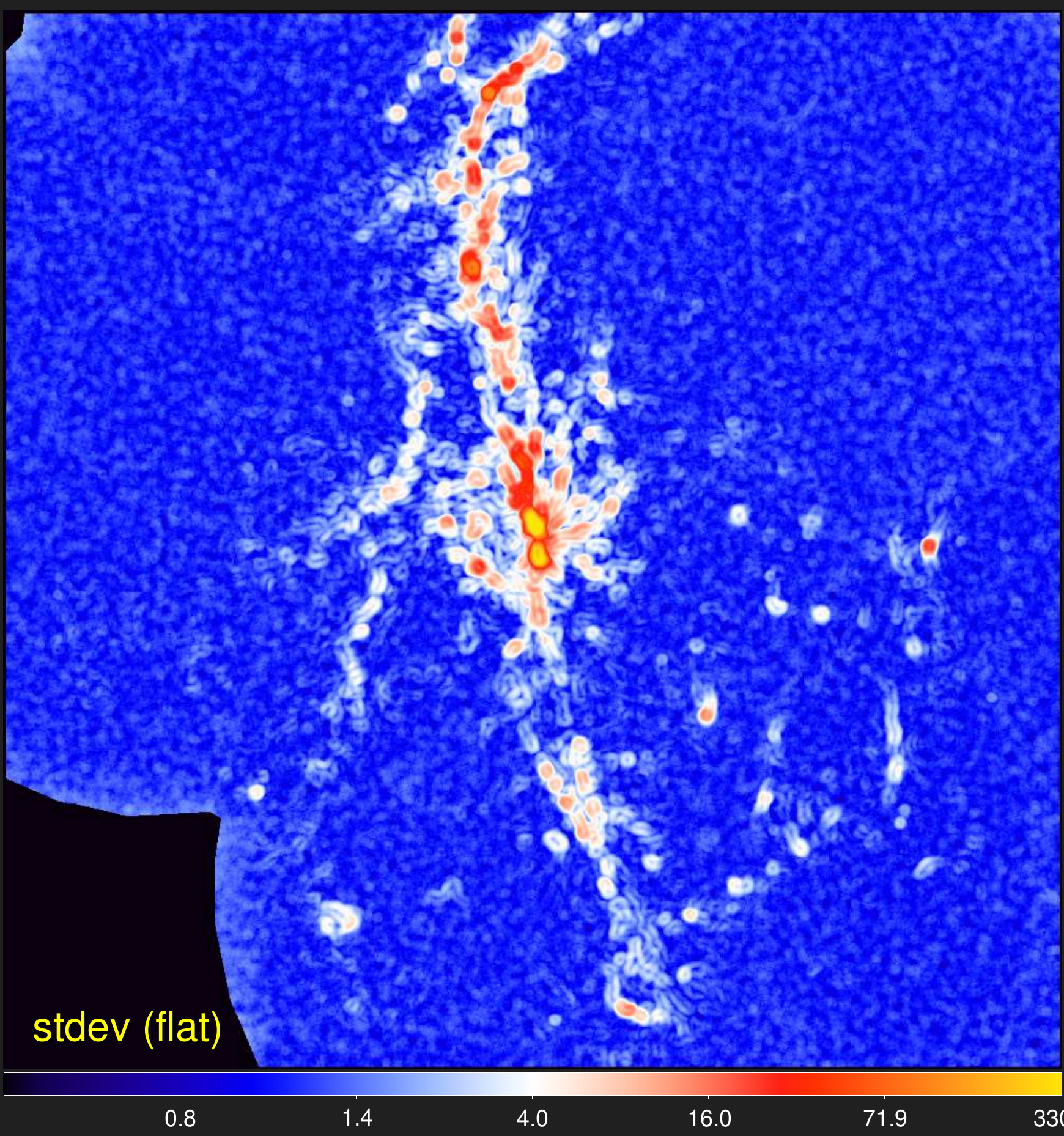}}}
\caption
{ 
Application of \textsl{getsf} to the \emph{JCMT} $\lambda{\,=\,}850$\,{${\mu}$m} image ($14.6${\arcsec\!} resolution) of the
\object{Orion\,A} star-forming cloud, adopting $\{X|Y\}_{\lambda}{\,=\,}30${\arcsec}. The \emph{top} row shows the original image
$\mathcal{I}_{\!\lambda}$ and the backgrounds $\mathcal{B}_{{\lambda}{\{X|Y\}}}$ of sources and filaments. The \emph{middle} row
shows the component $\mathcal{S}_{{\lambda}}$, the footprint ellipses of $212$ acceptably good sources on
$\mathcal{S}_{{\lambda}{\rm D}}$, and the component $\mathcal{F}_{{\lambda}{\rm D}}$ with $199$ skeletons $\mathcal{K}_{{k}{2}}$
corresponding to the scales $S_{\!k}{\,\approx\,}39${\arcsec}. The \emph{bottom} row shows the standard deviations
$\mathcal{U}_{\lambda}$ in the regularized component $\mathcal{S}_{{\lambda}{\rm R}}$, the flattening image
$\mathcal{Q}_{\lambda}$, and the standard deviations in the flattened component $\mathcal{S}_{{\lambda}{\rm
R}}\mathcal{Q}_{\lambda}^{-1}$. Some skeletons may only appear to have branches because they were widened for this presentation.
Intensities (in MJy\,sr$^{-1}$) are limited in range with logarithmic color mapping, except for $\mathcal{Q}_{\lambda}$, which is
shown with linear mapping.
} 
\label{oriona}
\vspace{20mm}
\end{figure*}

\object{Orion\,A} was observed with \emph{JCMT} at $450$ and $850$\,$\mu$m with the SCUBA-2 camera \citep{Lane_etal2016} with
angular resolutions of $9.8$ and $14.6${\arcsec}, respectively. The $0.86{\degr}{\times\,}0.86${\degr}
image\footnote{\url{https://www.canfar.net/storage/list/AstroDataCitationDOI/CISTI.CANFAR/16.0008/data}} at $850$\,$\mu$m in
Fig.~\ref{oriona} displays the northern part of the integral-shaped filament (ISF). Like with other ground-based submillimeter
observations that must subtract sky background, large-scale emission in the images has been filtered out \citep{Kirk_etal2018}. A
visual estimate suggests that the image contains substantial signal on spatial scales of up to ${\sim\,}100${\arcsec}. For the
two-wavelength \textsl{getsf} extraction, employing both $450$ and $850$\,$\mu$m images, maximum sizes
$\{X|Y\}_{\{450|850\}}{\,=\,}\{20,30,30,45\}${\arcsec\!} were adopted (Sect.~\ref{maxsizes}).

The $850$\,$\mu$m image of the ISF in Fig.~\ref{oriona} reveals the small-scale structures of the area most clearly because of the
spatial filtering effect of the observational technique. However, the central bright part of the ISF remains blended, and the
spatial decomposition by \textsl{getsf} helps isolate the sources $\mathcal{S}_{{\lambda}}$ in that area from the filaments
$\mathcal{F}_{{\lambda}}$ and their backgrounds $\mathcal{B}_{{\lambda}{\{X|Y\}}}$. The background of filaments is found to be
slightly negative, except in its central bright round area. In comparison with an average value of small-scale fluctuations in
$\mathcal{S}_{{\lambda}}$, they are larger by a factor of $2.7$ in the central zone and lower by a factor of $1.7$ in the
lower right corner. The standard deviations $\mathcal{U}_{{\lambda}}$ reveal imprints of the five overlapping scans from the
observations. The flattening algorithm of \textsl{getsf} effectively equalizes them and creates the flat detection images
$\mathcal{\{S|F\}}_{{\lambda}{\rm D}}$ of sources and filaments, improving their detection reliability.

The two-band source extraction in ISF with \textsl{getsf} cataloged $344$ sources, detected and measured in both wavebands
simultaneously. Only $257$ and $212$ sources at $450$ and $850$\,$\mu$m, respectively, are selected as acceptably good by
Eq.~(\ref{acceptable}); the S/N for the remaining detections is too low or they have other defects that are identified by the
measurements. Two additional \textsl{getsf} extractions, done on each image independently, resulted in catalogs with $319$ and
$283$ sources at $450$ and $850$\,$\mu$m, respectively. Independent extractions ignore the valuable information from the other
image, hence there are higher chances of spurious sources. With the additional condition that cataloged sources must be detected in
both images, the combined extraction catalog contains $223$ sources;$196$ and $183$ of these sources at $450$ and $850$\,$\mu$m,
respectively, are acceptably good. They represent the most reliable sources in the images, hence it is highly unlikely that there
are spurious sources among them.

The missing large-scale emission of the SCUBA-2 image helped \textsl{getsf} expose the many relatively faint, narrow
filaments within the wide, massive ISF. In the $\mathcal{F}_{{\lambda}}$ component, \textsl{getsf} identified $267$ and $199$
simple, non-branching segments of the filaments (Sect.~\ref{fildetection}) at $450$ and $850$\,$\mu$m, respectively, on transverse
scales of $28$ and $39${\arcsec}. This is similar to the existence of narrow sub-filaments on small scales within the resolved
\object{Taurus}, \object{Aquila}, and \object{IC\,5146} filaments (Fig.~\ref{ssfilams}) and consistent with the recent \emph{ALMA}
observations of ISF \citep{Hacar_etal2018}.




\subsection{Star-forming cloud \object{W\,43-MM1}}
\label{alma}

\begin{figure*}                                                               
\centering
\centerline{
  \resizebox{0.328\hsize}{!}{\includegraphics{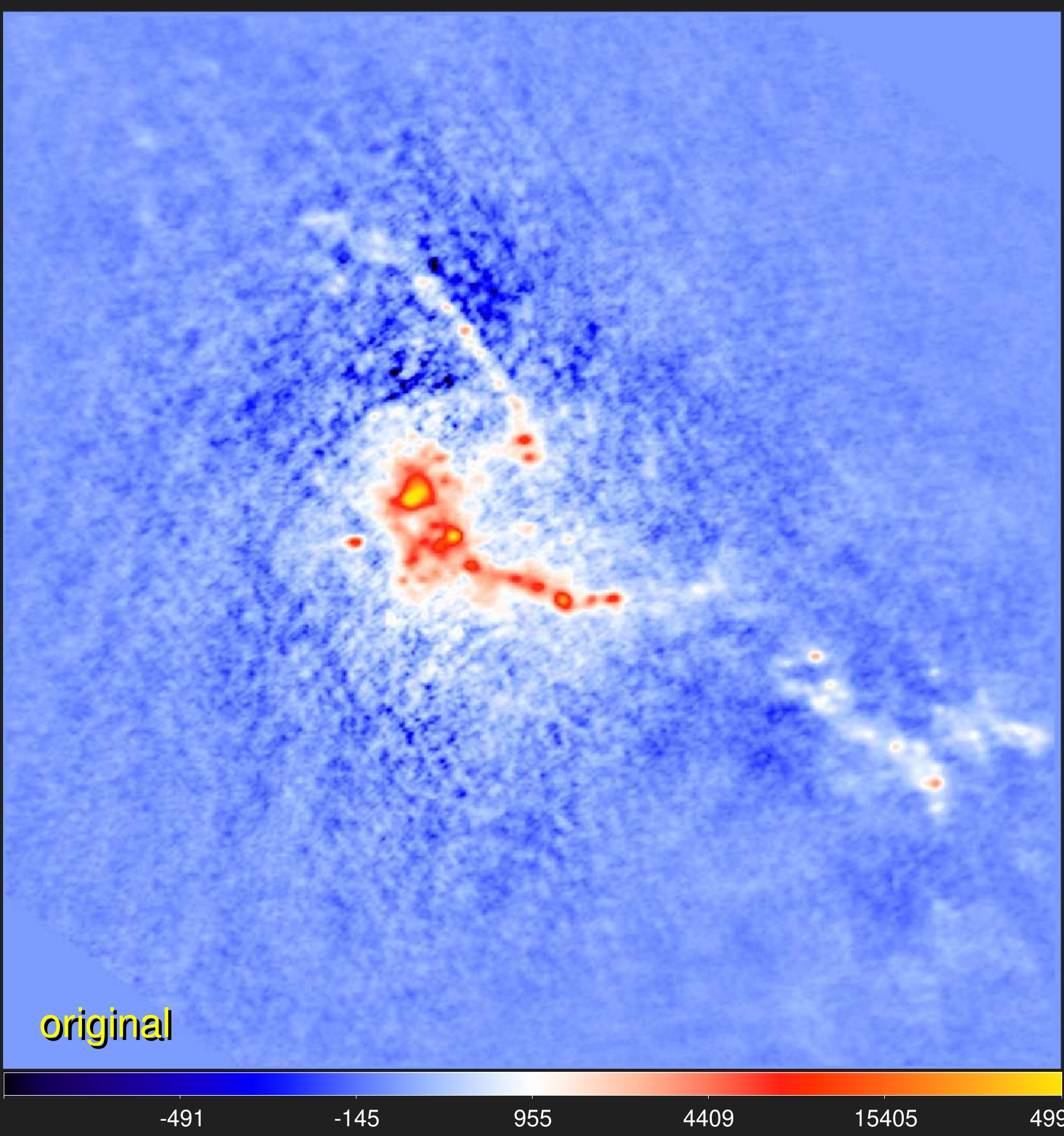}}
  \resizebox{0.328\hsize}{!}{\includegraphics{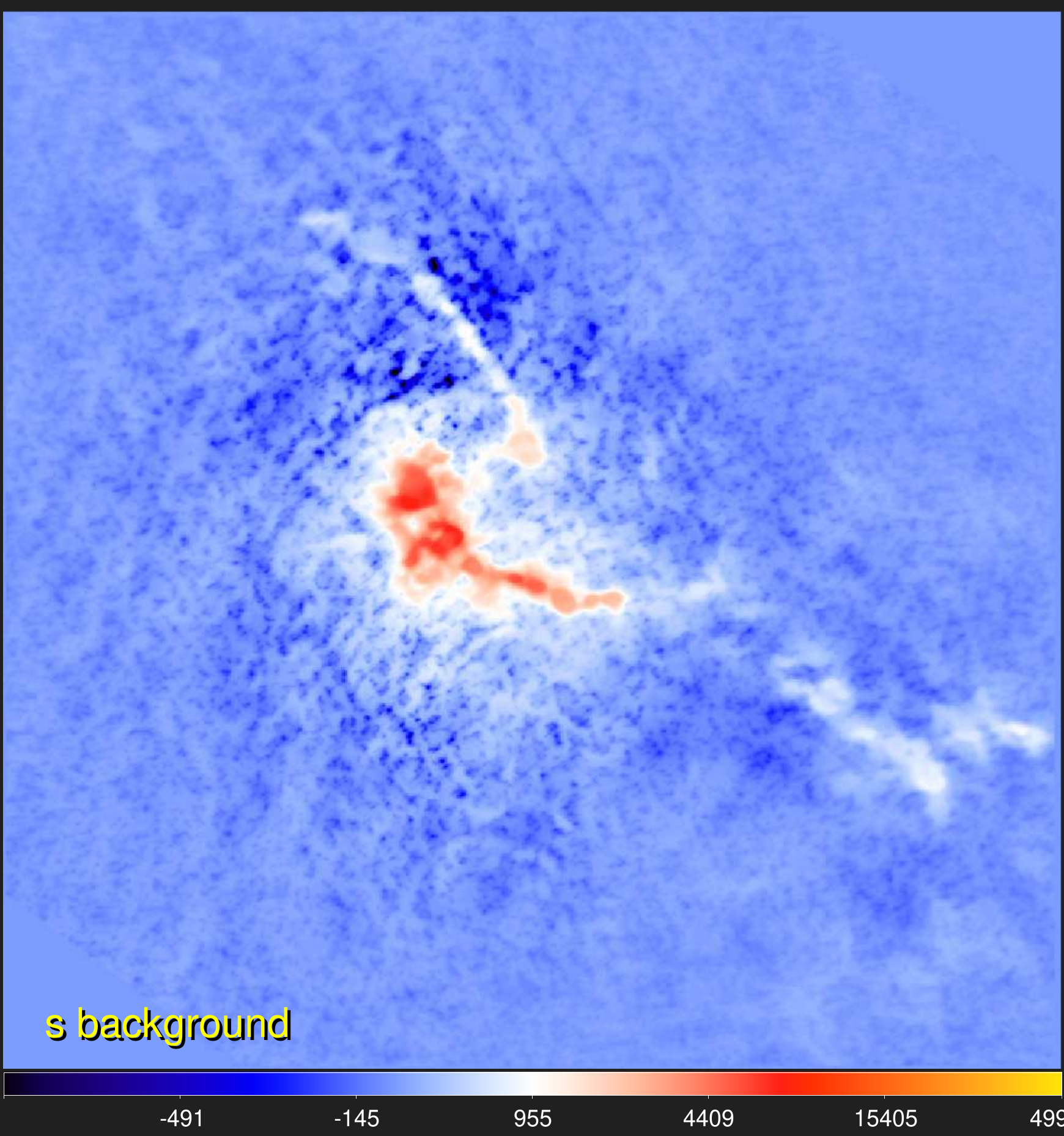}}
  \resizebox{0.328\hsize}{!}{\includegraphics{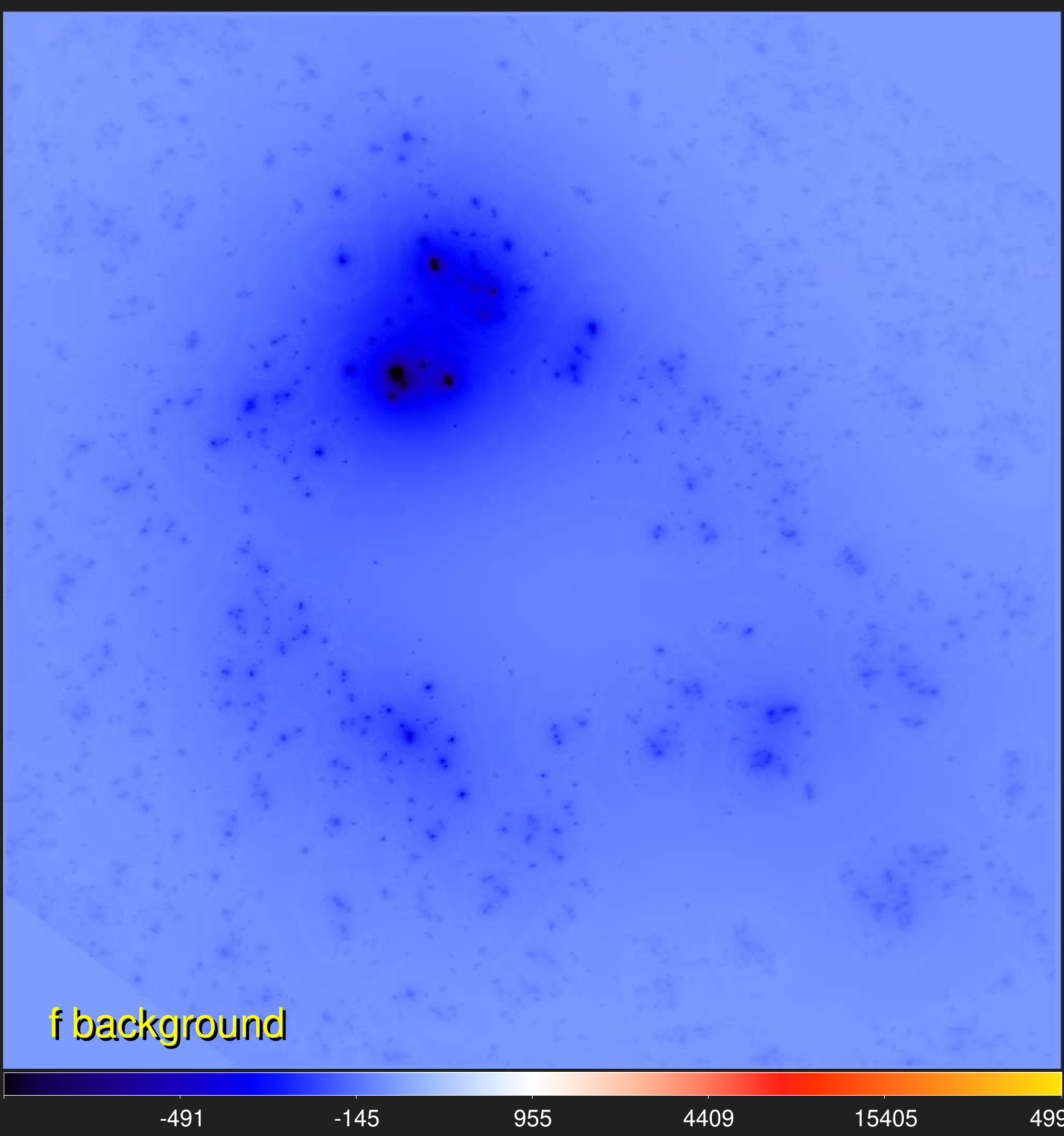}}}
\vspace{0.5mm}
\centerline{
  \resizebox{0.328\hsize}{!}{\includegraphics{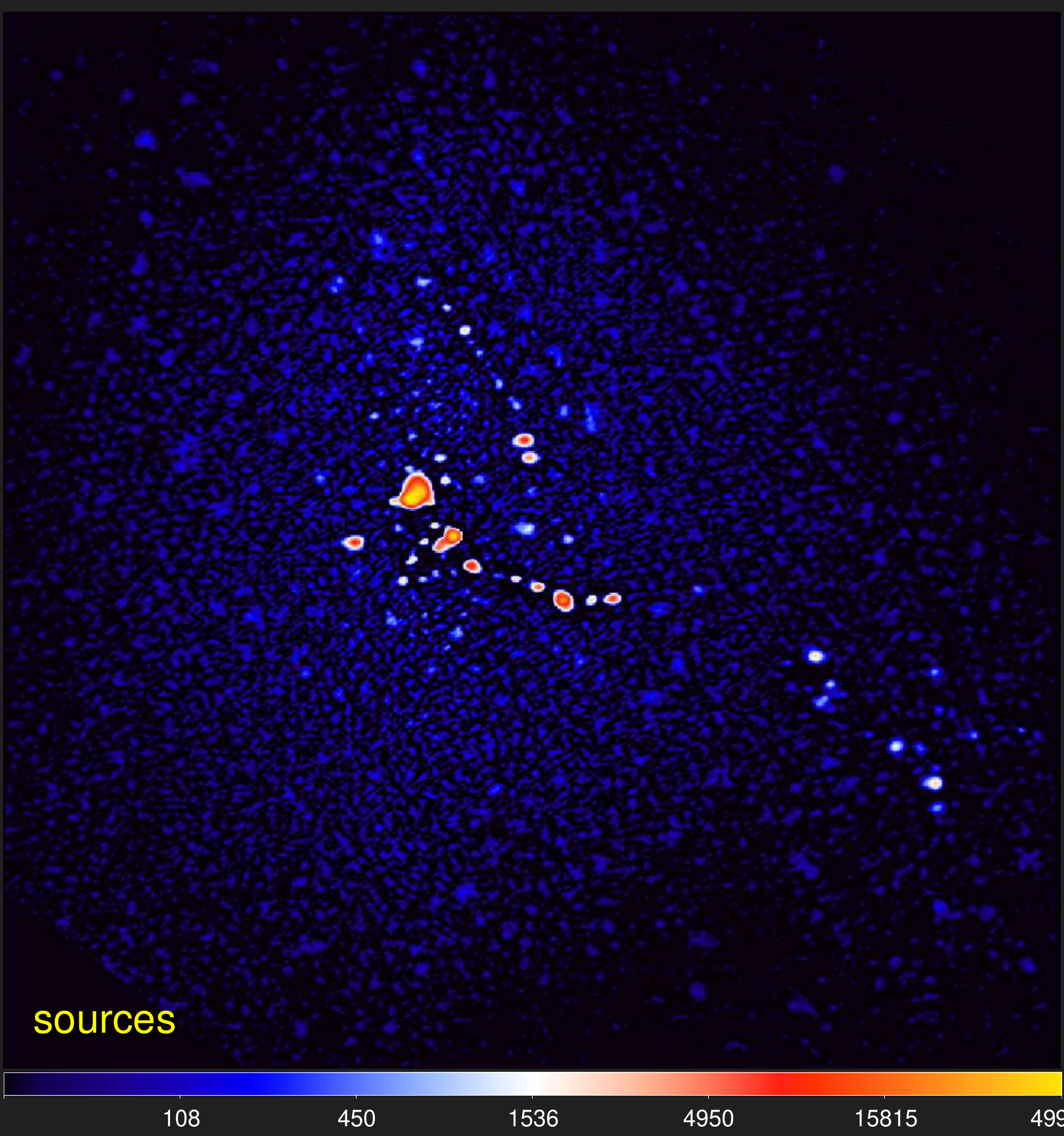}}
  \resizebox{0.328\hsize}{!}{\includegraphics{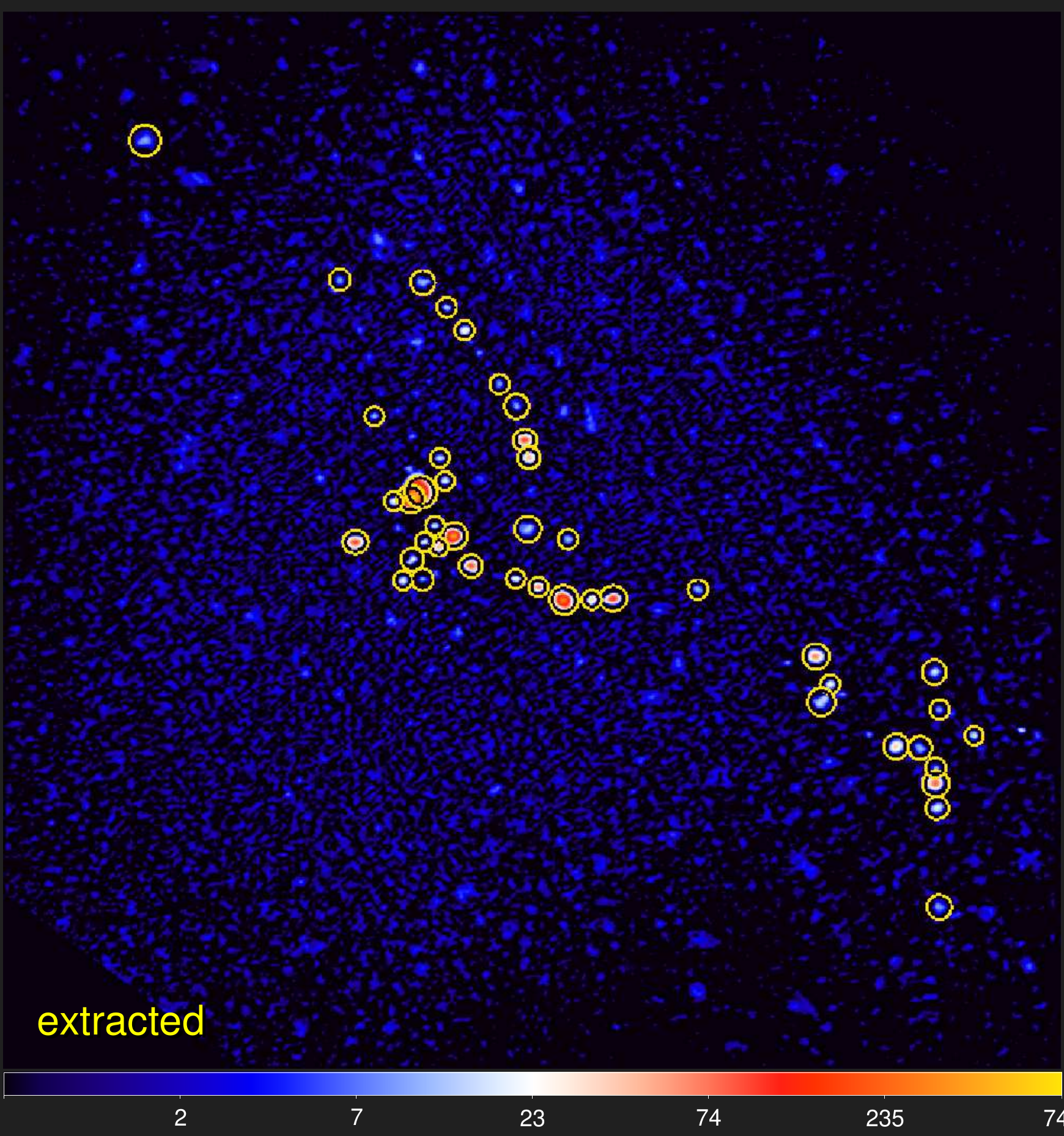}}
  \resizebox{0.328\hsize}{!}{\includegraphics{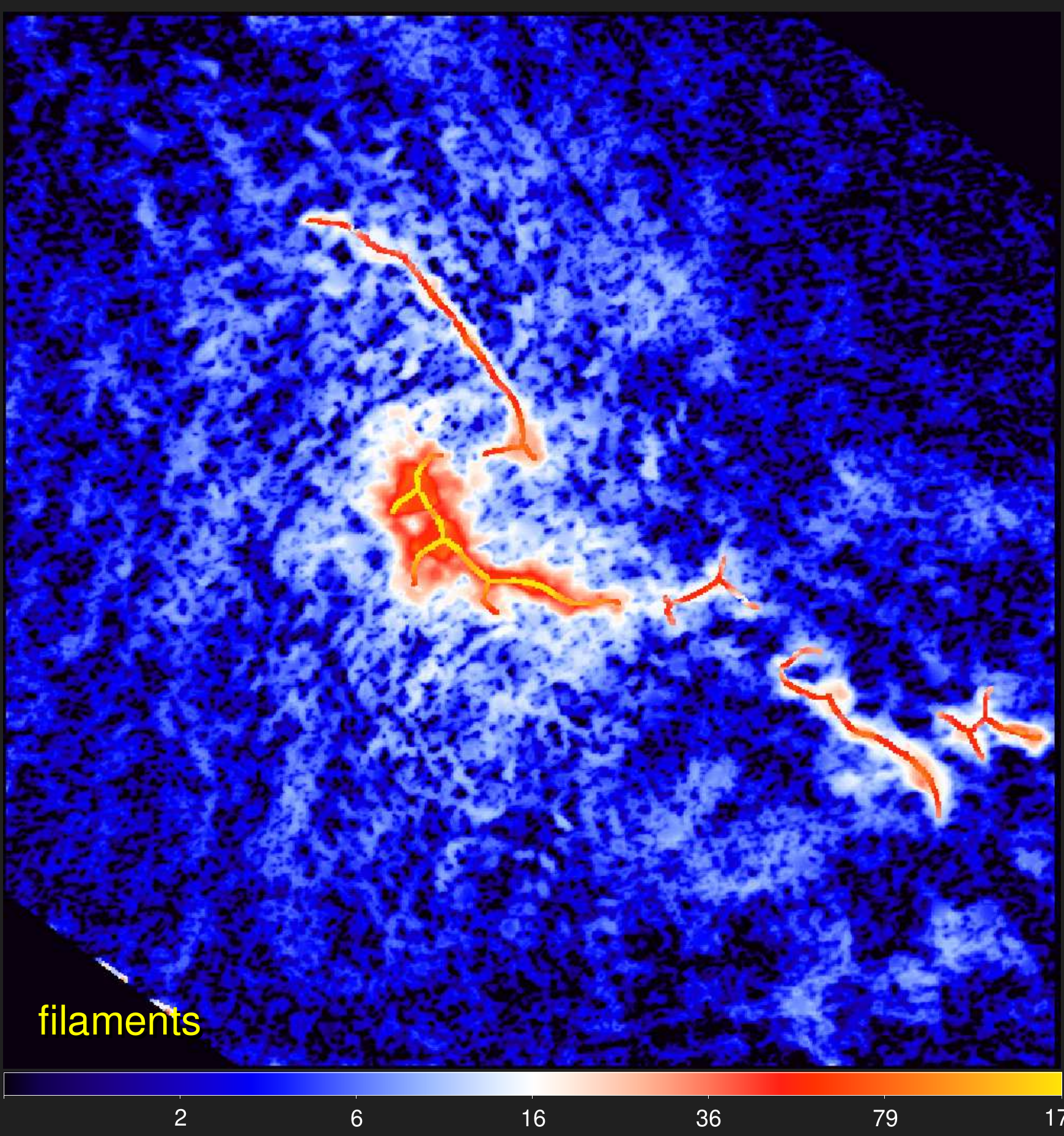}}}
\vspace{0.5mm}
\centerline{
  \resizebox{0.328\hsize}{!}{\includegraphics{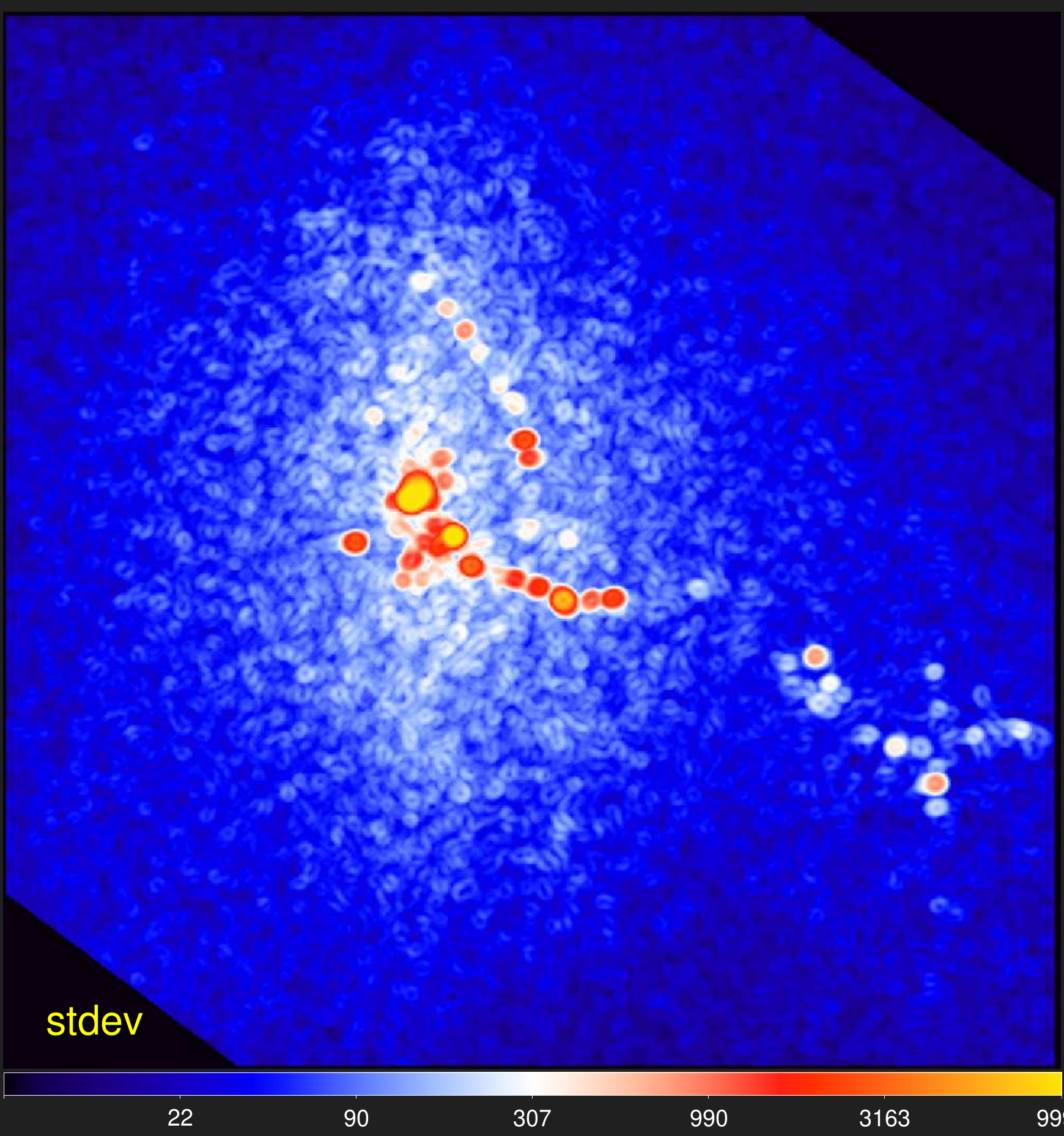}}
  \resizebox{0.328\hsize}{!}{\includegraphics{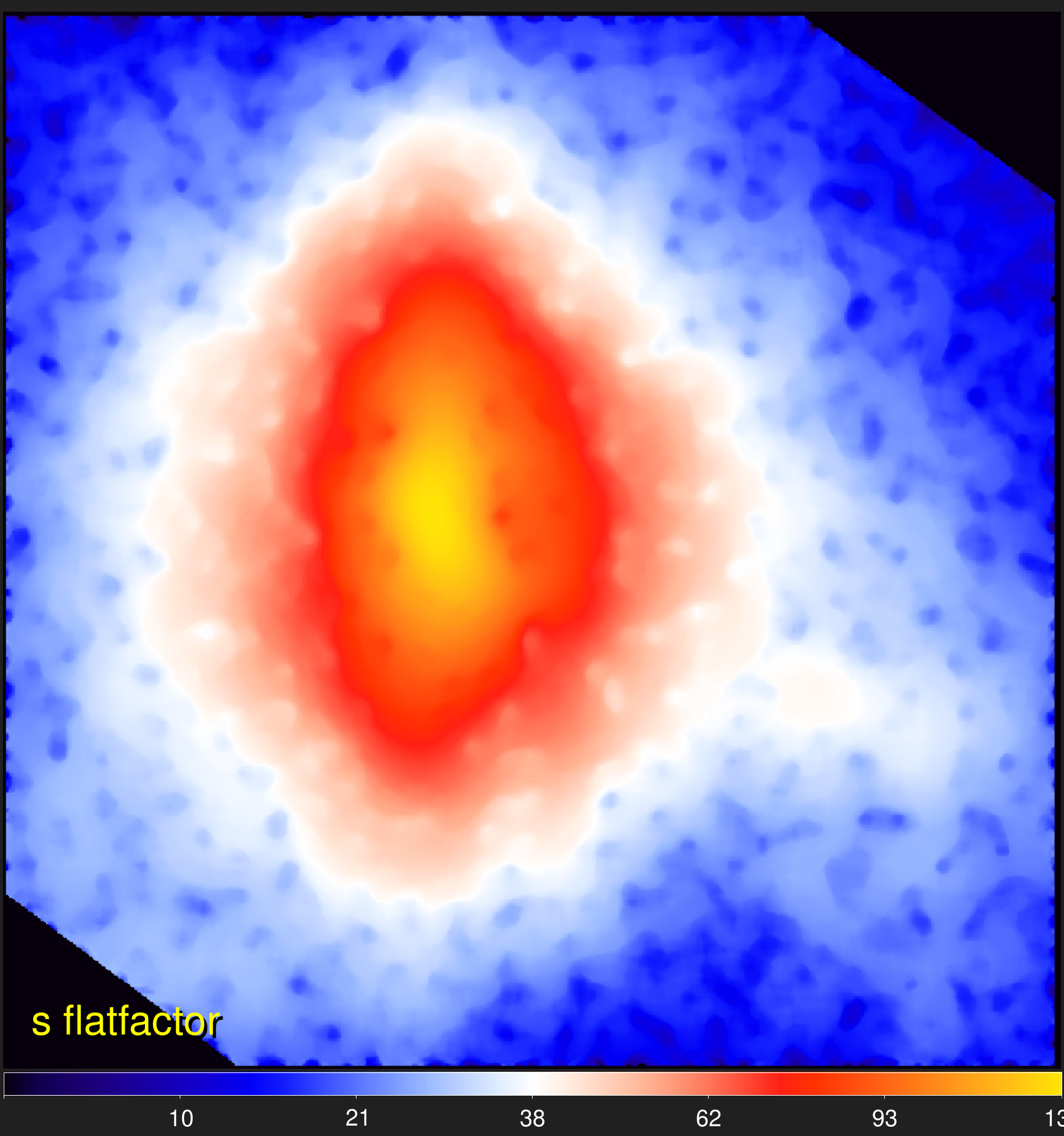}}
  \resizebox{0.328\hsize}{!}{\includegraphics{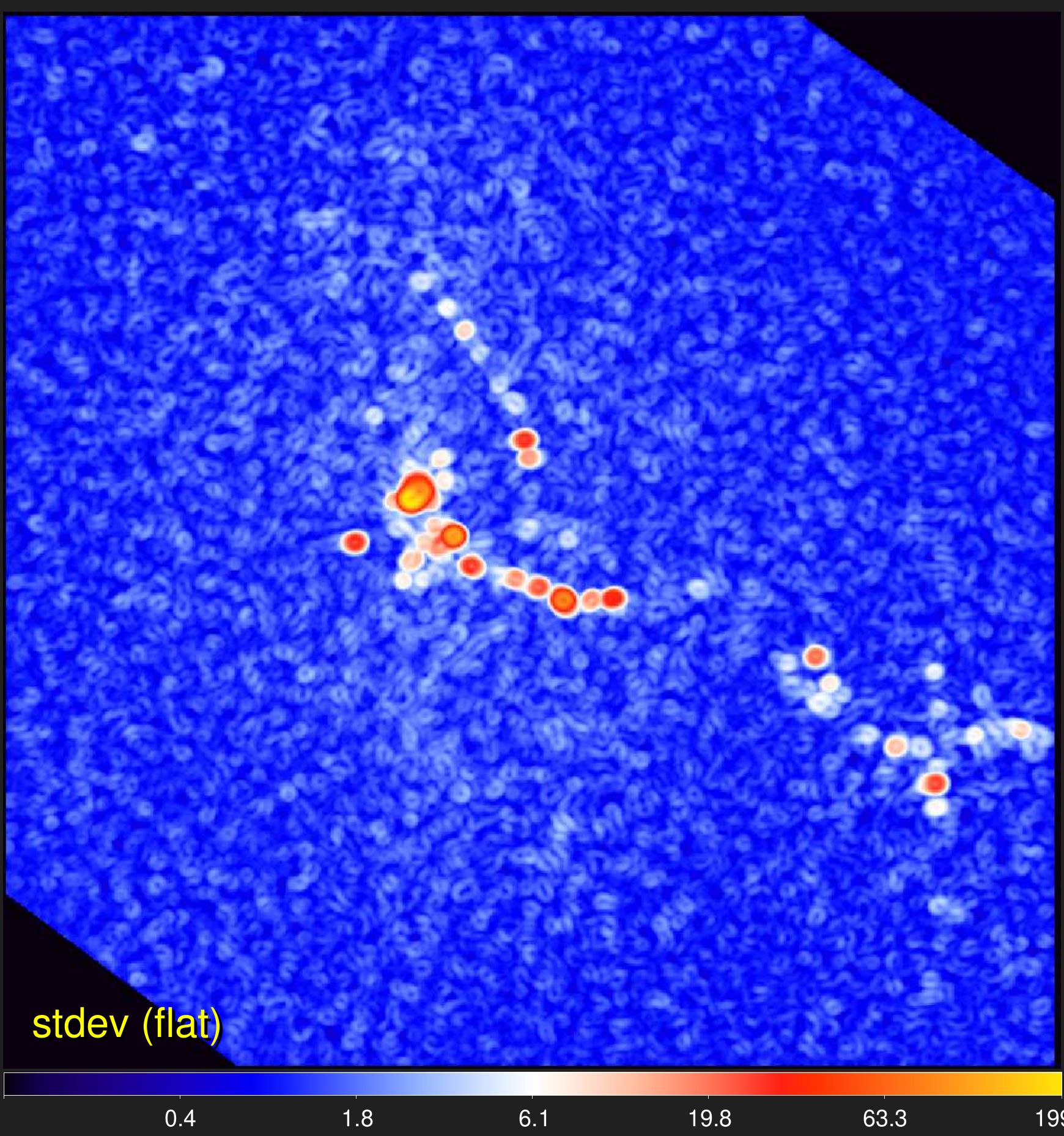}}}
\caption
{ 
Application of \textsl{getsf} to the \emph{ALMA} $\lambda{\,=}1300$\,{${\mu}$m} image ($0.44${\arcsec\!} resolution) of the
\object{W43-MM1} star-forming cloud, adopting $\{X|Y\}_{\lambda}{\,=\,}\{0.8,1.3\}${\arcsec\!}. The \emph{top} row shows the
original image $\mathcal{I}_{\!\lambda}$ and the backgrounds $\mathcal{B}_{{\lambda}{\{X|Y\}}}$ of sources and filaments. The
\emph{middle} row shows the component $\mathcal{S}_{{\lambda}}$, the footprint ellipses of $44$ acceptably good sources on
$\mathcal{S}_{{\lambda}{\rm D}}$, and the component $\mathcal{F}_{{\lambda}{\rm D}}$ with $15$ skeletons $\mathcal{K}_{{k}{2}}$
corresponding to the scales $S_{\!k}{\,\approx\,}2${\arcsec}. The \emph{bottom} row shows the standard deviations
$\mathcal{U}_{\lambda}$ in the regularized component $\mathcal{S}_{{\lambda}{\rm R}}$, the flattening image
$\mathcal{Q}_{\lambda}$, and the standard deviations in the flattened component $\mathcal{S}_{{\lambda}{\rm
R}}\mathcal{Q}_{\lambda}^{-1}$. Some skeletons may only appear to have branches because they were widened for this presentation.
Intensities (in MJy\,sr$^{-1}$) are limited in range with logarithmic color mapping, except for $\mathcal{Q}_{\lambda}$, which is
shown with square-root mapping.
} 
\label{w43mm1}
\vspace{20mm}
\end{figure*}

\object{W\,43-MM1} was observed with the $12$\,m array of the \emph{ALMA} interferometer (baselines $13{-}1045$\,m) in the
$233$\,GHz band centered at $1300$\,$\mu$m \citep{Motte_etal2018,Nony_etal2020}. The small $68{\arcsec}{\times\,}68${\arcsec} image
in Fig.~\ref{w43mm1} with an angular resolution of $0.44${\arcsec} contains spatial scales of up to $12${\arcsec}, beyond which the
interferometer was insensitive to the emission. For this source and filament extraction with \textsl{getsf}, the maximum size
$\{X|Y\}_{\lambda}{\,=\,}\{0.8,1.3\}${\arcsec} was adopted (Sect.~\ref{maxsizes}).

The interferometric image of \object{W\,43-MM1} (Fig.~\ref{w43mm1}) shows a cluster of relatively bright sources, some of them
blended, and three faint filamentary structures that appear to connect them. Separation of the components of sources
$\mathcal{S}_{{\lambda}}$ and filaments $\mathcal{F}_{{\lambda}}$ confirms that most sources are concentrated on (or near) the
faint continuous filaments. Almost the entire background $\mathcal{B}_{{\lambda}{Y}}$ of the filamentary structures is negative,
which is caused by the missing large scales in the observed images.

The small-scale fluctuation levels steeply increase toward the image center by more than an order of magnitude
(Fig.~\ref{w43mm1}), as evidenced by the standard deviations $\mathcal{U}_{{\lambda}}$. The small-scale structured noise from the
interferometric observations have both round or irregular, elongated shapes. Consequently, the separation of structural components
by \textsl{getsf} leads to many faint peaks in $\mathcal{S}_{{\lambda}}$ and $\mathcal{F}_{{\lambda}}$. The flattening algorithm
equalizes the fluctuation levels very effectively, providing reliable detection of sources in the flat
$\mathcal{\{S|F\}}_{{\lambda}{\rm D}}$. If not suppressed, such highly variable structured noise would produce many spurious
sources and filaments in the central area of the image.

This \emph{ALMA} image of \object{W\,43-MM1} contains only moderate numbers of sources and filaments. The extraction catalog
contains measurements of $44$ sources, and all of them are selected as acceptably good by Eq.~(\ref{acceptable}). This simple field
allows a visual verification that they all are the true sources and are not contaminated by the noise fluctuations. In the filament
component, \textsl{getsf} identified $15$ skeletons, tracing the simple, non-branching segments of the filaments
(Sect.~\ref{fildetection}) on spatial scales of ${\sim\,}2${\arcsec}, similar to the maximum width $Y_{\lambda}$ adopted for the
extraction.



\section{Strengths and limitations}
\label{strenlimits}


\subsection{Strengths}
\label{strengths}

In contrast to the other methods, \textsl{getsf} extracts sources and filaments simultaneously by combining available information
from all wavebands. Its flexible multiwavelength design enables handling of up to 99 images, not necessarily all of them observed
in different wavebands. The maximum number of images is arbitrary, representing the largest two-digit integer number used in the
output file names; the code can be updated to use a higher value if required for some applications. Any subset of the input images
that is deemed beneficial for the detection purposes can be used to detect the sources and filaments, whereas measurements of the
identified structures are provided for all input images. In a nonstandard application, the method can also be employed with the
position-velocity cubes if they are split into separate images along the velocity axis \citep[\textsl{getold} was used in this way
by][]{Shimajiri_etal2019}.

The images that are selected for detection are spatially decomposed to isolate the contributions of similar scales
(Appendix~\ref{decomposition}) and are then combined in a wavelength-independent set of single-scale detection images
(Sect.~\ref{combining}). This eliminates the necessity of associating independent detections across wavelengths in images with
greatly different angular resolutions and improves the detection and measurement accuracy. For example, positional association of
nearby sources detected at $160$\,$\mu$m and completely blended into a single clump at $500$\,$\mu$m does not make sense.

Separation of structural components in the images of highly structured observed regions in space provides independent images of
sources, filaments, and their backgrounds (Sect.~\ref{deriveback}), which is highly beneficial for the analysis and interpretation
of observations. Flattening of detection images equalizes the (nonuniform) small-scale background and noise fluctuations
(Sect.~\ref{flattening}). This greatly simplifies the images and allows reliable detection of sources and filaments in decomposed
single-scale images using a constant threshold, with a very low rate of spurious sources.

Sources and filaments of any size and width can be extracted by \textsl{getsf} provided that they are significantly smaller than
the image. Only the maximum size of the structures of interest must be specified for each image in order to limit the range of
spatial scales considered and the size of the structures to be measured and cataloged. The single parameter of the observed images
that \textsl{getsf} needs to know is the maximum size, which is determined from the images by users (Sect.~\ref{maxsizes}) on the
basis of their research interests. This single constrained parameter reduces the dependence of the extraction results on the human
factor to a minimum and makes their analysis and derived conclusions as objective as possible.

The numerical code is designed to be user-friendly and easy to run, providing diagnostics to help users avoid common problems. It
verifies the \textsl{getsf} configuration, input images, and masks for consistency, and it suggests solutions in various
circumstances during extractions. The software includes 21 utilities and scripts (Appendix~\ref{getsfdetails}), providing all kinds
of image processing necessary for \textsl{getsf} to run and more. They include the \textsl{fitfluxes} utility for spectral energy
distribution fitting of source fluxes or image pixels (and mass derivation) and the \textsl{hires} script that computes the
high-resolution surface density images (Sect.~\ref{hiresimages}). Most of the utilities are very useful for command-line image
manipulations, even without source and filament extractions.


\subsection{Limitations}
\label{limiations}

The method is designed and expected to work for the images that are not very sparse: most pixels must contain detectable signals
(measurable data). Examples of the images for which \textsl{getsf} might not produce reliable results are some extremely faint
X-ray or UV low-count images with isolated spiking pixels that are surrounded by large areas of pixels that were not assigned any
detectable signal. For such nonstandard images, \textsl{getsf} would still work and complete extractions, but its results might not
be reliable because the method relies on the standard deviations of the background or noise fluctuations outside structures, whose
values may not correctly represent the statistics of the observed data in these images. On the other hand, the images for
\textsl{getsf} extractions must not be extremely smooth: they must have some variations on scales of about the angular resolution.
However, such smooth images can easily be made perfectly suitable for \textsl{getsf} just by adding Gaussian noise at some faint
level that does not alter the structures of interest.

Separation of sources from filaments is not (and cannot be) perfect. It leaves very faint residuals of sources that end up in the
filament component. In practice, this is not important because most of the residuals are too faint (Fig.~\ref{flatfilcomb}) to
affect the filament properties. The background of very wide and/or overlapping filaments is likely to be derived less accurately
than that of the narrower and/or isolated filaments because the filaments are separated from the wider background areas. Filaments
that are separated from wider background peaks of comparable widths are likely to receive some contribution from the background
(Fig.~\ref{bgderivation}). In very rare cases, the footprint of a bright power-law peak might not be sufficiently expanded, which
leads to an underestimated flux.

The method takes quite considerable time to complete extractions, although \textsl{getsf} was optimized to run as fast as possible.
The aim of its design was to produce extraction results that are as reliable as possible because completeness and accuracy, not
speed, are of prime importance in astrophysical research. The runtime for the \textsl{getsf} applications presented in
Sect.~\ref{applications} is in the range of three hours to a week (the images with $430^{2}$ to $2000^{2}$ pixels and file sizes of
$800$\,KB to $16$\,MB). The two-wavelength extraction of sources and filaments for the subfield of \object{Orion\,A} described in
Sect.~\ref{scuba2} took 43 hours and required ${\sim\,}10$\,GB of disk space. The total processing time with \textsl{getsf} depends
on the numbers of pixels, wavelengths, iterations, detected sources and filaments, and on the processor and file system speed and
load. A source extraction run on $\text{eight}$ large images, each with $4800^{2}$ pixels ($92$\,MB file size), that detects and
measures ${\sim\,}3000$ sources, may need${\,}\text{about three}$ weeks and ${\sim\,}200$\,GB of disk space. Most of the time
\textsl{getsf} spends in the iterative separation of structural components: the actual extraction of sources and filaments takes
less than $10${\%} of the runtime. For the source extraction alone, the execution time is halved. In a properly planned research,
the processing time is almost never a limiting factor: much more time is usually spent on the analysis and interpretation of the
information delivered by the extraction and on describing the findings in a paper.

Many intermediate images are produced in the \textsl{getsf} extractions at each wavelength (for spatial decomposition, iterations,
etc.), hence they require large storage space. Between hundreds of MB and GB may be necessary for an extraction, depending on the
image size and the numbers of wavebands and iterations. It is necessary to keep many images until the end of the extraction
process; however, most of them may be deleted by \textsl{getsf} after the extraction has finished. The extraction results
themselves represent only ${\sim}\,20${\%} of the total size of the extraction directory. Computers with sufficiently large random
access memory are required to run \textsl{getsf} extractions on very large images. For the above range of image sizes, between $8$
and $64$\,GB may be necessary (the more memory, the better). The actual memory usage strongly depends on the number of sources
being detected and measured. Numbers of sources up to ${\sim\,}15000$ do not pose any problems to \textsl{getsf}, but substantially
larger numbers of detected sources require very large memory and long execution time.


\section{Conclusions}
\label{conclusions}

This paper presented \textsl{getsf}, the new multiscale method for extracting both sources and filaments in astronomical images
using separation of their structural components. It is specifically designed to handle multiwavelength sets of images and
extremely complex filamentary backgrounds, but it is perfectly applicable to a single image or very simple backgrounds. The new
code is freely downloadable from its website\footnote{\url{http://irfu.cea.fr/Pisp/alexander.menshchikov/}}, from the Astrophysics
Source Code Library\footnote{\url{https://ascl.net/2012.001}}, and also available from the author.

The main processing steps of \textsl{getsf} include (1) preparation of a complete set of images and derivation of high-resolution
surface densities, (2) spatial decomposition of the original images and separation of the structural components of sources and
filaments from each other and from their backgrounds, (3) flattening of the residual noise and background fluctuations in the
separate images of sources and filaments, (4) spatial decomposition of the flattened components of sources and filaments and their
combination of the over wavelengths, (5) detection of sources (positions) and filaments (skeletons) in the combined images of the
components, and (6) measurements of the properties of the detected sources and filaments and creation of the output catalogs and
images. Like its predecessor (\textsl{getold}, Papers I--III), \textsl{getsf} has a single user-definable parameter (per
wavelength), the maximum size of the structures of interest to extract. All internal parameters of \textsl{getsf} have been
calibrated and verified in numerous tests using various images from simulations and observations to ensure that the method
works well in all cases. 

This paper formulated \textsl{hires}, the algorithm for the derivation of high-resolution surface densities and temperatures from
the diffraction-limited multiwavelength far-infrared and submillimeter continuum observations, such as those obtained with
\emph{Herschel}. A substantial improvement over the original algorithm \citep{Palmeirim_etal2013} is the angular resolution of the
derived surface densities that may become as high as that of the shortest-wavelength image of a sufficient quality. In the case of
the \emph{Herschel} observations, the resolution may be as high as $5.6{\arcsec}$ for the slow scanning speed
($20{\arcsec}$s$^{-1}$) or $8.4{\arcsec}$ for the fast parallel mode ($60{\arcsec}$s$^{-1}$). If the $70$\,$\mu$m image appears too
noisy, excessively contaminated by the emission of polycyclic aromatic hydrocarbons or transiently heated very small dust grains,
or if it cannot be used for other reasons, then the highest resolution of surface densities is limited to that of the $100$ or
$160$\,$\mu$m images, that is, to $6.8{-}11.3{\arcsec}$ or $8.4{-}13.5{\arcsec}$, for the slow- or fast-scanning modes,
respectively. These high-resolution surface density images are especially useful for the detailed studies of the highly complex
structural diversity in the observed images and for deeper understanding of the physical processes within the heavily substructured
filaments and their relation to the formation of stars.

This paper described the set of simulated multiwavelength benchmark images for testing and comparing the source and filament
extraction methods to allow the researchers who need to perform such extractions to choose the most accurate algorithm for their
projects. Although the benchmark was designed to resemble the \emph{Herschel} observations of star-forming regions, the images are
suitable for testing and evaluating extraction methods for any astronomical projects and applications. It consists of the complex
fluctuating background cloud, the long dense filament, and many starless and protostellar cores with wide ranges of sizes, masses,
and intensity profiles, computed with a radiative transfer code. A separate paper (Men'shchikov 2021, submitted) presents a series
of the multiwavelength source extractions with \textsl{getsf} using three variants of the new benchmark with increasing complexity
levels and compares their results with those produced by \textsl{getold}. All benchmark images, the truth catalogs containing the
model parameters, and the reference extraction catalogs obtained by the author with \textsl{getsf} are available on its website.

The new extraction method can be used to conduct consistent and comparable studies of sources and filaments in various projects:
\textsl{getsf} is designed to work for all images with nonzero background or noise fluctuations, where most pixels carry nonzero
measured signal. The method is not limited to any particular area of astronomical research nor to the type of the telescopes or
instruments used, as demonstrated by its applications to the images obtained with \emph{XMM-Newton}, \emph{GALEX}, \emph{Hubble},
\emph{Spitzer}, \emph{Herschel}, \emph{APEX}, and \emph{ALMA}. Although no finite numbers of specific examples can prove that
\textsl{getsf} is universally applicable, they confirm a remarkably wide applicability of the method.


\begin{acknowledgements} 
This study used the \textsl{cfitsio} library \citep{Pence1999}, developed at HEASARC NASA (USA), \textsl{saoimage ds9}
\citep{JoyeMandel2003} and \textsl{wcstools} \citep{Mink2002}, developed at the Smithsonian Astrophysical Observatory (USA), and
the \textsl{stilts} library (by Mark Taylor), developed at Bristol University (UK). The \textsl{plot} utility and \textsl{ps12d}
library, used in this work to draw figures directly in the PostScript language, were written by the author using the
\textsl{psplot} library (by Kevin E. Kohler), developed at Nova Southeastern University Oceanographic Center (USA), and the
plotting subroutines from the MHD code \textsl{azeus} \citep{Ramsey2012}, developed by David Clarke and the author at Saint Mary's
University (Canada). This work used observations obtained with \emph{XMM–Newton}, an ESA science mission with instruments and
contributions directly funded by ESA Member States and NASA. This work used observations made with the \emph{Spitzer} Space
Telescope, which is operated by the Jet Propulsion Laboratory, California Institute of Technology, under a contract with NASA. This
work used observations made with the NASA/ESA \emph{Hubble} Space Telescope, and obtained from the \emph{Hubble} Legacy Archive,
which is a collaboration between the Space Telescope Science Institute (STScI/NASA), the Space Telescope European Coordinating
Facility (ST-ECF/ESA) and the Canadian Astronomy Data Centre (CADC/NRC/CSA). This paper used the SCUBA-2 data obtained at
\emph{JCMT} under program MJLSG31. The James Clerk Maxwell Telescope is operated by the East Asian Observatory on behalf of The
National Astronomical Observatory of Japan; Academia Sinica Institute of Astronomy and Astrophysics; the Korea Astronomy and Space
Science Institute; Center for Astronomical Mega-Science (as well as the National Key R\&D Program of China with No.
2017YFA0402700). Additional funding support is provided by the Science and Technology Facilities Council of the United Kingdom and
participating universities and organizations in the United Kingdom and Canada. Additional funds for the construction of SCUBA-2
were provided by the Canada Foundation for Innovation. This paper used the following ALMA data: ADS/JAO.ALMA\#2013.1.01365.S. ALMA
is a partnership of ESO (representing its member states), NSF (USA) and NINS (Japan), together with NRC (Canada), NSC and ASIAA
(Taiwan), and KASI (Republic of Korea), in cooperation with the Republic of Chile. The Joint ALMA Observatory is operated by ESO,
AUI/NRAO and NAOJ. The simulated surface density background $\mathcal{B}$ was derived from a synthetic scale-free background image
created by Ph.\,Andr{\'e}. A large set of images, used for testing and validation of \textsl{getsf}, includes those obtained in the
\emph{Herschel} Gould Belt Survey\footnote{\url{http://gouldbelt-herschel.cea.fr}} (HGBS, PI Ph.\,Andr{\'e}),
HOBYS\footnote{\url{http://hobys-herschel.cea.fr}} (PIs F.\,Motte, A.\,Zavagno, S.\,Bontemps), and \emph{ALMA}-IMF (PIs F.\,Motte,
A.\,Ginsburg, F.\,Louvet, P.\,Sanhoueza). HGBS and HOBYS are the \emph{Herschel} Key Projects jointly carried out by SPIRE
Specialist Astronomy Group 3 (SAG3), scientists of several institutes in the PACS Consortium (e.g., CEA Saclay, INAF-IAPS Rome,
LAM/OAMP Marseille), and scientists of the \emph{Herschel} Science Center (HSC). The author appreciates the valuable feedback,
received from G.\,Zhang, F.\,Louvet, and N.\,Kumar, on the \textsl{getsf} extractions in the X-shaped nebula, MHD simulations, and
\object{Mon\,R2}, respectively. The author is grateful to A.\,Zavagno, T.\,Nony, Y.\,Shimajiri, Ph.\,Andr{\'e}, D.\,Arzoumanian,
and P.\,Palmeirim for their comments on the manuscript.
\end{acknowledgements} 


\begin{appendix}


\section{Inaccuracies of the derived surface densities and temperatures}
\label{hiresinacc}

\begin{figure*}                                                               
\centering
\centerline{
  \resizebox{0.328\hsize}{!}{\includegraphics{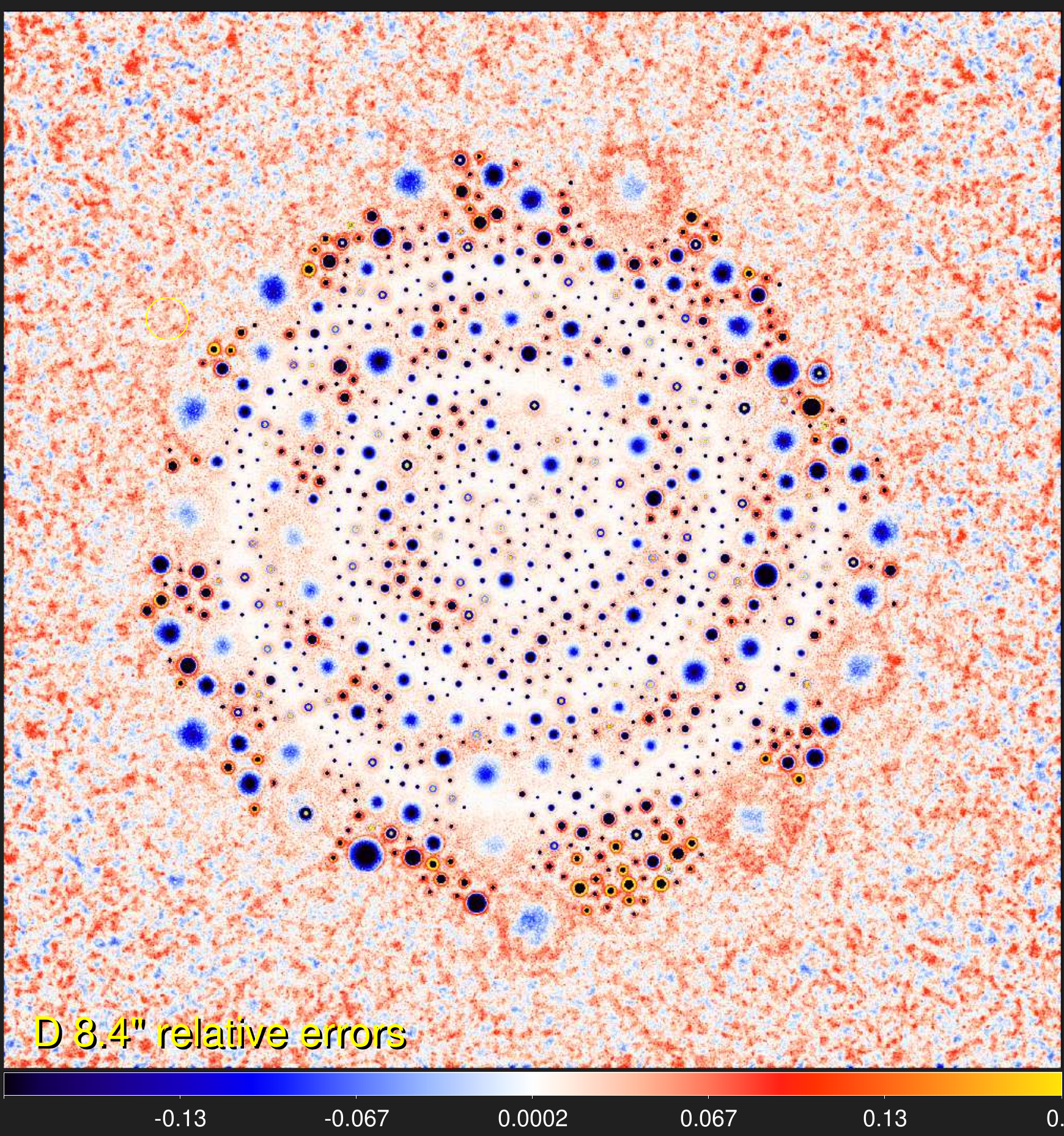}}
  \resizebox{0.328\hsize}{!}{\includegraphics{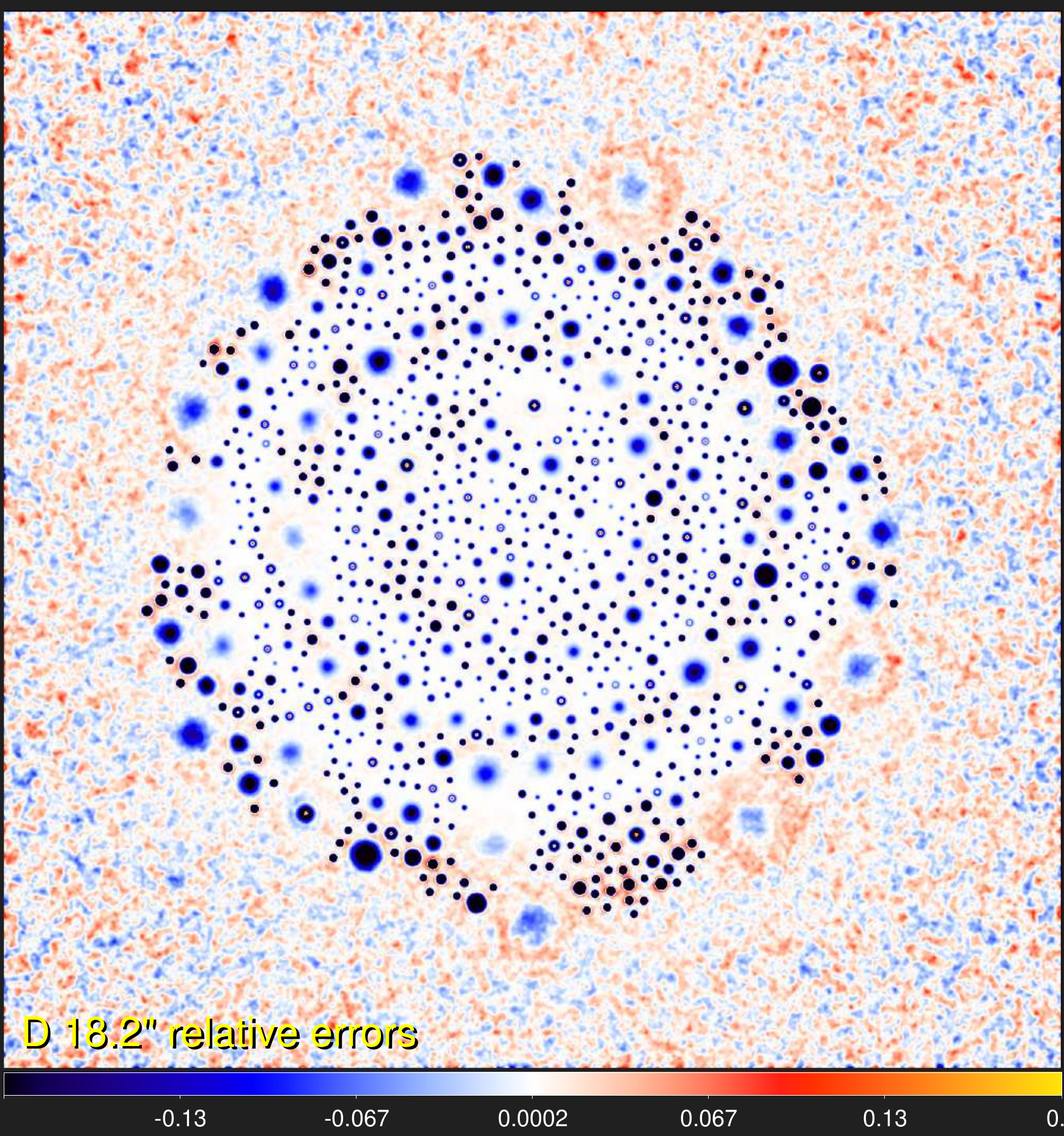}}
  \resizebox{0.328\hsize}{!}{\includegraphics{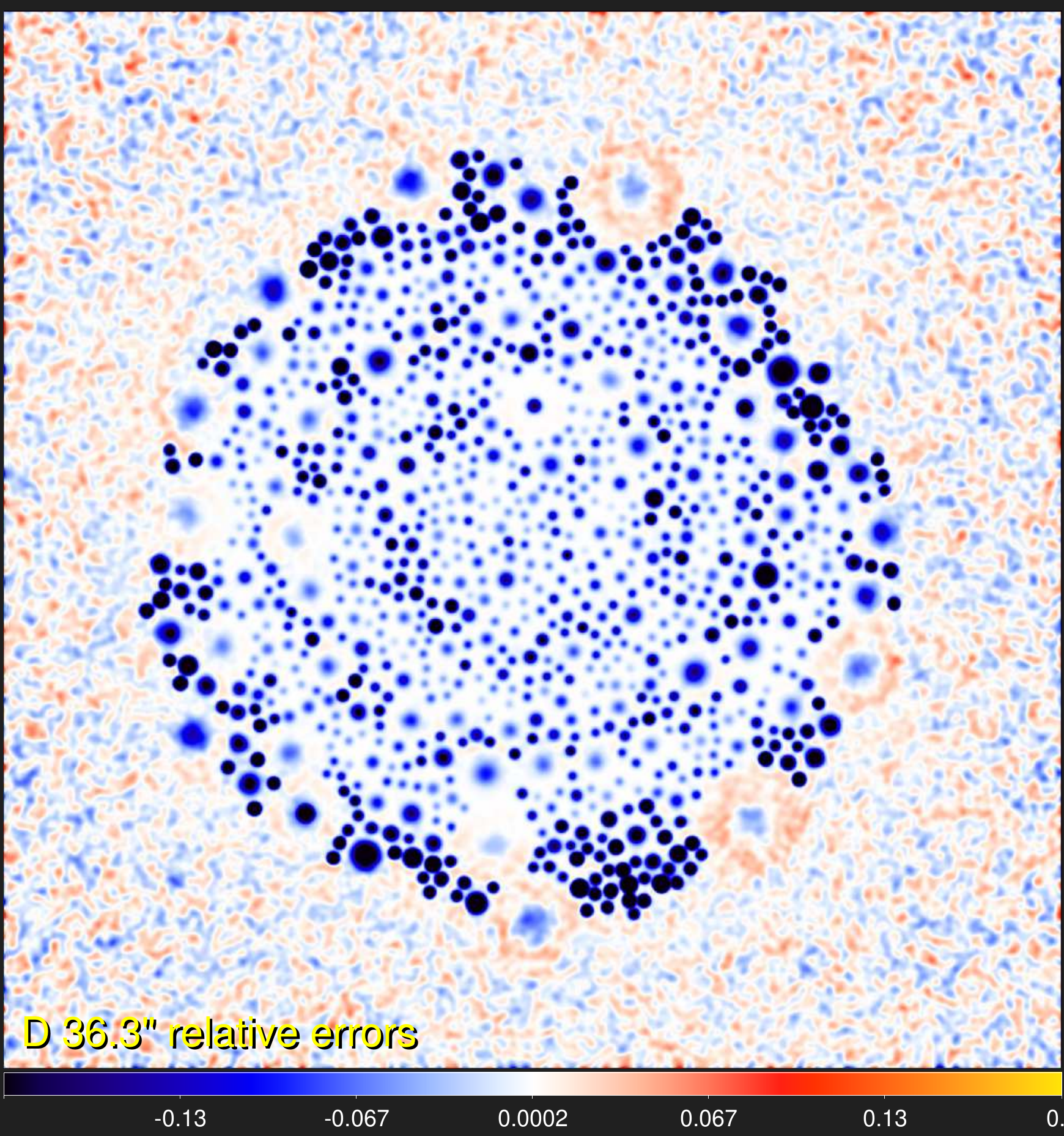}}}
\vspace{0.5mm}
\centerline{
  \resizebox{0.328\hsize}{!}{\includegraphics{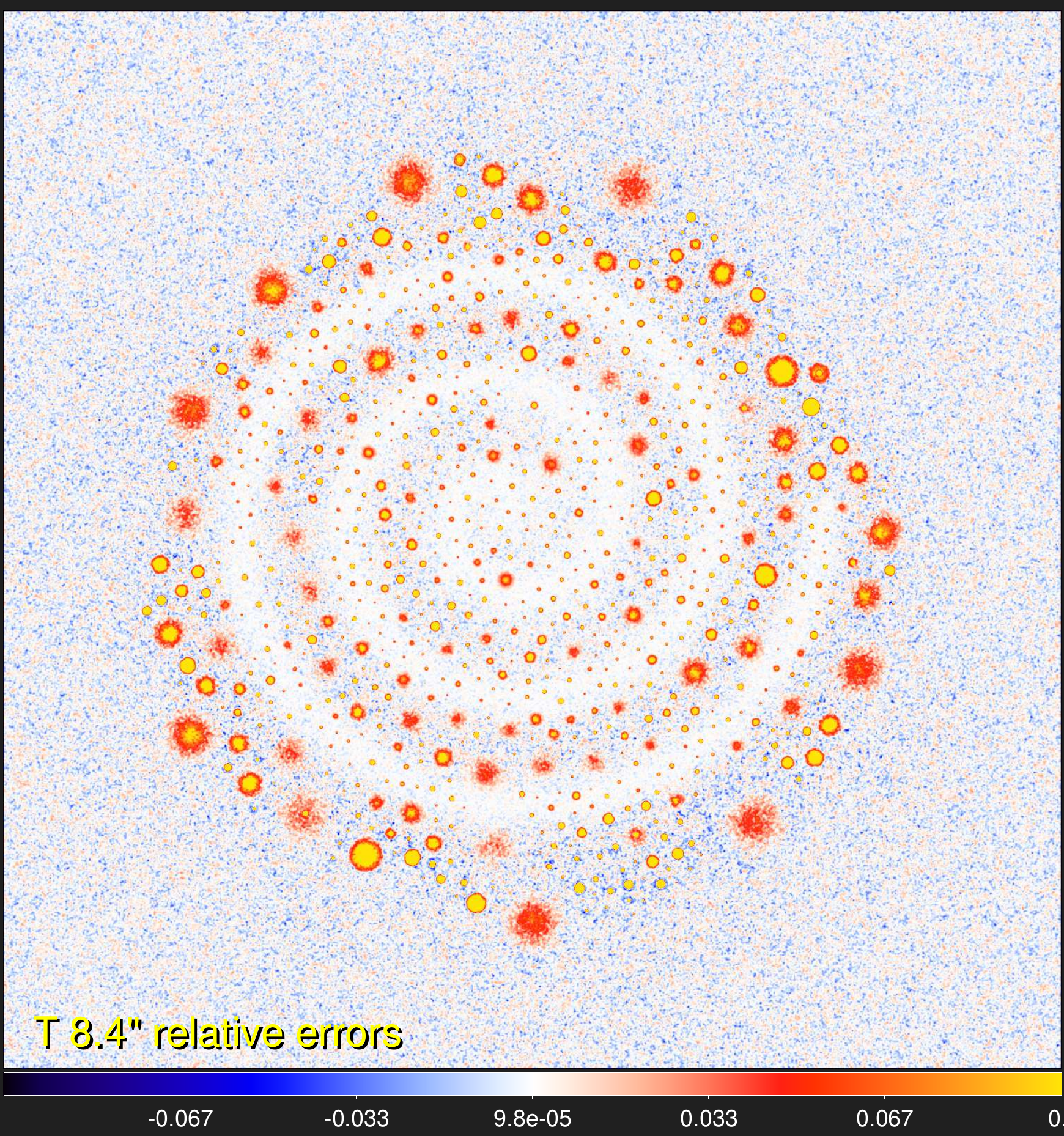}}
  \resizebox{0.328\hsize}{!}{\includegraphics{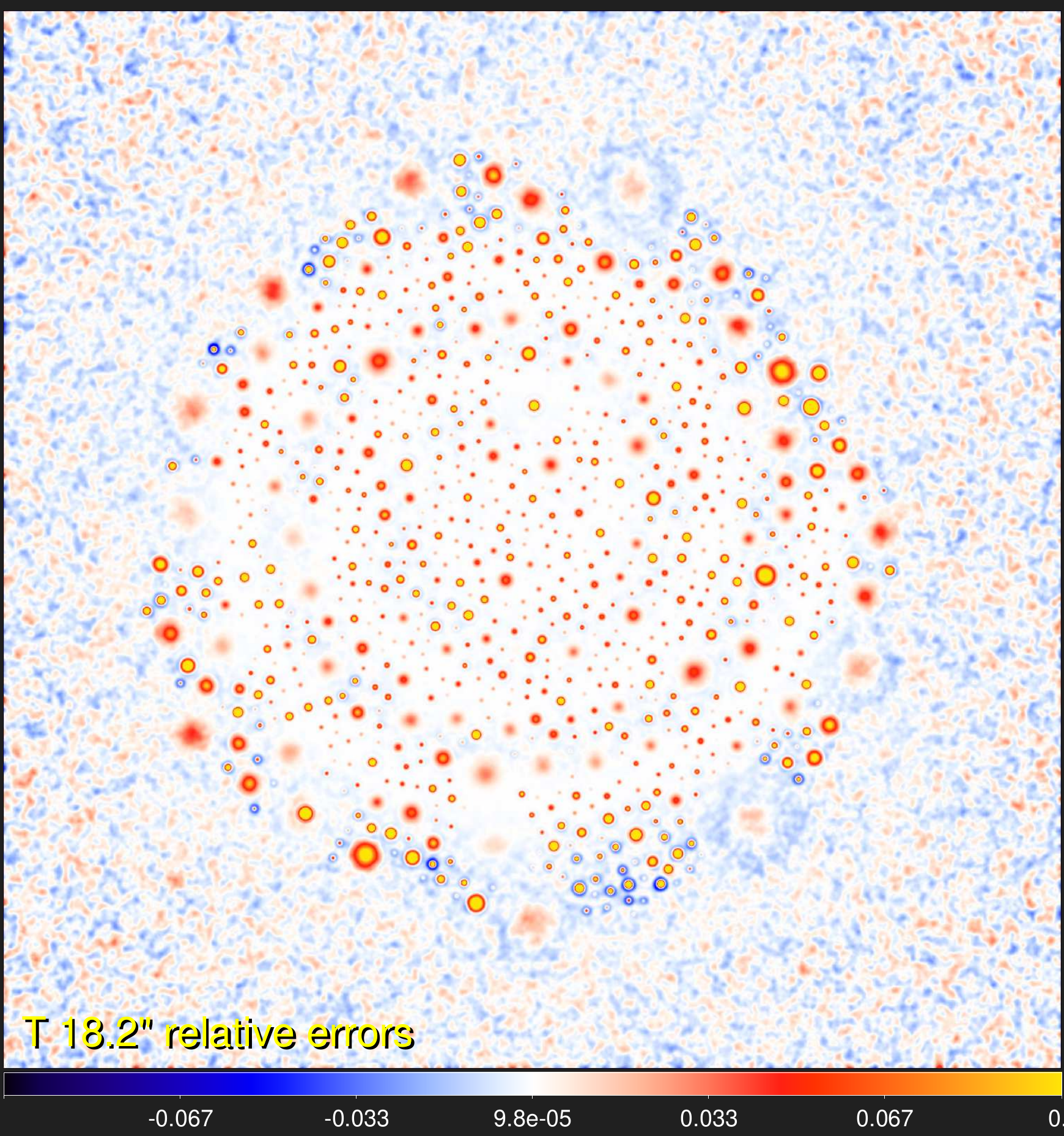}}
  \resizebox{0.328\hsize}{!}{\includegraphics{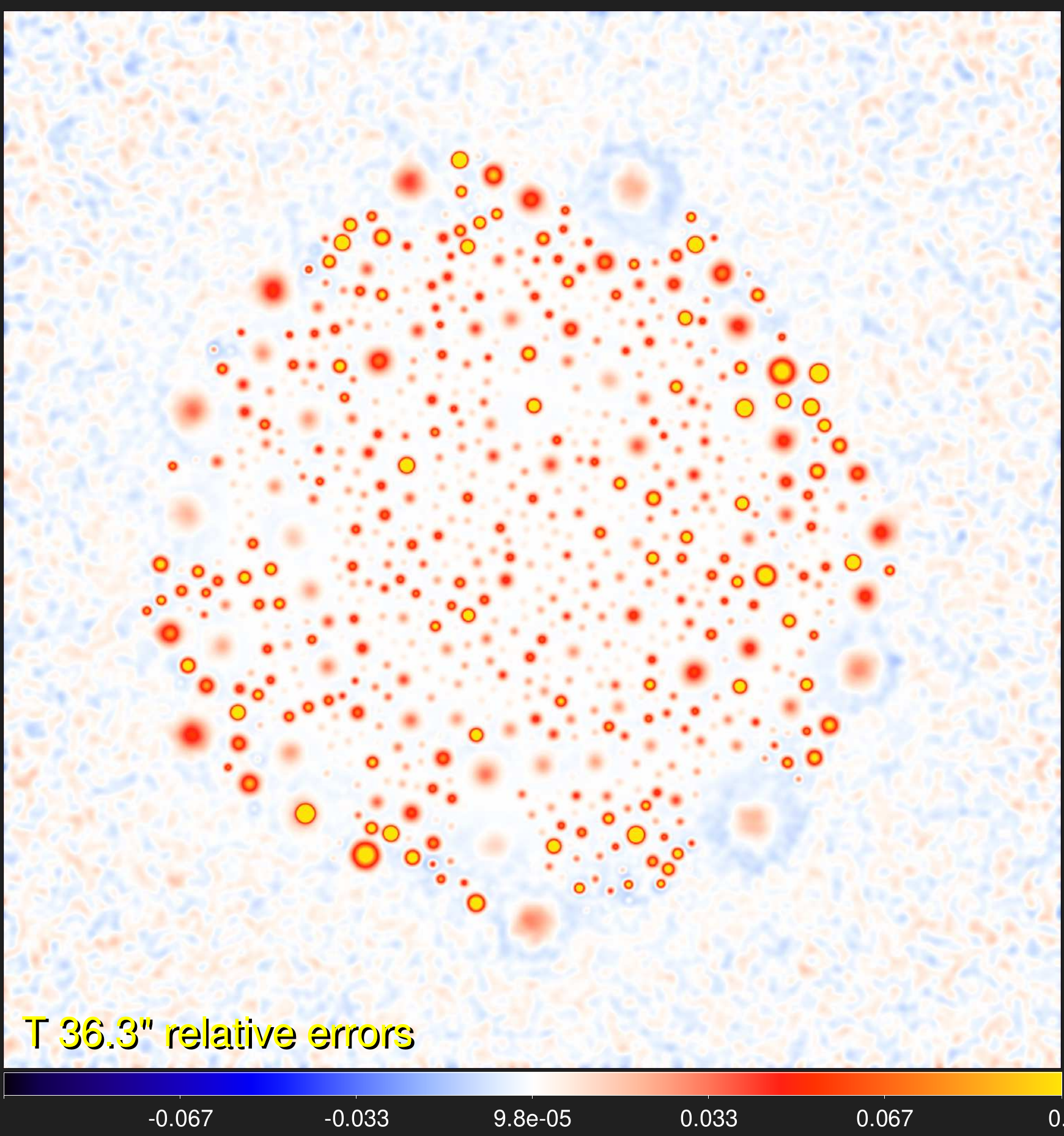}}}
\caption
{ 
Relative accuracies $\epsilon$ of the \textsl{hires} surface densities and temperatures derived from Eq.~(\ref{superdens})
(Sect.~\ref{hiresimages}) with respect to the true model images convolved to the same resolutions. The \emph{top} row shows the
errors in $\mathcal{D}_{8{\arcsec}}$\! ($\sigma{\,=\,}0.15$), $\mathcal{D}_{18{\arcsec}}$\! ($\sigma{\,=\,}0.06$), and
$\mathcal{D}_{36{\arcsec}}$\! ($\sigma{\,=\,}0.05$) and the \emph{bottom} row shows the errors in $\mathcal{T}_{8{\arcsec}}$\!
($\sigma{\,=\,}0.06$), $\mathcal{T}_{18{\arcsec}}$\! ($\sigma{\,=\,}0.05$), and $\mathcal{T}_{36{\arcsec}}$\!
($\sigma{\,=\,}0.05$). At the highest resolution of $8{\arcsec}$, the derived images are the most accurate, with the exception
of the unresolved protostellar peak surface densities (Fig.~\ref{simcores}), which  become strongly overestimated (up to a factor of
${\sim\,}10$) because the temperatures $\mathcal{T}_{\{2|3|4\}}$ along the lines of sight with large temperature
gradients are underestimated. The range of displayed values is reduced for better visibility. Linear color mapping.
} 
\label{hiresacc}
\end{figure*}

The algorithms described in Sect.~\ref{hiresimages} imply that the $160$, $250$, $350$, and $500$\,{${\mu}$m} images have
an accurate (consistent) intensity calibration. When we assume that the calibration inaccuracies can be described by constant
wavelength-dependent offsets, simple consistency checks and corrections can be made. Three independent estimates of
low-resolution temperatures ($\mathcal{T}_{\rm L1}$, $\mathcal{T}_{\rm L2}$, $\mathcal{T}_{\rm L3}$) are readily available from
fitting the images in three pairs of wavebands ($160{-}250$, $250{-}350$, and $350{-}500$\,{${\mu}$m}) with a low resolution of
$O_{500}$. If the median values of the three temperature maps differ by more than several percent, it would be necessary to adjust
some of the offsets and estimate $\mathcal{T}_{\rm L\{1|2|3\}}$ again. This iterative process is stopped when the three
temperatures become consistent.

The higher-resolution images are obtained at the cost of significantly stronger noise and greater chances of distortions and
spurious peaks. The quality of the resulting $\mathcal{\{D|T\}}_{\rm P}$ from Eq.~(\ref{hiresdentem}) strongly depends on the quality
of the original short-wavelength images. Higher levels of noise or map-making artifacts in the $250$ and $160$\,${\mu}$m images
would be amplified in the resulting maps in the process of fitting the spectral shapes $\Pi_{\lambda}$ of pixels, which is likely
to create significant small-scale distortions, predominantly in the pixels with strong line-of-sight temperature gradients that are usually
located over the dense sources or filaments. The differential terms $\delta\mathcal{\{D|T\}}_{\{3|2\}}$ that contain the
higher-resolution information are increasingly less accurate because they are obtained from fitting of only three and two
(noisier) images. It is very important to carefully inspect $\mathcal{\{D|T\}}_{\rm P}$ to ensure that they are
free of spurious small-scale structures before using them in any extraction. The \textsl{hires} images $\mathcal{\{D|T\}}_{O_{\rm
H}\!}$ from Eqs.~(\ref{superdens}) and (\ref{supertemp}) are much less affected by the problems because they use the contributions
$\delta\mathcal{\{D|T\}}_{\{4|3|2\}}$ from all three variants of the fitted temperatures for each of the six resolutions of the
original images.

The essential idea of the differential algorithm for improving the angular resolution of surface density was validated using the
benchmark images (Sect.~\ref{skybench}). The complete surface density $\mathcal{D}_{\rm C}{\,+\,}\mathcal{D}_{\rm S}$ was first
convolved to the resolutions of all \emph{Herschel} wavebands (Sect.~\ref{simcomplete}). The algorithm of Eq.~(\ref{hiresdentem}),
generalized to all six wavebands, was then applied to improve the lowest-resolution surface density using the unsharp masking of
Eq.~(\ref{diffterms}) and to successively recover each of the higher-resolution surface densities, all the way up to the highest
adopted resolution $O_{70}{\,=\,}8.4{\arcsec}$, with a resulting maximum error below $0.5${\%}. Although this is an excellent
accuracy of the scheme, real-life applications of the method involve fitting of the spectral pixel shapes $\Pi_{\lambda}$, hence
they inevitably suffer from larger inaccuracies (Fig.~\ref{hiresacc}).

The derived surface densities $\mathcal{D}_{\rm P}$ and $\mathcal{D}_{O_{\rm H}}$ (Sect.~\ref{hiresimages}) are not suitable for measuring dense structures, especially those with a central source of heating, because their inaccuracies
in the pixels with strong line-of-sight temperature gradients are too large \citep[e.g.,][]{Men'shchikov2016}. Comparisons with the true surface
densities in Fig.~\ref{hiresacc} show that the vast majority of the pixels outside bright sources are quite accurate, to better
than $0.5$\%. However, the inaccuracies become much larger in the places that are occupied by the sources with steep gradients of the
line-of-sight temperature. The starless cores and protostellar envelopes have markedly different radial temperature profiles,
therefore the errors that are induced in the derived surface densities are also very dissimilar in both their sign and magnitude.


\section{Single-scale spatial decomposition and standard deviations}
\label{decomposition}

\begin{figure}[!tbp]
\centering
\centerline{
  \resizebox{0.9\hsize}{!}{\includegraphics{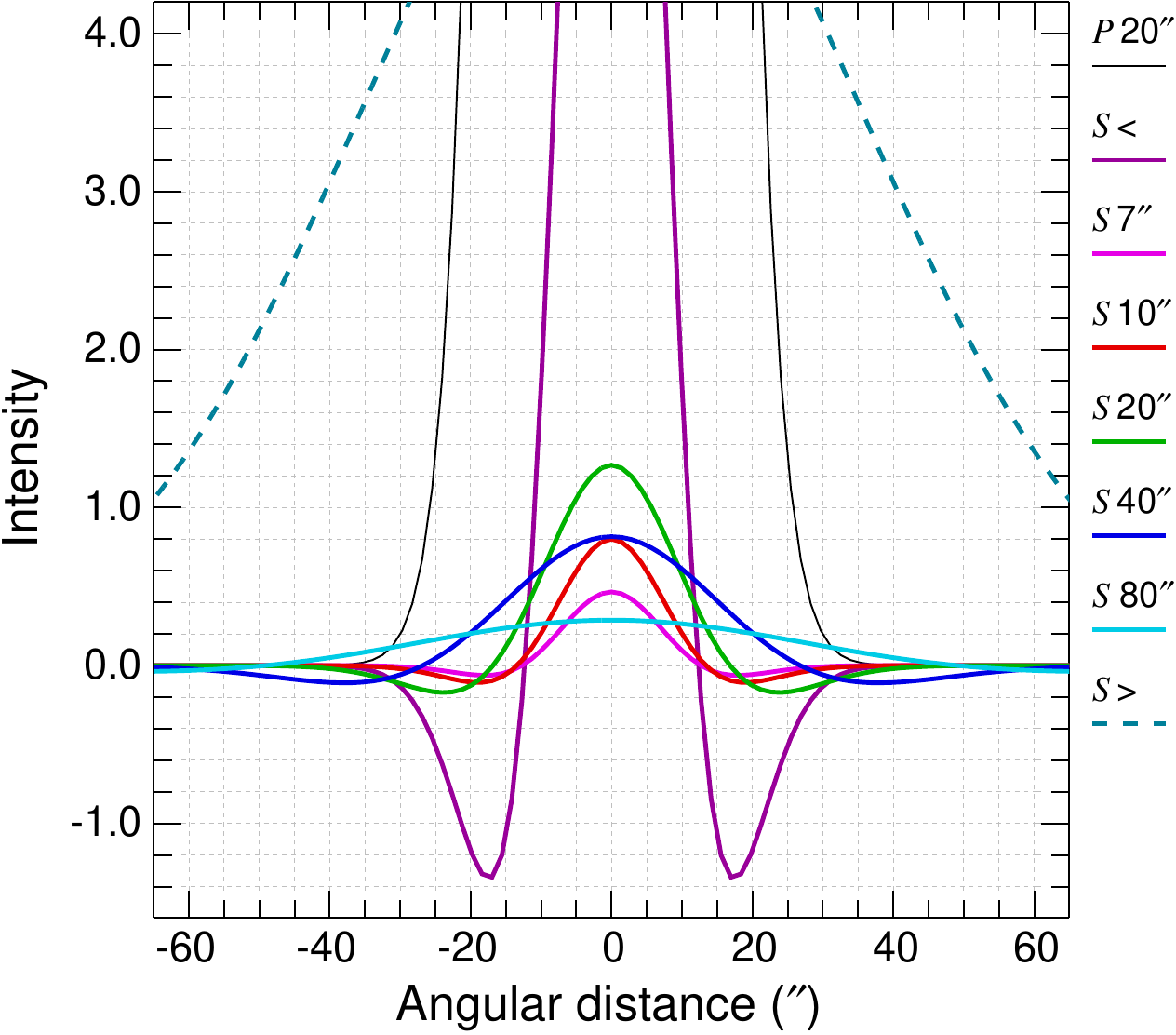}}}
\caption
{ 
Single-scale spatial decomposition for an unresolved source $\mathcal{P}$ with a peak value of 100 and resolution
${O_{\lambda}{\,=\,}20}${\arcsec} into 99 scales between ${S_{\!\rm min}{\,=\,}7}${\arcsec} and ${S_{\!\rm
max}{\,=\,}80}${\arcsec}, with the scale factor ${f{\,=\,}1.026}$. The profiles of the original Gaussian are shown for the six
selected single scales (from ${S{\,<\,}S_{\!\rm min}}$ to $S_{\!\rm max}$), and of the largest scales
(${\mathcal{G}_{99}{*\,}\mathcal{P}}$), outside the decomposition range (${S{\,>\,}S_{\!\rm max}}$).
} 
\label{gaussdecomposed}
\end{figure}

Following the \textsl{getold} general approach, \textsl{getsf} employs successive unsharp masking to decompose the prepared
original images $\mathcal{I}_{{\!\lambda}}$ (Sect.~\ref{obsimages}) into $N_{\rm S}$ single scales,
\begin{equation} 
\mathcal{I}_{{\!\lambda}{j}} = {\mathcal{G}_{j-1\,}{*\,}\mathcal{I}_{{\!\lambda}}} - 
{\mathcal{G}_{j\,}{*\,}\mathcal{I}_{{\!\lambda}}}, \,\,\, {j{\,=\,}1, 2,\dots, N_{\rm S}},
\label{successive}
\end{equation} 
where $\mathcal{G}_{j}$ are the circular Gaussian convolution kernels ($\mathcal{G}_{0}$ is to be regarded as the delta function)
with progressively increasing half-maximum sizes,
\begin{equation} 
{S_{\!j} = f\,S_{\!j-1}}, \,\,\, {S_{\!0} = S_{\!1} f^{\,-1}}, \,\,\, {S_{\!\rm min} \le S_{\!j} \le S_{\!\rm max}},
\label{bounds}
\end{equation} 
where ${f{\,>\,}1}$ is the discretization factor (typically ${f{\,\approx\,}1.05}$) and the limiting scales of the decomposition
range are
\begin{eqnarray} 
\left.\begin{aligned}
S_{\!\rm min}\!&= \max\left(2 \Delta, 0.33\,{\rm min}_{\lambda} \left( O_{\lambda} \right)\right),\\
S_{\!\rm max}\!&= {\rm max}_{\lambda} \left(\max\left( 4 X_{\lambda}, 4 Y_{\!\lambda} \right)\right),
\end{aligned}\right.
\label{decomrange}
\end{eqnarray} 
where $\Delta$ is the pixel size. The first image $\mathcal{I}_{{\!\lambda}{1}}$ contains the contribution from all scales below
$S_{\!\rm min}$, whereas the last image $\mathcal{I}_{{\!\lambda}{N_{\rm S}}}$ does not contain the signals from the scales above
$S_{\!\rm max}$, they are outside the range of scales being analyzed. The convolution is done with rescaling to conserve the total
flux, hence the originals $\mathcal{I}_{{\!\lambda}}$ can be recovered by summation of the $N_{\rm S}$ scales and all remaining
largest spatial scales,
\begin{equation} 
\mathcal{I}_{{\!\lambda}} = \sum\limits_{j=1}^{N_{\rm S}}\mathcal{I}_{{\!\lambda}{j}} + 
{\mathcal{G}_{N_{\rm S}}{*\,}\mathcal{I}_{{\!\lambda}}}.
\label{recovered}
\end{equation} 

The spatial decomposition is illustrated in Fig.~\ref{gaussdecomposed} using an example of a simple two-dimensional Gaussian shape
$\mathcal{P}$. As demonstrated in Papers I and II, the spatial decomposition has many useful properties. The filtered single-scale
images contain signals from a relatively narrow range of spatial scales, and their properties resemble the Gaussian statistics much
better than those of the originals, which are blends of all spatial scales. On the scales much smaller than the image size, the
decomposed images are well described by the global value of the standard deviation $\sigma_{{\!\lambda}{j}}$. Significant
departures from the Gaussian distribution in single scales above a certain threshold (e.g.,
$I_{{\lambda}{j}}{\,\ga\,}5\sigma_{{\!\lambda}{j}}$) indicate the presence of the real structures. The decomposition highlights the
structures of a specific width in the decomposed images on a matching scale. For example, a resolved isolated circular source with
a half-maximum size ${H_{\lambda}}$ has its maximum brightness in $\mathcal{I}_{{\!\lambda}{j}}$ on the scale
${S_{\!j}{\,\approx\,}H_{\lambda}}$ and a completely unresolved source produces the brightest signal on the smallest spatial scales
${S_{\!j}{\,\la\,}O_{\lambda}}$.

Following the \textsl{getold} approach (Papers I and II), \textsl{getsf} employs an iterative algorithm to determine the
single-scale $\sigma_{{\!\lambda}{j}}$ over the entire usable area ${\mathcal{I}_{{\!\lambda}{j}}\mathcal{M}_{\lambda}}$ of the
image to separate the real structures from other insignificant background or noise fluctuations. Before the iterations, the global
$\sigma_{{\!\lambda}{j0}}$ and the threshold ${\varpi_{{\lambda}{j0}}{\,=\,}3\sigma_{{\!\lambda}{j0}}}$ are computed over all
pixels. At the first and all subsequent iterations ($i{\,=\,}1, 2,\dots, N_{\rm I}$), significant peaks and hollows with
${|I_{{\lambda}{j}}|{\,\ge\,}\varpi_{{\lambda}{{j}{i-1}}}}$ are masked. The absolute value is taken, because structures have both
positive and negative counterparts in the decomposed images. Then \textsl{getsf} calculates a new (lower)
$\sigma_{{\!\lambda}{{j}{i}}}$ value outside the masked areas and all structures with
${{|I_{{\lambda}{j}}|{\,\ge\,}\varpi_{{\lambda}{{j}{i}}}}}$ are masked again. The iterations continue until the threshold converges
to a stable value of $\varpi_{{\lambda}{{j}{i}}}$, with corrections ${\delta\varpi_{{\lambda}{{j}{i}}}{\,<\,}1{\%}}$. The final
single-scale standard deviation is obtained as ${\sigma_{{\!\lambda}{j}}{\,=\,}\varpi_{{\lambda}{j}}/3}$ and its total value as
${{\sigma_{\!\lambda}}^{\!\!2} = \sum_{j}{\sigma_{\!{\lambda}{j}}}^{\!\!\!2}}$. The constant $3$, chosen empirically, provides both
suitable values of the resulting $\sigma_{{\!\lambda}{j}}$ values and good convergence of the iterations.

A notable difference with \textsl{getold} is that \textsl{getsf} does not need to correct the iterated thresholds using the
higher-order statistical moments (skewness and kurtosis) because significant structures are detected in accurately flattened
detection images (Sect.~\ref{extraction}), which ensures that the majority of pixels resemble a normal distribution. Furthermore,
precise $\sigma_{{\!\lambda}{j}}$ values are of relatively minor importance for the separation of structural components because the
separation is done in iterations and is based on the shapes that are removed from the single-scale slices (Sect.~\ref{clipping}),
not on the $\sigma_{{\!\lambda}{j}}$ value itself.


\section{Software suite}
\label{getsfdetails}

The method has been developed as a \textsl{bash} script \textsl{getsf} that executes a number of FORTRAN utilities, doing all
numerical computations. Linux or macOS systems with the \textsl{ifort} or \textsl{gfortran} compilers can be used to install the
code. For reading and writing images, \textsl{getsf} uses the \textsl{cfitsio} library \citep{Pence1999}; for resampling and
reprojecting images, it calls \textsl{swarp} \citep{Bertin_etal2002}; for convolving images, it uses the fast Fourier transform
routine \textsl{rlft3} \citep{Press_etal1992}; for determining the source coordinates $\alpha$ and $\delta$, it applies
\textsl{xy2sky} from \textsl{wcstools} \citep{Mink2002}; and for a colored screen output, it uses the \textsl{highlight} utility
(by Andr{\'e} Simon)\footnote{\url{http://www.andre-simon.de/}}, if the latter is installed.

The following list of the \textsl{getsf} utilities and scripts explains their purpose and functions. They are quite useful for
command-line image manipulations, even if there is no need in a source or filament extraction. Their usage information is displayed
when a utility is run without any parameter. The utilities are sorted in the decreasing sequence of their general usability outside
\textsl{getsf}.
\begin{tabbing} 
Utility \,\,\,\,\,\,\, \= Purpose \kill
\textsl{modfits}   \> modify an image or its header in various ways:\\
                   \> math transformations; profiling an image along a\\
                   \> line; image segmentation; filament skeletonization;\\
                   \> removal of connected clusters of pixels; addition or\\
                   \> removal of border areas; correction of saturated or\\
                   \> bad pixel areas; conversion of intensity units;\\
                   \> changes of the header keywords; \emph{etc}.\\
\textsl{operate}   \> operate on two input images: addition, subtraction,\\
                   \> multiplication, division; relative differencing;\\
                   \> minimization or maximization; extension or\\
                   \> expansion of masks; copying of an image header;\\
                   \> computation of surface densities, temperatures, or\\
                   \> intensities; \emph{etc}.\\
\textsl{imgstat}   \> display and/or save image statistical quantities;\\
                   \> produce mode-, mean-, or median-filtered images;\\
                   \> compute images of standard deviations, skewness,\\
                   \> kurtosis; \emph{etc}.\\
\textsl{fftconv}   \> fast Fourier transform or convolve image with few\\
                   \> predefined kernels or an external kernel image\\
\textsl{fitfluxes} \> fit spectral shapes of source fluxes or image pixel\\
                   \> intensities to derive masses or surface densities\\
\textsl{convolve}  \> convolve an image to a desired lower resolution\\
\textsl{resample}  \> resample and reproject an image with rotation\\
\textsl{hires}     \> high-resolution surface densities and temperatures\\
\textsl{prepobs}   \> convert observed images into the same pixel grid\\
\textsl{installg}  \> install \textsl{getsf} on a computer (macOS, Linux)\\
\textsl{iospeed}   \> test I/O speed of a hard drive for a specific image\\
\textsl{readhead}  \> display an image header or save selected keywords\\
\textsl{cleanbg}   \> interpolate background below source footprints\\
\textsl{ellipses}  \> overlay an image with ellipses of extracted sources\\
\textsl{sfinder}   \> detect sources in combined single-scale images\\
\textsl{smeasure}  \> measure and catalog properties of detected sources\\
\textsl{fmeasure}  \> measure and catalog properties of detected filaments\\
\textsl{finalcat}  \> produce the final catalogs of detected sources\\
\textsl{expanda}   \> expand masked areas of an image to its edges\\
\textsl{extractx}  \> extract all image extensions in separate images\\
\textsl{splitcube} \> split a data cube into separate images
\end{tabbing} 

The code is automated, flexible, and user-friendly; it can be downloaded from the
website\footnote{\url{http://irfu.cea.fr/Pisp/alexander.menshchikov/}}, the Astrophysics Source Code
Library\footnote{\url{https://ascl.net/2012.001}}, and it is also available from the author upon request. The website also contains
a detailed User's Guide and a complete validation extraction of sources and filaments in a small image for those who would like to
verify that their installed \textsl{getsf} produces correct results.

\end{appendix}


\bibliographystyle{aa}
\bibliography{aamnem99,getsf}

\end{document}